\documentclass[runningheads,traditabstract,times]{aa} 
\usepackage{epsfig}
\usepackage{graphics}
\usepackage{subfigure}
\usepackage{natbib}
\usepackage{amsmath}
\usepackage{enumerate}
\usepackage{fancyhdr}
\usepackage{afterpage} 
\usepackage{makecell}
\usepackage{caption}
\usepackage{afterpage} 
\usepackage{multicol}
\usepackage{supertabular}
\usepackage{longtable, tabu}
\usepackage{threeparttablex}
\usepackage{tablefootnote}
\usepackage{placeins}

\begin{document}

\title{C-GOALS II. Chandra Observations of the Lower Luminosity Sample of Nearby Luminous Infrared Galaxies in GOALS}

\author{N. Torres-Alb\`a\inst{1}, K. Iwasawa \inst{1,2}, T. D\'iaz-Santos \inst{3}, V. Charmandaris \inst{4,5}, C. Ricci \inst{3,6,7}, J.K. Chu \inst{8}, D. B. Sanders \inst{9},  L. Armus \inst{10}, L. Barcos-Mu\~noz \inst{11,12}, A.S. Evans \inst{11,13}, J.H. Howell \inst{10}, H. Inami \inst{14}, S.T. Linden \inst{13},  A.M. Medling \inst{15,16}, G.C. Privon \inst{17}, V. U \inst{18} and I. Yoon \inst{11}}
  
\institute{Institut de Ci\`encies del Cosmos (ICCUB), Universitat de Barcelona (IEEC-UB), Mart\'i i Franqu\`es, 1, 08028 Barcelona, Spain\\ \and ICREA, Pg. Llu\'is Companys 23, 08010 Barcelona, Spain \and N\'ucleo de Astronom\'ia de la Facultad de Ingenier\'ia y Ciencias, Universidad Diego Portales, Av. Ej\'ercito Libertador 441, Santiago, Chile \and Institute for Astronomy, Astrophysics, Space Applications \& Remote Sensing, National Observatory of Athens, GR-15236, Penteli, Greece \and University of Crete, Department of Physics, GR-71003, Heraklion, Greece \and Kavli Institute for Astronomy and Astrophysics, Peking University, Beijing 100871, China \and Chinese Academy of Sciences South America Center for Astronomy, Camino El Observatorio 1515, Las Condes, Santiago, Chile \and Gemini North Observatory, 670 N. A`ohoku Place, Hilo, HI 96720
 \and Institute for Astronomy, University of Hawaii, 2680 Woodlawn Drive, Honolulu, HI 96822, USA \and Infrared Processing and Analysis Center, MC 314-6, Caltech, 1200 E. California Blvd., Pasadena, CA 91125, USA \and National Radio Astronomy Observatory, 520 Edgemont Road, Charlottesville, VA 22903, USA \and Joint ALMA Observatory, Alonso de C\'ordova 3107, Vitacura, Santiago, Chile  \and Department of Astronomy, University of Virginia, P.O. Box 400325, Charlottesville, VA 22904, USA \and Univ Lyon, Univ Lyon1, Ens de Lyon, CNRS, Centre de Recherche Astrophysique de Lyon (CRAL) UMR5574, F-69230, SaintGenis-Laval, France \and Cahill Center for Astronomy and Astrophysics, California Institute of Technology, MS 249-17, Pasadena, CA 91125, USA \and Research School of Astronomy \& Astrophysics, Australian National University, Canberra, ACT 2611, Australia \and Department of Astronomy, University of Florida, 211 Bryant Space Sciences Center, Gainesville, 32611 FL, USA \and Department of Physics and Astronomy, 4129 Frederick Reines Hall, University of California, Irvine, CA 92697, USA}

\offprints{N. Torres-Albà \\ \email{ntorresalba@fqa.ub.es}}

\titlerunning{C-GOALS II}

\authorrunning{N. Torres-Alb\`a et al. }

\abstract{We analyze \textit{Chandra} X-ray observatory data for a sample of 63 luminous infrared galaxies (LIRGs), sampling the lower-infrared luminosity range of the Great Observatories All-Sky LIRG survey (GOALS), which includes the most luminous infrared selected galaxies in the local universe. X-rays are detected for 84 individual galaxies within the 63 systems, for which arcsecond resolution X-ray images, fluxes, infrared and X-ray luminosities, spectra and radial profiles are presented. Using X-ray and MIR selection criteria, we find AGN in (31$\pm$5)\% of the galaxy sample, compared to the (38$\pm$6)\% previously found for GOALS galaxies with higher infrared luminosities (C-GOALS I). Using mid-infrared data, we find that (59$\pm$9)\% of the X-ray selected AGN in the full C-GOALS sample do not contribute significantly to the bolometric luminosity of the host galaxy. Dual AGN are detected in two systems, implying a dual AGN fraction in systems that contain at least one AGN of (29$\pm$14)\%, compared to the (11$\pm$10)\% found for the C-GOALS I sample. Through analysis of radial profiles, we derive that most sources, and almost all AGN, in the sample are compact, with half of the soft X-ray emission generated within the inner $\sim 1$ kpc. For most galaxies, the soft X-ray sizes of the sources are comparable to those of the MIR emission. We also find that the hard X-ray faintness previously reported for the bright C-GOALS I sources is also observed in the brightest LIRGs within the sample, with $L_{\rm FIR}>8\times10^{10}$ L$_{\odot}$. \newline \newline}

\maketitle

\section{Introduction}\label{intro}

Luminous and Ultra-luminous Infrared Galaxies (LIRGs and ULIRGs) are galaxies with infrared (IR) luminosities exceeding $10^{11} \ L_\odot$ and $10^{12} \ L_\odot$ respectively. U/LIRGs are normally found to be gas-rich galaxy mergers, as tidal torques can funnel material from kpc-scales to the innermost regions of the galaxy and trigger intense star formation and/or AGN activity \citep[e.g.][]{Her1989,San1999,DimSpr2005}, the latter more significant with increasing infrared luminosity \citep[e.g.][]{ValLut2009,PetArm2011,AloPer2012}. 

These objects, common at redshifts $1-3$ where the peak of star formation in the universe is observed, represent a very important stage in galaxy evolution \citep[e.g.][]{CasNar2014}. The scenario proposed by e.g. \citealp{SanSoi1988,HopHer2005}, indicates that after a complete obscuration phase of the merger,  ULIRGs in a late stage of the interaction would later disperse or consume the gas and probably evolve into an obscured, type II quasar (QSO), and eventually into an exposed QSO. This process will ultimately lead to the formation of an elliptical galaxy, and accounts for the growth of the central supermassive black hole (e.g. \citealp{SanSoi1988}, \citealp{HopBun2009}). 

\begin{figure*}
\centering
\includegraphics[width=\textwidth,keepaspectratio]{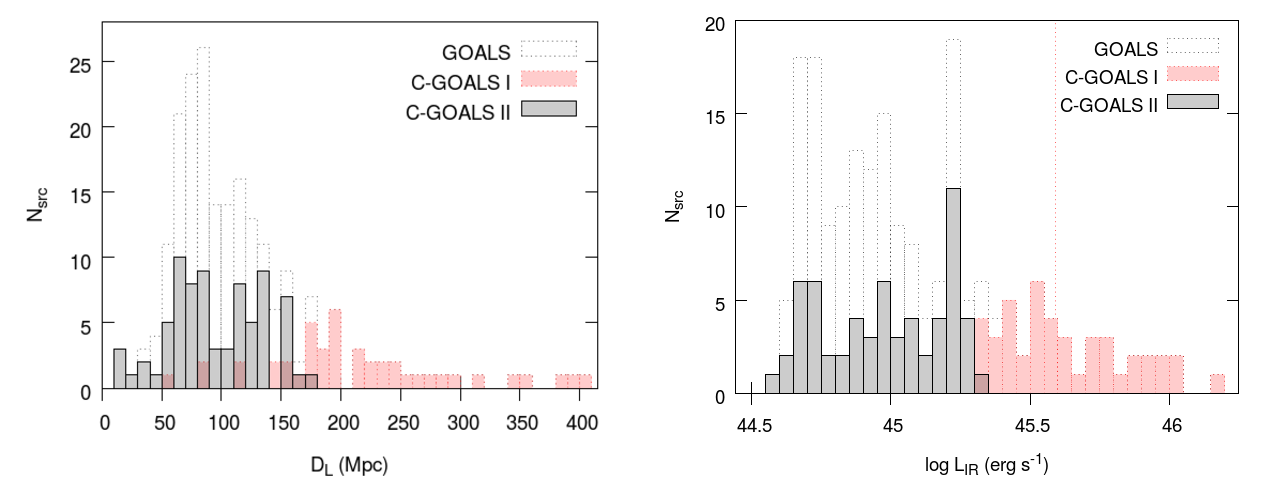}
\caption{The distribution of luminosity distance (left) and infrared luminosity $L_{\rm IR}(8-1000 \mu \rm m)$ (right) for the 44 objects of C-GOALS I \citep{IwaSan2011}, the 63 objects of C-GOALS II, and the 201 systems of the full GOALS sample \citep{ArmMaz2009}. The vertical dashed line represents $L_{\rm IR}=10^{12}$ L$_\odot$, the boundary between LIRGs and ULIRGs.}
\label{fig:SampleComp}
\end{figure*}

In agreement with this scenario, recent studies of pairs of galaxies have found that the fraction of dual AGN grows with decreasing separation between companions \citep[e.g.][]{EllPat2011,SatEll2014,SatSec2017,SilKam2011,KosMus2012}. More specifically, in a sample of U/LIRGs, \citet{StiArm2013} find an increase in the fraction of composite systems with merger stage. \citet{SatEll2014} find larger fractions of IR-selected AGN with respect to optically-selected AGN in mergers, which is likely due to the increase of obscuration. Evidences of and excess of AGN with large obscuring column densities in mergers are also found in recents works \citep[e.g.][]{DiaCha2010,KocBri2015,DelAle2016,LanRan2015,RicBau2017}. 

X-ray observations are an ideal tool to analyze properties of the inner regions of such obscured objects, due to the higher transparency of the gas and dust with respect to larger wavelengths. Previous studies of small  \citep[e.g.][]{FraBra2003,PtaHec2003, TenWil2005} and larger \citep[e.g.][]{TenVei2010,IwaSan2011,RicBau2017} samples of ULIRGs highlight the potential of X-rays in disentangling the AGN and starburst contribution and ability to detect enshrouded AGN.

Amongst the recent works, C-GOALS I \citep[\textit{Chandra}-GOALS I,][]{IwaSan2011}, is an X-ray study performed with the \textit{Chandra} X-ray Observatory (\textit{Chandra}, hereafter, \citealp{WeiTan2000}) of a complete sample of luminous infrared galaxies within the Great Observatories All-Sky LIRG Survey \citep[GOALS,][]{ArmMaz2009}. GOALS is a multi-wavelength study of the brightest infrared galaxies in the local Universe, a low-redshift subsample of the 60 $\mu$m flux selected \textit{IRAS} Revised Bright Galaxy Sample \citep[RBGS,][]{SanMaz2003}. The GOALS galaxies, all at z$<$0.088, are perfect laboratories for multi-wavelength studies of luminous infrared galaxies with a level of detail that only the observation of local galaxies allows. The arcsecond resolution provided by \textit{Chandra} can offer information of individual galaxies within mergers, and help disentangle previously undetected or unresolved AGN, in particular complementing studies of U/LIRGs at harder X-rays \citep[e.g.][]{RicBau2017}. 

The C-GOALS I paper presents data obtained by us and others with \textit{Chandra} and represents the X-ray component of the multi-wavelength survey for the most luminous infrared GOALS sources. This work, C-GOALS II, extends the X-ray study to a subsample of the lower luminosity range of GOALS galaxies. These data were obtained during \textit{Chandra} cycle 13 (PI: Sanders), combined with available archival data. The extension of the X-ray sample is motivated by the interest in reaching completeness in all wavelengths for the GOALS sample, and by the opportunity of comparing results derived at different infrared luminosity ranges. The sample contains galaxies at earlier merger stages, contributing to the expansion of previous studies into the domain of the less luminous LIRGS. In particular, \citet{IwaSan2011} observed a deviation in the correlation between infrared and X-ray luminosities in nearby star-forming galaxies \citep[e.g.][]{RanCom2003,GriGil2003,MinGil2014} for the galaxies in the C-GOALS I sample. Galaxies in the C-GOALS II sample have infrared luminosities that fall into the range where this change of behavior should occur, and are ideal to further study the reasons and possible implications of such deviation. 

The C-GOALS II sample is described and compared to the C-GOALS I sample in Sect. \ref{sample}. The observations and data reduction are described in Sect. \ref{ObsAndDR}. Results, including all X-ray images, fluxes, spectra and radial surface brightness profiles, are presented in Sect. \ref{results}, while derived properties and discussion of the X-ray and infrared luminosity correlation are presented in Sect. \ref{discussion}. Finally, we summarize our conclusions in Sect. \ref{conclusions}. Notes on individual objects can be found in Appendix \ref{NotesIO}, and X-ray contours, detailed images in the $0.5-2$ and $2-7$ keV bands, along with radial surface brightness profiles for each source, can be found in Appendix \ref{MWimages}.

\section{The Sample}\label{sample}

\begin{table*}
\footnotesize
\caption{Basic parameters of the objects in the C-GOALS II sample}
\label{table:sample}
\begin{tabular}{rllccccc}
\hline \hline
\begin{tabular}{c} No. \\ \\ (1) \end{tabular} & \begin{tabular}{c} \textit{IRAS} Name \\ \\ (2) \end{tabular} & \multicolumn{1}{l}{\begin{tabular}{c} Optical ID \\ \\ (3) \end{tabular}} & \begin{tabular}{c} RA (NED)\\ (J2000)\\ (4) \end{tabular} & \begin{tabular}{c} DEC (NED) \\ (J2000)  \\ (5) \end{tabular} &  \begin{tabular}{c} z \\   \\ (6) \end{tabular} & \begin{tabular}{c} $D_L$\\ (Mpc) \\ (7) \end{tabular} & \begin{tabular}{c} log$(L_{\rm IR})$\\ ($L_\odot$) \\ (8) \end{tabular} \\ \hline

45 & F13182+3424  & UGC 08387 & 13h 20m 35.34s   & +34d 08m 22.2s   & 0.0233 & 110.0 & 11.73 \\ 
47 & F01173+1405  & CGCG 436$-$030 & 01h 20m 02.72s   & +14d 21m 42.9s   & 0.0312 & 134.0 & 11.69 \\ 
49 & F01484+2220  & NGC 0695 & 01h 51m 14.24s   & 22d 34m 56.5s   & 0.0325 & 139.0 & 11.69 \\ \
50 & F12592+0436  & CGCG 043$-$099 & 13h 01m 50.80s   & +04d 20m 00.0s   & 0.0375 & 175.0 & 11.68 \\ 
51 & F11011+4107  & MCG+07$-$23$-$019 & 11h 03m 53.20s   & +40d 50m 57.0s   & 0.0345 & 158.0 & 11.62 \\ 
52 & F18329+5950  & NGC 6670 & 18h 33m 35.91s  & +59d 53m 20.2s   & 0.0289 & 129.5 & 11.65 \\ 
53 & F02512+1446  & UGC 02369 & 02h 54m 01.78s   & +14d 58m 24.9s   & 0.0312 & 136.0 & 11.67 \\ 
54 & F04315$-$0840  & NGC 1614 & 04h 33m 59.85s   & $-$08d 34m 44.0s   & 0.0159 & 67.8 & 11.65 \\ 
56 & F13497+0220  & NGC 5331 & 13h 52m 16.29s  & +02d 06m 17.0s   & 0.0330 & 155.0 & 11.66 \\ 
57 & F06076$-$2139 & IRAS F06076$-$2139 & 06h 09m 45.81s   & $-$21d 40m 23.7s   & 0.0374 & 165.0 & 11.65 \\ 
60 & F11231+1456  & IC 2810 & 11h 25m 47.30s   & +14d 40m 21.1s   & 0.0342 & 157.0 & 11.64 \\ 
63 & 18090+0130 & IRAS 18090+0130 & 18h 11m 35.91s  & +01d 31m 41.3s   & 0.0342 & 134.0 & 11.65 \\ 
64 & F01417+1651 & III Zw 035 & 01h 44m 30.45s   & +17d 06m 05.0s   & 0.0274 & 119.0  & 11.64 \\ 
65 & F10257$-$4339  & NGC 3256 & 10h 27m 51.27s   & $-$43d 54m 13.8s   & 0.0094 & 38.9 & 11.64 \\ 
67 & F16399$-$0937 & IRAS F16399$-$0937 & 16h 42m 40.21s  & $-$09d 43m 14.4s   & 0.0270 & 128.0 & 11.63 \\ 
68 & F16164$-$0746 & IRAS F16164$-$0746 & 16h 19m 11.79s  & $-$07d 54m 02.8s   & 0.0272 & 128.0 & 11.62 \\ 
69 & F18093$-$5744 & IC 4686/7 & 18h 13m 39.63s  & $-$57d 43m 31.3s  & 0.0173 & 81.9 & 11.62 \\ 
71 & F08354+2555  & NGC 2623 & 08h 38m 24.08s   & +25d 45m 16.6s   & 0.0185 & 84.1 & 11.60 \\ 
72 & F23135+2517 & IC 5298 & 23h 16m 00.70s   & +25d 33m 24.1s   & 0.0274 & 119.0  & 11.60 \\ 
73 & 20351+2521 & IRAS 20351+2521 & 20h 37m 17.72s   & +25d 31m 37.7s   & 0.0337 & 151.0 & 11.61 \\ 
75 & F16104+5235  & NGC 6090 & 16h 11m 40.70s  & +52d 27m 24.0s   & 0.0293 & 137.0 & 11.58 \\ 
79 & F13362+4831  & NGC 5256 & 13h 38m 17.52s  & +48d 16m 36.7s   & 0.0279 & 129.0 & 11.56 \\ 
80 & F03359+1523 & IRAS F03359+1523 & 03h 38m 46.70s   & +15d 32m 55.0s   & 0.0354 & 152.0 & 11.55 \\ 
81 & F04191$-$1855  & ESO 550$-$IG025 & 04h 21m 20.02s   & $-$18d 48m 47.6s   & 0.0322 & 135.8 & 11.51 \\ 
82 & F00085$-$1223  & NGC 0034 & 00h 11m 06.55s   & $-$12d 06m 26.3s   & 0.0196 & 84.1 & 11.49 \\ 
83 & F00506+7248  & MCG+12$-$02$-$001 & 00h 54m 03.61s   & +73d 05m 11.8s   & 0.0157 & 69.8 & 11.50 \\ 
85 & F17138$-$1017 & IRAS F17138$-$1017 & 17h 16m 35.79s  & $-$10d 20m 39.4s   & 0.0173 & 84.0 & 11.49 \\ 
95 & F12043$-$3140  & ESO 440$-$IG058 & 12h 06m 51.82s   & $-$31d 56m 53.1s   & 0.0232 & 112.0 & 11.43 \\ 
100 & F21453$-$3511 & NGC 7130 & 21h 48m 19.50s   & $-$34d 57m 04.7s   & 0.0162 & 72.7 & 11.42 \\ 
104 & F23488+1949 & NGC 7771 & 23h 51m 24.88s   & +20d 06m 42.6s   & 0.0143 & 61.2 & 11.40 \\ 
105 & F23157$-$0441 & NGC 7592 & 23h 18m 22.20s   & $-$04d 24m 57.6s   & 0.0244 & 106.0 & 11.40 \\ 
106 & F16577+5900  & NGC 6286 & 16h 58m 31.38s  & +58d 56m 10.5s   & 0.0183 & 85.7 & 11.37 \\ 
107 & F12590+2934  & NGC 4922 & 13h 01m 24.89s   & +29d 18m 40.0s   & 0.0236 & 111.0 & 11.38 \\ 
110 & F10015$-$0614  & NGC 3110 & 10h 04m 02.11s   & $-$06d 28m 29.2s   & 0.0169 & 79.5 & 11.37 \\ 
114 & F00402$-$2349  & NGC 0232 & 00h 42m 45.82s   & $-$23d 33m 40.9s   & 0.0227 & 95.2 & 11.44 \\ 
117 & F09333+4841  & MCG+08$-$18$-$013 & 09h 36m 37.19s   & +48d 28m 27.7s   & 0.0259 & 117.0 & 11.34 \\ 
120 & F15107+0724  & CGCG 049$-$057 & 15h 13m 13.09s  & +07d 13m 31.8s   & 0.0130 & 65.4  & 11.35 \\ 
121 & F02401$-$0013  & NGC 1068 & 02h 42m 40.71s   & $-$00d 00m 47.8s   & 0.0038 & 15.9 & 11.40 \\ 
123 & F02435+1253  & UGC 02238 & 02h 46m 17.49s   & +13d 05m 44.4s   & 0.0219 & 92.4 & 11.33 \\ 
127 & F13197$-$1627  & MCG$-$03$-$34$-$064 & 13h 22m 24.46s   & $-$16d 43m 42.9s   & 0.0165 & 82.2 & 11.28 \\ 
134 & F00344$-$3349  & ESO 350$-$IG038 & 00h 36m 52.25s   & $-$33d 33m 18.1s   & 0.0206 & 89.0 & 11.28 \\ 
136 & F23394$-$0353 & MCG$-$01$-$60$-$022 & 23h 42m 00.85s   & $-$03d 36m 54.6s   & 0.0232 & 100.0 & 11.27 \\ 
141 & F09437+0317 & IC 0563/4 & 09h 46m 20.71s  & +03d 03m 30.5s  & 0.0200 & 92.9 & 11.23 \\ 
142 & F13229$-$2934  & NGC 5135 & 13h 25m 44.06s  & $-$29d 50m 01.2s   & 0.0137 & 60.9 & 11.30 \\ 
144 & F13126+2453  & IC 0860 & 13h 15m 03.53s   & +24d 37m 07.9s   & 0.0112 & 56.8 & 11.14 \\ 
147 & F22132$-$3705 & IC 5179 & 22h 16m 09.10s   & $-$36d 50m 37.4s   & 0.0114 & 51.4 & 11.24 \\ \
148 & F03514+1546  & CGCG 465$-$012 & 03h 54m 16.08s   & +15d 55m 43.4s   & 0.0222 & 94.3 & 11.20 \\ 
157 & F12596$-$1529 & MCG$-$02$-$33$-$098/9 & 13h 02m 19.70s  & $-$15d 46m 03.0s  & 0.0159 & 78.7 & 11.17 \\ 
163 & F12243$-$0036  & NGC 4418 & 12h 26m 54.62s   & $-$00d 52m 39.2s   & 0.0073 & 36.5 & 11.19 \\ 
169 & F21330$-$3846 & ESO 343$-$IG013 & 21h 36m 10.83s   & $-$38d 32m 37.9s   & 0.0191 & 85.8 & 11.14 \\ 
170 & F06107+7822  & NGC 2146 & 06h 18m 37.71s   & +78d 21m 25.3s   & 0.0030 & 17.5 & 11.12 \\ 
174 & F14280+3126  & NGC 5653 & 14h 30m 10.42s  & +31d 12m 55.8s   & 0.0119 & 60.2 & 11.13 \\ 
178 & F12116+5448  & NGC 4194 & 12h 14m 09.47s   & +54d 31m 36.6s   & 0.0083 & 43.0 & 11.10 \\ 
179 & F23157+0618 & NGC 7591 & 23h 18m 16.28s   & +06d 35m 08.9s   & 0.0165 & 71.4 & 11.12 \\ 
182 & F00073+2538  & NGC 0023 & 00h 09m 53.41s   & +25d 55m 25.6s   & 0.0152 & 65.2 & 11.12 \\
188 & F23133$-$4251 & NGC 7552 & 23h 16m 10.77s   & $-$42d 35m 05.4s   & 0.0054 & 23.5 & 11.11 \\ 
191 & F04118$-$3207  & ESO 420$-$G013 & 04h 13m 49.69s   & $-$32d 00m 25.1s   & 0.0119 & 51.0 & 11.07 \\ 
194 & 08424$-$3130  & ESO 432$-$IG006 & 08h 44m 28.07s   & $-$31d 41m 40.6s   & 0.0162 & 74.4 & 11.08 \\ 
195 & F05365+6921  & NGC 1961 & 05h 42m 04.65s   & +69d 22m 42.4s   & 0.0131 & 59.0 & 11.06 \\ 
196 & F23444+2911 & NGC 7752/3 & 23h 47m 01.70s   & +29d 28m 16.3s   & 0.0162 & 73.6 & 11.07 \\ 
198 & F03316$-$3618  & NGC 1365 & 03h 33m 36.37s   & $-$36d 08m 25.4s   & 0.0055 & 17.9 & 11.00 \\ 
199 & F10196+2149  & NGC 3221 & 10h 22m 19.98s   & +21d 34m 10.5s   & 0.0137 & 65.7 & 11.09 \\ 
201 & F02071$-$1023  & NGC 0838 & 02h 09m 38.58s   & $-$10d 08m 46.3s   & 0.0128 & 53.8 & 11.05 \\ \hline\end{tabular}
 \begin{tablenotes}
\item \textbf{Notes:} Column(1): object number, also used in other tables. Column (2): original \textit{IRAS} source, where an ``F" prefix indicates the Faint Source Catalog and no prefix indicates the Point Source Catalog. Column (3): optical cross-identification, when available from NED. Columns (4), (5) and (6): the best available right ascension (J2000), declination and heliocentric redshift from NED as of October 2008. Column (7): the luminosity distance derived by correcting the heliocentric velocity for the 3-attractor flow model of \citet{MouHuc2000} and adopting cosmological parameters $H_0$ = 73 km s$^{-1}$ Mpc$^{-2}$, $\Omega_V$ = 0.73, and $\Omega_M$= 0.27, as provided by NED. Column (8): the total (8-1000) $\mu$m luminosity in log$_{10}$ Solar units as in \citep{ArmMaz2009}.\newline
  \end{tablenotes}
\end{table*}

GOALS \citep{ArmMaz2009} is a comprehensive study of 201 of the most luminous infrared-selected galaxies in the local Universe. The sample consists of 179 LIRGs and 22 ULIRGs, 85 of which are systems that contain multiple galaxies. GOALS is drawn from the \textit{IRAS} Revised Bright Galaxy Sample \citep[RBGS,][]{SanMaz2003}, with a luminosity threshold of $L_{\rm IR} \geq 10^{11} L_\odot$. The RBGS is a complete sample of galaxies, covering the whole sky, that have \textit{IRAS} 60 $\mu$m flux densities above 5.24 Jy and Galactic latitude $\vert b \vert \geq 5$º. 

\citet{IwaSan2011} studied a subsample of GOALS, C-GOALS I (hereafter, also CGI), which is complete in the higher infrared luminosity end of the GOALS sample (${\rm log} (L_{\rm IR}/L_\odot)= 11.73 - 12.57$). It contains 44 systems in the redshift range $z= 0.010-0.088$. The new sample, C-GOALS II (hereafter, also CGII), is an incomplete subsample of the lower luminosity section of GOALS, and includes all sources in the ${\rm log} (L_{\rm IR}/L_\odot)= 11.00 - 11.73$ range with available \textit{Chandra} data, as of January 2016. It is comprised of 63 systems, 30 of which contain multiple galaxies. The redshift range of the new sample is $z = 0.003-0.037$. The distribution of infrared luminosities and distances of the two samples is shown in Fig. \ref{fig:SampleComp}. Table \ref{table:sample} gives basic parameters for all the objects in the C-GOALS II sample. Note that names and positions refer to the infrared detected systems. Decomposition into individual galaxies is taken into account in Sect. \ref{results}. 

Figure \ref{fig:SampleComp} also evidences the incompleteness of CGII, comparing it with the full GOALS distribution of distances and luminosities. Of the 63 systems within CGII, 31 were observed through the same proposal, which was drawn to be representative of all possible merger stages. For the remaining 32 systems, data were taken from the archive according to availability. The proposal for which observing time was awarded varies in each case, and all targeting different scientific goals (e.g. study of AGN, SFR, X-ray binaries). For this reason, we do not expect our subsample to be biased toward a certain type of object, merger stage nor luminosity within the parent GOALS sample.

\section{Observations and Data reduction}\label{ObsAndDR}

Thirty-one systems were observed with \textit{Chandra} in cycle-13 (PI: Sanders) with a 15ks exposure on each target, carried out in imaging mode with the ACIS-S detector in VFAINT mode \citep{GarBau2003}. For the remaining 32 objects studied in this work, \textit{Chandra} data were obtained from the Archive. Exposure times for these targets vary from 4.88 to 58.34 ks, all taken with the ACIS-S detector in either FAINT or VFAINT mode. Table \ref{table:log} shows the observation log for the whole CGII sample, as well as the total source counts in the $0.5-7$ keV band for each object, obtained from the data analysis. The counts are derived for individual galaxies, and summed together when an object within the CGII sample contains more than one galaxy.

The data reduction was performed using the \textit{Chandra} data analysis package CIAO version 4.7 \citep{FruMcd2006}, and HEASARC's FTOOLS \citep{Bla1995}. The cosmology adopted here is consistent with that adopted by \citet{ArmMaz2009} and \citet{IwaSan2011}. Cosmological distances were computed by fist correcting for the 3-attractor flow model of \citet{MouHuc2000} and adopting $H_0=70$ km s$^{-1}$ Mpc$^{-1}$, $\Omega_V=0.72$, and $\Omega_M=0.28$ based on the 5-year WMAP results \citep{HinWei2009}, as provided by the NASA/IPAC Extragalactic Database (NED).

\begin{table*}
\footnotesize
\caption{\textit{Chandra} observation log for the objects in the CGII sample}
\label{table:log}
\begin{tabular}{rlcccccc}
\hline \hline
\begin{tabular}{c} No.\\ $\ $ \end{tabular}  & \begin{tabular}{c} Galaxy \\  $\ $ \end{tabular} & \begin{tabular}{c} Obs ID \\ $\ $ \end{tabular} & \begin{tabular}{c} Date   \\ $\ $ \end{tabular} & \begin{tabular}{c} Mode \\ $\ $  \end{tabular}  & \begin{tabular}{c} Exp. Time \\ (ks) \end{tabular} & \begin{tabular}{c} 0.5$-$7.0 keV$^a$  \\ (cts) \end{tabular}& \begin{tabular}{c} ${N_{\rm H,Gal}}^b$  \\ (10$^{20}$ cm$^{-2}$) \end{tabular} \\ \hline

45 & UGC 08387 & 7811 & 2007$-$02$-$19 & VFAINT & 14.07 & 277.9 $\pm$ 17.1 & 1.0 \\ 
47 & CGCG 436$-$030 & 15047 & 2012$-$11$-$24 & VFAINT & 13.82 & 168.6 $\pm$ 13.8  & 3.4 \\ 
49 & NGC 0695 & 15046 & 2013$-$01$-$01 & VFAINT & 14.78 & 312.9 $\pm$ 18.5 & 6.9 \\ 
50 & CGCG 043$-$099 & 15048 & 2012$-$11$-$23 & VFAINT & 14.78 & 71.9 $\pm$ 8.1 & 1.9 \\ 
51 & MCG+07$-$23$-$019 & 12977 & 2011$-$02$-$07 & VFAINT & 52.34 & 506.9 $\pm$ 26.8 & 1.0 \\ 
52 & NGC 6670 & 15049 & 2013$-$02$-$08 & VFAINT & 14.77 & 252.7 $\pm$ 16.8 & 3.9 \\ 
53 & UGC 02369 & 4058 & 2002$-$12$-$14 & FAINT & 9.68 & 120.6 $\pm$ 12.0 & 7.9 \\ 
54 & NGC 1614 & 15050 & 2012$-$11$-$21 & VFAINT & 15.76 & 800.0 $\pm$ 28.9 & 6.3 \\ 
56 & NGC 5331 & 15051 & 2013$-$05$-$12 & VFAINT & 14.78 & 121.9 $\pm$ 12.4 & 2.0 \\ 
57 & IRAS F06076$-$2139 & 15052 & 2012$-$12$-$12 & VFAINT & 14.78 & 52.4 $\pm$ 8.2 & 7.6 \\ 
60 & IC 2810 & 15053 & 2013$-$10$-$27 & VFAINT & 14.78 & 93.2 $\pm$ 11.7 & 2.5 \\ 
63 & IRAS 18090+0130 & 15054 & 2013$-$02$-$10 & VFAINT & 14.77 & 98.9 $\pm$ 11.3 & 20.2 \\ 
64 & III Zw 035 & 6855 & 2006$-$02$-$24 & FAINT & 14.98 & 81.4 $\pm$ 9.0 & 4.8 \\ 
65 & NGC 3256 & 835 & 2000$-$01$-$05 & FAINT & 27.80 & 8117.2 $\pm$ 102.3 & 9.1 \\ 
67 & IRAS F16399$-$0937 & 15055 & 2013$-$06$-$30 & VFAINT & 14.87 & 161.9 $\pm$ 14.4 & 13.0 \\ 
68 & IRAS F16164$-$0746 & 15057 & 2013$-$01$-$19 & VFAINT & 14.78 & 99.2 $\pm$ 11.3 & 11.3 \\ 
69 & IC 4686/7 & 15056 & 2012$-$11$-$19 & VFAINT & 14.48 & 519.7 $\pm$ 23.8 & 11.5 \\ 
71 & NGC 2623 & 4059 & 2003$-$01$-$03 & FAINT & 19.79 & 171.0 $\pm$ 14.1 & 3.1 \\ 
72 & IC 5298 & 15059 & 2013$-$02$-$04 & VFAINT & 14.78 & 222.8 $\pm$ 16.0 & 5.7 \\ 
73 & IRAS 20351+2521 & 15058 & 2012$-$12$-$13 & VFAINT & 13.56 & 146.8 $\pm$ 14.0 & 13.1 \\ 
75 & NGC 6090 & 6859 & 2006$-$05$-$14 & FAINT & 14.79 & 347.5 $\pm$ 19.3 & 1.6 \\ 
79 & NGC 5256 & 2044 & 2001$-$11$-$02 & FAINT & 19.69 & 1451.2 $\pm$ 43.5 & 1.7 \\ 
80 & IRAS F03359+1523 & 6856 & 2005$-$12$-$17 & FAINT & 14.76 & 108.2 $\pm$ 11.4 & 13.8 \\ 
81 & ESO 550$-$IG025 & 15060 & 2012$-$11$-$24 & VFAINT & 14.78 & 72.2 $\pm$ 10.6 & 3.2 \\ 
82 & NGC 0034 & 15061 & 2013$-$06$-$05 & VFAINT & 14.78 & 329.0 $\pm$ 19.5 & 2.1 \\ 
83 & MCG+12$-$02$-$001 & 15062 & 2012$-$11$-$22 & VFAINT & 14.31 & 311.0 $\pm$ 19.3 & 22.0 \\ 
85 & IRAS F17138$-$1017 & 15063 & 2013$-$07$-$12 & VFAINT & 14.78 & 207$-$3 $\pm$ 15.4 & 17.0 \\ 
95 & ESO 440$-$IG058 & 15064 & 2013$-$03$-$20 & VFAINT & 14.78 & 187.0 $\pm$ 16.1 & 5.6 \\ 
100 & NGC 7130 & 2188 & 2001$-$10$-$23 & FAINT & 38.64 & 3327.1 $\pm$ 59.3 & 1.9 \\ 
104 & NGC 7771 & 10397 & 2009$-$05$-$22 & VFAINT & 16.71 & 904.6 $\pm$ 34.6 & 4.0 \\ 
105 & NGC 7592 & 6860 & 2006$-$10$-$15 & FAINT & 14.99 & 388.7 $\pm$ 21.9 & 3.8 \\ 
106 & NGC 6286 & 10566 & 2009$-$09$-$18 & FAINT & 14.00 & 544.8 $\pm$ 27.9 & 1.8 \\ 
107 & NGC 4922 & 15065 & 2013$-$11$-$02 & VFAINT & 14.86 & 202.9 $\pm$ 17.2 & 0.9 \\ 
110 & NGC 3110 & 15069 & 2013$-$02$-$02 & VFAINT & 14.87 & 396.3 $\pm$ 22.3 & 3.5 \\ 
114 & NGC 0232 & 15066 & 2013$-$01$-$04 & VFAINT & 14.78 & 193.5 $\pm$ 15.7 & 1.4 \\ 
117 & MCG+08$-$18$-$013 & 15067 & 2013$-$06$-$03 & VFAINT & 13.79 & 101.7 $\pm$ 11.1 & 1.7 \\ 
120 & CGCG 049$-$057 & 10399 & 2009$-$04$-$17 & VFAINT & 19.06 & 30.2 $\pm$ 7.6 & 2.6 \\ 
121 & NGC 1068 & 344 & 2000$-$02$-$21 & FAINT & 47.44 & 100828.1 $\pm$ 326.7 & 2.9 \\
123 & UGC 02238 & 15068 & 2012$-$12$-$02 & VFAINT & 14.87 & 132.1 $\pm$ 13.5 & 8.9 \\ 
127 & MCG$-$03$-$34$-$064 & 7373 & 2006$-$07$-$31 & FAINT & 7.09 & 1029.3 $\pm$ 32.9 & 5.0 \\ 
134 & ESO 350$-$IG038 & 8175 & 2006$-$10$-$28 & VFAINT & 54.00 & 1794.5 $\pm$ 45.8 & 2.4 \\ 
136 & MCG$-$01$-$60$-$022 & 10570 & 2009$-$08$-$13 & FAINT & 18.90 & 325.4 $\pm$ 21.7 & 3.6 \\
141 & IC 0563/4 & 15070 & 2013$-$01$-$19 & VFAINT & 14.96 & 252.5 $\pm$ 18.9 & 3.8 \\
142 & NGC 5135 & 2187 & 2001$-$09$-$04 & FAINT & 29.30 & 3975.9 $\pm$ 68.0 & 4.9 \\ 
144 & IC 0860 & 10400 & 2009$-$03$-$24 & VFAINT & 19.15 & 25.9 $\pm$ 7.2 & 1.0 \\ 
147 & IC 5179 & 10392 & 2009$-$06$-$21 & VFAINT & 11.96 & 555.5 $\pm$ 32.2 & 1.4 \\ \
148 & CGCG 465$-$012 & 15071 & 2012$-$12$-$17 & VFAINT & 14.87 & 134.0 $\pm$ 13.4 & 14.8 \\ 
157 & MCG$-$02$-$33$-$098/9 & 15072 &  2013$-$05$-$08 & VFAINT & 14.87 & 141.0 $\pm$ 12.4 & 3.7 \\ 
163 & NGC 4418 & 4060 & 2003$-$03$-$10 & FAINT & 19.81 & 59.6 $\pm$ 15.3 & 1.9 \\ 
169 & ESO 343$-$IG013 & 15073 & 2013$-$06$-$13 & VFAINT & 14.78 & 139.6 $\pm$ 13.9 & 2.8 \\ 
170 & NGC 2146 & 3135 & 2002$-$11$-$16 & FAINT & 10.02 & 2144.2 $\pm$ 50.4 & 7.1 \\ 
174 & NGC 5653 & 10396 & 2009$-$04$-$11 & VFAINT & 16.52 & 387.1 $\pm$ 22.8 & 1.3 \\
178 & NGC 4194 & 7071 & 2006$-$09$-$09 & FAINT & 35.50 & 2410.3 $\pm$ 51.4 & 1.5 \\ 
179 & NGC 7591 & 10264 & 2009$-$07$-$05 & FAINT & 4.88 & 26.3 $\pm$ 6.1 & 5.6 \\ 
182 & NGC 0023 & 10401 & 2008$-$10$-$27 & VFAINT & 19.45 & 753.1 $\pm$ 31.9 & 3.4 \\
188 & NGC 7552 & 7848 & 2007$-$03$-$31 & FAINT & 5.08 & 832.8 $\pm$ 30.2 & 1.2 \\ 
191 & ESO 420$-$G013 & 10393 & 2009$-$05$-$13 & VFAINT & 12.42 & 759.0 $\pm$ 29.2 & 2.1 \\ 
194 & ESO 432$-$IG006 & 15074 & 2013$-$06$-$24 & VFAINT & 16.05 & 280.7 $\pm$ 20.0 & 19.3 \\ 
195 & NGC 1961 & 10531 & 2009$-$05$-$08 & VFAINT & 32.83 & 723.3 $\pm$ 40.0 & 8.1 \\ 
196 & NGC 7752/3 & 10569 & 2009$-$08$-$30 & FAINT & 11.99 & 96.0 $\pm$ 12.7 & 5.4 \\ 
198 & NGC 1365 & 6869 & 2006$-$04$-$20 & FAINT & 15.54 & 4644.2 $\pm$ 72.7 & 1.3 \\ 
199 &  NGC 3221 & 10398 & 2009$-$03$-$19 & VFAINT & 19.03 & 323.5 $\pm$ 28.3 & 1.9 \\
201 & NGC 0838 & 15667 & 2013$-$07$-$21 & VFAINT & 58.34 & 1996.0 $\pm$ 49.6 & 2.6 \\ \hline

\end{tabular}
 \begin{tablenotes}
\item \textbf{Notes:} $^{(a)}$ The source counts are background corrected and measured in the 0.5$-$7.0 keV band. The counts from separate components in a single system are obtained separately and then summed together. $^{(b)}$ The Galactic absorption column density is taken from the LAB HI map by \citet{KalBur2005}.\newline
  \end{tablenotes}

\end{table*}

\section{Results}\label{results}

\addtocounter{table}{1}

\begin{table}[h]
\setlength{\tabcolsep}{0.1cm}
\centering
\footnotesize
\caption{IR fractions}
\label{table:IRfrac}
\begin{tabular}{r@{\hskip 0.2cm}lcc}
\hline \hline
No. & \begin{tabular}{c}Galaxy in system\\  (1) \end{tabular} & \begin{tabular}{c} \% \\  (2) \end{tabular} & \begin{tabular}{c}Ref. \\  (3) \end{tabular} \\ 
\hline
47 & CGCG 436$-$030 (W) & 100 & 5\\ 
52 & NGC 6670 (W) & 62 & 1 \\ 
53 & UGC 02369 (S) & 98 & 2  \\ 
56 & NGC 5331 (S) & 81 & 1 \\ 
57 & IRAS F06076$-$2139 (N) & 88 & 2 \\ 
60 & IC 2810 (NW) & 68 & 1 \\ 
63 & IRAS 18090+0130 (E) & 81 & 1 \\ 
64 & III Zw 035 (N) & 100 & 5 \\ 
67 & IRAS F16399$-$0937 (N) & 90 & 3 \\ 
69 & IC 4687 (N,S)* & 66,22 & 1 \\ 
75 & NGC 6090 (NE) & 90 & 4 \\ 
79 & NGC 5256 (SW) & 63 & 2  \\ 
80 & IRAS F03359+1523 (E) & 100 & 5 \\ 
81 & ESO 550$-$IG025 (N) & 59 & 1 \\ 
83 & MCG+12$-$02$-$001 (E,W)* & 90,10 & 2 \\ 
95 & ESO 440$-$IG058 (S) & 89 & 1 \\ 
104 & NGC 7771 & 90 & 1 \\ 
105 & NGC 7592 (E,S)* & 63,0 & 2  \\ 
106 & NGC 6286 & 87 & 1 \\ 
107 & NGC 4922 (N) & 99 & 2 \\ 
110 & NGC 3110 (NE) & 91 & 1 \\
117 & MCG+08$-$18$-$013 (E) & 97 & 1 \\ 
127 & MCG$-$03$-$34$-$064 & 75 & 1 \\
141 & IC 0564 & 54 & 1 \\
157 & MCG$-$02$-$33$-$098/9 (W) & 69 & 2 \\ 
163 & NGC 4418 & 99 & 1 \\  
169 & ESO 343$-$IG013 (N) & 78 & 2 \\ 
179 & NGC 7591 & 94 & 1 \\
194 & ESO 432$-$IG006 (SW) & 63 & 1 \\ 
196 & NGC 7753 & 64 & 1 \\ 
\hline
\end{tabular}
\begin{tablenotes}
\item \textbf{Notes:} Column (1): name of the galaxy (galaxies) which emits most of the IR luminosity in a double (triple) system. Column (2): percentage of IR emission originating in the dominant component. Column (3): Reference from which the contribution to the IR luminosity is derived. \newline
1: Derived from \textit{Herschel} data \citep[as in][]{ChuSan2017}. \newline
2: Derived from MIPS 24 data  \citep[as in][]{DiaCha2010}. \newline
3: MIR determination from \citet{HaaSur2011}. \newline
4: Predicted IR emission from radio continuum in \citet{HatYos2004}. \newline
5: Refer to Notes on individual objects in Appendix \ref{NotesIO}. \newline 
(*) Triple component galaxy.
  \end{tablenotes}
\end{table}

Results of the X-ray analysis of the \textit{Chandra} data are presented in Table 3. For each galaxy we present the background-corrected ACIS-S X-ray soft band ($S, 0.5-2$ keV) count rate and X-ray hard band ($H, 2-7$ keV) count rate, the hardness ratio or X-ray color, estimated X-ray fluxes and luminosities in both soft and hard band, and the logarithmic ratio of each X-ray band to the infrared luminosity listed in Table \ref{table:sample}, $L_{\rm IR}(8$-$1000 \mu {\rm m})$. X-ray color, or hardness ratio, is computed as $HR=(H-S)/(H+S)$, using the bands previously defined. 

Individual galaxies belonging to the same GOALS system (i.e. contributing to one single \textit{IRAS} source) are identified by using the same GOALS Number. in the first column. Source names shown in the second column are used throughout this work; see Appendix \ref{NotesIO} for clarification on the identification of each component. 

Note that the hard X-ray flux ($F_{\rm HX}$) listed in Table 3 is in the $2-7$ keV band, where \textit{Chandra} is more sensitive; and the listed hard X-ray luminosities ($L_{\rm HX}$) refer to the the $2-10$ keV band. Spectral fitting to derive the fluxes is performed in the $2-7$ keV range as described in Sect. \ref{SpectralFitting}, and the fitted models are posteriorly used to estimate the luminosity up to 10 keV, in order to compare the derived results to those of previous works, in which the $2-10$ keV band is used.

Although significant intrinsic absorption in dusty objects such as LIRGs is likely present, X-ray luminosities are estimated by correcting for galactic absorption only. The X-ray spectra of our galaxies are complex, containing multiple components, with different degrees of obscuration, as explained in Sect. \ref{SpectralFitting}. As the estimated absorbing column density values are heavily model dependent, we do not use them to correct the luminosities listed in Table 3. 

As mentioned in Sect. \ref{sample}, many of the LIRGs in the CGII sample are composed of multiple galaxies, which are associated to a single GOALS object, as \textit{IRAS} is unable to resolve them. All spatially resolved components in the \textit{Chandra} data are presented separately, their count rates, fluxes and luminosities computed individually. In order to obtain the X-ray to infrared ratios listed in Table 3, the \textit{IRAS} flux associated to each object must be appropriately separated into the corresponding contribution of each component. This separation is carried out according to the best possible estimate available for each source, as listed in Table \ref{table:IRfrac}. The most accurate estimation would be derived by obtaining the separate contribution of each component from the FIR emission. Whenever possible, this is done using \textit{Herschel} photometric data \citep{ChuSan2017}. However, 14 of the multiple systems in the sample are unresolved by \textit{Herschel}, and thus the mid-infrared Spitzer MIPS 24 data are used for this purpose. 

For four systems in the sample the individual components remain unresolved at  mid-infrared wavelengths, and other determinations are used, as specified in Table \ref{table:IRfrac} and Appendix \ref{NotesIO}. 

Only objects that are detected in X-rays and contribute to at least 10\% of the infrared luminosity of the \textit{IRAS} source are analyzed and presented in this work. This cut means that out of the 63 GOALS systems in the sample, 84 individual galaxies are studied in CGII. No galaxy contributing $<10\%$ to the IR has a strong X-ray emission, but in cases in which the source is detected in the \textit{Chandra} data, it is specified in Appendix \ref{NotesIO}. For all galaxies in pairs that are not included in the analysis, their contribution to the IR luminosity of the bright component is taken into account. 

\subsection{X-ray images}\label{Xraycon}

We show how the X-ray radiation is related to the optical and IR emission, by comparing the $0.4-7$ keV brightness contours with HST, SDSS or IRAC images according to availability, in this order of preference. Appendix \ref{MWimages} shows X-ray contours overlaid on HST-ACS F814W (I-band) images (Evans et al. in prep.) for 27 objects, overlaid on SDSS DR-12 i-band images \citep{AlaAlb2015} for 18 objects and overlaid on IRAC channel 1 images \citep[][Mazzarella et al. in prep.]{ArmMaz2009} for the remaining 18 objects.

The contours are taken from a $0.4-7$ keV image, smoothed using a Gaussian filter with a dispersion of 1 arcsec, with the exception of NGC 5135, shown in Appendix \ref{MWimages}, for which a smoothing of 0.5 arcsec was used in order to preserve the two X-ray central peaks.

Eleven contour levels were defined, divided into ten equal logarithmic intervals, in the four different surface brightness ranges shown in Table \ref{table:ConRange}. ``Interval 1" is used for the majority of the sample. In order to outline lower surface brightness features in some sources, eleven contour levels starting at a lower surface brightness values were taken, as ``Interval 2'' or ``Interval 3". For a few systems, a higher lower surface brightness limit was taken in order to eliminate noisy features in the contours, defined as ``Interval 4". For bright objects NGC 1068 and NGC 1365, 21 contour levels were used instead, in order to reflect the X-ray morphology appropriately. Appendix \ref{MWimages} contains information on both which optical or IR image was used to overlay the X-ray contours on, and the Interval used for X-ray contour ranges.

\begin{table}
\centering
\footnotesize
\caption{X-ray contour ranges}
\label{table:ConRange}
\begin{tabular}{cccc}
\hline \hline
Interval & \begin{tabular}{c} Low \\  (1) \end{tabular} & \begin{tabular}{c} High \\  (2) \end{tabular} & \begin{tabular}{c} Levels \\  (3) \end{tabular} \\  \hline

1 & $2.5 \times 10^{-5}$ & $7 \times 10^{-3}$ & 11\\
2 & $1.7 \times 10^{-5}$ & $7 \times 10^{-3}$ & 11\\
3 & $1.0 \times 10^{-5}$ & $7 \times 10^{-3}$ & 11\\
4 & $4.0 \times 10^{-5}$ & $7 \times 10^{-3}$ & 11\\

\hline \hline
Galaxy & \begin{tabular}{c} Low \\  (1) \end{tabular} & \begin{tabular}{c} High \\  (2) \end{tabular} & \begin{tabular}{c} Levels \\  (4) \end{tabular} \\  \hline
CGCG 465-012 & $2.5 \times 10^{-5}$ & $2.0 \times 10^{-4}$ & 11\\
NGC 1068 & $2.5 \times 10^{-5}$ & $4.0 \times 10^{-2}$ & 21\\
NGC 5135 & $4.0 \times 10^{-5}$ & $1.0 \times 10^{-2}$  & 11\\
NGC 1365 & $2.5 \times 10^{-5}$ & $2.0 \times 10^{-2}$ & 21\\
NGC 0838 & $7.0 \times 10^{-6}$ & $4.0 \times 10^{-4}$ & 11\\

\hline
\end{tabular}
 \begin{tablenotes}
\item \textbf{Notes:} Column (1) and (2): lower and higher contour in [counts s$^{-1}$ arcsec$^{-1}$] for the given interval, respectively. (4): number of logarithmic contour levels.
  \end{tablenotes}
\end{table}

\begin{figure*}
  \centering 
\includegraphics[width=0.95\textwidth,keepaspectratio]{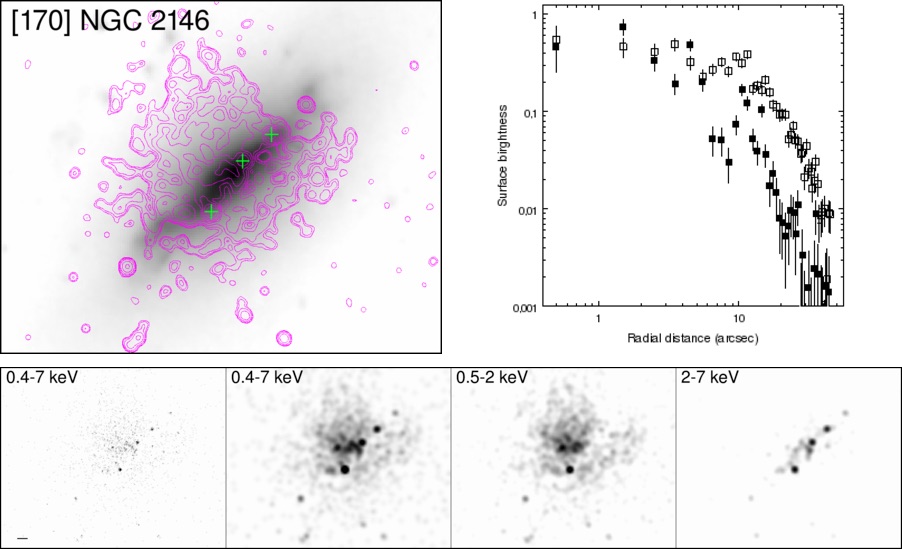}
  \caption{X-ray images and surface brightness profiles for NGC 2146. The orientation of the image is north up and east to the left. Similar figures for all 59 objects in the CGII sample are presented in Appendix \ref{MWimages}. \textit{Upper left:} X-ray (0.4$-$7 keV) brightness contours (magenta) with marked hard X-ray peaks (green crosses) overlaid on optical/infrared images. \textit{Upper right:} radial surface brightness profiles in the $0.5-2$ keV band (open squares) and the $2-7$ keV band (filled squares). Profiles have been centered using the brightness peak in the hard X-ray band, when clearly originating in the nucleus. Refer to Appendix \ref{NotesIO} for ambiguous objects. \textit{Bottom:} From left to right, unsmoothed and smoothed images in the 0.4$-$7 keV band, and smoothed images in the soft ($0.5-2$ keV) and hard ($2-7$ keV) bands. The pixel size is $\sim 0.5'' \times 0.5''$. The scale bar in the \textit{bottom left} image represents 5''.}
  \label{fig:NGC2146}
\end{figure*}

As the hard-band emission from all objects is generally more peaked and less intense than the soft-band emission, the contours mostly trace soft X-ray emission from the sources. For this reason, in sources for which one or more clear hard X-ray peaks are seen, these are marked with a green cross. We define a hard X-ray peak as point-source emission that clearly stands out from the rest of the photon distribution in the unsmoothed images. In cases where many point-like sources that are clearly not associated with any central source in the galaxy are present in an image, we opt not to mark them all individually. For a more detailed description of the X-ray emission in both bands, Appendix \ref{MWimages} also presents the smoothed and unsmoothed images in the $0.4-7$ keV band, and smoothed images in the soft ($0.5-2$ keV) and hard ($2-7$ keV) bands, for all objects. An example of one of these images is shown in Fig. \ref{fig:NGC2146}.

\subsection{X-ray spectra}\label{spectra} 

\begin{figure}
  \centering 
\includegraphics[width=0.95\columnwidth,keepaspectratio]{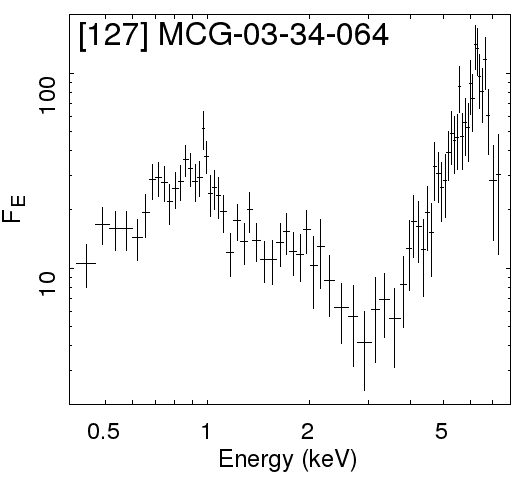}
  \caption{X-ray flux density spectra for MCG-03-34-064, obtained from the \textit{Chandra} ACIS. Flux density in units of $10^{-14}$ erg s$^{-1}$ cm$^{-2}$ keV$^{-1}$.}
  \label{fig:SampleSpectrum}
\end{figure}

Fig. \ref{fig:XraySpectra} in Appendix \ref{AllSpectra} presents the X-ray spectra for all sources. Spectral data is shown separately for each object with more than one resolved component. Instead of showing the usual count rate spectra, which are data folded through the detector response, we present the \textit{Chandra} ACIS spectra corrected for the detector response and converted into flux density units. This has the advantage of presenting the spectral properties without the need of spectral fitting, and facilitates comparison with other multi-wavelength data from GOALS. The flux density range for all spectra is set to be the same, 2 orders of magnitude, for consistent comparison. An example of one such spectra can be seen in Fig. \ref{fig:SampleSpectrum}.

Note that this presentation has the caveat of introducing some uncertainty, particularly when a spectral bin is large enough, within which the detector response varies rapidly; e.g. for galaxies with a low number of counts. It should also be taken into account that, even though these have been corrected for the detector effective area, the energy resolution of the detector is preserved, and therefore they are independent of any spectral model fitting; i.e. they are not unfolded spectra. 

It should also be noted that the spectra in Fig. \ref{fig:XraySpectra} are for display purposes only, and all physical quantities determined were obtained through spectral fitting of the count rate spectra, with the appropriate detector responses. 

\subsection{AGN selection}\label{AGNselection}

\begin{table}
\setlength{\tabcolsep}{0.05cm}
\centering
\footnotesize
\caption{Sources with an AGN signature in IR or X-rays.}
\label{table:AGNselection}
\begin{tabular}{rlcccc}
\hline \hline
No. & Galaxy & \begin{tabular}{c} AGN$_{bol}$ \\  (1) \end{tabular}& \begin{tabular}{c} [Ne v] \\  (2) \end{tabular} & \begin{tabular}{c} PAH \\  (3) \end{tabular} & \begin{tabular}{c} X \\  (4) \end{tabular}\\  \hline

45 & UGC 08387 & $0.03\pm0.01$ & Y & N & N \\ 
57 & IRAS F06076$-$2139 (S) & $-$ & N & N & C \\ 
68 & IRAS F16164$-$0746 & $0.05\pm0.01$ & Y & N & C \\ 
71 & NGC 2623 & $0.10\pm0.03$ & Y & N & C \\ 
72 & IC 5298 & $0.33\pm0.05$ & Y & N & A \\ 
79 & NGC 5256 (NE) & $0.23\pm0.07$(u) & Y(u) & N & A \\ 
79 & NGC 5256 (SW) & $0.23\pm0.07$(u) & Y(u) & N & L \\ 
82 & NGC 0034 & $0.04\pm0.02$ & N & N & A \\ 
85 & IRAS F17138$-$1017 &  $0.07\pm0.04$ & N & N & C \\ 
100 & NGC 7130 & $0.22\pm0.04$ & Y & N & LA \\ 
105 & NGC 7592 (W) & $0.20\pm0.06$ & Y(u) & N & N \\ 
106 & NGC 6286 & $0.11\pm0.06$ & N & N & A* \\
107 & NGC 4922 (N) & $0.17\pm0.05$ & Y & N & A \\ 
114 & NGC 0232 & $0.09\pm0.03$ & Y & N & N \\ 
120 & CGCG 049$-$057 & $0.04\pm0.02$ & N & N & C \\ 
121 & NGC 1068 & $1.00\pm0.01$& Y & $-$ & L \\ 
127 & MCG$-$03$-$34$-$064 & $0.88\pm0.04$& Y & Y & LA \\ 
136 & MCG $-$01$-$60$-$022 & $0.08\pm0.06$& N & $-$ & CA \\ 
142 & NGC 5135 & $0.24\pm0.06$& Y & N & L \\ 
144 & IC 0860 &  $0.06\pm0.03$& N & N & C \\ 
163 & NGC 4418 & $0.48\pm0.22$& N & Y & N \\ 
169 & ESO 343$-$IG013 (N) & $0.09\pm0.04$& N & N & C \\ 
191 & ESO 420$-$G013 & $0.25\pm0.04$& Y & N & N \\ 
194 & ESO 432$-$IG006 (NE) & $0.12\pm0.04$& N & N & A \\ 
194 & ESO 432$-$IG006 (SW) & $0.09\pm0.05$ & N & N & A \\ 
198 & NGC 1365 & $0.38\pm0.03$& N & N & CLA \\ 

\hline
\end{tabular}
 \begin{tablenotes}
\item \textbf{Notes:} Column (1): Contribution of the AGN into the bolometric luminosity of the galaxy \citep{DiaArm2017}. Column (2): Detection of the [Ne v] line \citep{PetArm2011}. Column (3): EW of the 6.2 $\mu$m PAH feature < 0.1 $\mu$m \citep{StiArm2013}. Column (4): X-ray AGN selection criteria. C: X-ray color ($HR > -0.3$); L: detection of 6.4 KeV line. A: absorbed AGN feature. Y: Source meets the criterion. N: Source does not meet the criterion. (u) unresolved detection in a multiple systems. (*) Only in the \textit{NuSTAR} data \citep{RicBau2016}. See Appendix \ref{NotesIO} for details on particular sources.
  \end{tablenotes}
\end{table}

\begin{figure}
\centering
\includegraphics[width=\columnwidth,keepaspectratio]{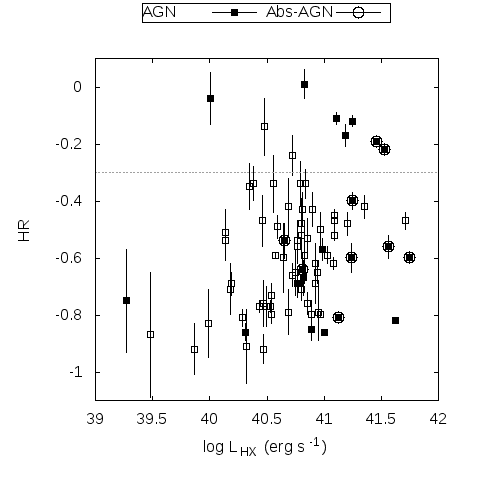}
\caption{Hardness ratio as a function of the $2-7$ keV luminosity for all sources in the CGII sample. All AGN from Table \ref{table:AGNselection} are plotted in filled squares, and those in which absorption features are fitted (labeled A in the table) are marked with an open circle. The dashed line shows the $-$0.3 boundary, above which sources are selected as AGN (unless evidence points toward lack of AGN presence, see Appendix \ref{NotesIO}).}
\label{fig:HRvsLHX}
\end{figure}

\begin{figure}
\centering
\includegraphics[width=\columnwidth,keepaspectratio]{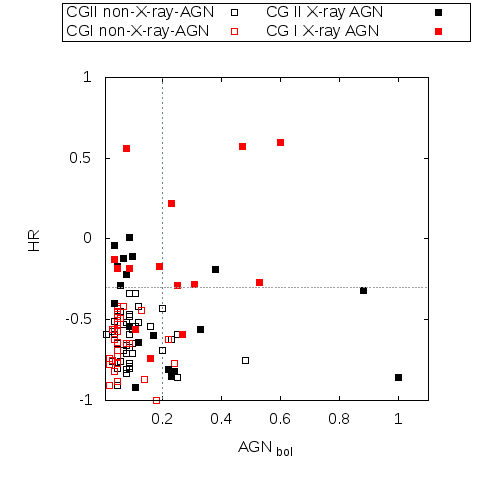}
\caption{Hardness ratio as a function of the fractional contribution of the AGN to the bolometric luminosity (as derived from MIR data, \citet{DiaArm2017}) of the source, in red for CGI sources and black for CGII. X-ray selected AGN from Table \ref{table:AGNselection} are plotted in filled squares.The horizontal dashed line shows the $HR=-0.3$ threshold. The vertical dashed line shows the value above which the AGN is energetically significant.}
\label{fig:HRvsAGNbol}
\end{figure}

AGN classification in the sample is performed using different criteria in both X-rays and infrared. Any galaxy that meets any of our selection criteria, described below, is classified as an AGN and listed in Table \ref{table:AGNselection}. 

The X-ray selection is performed using two different methods: X-ray color selection and detection of AGN spectral features. 

The X-ray color, or hardness ratio, gives the relative intensity of emission in hard and soft bands (in counts). A high $HR$ indicates strong emission above 2 keV, which is often associated with the presence of an obscured AGN, i.e. column density, $N_{\rm H}$ in the range of $10^{22}-10^{24}$ cm$^{-2}$. The threshold for AGN selection is chosen as $HR > -0.3$, as was for the CGI sample \citep[see][]{IwaSan2011}. 

Figure \ref{fig:HRvsLHX} shows the hardness ratio of all sources in the sample as a function of their hard X-ray luminosity. AGN selected through all criteria described in this section are plotted with filled squares, while all absorbed AGN are marked with open circles. Most AGN in the sample have a $HR$ below the threshold, as many are absorbed or not selected through X-rays. 

Some AGN are missed by this $HR$ selection because absorption in the nucleus is significant and soft X-ray emission coming from external starburst regions is strong. Such galaxies can still show a hard-band excess in their spectra, and if fitting them with an absorbed power-law with a fixed 1.8 photon index yields a high enough absorbing column density, we classify them as absorbed AGN (see Sect. \ref{HBfit}). Each of these cases is listed in table \ref{table:AGNselection} and discussed individually in Appendix \ref{NotesIO}. 

When absorption is even stronger, only reflected radiation can be observed in the hard band, and therefore sources appear weak, their $HR$ being even smaller. A clear signature of a highly obscured AGN is a the detection of a strong Fe K$\alpha$ line at 6.4 keV, which is also used as a criterion for AGN selection. We set a threshold of $2 \sigma$ for the detection of the iron line in order to classify a source as an AGN. Sources selected through this criterion are listed in Table \ref{table:AGNselection}, and details on the iron line fits can be found in Sect. \ref{SpectralFitting}, Table \ref{table:IronLines}.

The infrared selection is performed by means of the detection of the [Ne v] 14.32 $\mu$m line over kpc scales, which traces high-ionization gas. The ionization potential of [Ne v] is 96 eV, which is too high to be produced by OB stars. Therefore, detection of this line in the integrated spectra of galaxies is a good AGN indicator \citep[see][and references therein]{PetArm2011}. 

Another possible indicator is the equivalent width of the 6.2 $\mu$m PAH feature being lower than 0.1 $\mu$m.  Polycyclic Aromatic Hydro-carbons (PAHs) are either destroyed by the radiation originating from the AGN, or their features diluted in the spectra by the strong MIR continuum it creates; resulting in a low value of the EW \citep[see][and references therein]{StiArm2013}.

Using only the X-ray criteria, 21 galaxies have been found to host an AGN, which represents (25$\pm$5)\% of our sample. With the addition of IR criteria, 5 other galaxies are classified as AGN, resulting in a total AGN fraction of (31$\pm$5)\% for the 84 individually analyzed galaxies in CGII. Galaxies selected as AGN are presented in Table \ref{table:AGNselection}, along with optical classifications and whether or not they meet our X-ray and IR selection criteria. 

Two sources in the sample meet the selection criteria and yet we opt not to classify as AGN, for reasons explained in Appendix \ref{NotesIO}: IRAS F17138$-$1017 and IRAS F16399$-$0937 (S), which meet the $HR$ criterion.

Table \ref{table:AGNselection} also lists the contribution of the AGN to the bolometric luminosity for all sources classified as AGN. The contribution of the AGN to the MIR luminosity is derived by \citet{DiaArm2017}, for all GOALS galaxies, employing up to five \textit{Spitzer}/IRS diagnostics. Applying corrections based on SED templates of pure starbursts and AGN sources, they derive the fractional contribution of the AGN to the overall bolometric luminosity \citep[as in][]{VeiRup2009}. 

Figure \ref{fig:HRvsAGNbol} shows the $HR$ of all sources in the sample as a function of the fractional contribution of the AGN to the bolometric luminosity, $AGN_{\rm bol}$. Sources with a fraction larger than 0.2 can be considered to have an energetically significant AGN. X-ray selected AGN, through any of the three criteria mentioned above, are highlighted as filled symbols. All marked AGN below the $HR=-0.3$ threshold show signs of obscuration, as they have been selected through any of the other two X-ray criteria. In the full C-GOALS sample, 19/32 X-ray selected AGN lay below $AGN_{\rm bol} <0.2$. Therefore, more than half of the AGN detected through X-rays are not easily selected through the mentioned combination of MIR diagnostics. 

\subsection{X-ray spectral fitting}\label{SpectralFitting}

The $0.4-7$ keV \textit{Chandra} spectra of the CGII galaxies present similar properties to those of the CGI sample, analysed by \citet{IwaSan2011}; a mostly emission-line dominated soft X-ray band, and a hard X-ray power-law. As was already discussed in the mentioned work, both the spectral shape and the morphology of the emission (see images in Appendix \ref{MWimages}) suggest a different origin for the soft and hard X-rays, and therefore the two are analyzed separately.

A few objects present in the sample (IRAS 18090+0130 (W), IRAS F06076$-$2139 (S), IC 0860 and NGC 7591) have not been fitted due to an excessively low count number, of the order of $\lesssim 25$ cts, in the full $0.4-7$ keV band. For these sources, only the count rates and $HR$ have been computed, and results on fluxes and luminosities are not presented in Table 3.

For the majority of our sources, which have few counts, fitting is done through C statistic minimization instead of $\chi^2$ minimization.

\subsubsection{Soft band ($0.4-2$ keV) fitting}\label{SBfit}

Starburst galaxies, when not dominated by a luminous AGN, have soft X-ray emission originating in hot interstellar gas ($\sim 0.2-1$keV), which is shock-heated by supernovae explosions and stellar winds from massive stars. Emission from hot gas can generally be fitted with a standard thermal emission model, with a solar abundance pattern, e.g. \textit{mekal} \citep{MewGro1985, Kaa1992,LieOst1995}. However, in our data, such a simple model does not agree with many of the observed emission line strengths and provides an unsatisfactory fit in most cases. A better fit can be obtained either with a modified abundance pattern, richer in $\alpha$ elements; or through the overlap of more than one \textit{mekal} at different temperatures. 

The hot gas within a starburst region is expected to be enriched by $\alpha$ elements, produced in core-collapse supernovae. Metal abundances should deviate from a solar pattern, as has been found for star-forming knots in nearby galaxies, e.g. The Antennae \citep{FabBal2004}. At the same time, the extended soft X-ray emitting gas is expected to be multi-phase: the shocked gas swept away by a starburst wind seen at outer radii is free from absorption while the hotter gas at inner radii may have some absorption of the interstellar medium \citep[e.g.][]{StrSte2000}. A temperature gradient can be approximated at first order as two \textit{mekal} models with different temperatures. One would model the most external, colder gas component (at $T=T_1$) which is located far away from the nucleus, and therefore is less absorbed by the interstellar material. The other would model the inner, hotter gas (at $T=T_2$), obscured by the denser material present in the central region of the galaxy. 

Ideally, the data should be modeled using more than one \textit{mekal} component, with different temperatures and absorbing column densities, and with non-solar metal abundances. However, given the quality of the data, this would imply severely overfitting. As we are interested in probing the level of obscuration in the C-GOALS sources, we opt to model the data using two \textit{mekal} models as defined above, which both have solar abundance patterns. 

The results obtained through this fitting, the parameters of which can be seen in Table 7, show that it is possible to satisfactorily fit the sources with a high enough number counts using this model, which is to be expected if part of the emission truly originates in a denser, inner region. However, we note that this model is not clearly superior to a single \textit{mekal} component with non-solar abundances, as was used by \citet{IwaSan2011} on the CGI sample; and that most of the analyzed sources do not have good enough data quality to determine a clear best fit between the two models. 
\addtocounter{table}{1}

\begin{figure*}
\centering
\includegraphics[width=\textwidth,keepaspectratio]{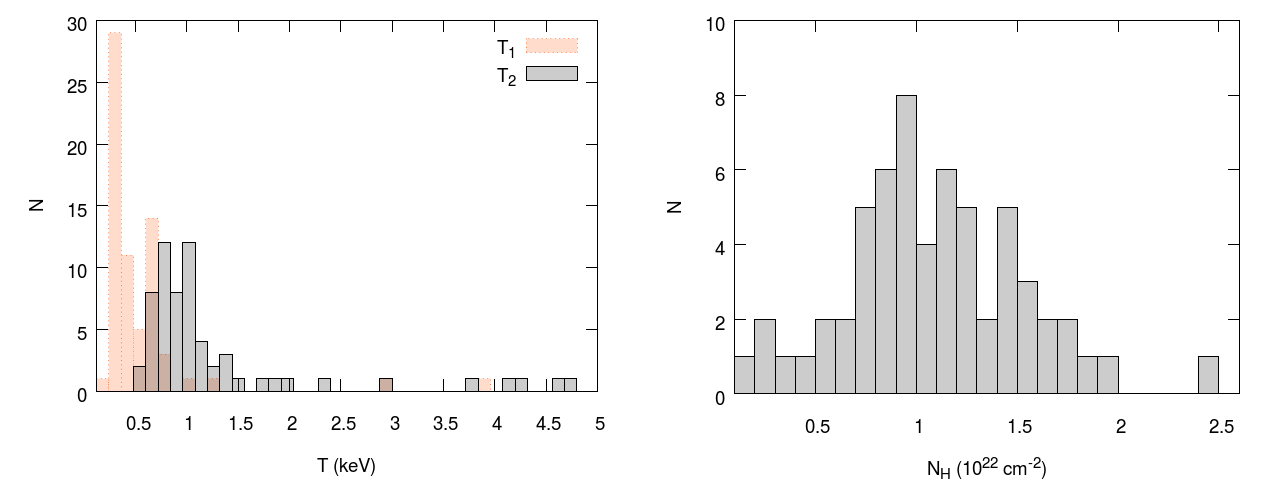}
\caption{\textit{Left}: The distribution of \textit{mekal} model temperatures, where $T_1$ is the temperature of the external, colder gas component and $T_2$ the temperature of the internal, hotter gas component, for the CGII sample. \textit{Right:} The distribution of absorbing column densities associated to the inner, hotter gas component, for the CGII sample.}
\label{fig:MekalParam}
\end{figure*}

The distribution of the obtained parameters for the full CGII sample can be seen in the histograms presented in Fig. \ref{fig:MekalParam}. The temperature associated to the colder \textit{mekal} component ($T_1$) used to model the $0.5-2$ keV emission of each source presents a more narrow distribution than that associated to the hotter component ($T_2$). The distribution of $T_1$ has a median value of $0.38 \pm 0.03$ keV, and an interquartile range of $0.32 - 0.63$ keV. The distribution of $T_2$ has, as expected, a higher median value, of $0.97\pm0.18$. The interquartile range is  $0.77-1.2$ keV range, with a long tail extending up to $T_2 \sim 4.5$ keV. Note that even though the two distributions overlap (i.e. some $T_1$ values are larger than some $T_2$ values), for each single source $T_2>T_1$. 

Figure \ref{fig:MekalParam} also shows the distribution of column densities, $N_{\rm H}$, absorbing the hottest \textit{mekal} component. The median of the distribution is $(1.1\pm0.2)\times 10^{22}$ cm$^{-2}$, with an interquartile range of $(0.8-1.4)\times 10^{22}$ cm$^{-2}$.

A few sources, named in Sect. \ref{HBfit}, have been modeled in the full $0.4-7$ keV band with a single power-law; and therefore no values for $T_1$, $T_2$ and $N_{\rm H}$ have been derived for them. Also, NGC 6285 and IC 2810 (SE) have been fitted with a single \textit{mekal} component in the $0.4-2$ keV range, and NGC 7752/3 (NE) and ESO 440-IG058 (N) have been fitted with a single \textit{mekal} component in the full $2-7$ keV range. These sources have not been included in the histograms shown in Fig. \ref{fig:MekalParam}, or the averages previously mentioned.

\subsubsection{Hard band ($2-7$ keV) fitting}\label{HBfit}

In the hard X-ray band, where the emission from hot interstellar gas and young stars significantly decreases, X-ray binaries dominate the emission in the absence of an AGN. Their emission can be fitted by a simple power-law. The photon index, $\Gamma$, is the slope of a power-law model that describes a photon spectrum, defined as ${\rm d}N/{\rm d}E \propto E^{-\Gamma}$ photons cm$^{-2}$ s$^{-1}$ keV$^{-1}$. Derived values for $\Gamma$ for all sources in the CGII sample can be found in Table 7. 

This fit is only performed for galaxies that have a minimum of 20 cts in the $2-7$ keV range. Galaxies with a lower count number have been fitted while imposing a fixed power-law photon index of 2 (average spectral slope found for a sample of local starburst galaxies, \citealp{RanCom2003}), leaving only the model normalization as a free parameter. This limit is set in order to obtain meaningful constraints for the spectral slope. It is lower than the one fixed for the CGI sample (50 cts) since many of the sources in the current sample are much fainter, as expected given their lower infrared luminosities.

A few objects within the sample show a clear, steep increase of flux at energies $\geq 3-4$ keV (see e.g. MCG$-$03$-$34$-$064 in Fig. \ref{fig:XraySpectra}), which is a sign of the presence of an absorbed AGN \citep[see e.g.][]{TurMil2009}. In such cases, which all have a count number higher than 20 cts, we fit an absorbed power-law imposing a fixed photon index of 1.8 (a typically expected value for the photon index of an AGN, see e.g. \citealp{NanPou1994}). This leaves the absorbing column density, $N_{\rm H}$, as a free parameter. This model is preferred if the fit yields values $N_{\rm H} \gtrsim 10^{23}$ cm$^{-2}$, and is statistically better than a simple power-law fit. In such cases, we classify the source as an absorbed AGN. 

A few sources in the sample (NGC 5331 (N), IRAS F16399$-$0937 (S), ESO 550$-$IG025 (S), MCG+12$-$02$-$001 (W), CGCG 049$-$057, UGC 02238, NGC 4418 and ESO 343$-$IG013 (N) and (S)) are clearly best-fitted with a single power-law in the full $0.4-7$ keV band, and the $\Gamma$ parameter shown in Table 7 corresponds to that fit. 

\subsubsection{Iron K$\alpha$ lines}

The Fe K$\alpha$ line is a frequently used, reliable diagnostic of heavily obscured AGN. As already mentioned in Sect. \ref{AGNselection}, we use it as one of our X-ray AGN selection criteria. The cold iron line seen in some of the CGII sources is fitted with a Gaussian model centered at 6.4 keV. Six sources in the CGII sample have such a line fitted with a significance above $2 \sigma$, which is the threshold we set to consider a detection. 

A more conservative and frequently used threshold to consider a line as detected in the data is a 3$\sigma$ significance. Imposing this more restrictive criterion, only NGC 1068 and MCG$-$01$-$34$-$064 would have detected Fe K$\alpha$ lines in the sample. The threshold is lowered because of the low signal-to-noise ratio for all sources in the CGII sample. However, note that lowering it to $2\sigma$ does not change the fraction of selected AGN within the sample, as all sources with a line detection also meet other selection criteria. We still consider the presence of this line at $2\sigma$ to be relevant information, which can give support to other AGN determinations, and therefore include it in the analysis. 

Parameters of the fit for these six sources are shown in Table \ref{table:IronLines}, including the line energy, intensity and equivalent width with respect to the continuum. The detection of these lines has been previously reported based on other X-ray observations \citep{KoyIno1989, BanKle1990, MazIwa2012, GilCom1999, LevKro2002,RicUed2014,RisMin2009}.

\begin{table}
\centering
\tabcolsep=0.10cm
\footnotesize
\caption{Fe K$\alpha$ line fits.}
\label{table:IronLines}
\begin{tabular}{rlccc}
\hline \hline
No. & Galaxy & \begin{tabular}{c} E \\  (keV) \end{tabular} & \begin{tabular}{c} I \\  ($10^{-6}$ s$^{-1}$ cm$^{-2}$)\end{tabular} & \begin{tabular}{c} EW \\  (keV) \end{tabular} \\  \hline

79 & NGC 5256 (SW) & 6.44$_{-0.05}^{+0.04}$ & 1.4$_{-0.6}^{+0.9}$ & 4.0$_{-1.8}^{+2.6}$ \\
100 & NGC 7130 & 6.42$_{-0.04}^{+0.03}$ & 3.3$_{-1.3}^{+1.6}$ & 0.8$_{-0.3}^{+0.4}$ \\
121 & NGC 1068 & 6.43$_{-0.04}^{+0.07}$ & 32.2$_{-9.0}^{+12.6}$ & 0.9$_{-0.3}^{+0.3}$ \\
127 & MCG$-$03$-$34$-$064 & 6.43$_{-0.08}^{+0.10}$ & 77.1$_{-25.4}^{+32.8}$ & 0.7$_{-0.2}^{+0.3}$ \\
142 & NGC 5135 & 6.41$_{-0.03}^{+0.03}$ & 7.2$_{-2.5}^{+2.9}$ & 1.1$_{-0.4}^{+0.4}$ \\
198 & NGC 1365 & 6.35$_{-0.04}^{+0.03}$ & 40.3$_{-14.0}^{+15.0}$ & 0.14$_{-0.04}^{+0.07}$ \\

\hline
\end{tabular}
 \begin{tablenotes}
\item \textbf{Notes:} Iron K$\alpha$ Line detections with a significance of $2 \sigma$ or higher. The line centroid energy is measured in the rest frame. Errors reported correspond to $1\sigma$ for one parameter of interest, leaving 5 parameters free.
  \end{tablenotes}
\end{table}

\subsection{X-ray luminosities and correlation with $L_{\rm IR}$}\label{XvsIR}

\begin{figure*}
\centering
\includegraphics[width=\textwidth,keepaspectratio]{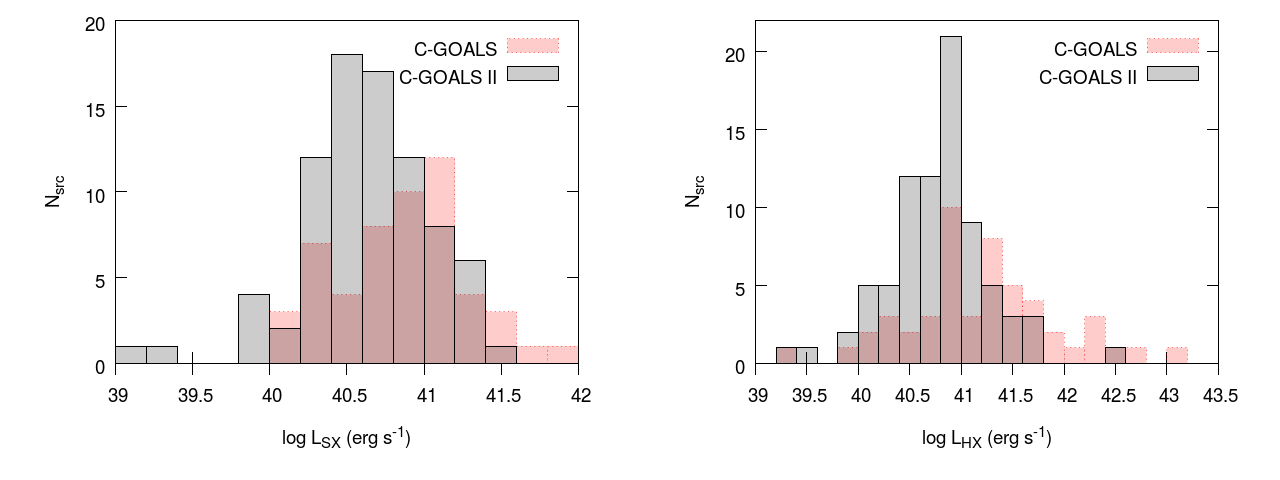}
\caption{The distributions of soft band X-ray luminosity, $0.5-2$ keV (left), and hard band X-ray luminosity, $2-10$ keV (right), for the individual galaxies of CGI and CGII.}
\label{fig:XrayLDist}
\end{figure*}

\begin{figure*}
\centering
\includegraphics[width=\textwidth,keepaspectratio]{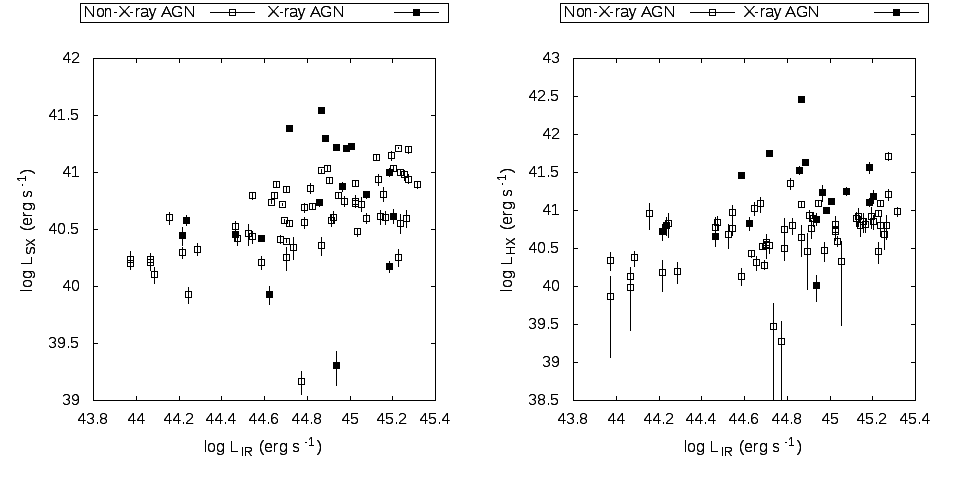}
\caption{Plots of soft (left, $0.5-2$ keV) and hard (right, $2-10$ keV) X-ray luminosity versus infrared luminosity, where the X-ray luminosity is corrected only for Galactic absorption. X-ray selected AGN, shown in Table \ref{table:AGNselection}, are shown in black. When multiple objects are present in a source, their infrared luminosity is divided as shown in Table \ref{table:IRfrac}.}
\label{fig:XvsIR}
\end{figure*}

\begin{figure*}
\centering
\includegraphics[width=\textwidth,keepaspectratio]{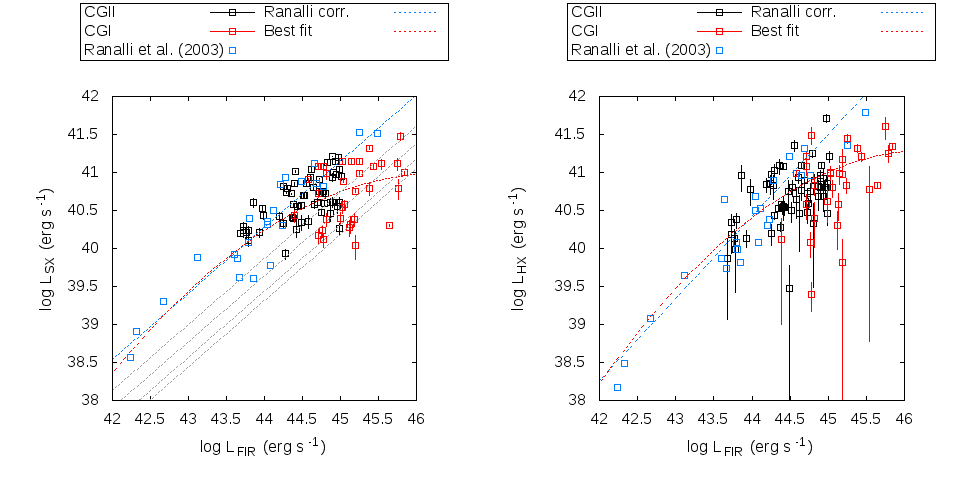}
\caption{Plots of soft (left, $0.5-2$ keV) and hard (right, $2-10$ keV) X-ray luminosity versus FIR luminosity derived as in Eq. \ref{eq:LFIR}, where the X-ray luminosity is corrected only for Galactic absorption. Data used by \citet{RanCom2003}, along with their derived correlation, are shown in blue. CGI and CGII data (for galaxies without an AGN) are plotted in red and black squares respectively. When multiple objects are present in a source, their infrared luminosity is divided as shown in Table \ref{table:IRfrac}. All sources containing AGN, as listed in Table \ref{table:AGNselection} or classified as AGN by \citet{IwaSan2011} have been removed both from the plot and from the fits. The red, dashed line shows our best quadratic fit for the C-GOALS + \citet{RanCom2003} data. Grey, dashed lines (left) show theoretical lines of obscuration for $N_{\rm H}=0.5,1.0,2.0, 5.0 \times 10^{22}$ cm$^{22}$, as described in Sect. \ref{XvsIR}.}
\label{fig:XvsFIR}
\end{figure*}

Figure \ref{fig:XrayLDist} shows the distribution of derived luminosities in soft and hard band, presented in Table 3, compared with that of those obtained for the CGI sample of \citet{IwaSan2011}. For CGII, the distributions peak at $\log(L_{\rm SX}) \sim 40.6$ erg s$^{-1}$ and $\log(L_{\rm HX}) \sim 40.9$ erg s$^{-1}$, which is slightly lower than the peak of both bands for CGI sample, at $\log(L_{\rm X}) \sim 41.1$ erg s$^{-1}$. The median logarithmic values for the soft and hard band luminosities are listed in Table \ref{table:Stats}. CGII has lower X-ray luminosity values than CGI, as expected, reflecting a correlation between Infrared and X-ray luminosity, seen in both the CGI and CGII samples (see Figs. \ref{fig:XvsIR} and \ref{fig:XvsFIR}).

\begin{table}
\centering
\footnotesize
\caption{Statistical X-ray properties of the sample.}
\label{table:Stats}
\begin{tabular}{lccc}
\hline \hline
 & CGI & CGII & C-GOALS \\  \hline
log$(L_{\rm SX})$ & $40.9\pm0.3$ & $40.6\pm0.2$ & $40.8\pm0.4$ \\
log$(L_{\rm HX})$  & $41.2\pm1.7$ & $40.8\pm0.8$ & $40.9\pm0.6$ \\
log$(L_{\rm SX}/L_{\rm IR})$ & $-4.53 \pm 0.34$ & $-4.16 \pm 0.42$ & $-4.26 \pm 0.48$ \\
log$(L_{\rm HX}/L_{\rm IR})$ & $-4.40 \pm 0.63$ & $-4.04 \pm 0.48$ & $-4.17 \pm 0.59$\\
log$(L_{\rm SX}/L_{\rm IR})*$ & $-4.6 \pm 0.1$ & $ -4.18 \pm 0.35$ &  $-4.26 \pm 0.45$\\
log$(L_{\rm HX}/L_{\rm IR})*$ & $-4.5 \pm 0.1$ & $ -4.18 \pm 0.37 $ & $-4.23 \pm 0.54$\\
log$(L_{\rm SX}/L_{\rm FIR})*$ & $-4.44 \pm 0.45$ & $ -3.82 \pm 0.32$ &  $-3.96 \pm 0.47$\\
log$(L_{\rm HX}/L_{\rm FIR})*$ & $-4.22 \pm 0.52$ & $ -3.81 \pm 0.45 $ & $-3.96 \pm 0.52$\\
\hline
\end{tabular}
 \begin{tablenotes}
\item \textbf{Notes:} Median values of the distribution of soft band and hard X-ray luminosities; and the distributions of the ratios of X-ray to IR luminosities for the C-GOALS I, C-GOALS II and full C-GOALS sample. (*) AGN removed from the sample.
  \end{tablenotes}
\end{table}

The origin of this correlation is in the presence of star formation in the galaxies. Far-infrared luminosity measurements detect the energy absorbed by the dust of the interstellar medium from young, bright stars; and thus are a good estimator of the total Star Formation Rate (SFR) \citep[e.g.][]{Ken1998}. In galaxies with a considerable amount of star formation, such as starburst galaxies, emission in other wavelengths can also be related to young and massive stars, such as X-ray luminosity (e.g. X-ray binaries emission, supernova remnants (SNRs)). Therefore, it has been suggested that if a good correlation between X-ray luminosity and IR luminosity exists in galaxies, the SFR can be directly inferred from the X-ray luminosity. Compatible correlations have been found in previous works for local star-forming galaxies with infrared luminosities lower than those of LIRGS \citep[e.g.][]{RanCom2003, GriGil2003, MinGil2014}. 

Figure \ref{fig:XvsIR} shows the X-ray luminosity as a function of the \textit{IRAS} infrared luminosity. The data show a moderate correlation, with a typical spread of more than one order of magnitude when only considering sources without detected X-ray AGN presence (open squares). Sources which contain X-ray AGN typically lay above the trend, adding scatter to the correlation. Sources with multiple components are separated into their respective contributions, and plotted separately, as it has been shown that when they are plotted summed into one single source, the correlation becomes less clear \citep{IwaSan2009}. Their total ($8-1000$) $\mu$m \textit{IRAS} luminosity, as in Table \ref{table:sample}, is separated into the different contributions using the percentages indicated in Table \ref{table:IRfrac}.

As was already seen in the CGI galaxies, X-ray selected AGN tend to be more luminous in X-rays than the rest of the sample, especially in the hard band. However, the values for $\log(L_{\rm X}/L_{\rm IR})$ are higher for the CGII sample than for CGI, as listed in Table \ref{table:Stats}. This result means that while our sample is less bright both in X-ray and in infrared when compared to the CGI sample, we find a higher relative X-ray to infrared luminosity. Removing X-ray selected AGN from both samples gives lower ratios, which we also list. 

Comparing this average values to the $\log(L_{\rm X}/L_{\rm IR}) \sim - 3.7$ found by \citep{RanCom2003} for local, star-forming galaxies with lower SFR, we find a large discrepancy. However, their IR luminosity does not include the 1$-$40$\mu$m range, which may contribute a non-negligible amount of power, in particular for IR warm, AGN dominated systems. Therefore direct comparison needs to introduce a correction. Furthermore, only at radio and FIR wavelengths are the most intense starbursts transparent \citep[e.g.][]{Con1992}, so that from their detected flux SFR can be accurately estimated. 

Infrared luminosities derived by \citet{RanCom2003}, hereafter $L_{\rm FIR}$, follow the expression
\begin{equation}\label{eq:LFIR}
L_{\rm FIR}=1.26 \times 10^{-11} (2.58\ S_{60\mu} + S_{100\mu})\  {\rm erg \ s}^{-1} {\rm cm}^{-2},
\end{equation}
from \citet{HelSoi1985}. 

We use this expression to derive $L_{\rm FIR}$ for all non-AGN objects in the CGI and CGII samples, again accounting for the contribution of multiple components following Table \ref{table:IRfrac}. As listed in Table \ref{table:Stats}, $\log(L_{\rm X}/L_{\rm FIR})$ is similar to -3.7 for the CGII objects, but not for the galaxies in CGI. A direct comparison between the distribution followed by objects analyzed by \citet{RanCom2003}, as well as their derived correlation, and GOALS objects is shown in Fig. \ref{fig:XvsFIR}. 

The best fit correlations derived by \citet{RanCom2003} are:
\begin{equation}\label{eq:SoftRanCorr}
log(L_{\rm SX})=(0.87 \pm 0.08) log(F_{\rm FIR})+2.0 \pm 3.7
\end{equation}
\begin{equation}\label{eq:HardRanCorr}
log(L_{\rm HX})=(1.08 \pm 0.09) log(F_{\rm FIR})-7.1 \pm 4.2
\end{equation}
which correspond to the blue dotted lines plotted in Fig. \ref{fig:XvsFIR}. 

Galaxies in the GOALS sample with $L_{\rm FIR} \lesssim 8 \times 10^{10}\ L_{\odot}$ follow Eqs. \ref{eq:SoftRanCorr} and \ref{eq:HardRanCorr}, but those with higher $L_{\rm IR}$ have systematically lower X-ray luminosity than predicted. 

This behavior suggests that a better fit would be obtained with a quadratic relation in log scale. Using the least-squares method, we obtain a best fit for the C-GOALS + \citet{RanCom2003} data:
\begin{equation}\label{eq:SoftBrokenPW}
      \log(L_{\rm SX})=(-0.17\pm0.04) x^2+(15\pm2) x + (-263\pm8)
\end{equation}
\begin{equation}\label{eq:HardBrokenPW}
      \log(L_{\rm HX})=(-0.17\pm0.06) x^2+(15.2\pm2) x + (-169\pm10)
\end{equation}
where $x=\log(L_{\rm FIR})$. This fit is plotted as a red dashed line in Fig. \ref{fig:XvsFIR}. Below FIR luminosities of $\sim 3\times 10^{44}$ erg s$^{-1}$ (i.e. $\sim 8 \times 10^{10}$ L$_\odot$) the quadratic fit overlaps with the linear correlation. Above this value, X-ray luminosity increases far more slowly with far-infrared luminosity. This effect is larger in soft X-rays. Note that fitting a single power-law to the full data sample we do not recover a relation that is compatible, within the errors, with Eq. \ref{eq:SoftRanCorr}. A power-law fit also yields a larger $\chi ^{2}$ value than the fits given by Eqs. \ref{eq:SoftBrokenPW} and \ref{eq:HardBrokenPW}.

As soft X-rays are easily absorbed by moderate column densities, we show the effect that obscuration could have on the \citet{RanCom2003} correlation. In order to do so, we take an average spectrum that is characteristic of the galaxies within our sample: a double component \textit{mekal} model with temperatures $T_1=0.38$ and $T_2=0.97$, the median values derived from our soft X-ray analysis. According to our model, the inner component of $T=T_2$ can have considerable absorption, and fitting yields values in the range $N_{\rm H}0.1-2.5 \times 10^{22}$ cm$^{-2}$. We assume different column densities, $N_{\rm H}=0.5,1.0,2.0, 5.0 \times 10^{22}$ cm$^{22}$, and absorb the hotter component. We use the model to calculate the decrease in flux caused by the different column densities, and considering that the linear correlation derived by \citet{RanCom2003} has no intrinsic absorption, we plot the ``absorbed'' equivalent correlations in Fig. \ref{fig:XvsFIR}. $N_{\rm H} = 5.0 \times 10^{22}$ cm$^{-2}$ actually absorbs more than 99\% of the emission of the inner component in the $0.5-2$ keV range, meaning larger column densities would result in no change in the emission i.e. only the emission of the outer, unabsorbed component remains.  

\subsection{Radial profiles}

\begin{table*}
\setlength{\tabcolsep}{0.05cm}
\centering
\footnotesize
\caption{Half-light radius}
\label{table:HalfRadius}
\begin{tabular}{rlcc|rlcc}
\hline \hline

No. & Galaxy & R$_{\rm H}^{\rm SX}$ ('') & R$_{\rm H}^{\rm SX}$ (kpc) & No. & Galaxy &  R$_{\rm H}^{\rm SX}$ ('') & R$_{\rm H}^{\rm SX}$ (kpc) \\ \hline
45	&	 UGC 08387 	&	 2.4 	&	 1.3 	&	104	&	 NGC 7771 	&	 3.0 	&	 0.9 \\ 
47	&	 CGCG 436$-$030 	&	 2.6 	&	 1.7 	&	104	&	 NGC 7770 	&	 3.7 	&	 1.1 \\ 
49	&	 NGC 0695 	&	 3.7 	&	 2.5 	&	105	&	 NGC 7592 (E) 	&	 3.1 	&	 1.6 \\ 
50	&	 CGCG 043$-$099 	&	 1.2 	&	 1.0 	&	105	&	 NGC 7592 (W) 	&	 0.9 	&	 0.5 \\ 
51	&	 MCG+07$-$23$-$019 	&	 4.3 	&	 3.3 	&	106	&	 NGC 6286 	&	 7.2 	&	 3.0 \\ 
52	&	 NGC 6670 (E) 	&	 3.0 	&	 1.9 	&	106	&	 NGC 6285 	&	 1.7 	&	 0.7 \\ 
52	&	 NGC 6670 (W) 	&	 4.0 	&	 2.5 	&	107	&	 NGC 4922 (N) 	&	 <0.5 	&	 <0.3 \\ 
53	&	 UGC 02369 (S) 	&	 1.5 	&	 1.0 	&	110	&	 NGC 3110 	&	 9.0 	&	 3.5 \\ 
54	&	 NGC 1614 	&	 1.4 	&	 0.5 	&	114	&	 NGC 0232 	&	 1.5 	&	 0.7 \\ 
56	&	 NGC 5331 (N) 	&	 2.1 	&	 1.6 	&	117	&	 MCG+08$-$18$-$013(E) 	&	 2.3 	&	 1.3 \\ 
56	&	 NGC 5331 (S) 	&	 3.0 	&	 2.3 	&	120	&	 CGCG 049$-$057 	&	 1.1 	&	 0.3 \\ 
57	&	 IRAS F06076$-$2139(N) 	&	 0.7 	&	 0.6 	&	121	&	 NGC 1068 	&	 7.8 	&	 0.6 \\ 
57	&	 IRAS F06076$-$2139(S) 	&	 $-$ 	&	 $-$ 	&	123	&	 UGC 02238 	&	 4.0 	&	 1.8 \\ 
60	&	 IC 2810(NW) 	&	 1.2 	&	 0.9 	&	127	&	 MCG$-$03$-$34$-$064 	&	 <0.5 	&	 <0.2 \\ 
60	&	 IC 2810 (SE) 	&	 0.7 	&	 0.5 	&	134	&	 ESO 350$-$IG038 	&	 2.4 	&	 1.0 \\ 
63	&	 IRAS 18090+0130 (E) 	&	 2.4 	&	 1.6 	&	136	&	 MCG $-$01$-$60$-$022 	&	 4.6 	&	 2.2 \\ 
63	&	 IRAS 18090+0130(W) 	&	 <0.5 	&	 <0.3 	&	141	&	 IC 0564 	&	 8.0 	&	 3.6 \\ 
64	&	 III Zw 035 (S) 	&	 1.2 	&	 0.7 	&	141	&	 IC 0563 	&	 8.1 	&	 3.6 \\ 
65	&	 NGC 3256 	&	 7.7 	&	 1.5 	&	142	&	 NGC 5135 	&	 2.1 	&	 0.6 \\ 
67	&	 IRAS F16399$-$0937(N) 	&	 3.5 	&	 2.2 	&	144	&	 IC 0860 	&	 6.0 	&	 1.7 \\ 
67	&	 IRAS F16399$-$0937(S) 	&	 5.3 	&	 3.3 	&	147	&	 IC 5179 	&	 14.6 	&	 3.6 \\ 
68	&	 IRAS F16164$-$0746 	&	 1.5 	&	 0.9 	&	148	&	 CGCG 465$-$012 	&	 4.6 	&	 2.1 \\ 
69	&	 IC 4687 	&	 3.5 	&	 1.4 	&	157	&	 MCG$-$02$-$33$-$099 	&	 1.3 	&	 0.5 \\ 
69	&	 IC 4686 	&	 <0.5 	&	 <0.2 	&	157	&	 MCG$-$02$-$33$-$098 	&	 1.2 	&	 0.5 \\ 
69	&	 IC 4689 	&	 3.5 	&	 1.4 	&	163	&	 NGC 4418 	&	 2.7 	&	 0.5 \\ 
71	&	 NGC 2623 	&	 1.0 	&	 0.4 	&	169	&	 ESO 343$-$IG013 (N) 	&	 1.2 	&	 0.5 \\ 
72	&	 IC 5298 	&	 <0.5 	&	 <0.3 	&	169	&	 ESO 343$-$IG013 (S) 	&	 5.7 	&	 2.4 \\ 
73	&	 IRAS 20351+2521 	&	 4.0 	&	 2.9 	&	170	&	 NGC 2146 	&	 15.5 	&	 1.3 \\ 
75	&	 NGC 6090 (NE) 	&	 1.6 	&	 1.1 	&	174	&	 NGC 5653 	&	 9.0 	&	 2.6 \\ 
75	&	 NGC6090 (SW) 	&	 1.4 	&	 0.9 	&	178	&	 NGC 4194 	&	 4.2 	&	 0.9 \\ 
79	&	 NGC 5256 (NE) 	&	 0.8 	&	 0.5 	&	179	&	 NGC 7591 	&	 0.8 	&	 0.3 \\ 
79	&	 NGC 5256 (SW) 	&	 2.2 	&	 1.4 	&	182	&	 NGC 0023 	&	 3.2 	&	 1.0 \\ 
80	&	 IRAS F03359+1523(E) 	&	 1.9 	&	 1.4 	&	188	&	 NGC 7552 	&	 3.2 	&	 0.4 \\ 
81	&	 ESO 550$-$IG025 (N) 	&	 2.2 	&	 1.4 	&	191	&	 ESO 420$-$G013 	&	 1.0 	&	 0.2 \\ 
81	&	 ESO 550$-$IG025 (S) 	&	 1.0 	&	 0.7 	&	194	&	 ESO 432$-$IG006 (NE) 	&	 2.3 	&	 0.8 \\ 
82	&	 NGC 0034 	&	 1.3 	&	 0.5 	&	194	&	 ESO 432$-$IG006 (SW) 	&	 2.5 	&	 0.9 \\ 
83	&	 MCG+12$-$02$-$001 (E) 	&	 2.1 	&	 0.7 	&	195	&	 NGC 1961 	&	 8.4 	&	 2.4 \\ 
83	&	 MCG+12$-$02$-$001 (W) 	&	 2.3 	&	 0.8 	&	196	&	 NGC 7752/3 (NE) 	&	 0.8 	&	 0.3 \\ 
85	&	 IRAS F17138$-$1017 	&	 2.8 	&	 1.1 	&	196	&	 NGC 7752/3 (SW) 	&	 5.6 	&	 2.0 \\ 
95	&	 ESO 440$-$IG058 (N) 	&	 0.6 	&	 0.3 	&	198	&	 NGC 1365 	&	 7.0 	&	 0.6 \\ 
95	&	 ESO 440$-$IG058 (S) 	&	 3.8 	&	 2.1 	&	199	&	 NGC 3221 	&	 13.2 	&	 4.2 \\ 
100	&	 NGC 7130 	&	 <0.5 	&	 <0.2 	&	201	&	 NGC 0838 	&	 5.1 	&	 1.3 \\ 

\hline
\end{tabular}
 \begin{tablenotes}
\item {\textbf{Notes:} Radius up to which half of the source counts in the $0.5-2$ keV band are emitted, for the 84 galaxies analyzed within the CGII sample.} 
  \end{tablenotes}
\end{table*}

\begin{figure}
\centering
\includegraphics[width=\columnwidth,keepaspectratio]{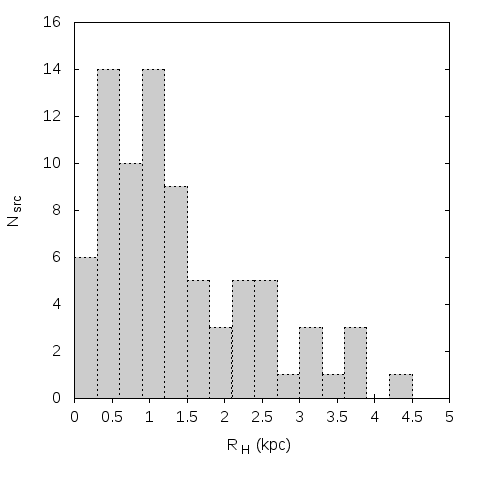}
\caption{Histogram of half-light radii for the sources in the CGII sample.}
\label{fig:RHhist}
\end{figure}

\begin{figure}
\centering
\includegraphics[width=\columnwidth,keepaspectratio]{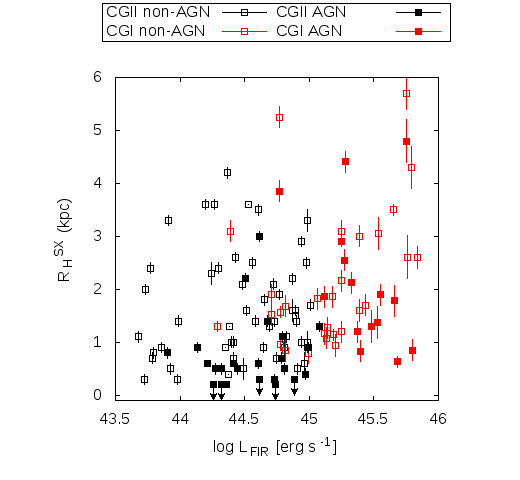}
\caption{Half-light radius as a function of  far-infrared luminosity (as in Eq. \ref{eq:LFIR}) for galaxies within the CGII (black) and CGI (red) sample. Sources without an AGN are plotted in open squares, while sources with an AGN are plotted in filled squares.}
\label{fig:RhalfvsLIR}
\end{figure}

Radial profiles for all sources (except IRAS F06076$-$2139 (S), for which not enough counts are detected in the \textit{Chandra} data) are characterized in two different ways. In the first method, we compute the soft X-ray half-light radius (R$_{\rm H}^{\rm SX}$) for the $0.5-2$ keV band as the radius within which half of the total number of counts is emitted. In the second method, surface brightness profiles are computed and provided in Appendix \ref{MWimages}, also shown in Fig. \ref{fig:NGC2146} for NGC 2146. These profiles are computed in the soft $0.5-2$ keV band, shown in open squares, and the hard $2-7$ keV band, in filled squares. 

Profiles are centered using the hard X-ray peak that corresponds to the nucleus of the galaxy, which typically corresponds to the near-infrared nucleus. In cases in which there is no clear central emission in the ($2-7$) keV band, the profiles are centered using infrared images. For all galaxies that have radial profiles centered using infrared images, a comment has been added in Appendix \ref{NotesIO}. 

The values for R$_{\rm H}^{\rm SX}$, both in arcsec and in kpc, are provided in Table \ref{table:HalfRadius}. While this value can give an idea of the size of the central, more intensely emitting region of a galaxy, note that for sources without an extensive diffuse emission (e.g. NGC 3221, which is mostly composed of point-sources), it might not have a physical meaning. Other sources show non-axisymmetric morphology, most likely associated to extended starburst winds (e.g. UGC 08387, NGC 6286 (SE), NGC 2146, NGC 4194, NGC 1365 and NGC 0838). See Appendix \ref{MWimages} for detailed images of the morphology of the X-ray emission in all sources

Due to the pixel size of the \textit{Chandra} CCD, the smallest radius within which counts can be computed is limited to $0.5$''. Very compact sources can have more than half of their detected counts within this region, making the estimation of R$_{\rm H}^{\rm SX}$ impossible. This is the case for six of our sources, for which an upper-limit is provided. It would also be the case for the vast majority of sources when considering the hard band emission, which is the reason we do not provide values of R$_{\rm H}^{\rm HX}$.

Figure \ref{fig:RHhist} shows a histogram of all half-light radii presented in Table \ref{table:HalfRadius}. Note that sources with only upper limits derived are included, as they all fall below R$_{\rm H}^{\rm SX}=0.3$ kpc, which is the bin size. The distribution of R$_{\rm H}^{\rm SX}$ has a median of $1.0\pm0.1$ kpc, with an interquartile range of $0.5-1.9$ kpc. This shows that most sources within CGII have a compact X-ray distribution, with half of the emission being generated within the inner $\sim 1$ kpc. VLA 33 GHz studies of the 22 brightest U/LIRGs in the C-GOALS sample find half-light radii below $1.7$ kpc for all sources, meaning that the emission is also compact in other wavelengths \citep{BarLer2017}.

Note that all values of R$_{\rm H}^{\rm SX}$ are actually upper limits, as any amount of obscuration in the sources (likely important in most, as seen in Fig. \ref{fig:XvsFIR}), which is concentrated in the inner regions, will result in an apparent decrease of compactness. 

Figure \ref{fig:RhalfvsLIR} shows the distribution of half-light radii as a function of the far-infrared luminosity. The size of the most intensely X-ray emitting region shows no clear correlation with the overall infrared luminosity, though CGI sources tend to have higher X-ray radii. Within the CGII sample, sources that contain an AGN, as listed in Table \ref{table:AGNselection}, plotted in filled squares, tend to be compact.  (83$\pm$7)\% of them have R$_{\rm H}^{\rm SX} \leq 1$ kpc. This is in agreement with previous results from a study of the extended MIR emission in GOALS using Spitzer/IRS spectroscopy, where it was found that progressively more AGN-dominated galaxies tend to show more compact MIR emission \citep{DiaCha2010}. 

\begin{figure}
\centering
\includegraphics[width=\columnwidth,keepaspectratio]{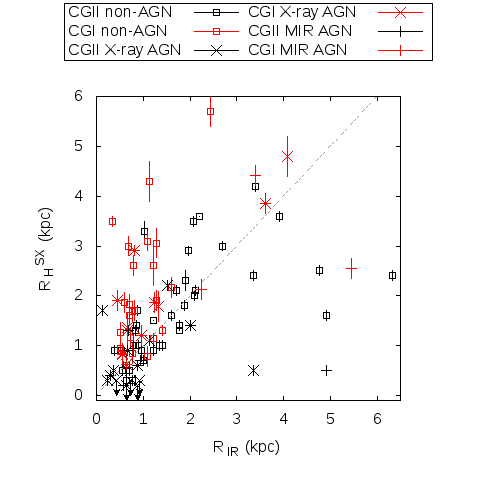}
\caption{Soft X-ray half-light radius as a function of 70 $\mu$m far-infrared radius, taken from \citet{DiaArm2017}. CGI data is in red, and CGII data in black. X-ray selected AGN and Mid-IR selected AGN (as specified in Table \ref{table:AGNselection}) are plotted separately. The dashed line shows $R_{\rm IR}=R_{\rm H}^{\rm SX}$, and is not a fit to the data. 11 systems within the whole C-GOALS sample are resolved into individual galaxies in X-rays but not at 70 $\mu$m, and are thus not plotted.}
\label{fig:RSXvsRIR}
\end{figure}

Figure \ref{fig:RSXvsRIR} shows a comparison between R$_{\rm H}^{\rm SX}$ and the IR radius at 70 $\mu$m taken from \citet{DiaArm2017}, for all C-GOALS sources. Most sources in the sample are placed close to the $R_{\rm IR}=R_{\rm H}^{\rm SX}$ line. CGI sources tend to have larger X-ray half-light radius for a given characteristic IR radius, when compared to CGII sources. 

Outliers with very compact soft X-ray emission and yet a large IR radius are X-ray selected AGN ESO 343$-$IG013 (N) and IR selected AGN NGC 7592 (W). Both sources show clear strong hard X-ray peaks in the nucleus (see images in Appendix \ref{MWimages}). Another extreme outlier is IRAS F12112+0305, with R$_{\rm H}^{\rm SX}=5.7\pm0.3$ kpc and yet much more compact IR emission. 

\subsection{Luminosity Surface Density}

Using the luminosities listed in Table 3 and the R$_{\rm H}^{\rm SX}$ in Table \ref{table:HalfRadius} we derive luminosity surface densities for all sources in the C-GOALS sample, as $\Sigma_{\rm SX} = L_{\rm SX}/\pi (R_{\rm H}^{\rm SX})^2$. $\Sigma_{\rm IR}$ is derived using 70 $\mu$m radii from \citet{DiaArm2017}.

\begin{figure*}
\centering
\includegraphics[width=\textwidth,keepaspectratio]{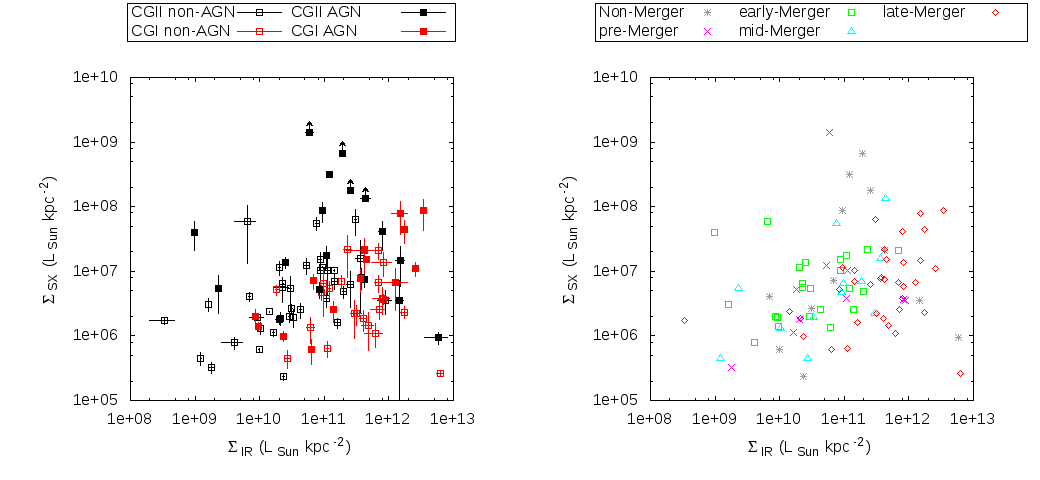}
\caption{Plots of X-ray luminosity surface density versus infrared luminosity surface density. Left: Red data corresponds to CGI sources and black data corresponds to CGII sources. AGN are highlighted as filled symbols. Right: Sources in the full C-GOALS sample are plotted in different symbols according to merger stage, as derived by \citet{StiArm2013}. 11 systems within the whole C-GOALS sample are resolved into individual galaxies in X-rays but not at 70 $\mu$m, and are thus not plotted.}
\label{fig:Sigma}
\end{figure*}

Figure \ref{fig:Sigma} shows $\Sigma_{\rm SX}$ as a function of $\Sigma_{\rm IR}$. X-ray surface density tends to increase with infrared surface density, though the correlation is broader than the one existing between luminosities. The left plot highlights AGN in filled symbols and separates the CGI and CGII samples. CGI sources, brighter in infrared, tend to have lower $\Sigma_{\rm SX}$ for a given $\Sigma_{\rm IR}$. Within a given sample, sources with AGN tend to have a larger $\Sigma_{\rm SX}$, which is to be expected given how they are both brighter in X-rays and also more compact. 
Figure \ref{fig:Sigma} also shows, on the right, the same figure but highlighting merger stage of the sources in the sample. Information on merger stage is taken from \citet{StiArm2013}, derived from visual inspection of IRAC 3.6 $\mu$m (Channel 1) images. Classification is Non-merger, pre-merger (galaxy pairs prior to a first encounter), early-stage merger (post-first-encounter with galaxy disks still symmetric and in tact but with signs of tidal tails), mid-stage merger (showing amorphous disks, tidal tails, and other signs of merger activity) and late-stage merger (two nuclei in a common envelope). Sources in late-stage mergers tend to have higher $\Sigma_{\rm IR}$, but lower $\Sigma_{\rm SX}$ for the same $\Sigma_{\rm IR}$. 
\begin{table}
\centering
\footnotesize
\caption{Merger stage.}
\label{table:MergerStage}
\begin{tabular}{lccc}
\hline \hline
 & CGI & CGII & C-GOALS \\  \hline
Non-merger & 4\% & 18\% & 12\% \\
Pre-merger & 4\% & 14\% & 10\% \\
early-stage merger & 30\% & 31\% & 30\% \\
mid-stage merger & 17\% & 18\% & 18\% \\
late-stage merger & 44\% & 18\% & 27\% \\
\hline
\end{tabular}
 \begin{tablenotes}
\item \textbf{Notes:} Percentage of sources classified as different merger stages by \citet{StiArm2013} for the C-GOALS I, C-GOALS II and full C-GOALS sample.
  \end{tablenotes}
\end{table}

Percentages of merger stages in the sample are listed in Table \ref{table:MergerStage}. Many CGI sources are late mergers, hence their higher infrared luminosity. This can be seen also in Fig. \ref{fig:Sigma}, as CGI and late merger have the same behavior with respect to the rest of the sources in the sample. 

\section{Discussion}\label{discussion} 

\subsection{X-ray to infrared luminosity relation}

The X-ray to infrared luminosity (or SFR) correlation has been studied in numerous previous works \citep[e.g.][]{FabTri1985,FabGio1988,Fab1989,BauAle2002,GriGil2003,RanCom2003,GilGri2004,PerRep2007,MinGil2014}. For soft X-rays, it originates in starburst-wind shock-heated gas. For hard X-rays ($2-10$ keV) the  relation is thought to be originated in High Mass X-ray Binaries (HMXB), end products of star formation. At low star forming rates, i.e. for local starburst galaxies with $L_{\rm IR}\ll 10^{11}$ L$\odot$, low-mass X-ray binaries (LMXBs) can significantly contribute to the X-ray luminosity. Luminosity from LMXBs correlates with galaxy stellar mass ($M_{\rm *}$), and this dependence must be considered, along with the contribution of SFR \citep[e.g.][]{ColHec2004,Gil2004,LehBra2008,LehAle2010}. 

Figure \ref{fig:XvsFIR} shows a comparison between C-GOALS data and the correlation derived by \citet{RanCom2003} for a sample of nearby starbursting galaxies. Works which include the LMXB contribution at low luminosities, show a slight decrease of the slope at the high luminosity end. Therefore, their correlation can be used as a point-of-reference against which to plot the U/LIRG data, though a rigorous comparison would require inclusion of all previously mentioned works, and is beyond the scope of this work. 

It is clear that at higher infrared luminosities the correlation breaks down in an apparent deficit of X-ray flux, more extreme in the $0.5-2$ keV band. This has been observed since the inclusion of the C-GOALS ULIRGs and high-luminosity LIRGs into the mentioned correlations \citep[e.g.][]{IwaSan2009,LehAle2010,IwaSan2011}. The inclusion of our CGII data provides more information on the transition between low-IR-luminosity galaxies and ULIRGs. 

This underluminosity, or X-ray quietness, is explained in many works as an effect of obscuration. U/LIRGs have extremely high concentrations of gas and dust in their inner regions, resulting in compact starbursts. High gas column densities can easily absorb soft X-rays, and in the most extreme cases, even hard X-rays. 

Galaxies in the CGII sample, which are less IR-luminous than those in the CGI sample, are generally found in less-advanced mergers (see Table \ref{table:MergerStage}). The concentrations of gas and dust in the inner regions of the galaxies is larger in the more advanced mergers \citep[e.g.][]{RicBau2017}, implying that the contribution of obscuration is more important at higher IR luminosities. From an IR point of view, \citet{DiaCha2010} observe that late-stage mergers are much more compact, hinting also at larger column densities. 

\subsubsection{Soft X-ray faintness}

As shown in Fig. \ref{fig:XvsFIR}, the obscuring column densities necessary to dim the soft X-ray emission in most of the sources are compatible with those derived from the two-component model, plotted in Fig. \ref{fig:MekalParam} and listed in Table 7. The derived values of $N_{\rm H}$ are lower limits, as any gas phase with higher obscuration contributes less significantly to the X-ray emission, or is even completely absorbed, and therefore cannot be fitted.

We note, that our spectral model is based on the existence of two distinct phases in the galaxy ISM. Emission is likely to come from a complex phase-continuum of gas, and thus individual estimates of properties based on the spectral fitting should be taken with caution \citep[see e.g.][for a discussion]{StrSte2000}. However, the simple two-phase model is the most complex we can fit given our data, and it shows that the column densities can at least explain the data. 

Figure \ref{fig:Sigma} shows lower $\Sigma_{\rm SX}$ for CGI galaxies, which reflects both their X-ray faintness and larger soft X-ray sizes (Fig. \ref{fig:RSXvsRIR}). We define the size of the emission as the half-light radius, meaning
that larger sizes indicate a less compact source. This implies that the faintness is most likely originating in the center of the source. As CGI galaxies are in more advanced merger stages, and should have higher column densities, this is likely to be an effect of obscuration. Another likely contribution to the larger soft X-ray sizes is the strongest starbursts in CGI galaxies, which generate larger soft X-ray nebulae.

We note that the \textit{Chandra} resolution is much better than that of \textit{Spitzer}, which should be taken into account in any direct comparison between characteristics sizes or luminosity surface densities. Higher resolution should imply a tendency toward deriving higher compactness, while Fig. \ref{fig:RSXvsRIR} shows the opposite: X-ray sizes are generally similar, or even larger than IR sizes. However, we do not know what the IR emission would look like at similar resolutions. This difference between the datasets could explain the presence of outliers below the $R_{\rm IR} = R_{\rm H}^{\rm SX}$ line, and add to the dispersion of the data. Future JWST observations would allow for better comparison. 

In conclusion, the soft X-ray faintness, and therefore the quadratic best fit curve given by Eq. \ref{eq:SoftBrokenPW}, can be explained through obscuration, as the necessary column densities are present in the galaxies within the C-GOALS sample.

\subsubsection{Hard X-ray faintness}

Attributing to extinction the observed faintness in hard X-rays, requires much higher gas column densities. While it is clear that more IR-luminous sources (i.e. late-stage mergers) are generally more heavily obscured \citep[e.g.][for a GOALS subsample]{RicBau2017}, the most extreme sources are $\sim 1-2$ dex below the correlation shown in Fig. \ref{fig:XvsFIR}. This implies that between $\sim 90-99$\% of the central starburst region must be covered in medium dense enough to suppress even hard X-rays. To obscure the emission in the $2-8$ keV band in which \textit{Chandra} is sensitive, the necessary column densities would be of the order of $\sim 10^{24}$ cm$^{-2}$. Sources in the sample that are undetected by \textit{NuSTAR} \citep{RicBau2017} would require even higher column densities, of the order of  $\sim 10^{25}$ cm$^{-2}$. 

In order to explain the observed faintness, regions of sizes of the order of the R$_{\rm H}^{\rm SX}$ listed in Table \ref{table:HalfRadius} would need to be covered in the mentioned high $N_{\rm H}$. A column density of $10^{25}$ cm$^{-2}$ could imply $H_2$ masses of the order of $\sim 10^{10}-10^{11}\ M_\odot$ for a nuclear star-forming region of 500 pc of radius. It is unclear if such large gas masses are truly concentrated in the inner regions of ULIRGs in the GOALS sample, and thus if this ``self-absorbed starburst'' scenario is feasible. 

To disentangle the origin of the faintness, \citet{IwaSan2009} stack spectra of non-AGN sources in CGI, recovering a high-ionization Fe K feature. This feature can be explained by the presence of an internally shocked, hot bubble produced by thermalizing the energy of supernovae and stellar winds \citep[e.g.][]{CheCle1985}, which, contrary to the SNe and HMXB emission, could be visible through the obscuring material. With high star formation rates, the luminosity and spectrum with strong Fe $_{\rm XXV}$ line can be reproduced \citep[e.g.][ for Arp 220]{IwaSan2005}. This high ionization line could also originate from low density gas, photoionized by a hidden AGN, \citep[e.g.][]{AntMil1985,KroKal1987} and has been observed as the dominant Fe K feature in some Compton-thick AGNs \citep[e.g.][]{NanIwa2007, BraRee2009}. 

Therefore, another explanation for the X-ray faintness could be the presence of a completely obscured AGN in the nucleus of these galaxies. This AGN would contribute to the infrared emission, while escaping X-ray detection. While the column densities needed to cover the AGN are as high as those needed to self-absorb a starburst, the obscured region would be much smaller. This would imply much smaller masses, easily found in the nuclei of GOALS galaxies. The hidden AGN versus extremely compact starburst scenario has been previously discussed for some of the C-GOALS sources that show the higher X-ray faintness. Cases like Arp 220 \citep[e.g.][]{ScoMur2017,BarAal2018} or NGC 4418 \citep[e.g.][]{CosAal2013,CosSak2015} are compatible with both scenarios. However, it is worth noting that the hidden AGN scenario requires the presence of an AGN with significant IR emission in order to explain the X-ray faintness, which means it is probably unlikely that MIR determinations would systematically fail to pick their signature. 

\citet{DiaArm2017} suggest, from interpretation of Herschel Far Infrared data on the full GOALS sample, that the fraction of young, massive stars per star-forming region in ULIRGs might be higher than expected. This does not imply a change in the initial mass function, but the presence of very young star-forming regions, in which most massive stars still have not disappeared (age less than a few Myr \citealp{InaArm2013}). In such a case, the massive stars can contribute to the infrared emission, but the number of HMXB and SNe associated to the region will be low, as those are end-products of the star formation. This would result in a lower-than-expected X-ray luminosity for a given infrared luminosity. Furthermore, in such a scenario, the winds of very massive stars could generate the hot gas that explains the Fe $_{\rm XXV}$ line, without the need to invoke extreme obscuration over a large population of HMXB. 

In order to truly understand the origin of the X-ray faintness, further observations are needed that provide information on the obscuration within the sources (e.g. ALMA or NuSTAR observations), or on the unobscured star formation rates (e.g. through radio observations). 

\subsection{AGN and double AGN fraction}

In Sect. \ref{AGNselection} we have shown that (38$\pm$6)\% of systems (24/63) within CGII contain an AGN, (31$\pm$5)\% (26/84) of individual galaxies being classified as AGN, according to mid-IR and/or X-ray criteria. This fraction can be compared to the (50$\pm$7)\% of systems, or (38$\pm$7)\% (21/55) analyzed individual galaxies classified by \citet{IwaSan2011} when analyzing the more IR-luminous objects in the CGI sample. This result may hint toward a slight increase of AGN presence with IR luminosity, though the fractions in the two samples are compatible within the statistical errors. Although the increase of AGN fraction as a function of luminosity found here is not statistically significant, it is consistent with previous findings in optical and infrared spectroscopy \citep[e.g.][]{VeiKim1995,KimSan1995,YuaKew2010, StiArm2013}.

Double AGN are detected in two interacting systems, NGC 5256 and ESO 432-IG006, of the 30 multiple systems analyzed here (7$\pm$4)\%. In the C-GOALS sample one double AGN system, NGC 6240, was detected, out of 24 multiple systems analyzed (4$\pm$4)\%.

Theoretical estimates derived from merger simulations performed by \citet{CapDot2017}, which take into account observational effects (e.g. observation angle, distance dimming of X-ray luminosity, obscuration of gas surrounding central BH), conclude that in a sample of major mergers hosting at least one AGN, the fraction of dual AGN should be $\sim 20-30$\%. \citet{KosMus2012} study a sample of 167 nearby ($z<0.05$) X-ray selected AGN, finding a fraction of dual AGN in multiple systems of 19.5\%. When separated into major pairs (mass ratio $\geq$ 0.25) and minor pairs, they find 37.1\% and 4.8\% respectively. Other studies \citep[e.g.][]{EllPat2011, SatSec2017} find a statistical excess of dual AGN that decreases with galaxy separation. Therefore, dual AGN activation is more likely in advanced merger stages. 

Within CGI and CGII, the fraction of double AGN in systems that host at least one AGN is (11$\pm$10)\% (1/9) and (29$\pm$14)\% (2/7) respectively. The fraction found in CGII falls well within the ranges found in the two previously mentioned works, while the dual AGN fraction is CGI is just barely compatible within the errors. Also, CGI galaxies are generally found in more advanced merger stages (see Table \ref{table:MergerStage}), meaning that according to predictions their dual AGN fraction should be closer to the mentioned 37.1\%, and not the lower 20\%.

The lack of dual AGN in the CGI sample could be explained with heavy obscuration, which is expected to be important for these sources, as discussed in the previous section. Compton-thick ($N_{\rm H} > 1.5 \times 10^{24}$ cm$^{-1}$)  AGN may be completely obscured in our \textit{Chandra} data (e.g. Mrk 273, \citealp{IwaU2018}), and their scattered continuum or Fe K$_\alpha$ lines too faint to be detected . Mid-IR criteria can be effective in such cases, even though they may also miss the most heavily buried AGN \citep[e.g.][]{SnyHay2013}. Recent simulations by \citet{BleSny2017} find that much of the AGN lifetime is still undetected with common mid-IR selection criteria, even in the late stages of gas-rich major mergers. This effect is incremented for AGN which don't contribute significantly to the bolometric luminosity, especially when considering that the presence of strong a starburst can help dilute the AGN signature. Fig. \ref{fig:HRvsAGNbol} shows up to 19/32 X-ray selected AGN in C-GOALS that contribute less than 20\% to the bolometric luminosity, most of which are missed by mid-IR selection criteria (see Table \ref{table:AGNselection}). 

Another likely contribution into the low fraction of dual AGN found in CGI is the inability to resolve individual nuclei in a late-stage merger. Many CGI galaxies are found in such a stage. While \textit{Chandra} has a high spatial resolution ($\sim$0.5''), very closely interacting nuclei, with separations of the order of $\lesssim 200-300$ pc, would remain unresolved in our sample. \textit{Spitzer} data, also used in this work for the mid-IR AGN selection, has a much lower resolution, and would not resolve double AGN with even further separations. 

However, as our sample sizes are small and therefore the statistical errors large, we cannot make any strong statements regarding a decreasing trend of double AGN presence with infrared luminosity.

\section{Conclusions}\label{conclusions}

We analyze \textit{Chandra}-ACIS data for a sample of 63 U/LIRGs, composed by 84 individual galaxies (CGII). These galaxies are a low-IR-luminosity subsample of GOALS, a complete flux-limited sample of the 60 $\mu$m selected, bright galaxies in the local universe (z$<$0.08). Arcsec-resolution images, spectra and radial surface brightness profiles are presented. We compare the observations with \textit{Spitzer} and \textit{Herschel} data to contrast their X-ray and infrared properties. We also compare our results to those found by \citet{IwaSan2011} for the high-IR-luminosity subsample of GOALS (CGI). We summarize our main findings. 

\begin{itemize}
\item Objects with AGN signature represent a (31$\pm$5)\% of the CGII sample, compared to the (38$\pm$7)\% reported for the CGI sample. Double AGN are detected in two interacting systems, implying that the fraction of of double AGN in systems that host at least one AGN is (29$\pm$14)\%, in contrast to the (11$\pm$10)\% found for the CGI sample. 

\item 19/32 of the X-ray selected AGN in the full C-GOALS sample (CGI+CGII) are not energetically significant, contributing less than $20\%$ to the bolometric luminosity of the galaxy, according to MIR determinations. 

\item The brightest LIRGs, at $L_{\rm FIR} > 8 \times 10^{10}$ L$_\odot$, show hard X-ray faintness with respect to the luminosities predicted by correlations found for nearby star-forming galaxies. This behavior is accentuated for the CGI ULIRGs. Possible explanations for the sources with most extreme deviations include a self-absorbed starburst, an obscured AGN or the presence of extremely young star-forming regions. 

\item The extended soft X-ray emission shows a spectrum consistent with thermal emission from a two-phase gas, with an inner, hotter and more heavily obscured component, and an outer, colder and unobscured component. 

\item According to our modeling, an obscuration of the inner component in the range of $N_{\rm H} = 1-5 \times 10^{22}$ cm$^{-2}$ can explain the soft X-ray faintness for the vast majority of the sources. 

\item Most sources within CGII have a compact soft X-ray morphology. (50$\pm$8)\% of the sources generate half of the emission within the inner $\sim 1$ kpc. This behavior is accentuated for AGN, with (83$\pm$7)\% of the sources with half-light radius below $\sim 1$ kpc. 

\item CGI sources are, in comparison, less compact, which is most likely an effect of obscuration in the inner regions. 

\item Most sources in CGII have similar soft X-ray and MIR sizes, though there is important dispersion in this relation. 

\end{itemize}

\section*{Acknowledgments}

We thank the anonymous referee for helpful comments and suggestions. We acknowledge support by the Spanish Ministerio de Econom\'{i}a y Competitividad (MINECO/FEDER, UE) under grants AYA2013-47447-C3-1-P and MDM-2014-0369 of ICCUB (Unidad de Excelencia `Mar\'{i}a de Maeztu'). N.T-A. acknowledges support from MINECO through FPU14/04887 grant. T.D.-S. acknowledges support from ALMA-CONICYT project 31130005 and FONDECYT regular project 1151239. G.C.P. acknowledges support from the University of Florida. This work was conducted in part at the Aspen Center for Physics, which is supported by NSF grant PHY-1607611; V.U., G.C.P, D.B.S., A.M.M., T.D-S. and A.S.E. thank the Center for its hospitality during the Astrophysics of Massive Black Holes Mergers workshop in June and July 2018. The scientific results reported in this article are based on observations made by Chandra X-ray Observatory, and has made use of the NASA/IPAC Extragalactic Database (NED) which is operated by the Jet Propulsion Laboratory, California Institute of Technology under contract with NASA. We acknowledge the use of the software packages CIAO and HEASoft.

\bibliographystyle{aa}
\bibliography{Referencies}{}  

\begin{thebibliography}{157}
\expandafter\ifx\csname natexlab\endcsname\relax\def\natexlab#1{#1}\fi

\bibitem[{{Alam} {et~al.}(2015){Alam}, {Albareti}, {Allende Prieto}, {Anders},
  {Anderson}, {Anderton}, {Andrews}, {Armengaud}, {Aubourg}, {Bailey}, \&
  et~al.}]{AlaAlb2015}
{Alam}, S., {Albareti}, F.~D., {Allende Prieto}, C., {et~al.} 2015, \apjs, 219,
  12

\bibitem[{{Alonso-Herrero} {et~al.}(2012){Alonso-Herrero}, {Pereira-Santaella},
  {Rieke}, \& {Rigopoulou}}]{AloPer2012}
{Alonso-Herrero}, A., {Pereira-Santaella}, M., {Rieke}, G.~H., \& {Rigopoulou},
  D. 2012, \apj, 744, 2

\bibitem[{{Antonucci} \& {Miller}(1985)}]{AntMil1985}
{Antonucci}, R.~R.~J. \& {Miller}, J.~S. 1985, \apj, 297, 621

\bibitem[{{Armus} {et~al.}(2009){Armus}, {Mazzarella}, {Evans}, {Surace},
  {Sanders}, {Iwasawa}, {Frayer}, {Howell}, {Chan}, {Petric}, {Vavilkin},
  {Kim}, {Haan}, {Inami}, {Murphy}, {Appleton}, {Barnes}, {Bothun}, {Bridge},
  {Charmandaris}, {Jensen}, {Kewley}, {Lord}, {Madore}, {Marshall},
  {Melbourne}, {Rich}, {Satyapal}, {Schulz}, {Spoon}, {Sturm}, {U}, {Veilleux},
  \& {Xu}}]{ArmMaz2009}
{Armus}, L., {Mazzarella}, J.~M., {Evans}, A.~S., {et~al.} 2009, \pasp, 121,
  559

\bibitem[{{Atek} {et~al.}(2008){Atek}, {Kunth}, {Hayes}, {{\"O}stlin}, \&
  {Mas-Hesse}}]{AteKun2008}
{Atek}, H., {Kunth}, D., {Hayes}, M., {{\"O}stlin}, G., \& {Mas-Hesse}, J.~M.
  2008, \aap, 488, 491

\bibitem[{{Baan} \& {Kl{\"o}ckner}(2006)}]{BaaKlo2006}
{Baan}, W.~A. \& {Kl{\"o}ckner}, H.-R. 2006, \aap, 449, 559

\bibitem[{{Band} {et~al.}(1990){Band}, {Klein}, {Castor}, \&
  {Nash}}]{BanKle1990}
{Band}, D.~L., {Klein}, R.~I., {Castor}, J.~I., \& {Nash}, J.~K. 1990, \apj,
  362, 90

\bibitem[{{Barcos-Mu{\~n}oz} {et~al.}(2018){Barcos-Mu{\~n}oz}, {Aalto},
  {Thompson}, {Sakamoto}, {Mart{\'{\i}}n}, {Leroy}, {Privon}, {Evans}, \&
  {Kepley}}]{BarAal2018}
{Barcos-Mu{\~n}oz}, L., {Aalto}, S., {Thompson}, T.~A., {et~al.} 2018, \apjl,
  853, L28

\bibitem[{{Barcos-Mu{\~n}oz} {et~al.}(2017){Barcos-Mu{\~n}oz}, {Leroy},
  {Evans}, {Condon}, {Privon}, {Thompson}, {Armus}, {D{\'{\i}}az-Santos},
  {Mazzarella}, {Meier}, {Momjian}, {Murphy}, {Ott}, {Sanders}, {Schinnerer},
  {Stierwalt}, {Surace}, \& {Walter}}]{BarLer2017}
{Barcos-Mu{\~n}oz}, L., {Leroy}, A.~K., {Evans}, A.~S., {et~al.} 2017, \apj,
  843, 117

\bibitem[{{Bauer} {et~al.}(2002){Bauer}, {Alexander}, {Brandt},
  {Hornschemeier}, {Vignali}, {Garmire}, \& {Schneider}}]{BauAle2002}
{Bauer}, F.~E., {Alexander}, D.~M., {Brandt}, W.~N., {et~al.} 2002, \aj, 124,
  2351

\bibitem[{{Blackburn}(1995)}]{Bla1995}
{Blackburn}, J.~K. 1995, in Astronomical Society of the Pacific Conference
  Series, Vol.~77, Astronomical Data Analysis Software and Systems IV, ed.
  R.~A. {Shaw}, H.~E. {Payne}, \& J.~J.~E. {Hayes}, 367

\bibitem[{{Blecha} {et~al.}(2017){Blecha}, {Snyder}, {Satyapal}, \&
  {Ellison}}]{BleSny2017}
{Blecha}, L., {Snyder}, G.~F., {Satyapal}, S., \& {Ellison}, S.~L. 2017, ArXiv
  e-prints [\eprint[arXiv]{1711.02094}]

\bibitem[{{Braito} {et~al.}(2009){Braito}, {Reeves}, {Della Ceca}, {Ptak},
  {Risaliti}, \& {Yaqoob}}]{BraRee2009}
{Braito}, V., {Reeves}, J.~N., {Della Ceca}, R., {et~al.} 2009, \aap, 504, 53

\bibitem[{{Brightman} \& {Nandra}(2011)}]{BriNan2011}
{Brightman}, M. \& {Nandra}, K. 2011, \mnras, 413, 1206

\bibitem[{{Capelo} {et~al.}(2017){Capelo}, {Dotti}, {Volonteri}, {Mayer},
  {Bellovary}, \& {Shen}}]{CapDot2017}
{Capelo}, P.~R., {Dotti}, M., {Volonteri}, M., {et~al.} 2017, \mnras, 469, 4437

\bibitem[{{Casey} {et~al.}(2014){Casey}, {Narayanan}, \& {Cooray}}]{CasNar2014}
{Casey}, C.~M., {Narayanan}, D., \& {Cooray}, A. 2014, \physrep, 541, 45

\bibitem[{{Chapman} {et~al.}(1990){Chapman}, {Staveley-Smith}, {Axon}, {Unger},
  {Cohen}, {Pedlar}, \& {Davies}}]{ChaSta1990}
{Chapman}, J.~M., {Staveley-Smith}, L., {Axon}, D.~J., {et~al.} 1990, \mnras,
  244, 281

\bibitem[{{Chevalier} \& {Clegg}(1985)}]{CheCle1985}
{Chevalier}, R.~A. \& {Clegg}, A.~W. 1985, \nat, 317, 44

\bibitem[{{Chu} {et~al.}(2017){Chu}, {Sanders}, {Larson}, {Mazzarella},
  {Howell}, {D{\'{\i}}az-Santos}, {Xu}, {Paladini}, {Schulz}, {Shupe},
  {Appleton}, {Armus}, {Billot}, {Chan}, {Evans}, {Fadda}, {Frayer}, {Haan},
  {Ishida}, {Iwasawa}, {Kim}, {Lord}, {Murphy}, {Petric}, {Privon}, {Surace},
  \& {Treister}}]{ChuSan2017}
{Chu}, J.~K., {Sanders}, D.~B., {Larson}, K.~L., {et~al.} 2017, \apjs, 229, 25

\bibitem[{{Clemens} {et~al.}(2008){Clemens}, {Vega}, {Bressan}, {Granato},
  {Silva}, \& {Panuzzo}}]{CleVeg2008}
{Clemens}, M.~S., {Vega}, O., {Bressan}, A., {et~al.} 2008, \aap, 477, 95

\bibitem[{{Colbert} {et~al.}(2004){Colbert}, {Heckman}, {Ptak}, {Strickland},
  \& {Weaver}}]{ColHec2004}
{Colbert}, E.~J.~M., {Heckman}, T.~M., {Ptak}, A.~F., {Strickland}, D.~K., \&
  {Weaver}, K.~A. 2004, \apj, 602, 231

\bibitem[{{Condon}(1992)}]{Con1992}
{Condon}, J.~J. 1992, \araa, 30, 575

\bibitem[{{Corbett} {et~al.}(2002){Corbett}, {Norris}, {Heisler}, {Dopita},
  {Appleton}, {Struck}, {Murphy}, {Marston}, {Charmandaris}, {Kewley}, \&
  {Zezas}}]{CorNor2002}
{Corbett}, E.~A., {Norris}, R.~P., {Heisler}, C.~A., {et~al.} 2002, \apj, 564,
  650

\bibitem[{{Costagliola} {et~al.}(2013){Costagliola}, {Aalto}, {Sakamoto},
  {Mart{\'{\i}}n}, {Beswick}, {Muller}, \& {Kl{\"o}ckner}}]{CosAal2013}
{Costagliola}, F., {Aalto}, S., {Sakamoto}, K., {et~al.} 2013, \aap, 556, A66

\bibitem[{{Costagliola} {et~al.}(2015){Costagliola}, {Sakamoto}, {Muller},
  {Mart{\'{\i}}n}, {Aalto}, {Harada}, {van der Werf}, {Viti}, {Garcia-Burillo},
  \& {Spaans}}]{CosSak2015}
{Costagliola}, F., {Sakamoto}, K., {Muller}, S., {et~al.} 2015, \aap, 582, A91

\bibitem[{{Del Moro} {et~al.}(2016){Del Moro}, {Alexander}, {Bauer}, {Daddi},
  {Kocevski}, {McIntosh}, {Stanley}, {Brandt}, {Elbaz}, {Harrison}, {Luo},
  {Mullaney}, \& {Xue}}]{DelAle2016}
{Del Moro}, A., {Alexander}, D.~M., {Bauer}, F.~E., {et~al.} 2016, \mnras, 456,
  2105

\bibitem[{{Di Matteo} {et~al.}(2005){Di Matteo}, {Springel}, \&
  {Hernquist}}]{DimSpr2005}
{Di Matteo}, T., {Springel}, V., \& {Hernquist}, L. 2005, \nat, 433, 604

\bibitem[{{D{\'{\i}}az-Santos} {et~al.}(2017){D{\'{\i}}az-Santos}, {Armus},
  {Charmandaris}, {Lu}, {Stierwalt}, {Stacey}, {Malhotra}, {van der Werf},
  {Howell}, {Privon}, {Mazzarella}, {Goldsmith}, {Murphy}, {Barcos-Mu{\~n}oz},
  {Linden}, {Inami}, {Larson}, {Evans}, {Appleton}, {Iwasawa}, {Lord},
  {Sanders}, \& {Surace}}]{DiaArm2017}
{D{\'{\i}}az-Santos}, T., {Armus}, L., {Charmandaris}, V., {et~al.} 2017, \apj,
  846, 32

\bibitem[{{D{\'{\i}}az-Santos} {et~al.}(2010){D{\'{\i}}az-Santos},
  {Charmandaris}, {Armus}, {Petric}, {Howell}, {Murphy}, {Mazzarella},
  {Veilleux}, {Bothun}, {Inami}, {Appleton}, {Evans}, {Haan}, {Marshall},
  {Sanders}, {Stierwalt}, \& {Surace}}]{DiaCha2010}
{D{\'{\i}}az-Santos}, T., {Charmandaris}, V., {Armus}, L., {et~al.} 2010, \apj,
  723, 993

\bibitem[{{Dopita} {et~al.}(2002){Dopita}, {Pereira}, {Kewley}, \&
  {Capaccioli}}]{DopPer2002}
{Dopita}, M.~A., {Pereira}, M., {Kewley}, L.~J., \& {Capaccioli}, M. 2002,
  \apjs, 143, 47

\bibitem[{{Dudik} {et~al.}(2009){Dudik}, {Satyapal}, \& {Marcu}}]{DudSat2009}
{Dudik}, R.~P., {Satyapal}, S., \& {Marcu}, D. 2009, \apj, 691, 1501

\bibitem[{{Ellison} {et~al.}(2011){Ellison}, {Patton}, {Mendel}, \&
  {Scudder}}]{EllPat2011}
{Ellison}, S.~L., {Patton}, D.~R., {Mendel}, J.~T., \& {Scudder}, J.~M. 2011,
  \mnras, 418, 2043

\bibitem[{{Evans} {et~al.}(2008){Evans}, {Vavilkin}, {Pizagno}, {Modica},
  {Mazzarella}, {Iwasawa}, {Howell}, {Surace}, {Armus}, {Petric}, {Spoon},
  {Barnes}, {Suer}, {Sanders}, {Chan}, \& {Lord}}]{EvaVav2008}
{Evans}, A.~S., {Vavilkin}, T., {Pizagno}, J., {et~al.} 2008, \apjl, 675, L69

\bibitem[{{Fabbiano}(1989)}]{Fab1989}
{Fabbiano}, G. 1989, \araa, 27, 87

\bibitem[{{Fabbiano} {et~al.}(2004){Fabbiano}, {Baldi}, {King}, {Ponman},
  {Raymond}, {Read}, {Rots}, {Schweizer}, \& {Zezas}}]{FabBal2004}
{Fabbiano}, G., {Baldi}, A., {King}, A.~R., {et~al.} 2004, \apjl, 605, L21

\bibitem[{{Fabbiano} {et~al.}(1988){Fabbiano}, {Gioia}, \&
  {Trinchieri}}]{FabGio1988}
{Fabbiano}, G., {Gioia}, I.~M., \& {Trinchieri}, G. 1988, \apj, 324, 749

\bibitem[{{Fabbiano} \& {Trinchieri}(1985)}]{FabTri1985}
{Fabbiano}, G. \& {Trinchieri}, G. 1985, \apj, 296, 430

\bibitem[{{Falstad} {et~al.}(2015){Falstad}, {Gonz{\'a}lez-Alfonso}, {Aalto},
  {van der Werf}, {Fischer}, {Veilleux}, {Mel{\'e}ndez}, {Farrah}, \&
  {Smith}}]{FalGon2015}
{Falstad}, N., {Gonz{\'a}lez-Alfonso}, E., {Aalto}, S., {et~al.} 2015, \aap,
  580, A52

\bibitem[{{Franceschini} {et~al.}(2003){Franceschini}, {Braito}, {Persic},
  {Della Ceca}, {Bassani}, {Cappi}, {Malaguti}, {Palumbo}, {Risaliti},
  {Salvati}, \& {Severgnini}}]{FraBra2003}
{Franceschini}, A., {Braito}, V., {Persic}, M., {et~al.} 2003, \mnras, 343,
  1181

\bibitem[{{Fruscione} {et~al.}(2006){Fruscione}, {McDowell}, {Allen},
  {Brickhouse}, {Burke}, {Davis}, {Durham}, {Elvis}, {Galle}, {Harris},
  {Huenemoerder}, {Houck}, {Ishibashi}, {Karovska}, {Nicastro}, {Noble},
  {Nowak}, {Primini}, {Siemiginowska}, {Smith}, \& {Wise}}]{FruMcd2006}
{Fruscione}, A., {McDowell}, J.~C., {Allen}, G.~E., {et~al.} 2006, in
  \procspie, Vol. 6270, Society of Photo-Optical Instrumentation Engineers
  (SPIE) Conference Series, 62701V

\bibitem[{{Garmire} {et~al.}(2003){Garmire}, {Bautz}, {Ford}, {Nousek}, \&
  {Ricker}}]{GarBau2003}
{Garmire}, G.~P., {Bautz}, M.~W., {Ford}, P.~G., {Nousek}, J.~A., \& {Ricker},
  Jr., G.~R. 2003, in \procspie, Vol. 4851, X-Ray and Gamma-Ray Telescopes and
  Instruments for Astronomy., ed. J.~E. {Truemper} \& H.~D. {Tananbaum}, 28--44

\bibitem[{{Gilfanov}(2004)}]{Gil2004}
{Gilfanov}, M. 2004, \mnras, 349, 146

\bibitem[{{Gilfanov} {et~al.}(2004){Gilfanov}, {Grimm}, \&
  {Sunyaev}}]{GilGri2004}
{Gilfanov}, M., {Grimm}, H.-J., \& {Sunyaev}, R. 2004, Nuclear Physics B
  Proceedings Supplements, 132, 369

\bibitem[{{Gilli} {et~al.}(1999){Gilli}, {Comastri}, {Brunetti}, \&
  {Setti}}]{GilCom1999}
{Gilli}, R., {Comastri}, A., {Brunetti}, G., \& {Setti}, G. 1999, \na, 4, 45

\bibitem[{{Goldader} {et~al.}(1997){Goldader}, {Joseph}, {Doyon}, \&
  {Sanders}}]{GolJos1997}
{Goldader}, J.~D., {Joseph}, R.~D., {Doyon}, R., \& {Sanders}, D.~B. 1997,
  \apjs, 108, 449

\bibitem[{{Grimm} {et~al.}(2003){Grimm}, {Gilfanov}, \& {Sunyaev}}]{GriGil2003}
{Grimm}, H.-J., {Gilfanov}, M., \& {Sunyaev}, R. 2003, \mnras, 339, 793

\bibitem[{{Haan} {et~al.}(2011){Haan}, {Surace}, {Armus}, {Evans}, {Howell},
  {Mazzarella}, {Kim}, {Vavilkin}, {Inami}, {Sanders}, {Petric}, {Bridge},
  {Melbourne}, {Charmandaris}, {Diaz-Santos}, {Murphy}, {U}, {Stierwalt}, \&
  {Marshall}}]{HaaSur2011}
{Haan}, S., {Surace}, J.~A., {Armus}, L., {et~al.} 2011, \aj, 141, 100

\bibitem[{{Hattori} {et~al.}(2004){Hattori}, {Yoshida}, {Ohtani}, {Sugai},
  {Ishigaki}, {Sasaki}, {Hayashi}, {Ozaki}, {Ishii}, \& {Kawai}}]{HatYos2004}
{Hattori}, T., {Yoshida}, M., {Ohtani}, H., {et~al.} 2004, \aj, 127, 736

\bibitem[{{Helou} {et~al.}(1985){Helou}, {Soifer}, \&
  {Rowan-Robinson}}]{HelSoi1985}
{Helou}, G., {Soifer}, B.~T., \& {Rowan-Robinson}, M. 1985, \apjl, 298, L7

\bibitem[{{Hernquist}(1989)}]{Her1989}
{Hernquist}, L. 1989, \nat, 340, 687

\bibitem[{{Hill} {et~al.}(2001){Hill}, {Heisler}, {Norris}, {Reynolds}, \&
  {Hunstead}}]{HilHei2001}
{Hill}, T.~L., {Heisler}, C.~A., {Norris}, R.~P., {Reynolds}, J.~E., \&
  {Hunstead}, R.~W. 2001, \aj, 121, 128

\bibitem[{{Hinshaw} {et~al.}(2009){Hinshaw}, {Weiland}, {Hill}, {Odegard},
  {Larson}, {Bennett}, {Dunkley}, {Gold}, {Greason}, {Jarosik}, {Komatsu},
  {Nolta}, {Page}, {Spergel}, {Wollack}, {Halpern}, {Kogut}, {Limon}, {Meyer},
  {Tucker}, \& {Wright}}]{HinWei2009}
{Hinshaw}, G., {Weiland}, J.~L., {Hill}, R.~S., {et~al.} 2009, \apjs, 180, 225

\bibitem[{{Hopkins} {et~al.}(2009){Hopkins}, {Bundy}, {Murray}, {Quataert},
  {Lauer}, \& {Ma}}]{HopBun2009}
{Hopkins}, P.~F., {Bundy}, K., {Murray}, N., {et~al.} 2009, \mnras, 398, 898

\bibitem[{{Hopkins} {et~al.}(2005){Hopkins}, {Hernquist}, {Cox}, {Di Matteo},
  {Martini}, {Robertson}, \& {Springel}}]{HopHer2005}
{Hopkins}, P.~F., {Hernquist}, L., {Cox}, T.~J., {et~al.} 2005, Astrophysical
  Journal, 630, 705

\bibitem[{{Howell} {et~al.}(2007){Howell}, {Mazzarella}, {Chan}, {Lord},
  {Surace}, {Frayer}, {Appleton}, {Armus}, {Evans}, {Bothun}, {Ishida}, {Kim},
  {Jensen}, {Madore}, {Sanders}, {Schulz}, {Vavilkin}, {Veilleux}, \&
  {Xu}}]{HowMaz2007}
{Howell}, J.~H., {Mazzarella}, J.~M., {Chan}, B.~H.~P., {et~al.} 2007, \aj,
  134, 2086

\bibitem[{{Hutchings} {et~al.}(1990){Hutchings}, {Lo}, {Neff}, {Stanford}, \&
  {Unger}}]{HutLo1990}
{Hutchings}, J.~B., {Lo}, E., {Neff}, S.~G., {Stanford}, S.~A., \& {Unger},
  S.~W. 1990, \aj, 100, 60

\bibitem[{{Inami} {et~al.}(2013){Inami}, {Armus}, {Charmandaris}, {Groves},
  {Kewley}, {Petric}, {Stierwalt}, {D{\'{\i}}az-Santos}, {Surace}, {Rich},
  {Haan}, {Howell}, {Evans}, {Mazzarella}, {Marshall}, {Appleton}, {Lord},
  {Spoon}, {Frayer}, {Matsuhara}, \& {Veilleux}}]{InaArm2013}
{Inami}, H., {Armus}, L., {Charmandaris}, V., {et~al.} 2013, \apj, 777, 156

\bibitem[{{Inui} {et~al.}(2005){Inui}, {Matsumoto}, {Tsuru}, {Koyama},
  {Matsushita}, {Peck}, \& {Tarchi}}]{InuMat2005}
{Inui}, T., {Matsumoto}, H., {Tsuru}, T.~G., {et~al.} 2005, \pasj, 57, 135

\bibitem[{{Iwasawa} {et~al.}(2009){Iwasawa}, {Sanders}, {Evans}, {Mazzarella},
  {Armus}, \& {Surace}}]{IwaSan2009}
{Iwasawa}, K., {Sanders}, D.~B., {Evans}, A.~S., {et~al.} 2009, \apjl, 695,
  L103

\bibitem[{{Iwasawa} {et~al.}(2005){Iwasawa}, {Sanders}, {Evans}, {Trentham},
  {Miniutti}, \& {Spoon}}]{IwaSan2005}
{Iwasawa}, K., {Sanders}, D.~B., {Evans}, A.~S., {et~al.} 2005, \mnras, 357,
  565

\bibitem[{{Iwasawa} {et~al.}(2011){Iwasawa}, {Sanders}, {Teng}, {U}, {Armus},
  {Evans}, {Howell}, {Komossa}, {Mazzarella}, {Petric}, {Surace}, {Vavilkin},
  {Veilleux}, \& {Trentham}}]{IwaSan2011}
{Iwasawa}, K., {Sanders}, D.~B., {Teng}, S.~H., {et~al.} 2011, \aap, 529, A106

\bibitem[{{Iwasawa} {et~al.}(2018){Iwasawa}, {U}, {Mazzarella}, {Medling},
  {Sanders}, \& {Evans}}]{IwaU2018}
{Iwasawa}, K., {U}, V., {Mazzarella}, J.~M., {et~al.} 2018, \aap, 611, A71

\bibitem[{{Joy} \& {Harvey}(1987)}]{JoyHar1987}
{Joy}, M. \& {Harvey}, P.~M. 1987, \apj, 315, 480

\bibitem[{{Kaaret} \& {Alonso-Herrero}(2008)}]{KaaAlo2008}
{Kaaret}, P. \& {Alonso-Herrero}, A. 2008, \apj, 682, 1020

\bibitem[{{Kaastra}(1992)}]{Kaa1992}
{Kaastra}, J. 1992, An X-Ray Spectral Code for Optically Thin Plasmas (Internal
  SRON-Leiden Report, updated version 2.0)

\bibitem[{{Kalberla} {et~al.}(2005){Kalberla}, {Burton}, {Hartmann}, {Arnal},
  {Bajaja}, {Morras}, \& {P{\"o}ppel}}]{KalBur2005}
{Kalberla}, P.~M.~W., {Burton}, W.~B., {Hartmann}, D., {et~al.} 2005, \aap,
  440, 775

\bibitem[{{Kennicutt}(1998)}]{Ken1998}
{Kennicutt}, Jr., R.~C. 1998, \apj, 498, 541

\bibitem[{{Kewley} {et~al.}(2001){Kewley}, {Dopita}, {Sutherland}, {Heisler},
  \& {Trevena}}]{KewDop2001}
{Kewley}, L.~J., {Dopita}, M.~A., {Sutherland}, R.~S., {Heisler}, C.~A., \&
  {Trevena}, J. 2001, \apj, 556, 121

\bibitem[{{Kewley} {et~al.}(2006){Kewley}, {Groves}, {Kauffmann}, \&
  {Heckman}}]{KewGro2006}
{Kewley}, L.~J., {Groves}, B., {Kauffmann}, G., \& {Heckman}, T. 2006, \mnras,
  372, 961

\bibitem[{{Kim} {et~al.}(1995){Kim}, {Sanders}, {Veilleux}, {Mazzarella}, \&
  {Soifer}}]{KimSan1995}
{Kim}, D.-C., {Sanders}, D.~B., {Veilleux}, S., {Mazzarella}, J.~M., \&
  {Soifer}, B.~T. 1995, \apjs, 98, 129

\bibitem[{{Kocevski} {et~al.}(2015){Kocevski}, {Brightman}, {Nandra},
  {Koekemoer}, {Salvato}, {Aird}, {Bell}, {Hsu}, {Kartaltepe}, {Koo}, {Lotz},
  {McIntosh}, {Mozena}, {Rosario}, \& {Trump}}]{KocBri2015}
{Kocevski}, D.~D., {Brightman}, M., {Nandra}, K., {et~al.} 2015, \apj, 814, 104

\bibitem[{{Koss} {et~al.}(2012){Koss}, {Mushotzky}, {Treister}, {Veilleux},
  {Vasudevan}, \& {Trippe}}]{KosMus2012}
{Koss}, M., {Mushotzky}, R., {Treister}, E., {et~al.} 2012, \apjl, 746, L22

\bibitem[{{Koyama} {et~al.}(1989){Koyama}, {Inoue}, {Tanaka}, {Awaki},
  {Takano}, {Ohashi}, \& {Matsuoka}}]{KoyIno1989}
{Koyama}, K., {Inoue}, H., {Tanaka}, Y., {et~al.} 1989, \pasj, 41, 731

\bibitem[{{Kreckel} {et~al.}(2014){Kreckel}, {Armus}, {Groves}, {Lyubenova},
  {D{\'{\i}}az-Santos}, {Schinnerer}, {Appleton}, {Croxall}, {Dale}, {Hunt},
  {Beir{\~a}o}, {Bolatto}, {Calzetti}, {Donovan Meyer}, {Draine}, {Hinz},
  {Kennicutt}, {Meidt}, {Murphy}, {Smith}, {Tabatabaei}, \&
  {Walter}}]{KreArm2014}
{Kreckel}, K., {Armus}, L., {Groves}, B., {et~al.} 2014, \apj, 790, 26

\bibitem[{{Krolik} \& {Kallman}(1987)}]{KroKal1987}
{Krolik}, J.~H. \& {Kallman}, T.~R. 1987, \apjl, 320, L5

\bibitem[{{Kunth} {et~al.}(2003){Kunth}, {Leitherer}, {Mas-Hesse},
  {{\"O}stlin}, \& {Petrosian}}]{KunLei2003}
{Kunth}, D., {Leitherer}, C., {Mas-Hesse}, J.~M., {{\"O}stlin}, G., \&
  {Petrosian}, A. 2003, \apj, 597, 263

\bibitem[{{Kuo} {et~al.}(2008){Kuo}, {Lim}, {Tang}, \& {Ho}}]{KuoLim2008}
{Kuo}, C.-Y., {Lim}, J., {Tang}, Y.-W., \& {Ho}, P.~T.~P. 2008, \apj, 679, 1047

\bibitem[{{Lahuis} {et~al.}(2007){Lahuis}, {Spoon}, {Tielens}, {Doty}, {Armus},
  {Charmandaris}, {Houck}, {St{\"a}uber}, \& {van Dishoeck}}]{LahSpo2007}
{Lahuis}, F., {Spoon}, H.~W.~W., {Tielens}, A.~G.~G.~M., {et~al.} 2007, \apj,
  659, 296

\bibitem[{{Lanzuisi} {et~al.}(2015){Lanzuisi}, {Ranalli}, {Georgantopoulos},
  {Georgakakis}, {Delvecchio}, {Akylas}, {Berta}, {Bongiorno}, {Brusa},
  {Cappelluti}, {Civano}, {Comastri}, {Gilli}, {Gruppioni}, {Hasinger},
  {Iwasawa}, {Koekemoer}, {Lusso}, {Marchesi}, {Mainieri}, {Merloni},
  {Mignoli}, {Piconcelli}, {Pozzi}, {Rosario}, {Salvato}, {Silverman},
  {Trakhtenbrot}, {Vignali}, \& {Zamorani}}]{LanRan2015}
{Lanzuisi}, G., {Ranalli}, P., {Georgantopoulos}, I., {et~al.} 2015, \aap, 573,
  A137

\bibitem[{{Larson} {et~al.}(2016){Larson}, {Sanders}, {Barnes}, {Ishida},
  {Evans}, {U}, {Mazzarella}, {Kim}, {Privon}, {Mirabel}, \&
  {Flewelling}}]{LarSan2016}
{Larson}, K.~L., {Sanders}, D.~B., {Barnes}, J.~E., {et~al.} 2016, \apj, 825,
  128

\bibitem[{{Lehmer} {et~al.}(2010){Lehmer}, {Alexander}, {Bauer}, {Brandt},
  {Goulding}, {Jenkins}, {Ptak}, \& {Roberts}}]{LehAle2010}
{Lehmer}, B.~D., {Alexander}, D.~M., {Bauer}, F.~E., {et~al.} 2010, \apj, 724,
  559

\bibitem[{{Lehmer} {et~al.}(2008){Lehmer}, {Brandt}, {Alexander}, {Bell},
  {Hornschemeier}, {McIntosh}, {Bauer}, {Gilli}, {Mainieri}, {Schneider},
  {Silverman}, {Steffen}, {Tozzi}, \& {Wolf}}]{LehBra2008}
{Lehmer}, B.~D., {Brandt}, W.~N., {Alexander}, D.~M., {et~al.} 2008, \apj, 681,
  1163

\bibitem[{{Levenson} {et~al.}(2002){Levenson}, {Krolik}, {{\.Z}ycki},
  {Heckman}, {Weaver}, {Awaki}, \& {Terashima}}]{LevKro2002}
{Levenson}, N.~A., {Krolik}, J.~H., {{\.Z}ycki}, P.~T., {et~al.} 2002, \apjl,
  573, L81

\bibitem[{{Levenson} {et~al.}(2005){Levenson}, {Weaver}, {Heckman}, {Awaki}, \&
  {Terashima}}]{LevWea2005}
{Levenson}, N.~A., {Weaver}, K.~A., {Heckman}, T.~M., {Awaki}, H., \&
  {Terashima}, Y. 2005, \apj, 618, 167

\bibitem[{{Liedahl} {et~al.}(1995){Liedahl}, {Osterheld}, \&
  {Goldstein}}]{LieOst1995}
{Liedahl}, D.~A., {Osterheld}, A.~L., \& {Goldstein}, W.~H. 1995, \apjl, 438,
  L115

\bibitem[{{Lipovetsky} {et~al.}(1988){Lipovetsky}, {Neizvestny}, \&
  {Neizvestnaya}}]{LipNei1988}
{Lipovetsky}, V.~A., {Neizvestny}, S.~I., \& {Neizvestnaya}, O.~M. 1988,
  Soobshcheniya Spetsial'noj Astrofizicheskoj Observatorii, 55, 5

\bibitem[{{Lira} {et~al.}(2002){Lira}, {Ward}, {Zezas}, {Alonso-Herrero}, \&
  {Ueno}}]{LirWar2002}
{Lira}, P., {Ward}, M., {Zezas}, A., {Alonso-Herrero}, A., \& {Ueno}, S. 2002,
  \mnras, 330, 259

\bibitem[{{Lonsdale} {et~al.}(1993){Lonsdale}, {Smith}, \&
  {Lonsdale}}]{LonSmi1993}
{Lonsdale}, C.~J., {Smith}, H.~J., \& {Lonsdale}, C.~J. 1993, \apjl, 405, L9

\bibitem[{{Luangtip} {et~al.}(2015){Luangtip}, {Roberts}, {Mineo}, {Lehmer},
  {Alexander}, {Jackson}, {Goulding}, \& {Fischer}}]{LuaRob2015}
{Luangtip}, W., {Roberts}, T.~P., {Mineo}, S., {et~al.} 2015, \mnras, 446, 470

\bibitem[{{Magdziarz} \& {Zdziarski}(1995)}]{MagZdz1995}
{Magdziarz}, P. \& {Zdziarski}, A.~A. 1995, \mnras, 273, 837

\bibitem[{{Maia} {et~al.}(1996){Maia}, {Suzuki}, {da Costa}, {Willmer}, \&
  {Rite}}]{MaiSuz1996}
{Maia}, M.~A.~G., {Suzuki}, J.~A., {da Costa}, L.~N., {Willmer}, C.~N.~A., \&
  {Rite}, C. 1996, \aaps, 117, 487

\bibitem[{{Maiolino} {et~al.}(2003){Maiolino}, {Comastri}, {Gilli}, {Nagar},
  {Bianchi}, {B{\"o}ker}, {Colbert}, {Krabbe}, {Marconi}, {Matt}, \&
  {Salvati}}]{MaiCom2003}
{Maiolino}, R., {Comastri}, A., {Gilli}, R., {et~al.} 2003, \mnras, 344, L59

\bibitem[{{Mazzarella} {et~al.}(2012){Mazzarella}, {Iwasawa}, {Vavilkin},
  {Armus}, {Kim}, {Bothun}, {Evans}, {Spoon}, {Haan}, {Howell}, {Lord},
  {Marshall}, {Ishida}, {Xu}, {Petric}, {Sanders}, {Surace}, {Appleton},
  {Chan}, {Frayer}, {Inami}, {Khachikian}, {Madore}, {Privon}, {Sturm}, {U}, \&
  {Veilleux}}]{MazIwa2012}
{Mazzarella}, J.~M., {Iwasawa}, K., {Vavilkin}, T., {et~al.} 2012, \aj, 144,
  125

\bibitem[{{Mel{\'e}ndez} {et~al.}(2014){Mel{\'e}ndez}, {Heckman},
  {Mart{\'{\i}}nez-Paredes}, {Kraemer}, \& {Mendoza}}]{MelHec2014}
{Mel{\'e}ndez}, M., {Heckman}, T.~M., {Mart{\'{\i}}nez-Paredes}, M., {Kraemer},
  S.~B., \& {Mendoza}, C. 2014, \mnras, 443, 1358

\bibitem[{{Mewe} {et~al.}(1985){Mewe}, {Gronenschild}, \& {van den
  Oord}}]{MewGro1985}
{Mewe}, R., {Gronenschild}, E.~H.~B.~M., \& {van den Oord}, G.~H.~J. 1985,
  \aaps, 62, 197

\bibitem[{{Mineo} {et~al.}(2014){Mineo}, {Gilfanov}, {Lehmer}, {Morrison}, \&
  {Sunyaev}}]{MinGil2014}
{Mineo}, S., {Gilfanov}, M., {Lehmer}, B.~D., {Morrison}, G.~E., \& {Sunyaev},
  R. 2014, Monthly Notices of the RAS, 437, 1698

\bibitem[{{Mirabel} \& {Sanders}(1988)}]{MirSan1988}
{Mirabel}, I.~F. \& {Sanders}, D.~B. 1988, \apj, 335, 104

\bibitem[{{Moorwood} \& {Oliva}(1994)}]{MooOli1994}
{Moorwood}, A.~F.~M. \& {Oliva}, E. 1994, \apj, 429, 602

\bibitem[{{Mould} {et~al.}(2000){Mould}, {Huchra}, {Freedman}, {Kennicutt},
  {Ferrarese}, {Ford}, {Gibson}, {Graham}, {Hughes}, {Illingworth}, {Kelson},
  {Macri}, {Madore}, {Sakai}, {Sebo}, {Silbermann}, \& {Stetson}}]{MouHuc2000}
{Mould}, J.~R., {Huchra}, J.~P., {Freedman}, W.~L., {et~al.} 2000, \apj, 529,
  786

\bibitem[{{Mudd} {et~al.}(2014){Mudd}, {Mathur}, {Guainazzi}, {Piconcelli},
  {Bianchi}, {Komossa}, {Vignali}, {Lanzuisi}, {Nicastro}, {Fiore}, \&
  {Maiolino}}]{MudMat2014}
{Mudd}, D., {Mathur}, S., {Guainazzi}, M., {et~al.} 2014, \apj, 787, 40

\bibitem[{{Nandra} \& {Iwasawa}(2007)}]{NanIwa2007}
{Nandra}, K. \& {Iwasawa}, K. 2007, \mnras, 382, L1

\bibitem[{{Nandra} \& {Pounds}(1994)}]{NanPou1994}
{Nandra}, K. \& {Pounds}, K.~A. 1994, \mnras, 268, 405

\bibitem[{{Norris} \& {Forbes}(1995)}]{NorFor1995}
{Norris}, R.~P. \& {Forbes}, D.~A. 1995, \apj, 446, 594

\bibitem[{{Oda} {et~al.}(2018){Oda}, {Ueda}, {Tanimoto}, \&
  {Ricci}}]{OdaUed2018}
{Oda}, S., {Ueda}, Y., {Tanimoto}, A., \& {Ricci}, C. 2018, \apj, 855, 79

\bibitem[{{O'Sullivan} {et~al.}(2014){O'Sullivan}, {Zezas}, {Vrtilek},
  {Giacintucci}, {Trevisan}, {David}, {Ponman}, {Mamon}, \&
  {Raychaudhury}}]{OsuZez2014}
{O'Sullivan}, E., {Zezas}, A., {Vrtilek}, J.~M., {et~al.} 2014, \apj, 793, 73

\bibitem[{{Parra} {et~al.}(2010){Parra}, {Conway}, {Aalto}, {Appleton},
  {Norris}, {Pihlstr{\"o}m}, \& {Kewley}}]{ParCon2010}
{Parra}, R., {Conway}, J.~E., {Aalto}, S., {et~al.} 2010, \apj, 720, 555

\bibitem[{{Pereira-Santaella} {et~al.}(2015{\natexlab{a}}){Pereira-Santaella},
  {Alonso-Herrero}, {Colina}, {Miralles-Caballero}, {P{\'e}rez-Gonz{\'a}lez},
  {Arribas}, {Bellocchi}, {Cazzoli}, {D{\'{\i}}az-Santos}, \& {Piqueras
  L{\'o}pez}}]{PerAlo2015}
{Pereira-Santaella}, M., {Alonso-Herrero}, A., {Colina}, L., {et~al.}
  2015{\natexlab{a}}, \aap, 577, A78

\bibitem[{{Pereira-Santaella} {et~al.}(2015{\natexlab{b}}){Pereira-Santaella},
  {Colina}, {Alonso-Herrero}, {Usero}, {D{\'{\i}}az-Santos},
  {Garc{\'{\i}}a-Burillo}, {Alberdi}, {Gonzalez-Martin}, {Herrero-Illana},
  {Imanishi}, {Levenson}, {P{\'e}rez-Torres}, \& {Ramos Almeida}}]{PerCol2015}
{Pereira-Santaella}, M., {Colina}, L., {Alonso-Herrero}, A., {et~al.}
  2015{\natexlab{b}}, \mnras, 454, 3679

\bibitem[{{Persic} \& {Rephaeli}(2007)}]{PerRep2007}
{Persic}, M. \& {Rephaeli}, Y. 2007, \aap, 463, 481

\bibitem[{{Petric} {et~al.}(2011){Petric}, {Armus}, {Howell}, {Chan},
  {Mazzarella}, {Evans}, {Surace}, {Sanders}, {Appleton}, {Charmandaris},
  {D{\'{\i}}az-Santos}, {Frayer}, {Haan}, {Inami}, {Iwasawa}, {Kim}, {Madore},
  {Marshall}, {Spoon}, {Stierwalt}, {Sturm}, {U}, {Vavilkin}, \&
  {Veilleux}}]{PetArm2011}
{Petric}, A.~O., {Armus}, L., {Howell}, J., {et~al.} 2011, \apj, 730, 28

\bibitem[{{Ptak} {et~al.}(2003){Ptak}, {Heckman}, {Levenson}, {Weaver}, \&
  {Strickland}}]{PtaHec2003}
{Ptak}, A., {Heckman}, T., {Levenson}, N.~A., {Weaver}, K., \& {Strickland}, D.
  2003, \apj, 592, 782

\bibitem[{{Ranalli} {et~al.}(2003){Ranalli}, {Comastri}, \&
  {Setti}}]{RanCom2003}
{Ranalli}, P., {Comastri}, A., \& {Setti}, G. 2003, \aap, 399, 39

\bibitem[{{Ricci} {et~al.}(2016){Ricci}, {Bauer}, {Treister},
  {Romero-Ca{\~n}izales}, {Arevalo}, {Iwasawa}, {Privon}, {Sanders},
  {Schawinski}, {Stern}, \& {Imanishi}}]{RicBau2016}
{Ricci}, C., {Bauer}, F.~E., {Treister}, E., {et~al.} 2016, \apj, 819, 4

\bibitem[{{Ricci} {et~al.}(2017){Ricci}, {Bauer}, {Treister}, {Schawinski},
  {Privon}, {Blecha}, {Arevalo}, {Armus}, {Harrison}, {Ho}, {Iwasawa},
  {Sanders}, \& {Stern}}]{RicBau2017}
{Ricci}, C., {Bauer}, F.~E., {Treister}, E., {et~al.} 2017, \mnras, 468, 1273

\bibitem[{{Ricci} {et~al.}(2014){Ricci}, {Ueda}, {Paltani}, {Ichikawa},
  {Gandhi}, \& {Awaki}}]{RicUed2014}
{Ricci}, C., {Ueda}, Y., {Paltani}, S., {et~al.} 2014, \mnras, 441, 3622

\bibitem[{{Risaliti} {et~al.}(2005){Risaliti}, {Elvis}, {Fabbiano}, {Baldi}, \&
  {Zezas}}]{RisElv2005}
{Risaliti}, G., {Elvis}, M., {Fabbiano}, G., {Baldi}, A., \& {Zezas}, A. 2005,
  \apjl, 623, L93

\bibitem[{{Risaliti} {et~al.}(2007){Risaliti}, {Elvis}, {Fabbiano}, {Baldi},
  {Zezas}, \& {Salvati}}]{RisElv2007}
{Risaliti}, G., {Elvis}, M., {Fabbiano}, G., {et~al.} 2007, \apjl, 659, L111

\bibitem[{{Risaliti} {et~al.}(2000){Risaliti}, {Gilli}, {Maiolino}, \&
  {Salvati}}]{RisGil2000}
{Risaliti}, G., {Gilli}, R., {Maiolino}, R., \& {Salvati}, M. 2000, \aap, 357,
  13

\bibitem[{{Risaliti} {et~al.}(2009){Risaliti}, {Miniutti}, {Elvis}, {Fabbiano},
  {Salvati}, {Baldi}, {Braito}, {Bianchi}, {Matt}, {Reeves}, {Soria}, \&
  {Zezas}}]{RisMin2009}
{Risaliti}, G., {Miniutti}, G., {Elvis}, M., {et~al.} 2009, \apj, 696, 160

\bibitem[{{Romero-Ca{\~n}izales} {et~al.}(2017){Romero-Ca{\~n}izales},
  {Alberdi}, {Ricci}, {Ar{\'e}valo}, {P{\'e}rez-Torres}, {Conway}, {Beswick},
  {Bondi}, {Muxlow}, {Argo}, {Bauer}, {Efstathiou}, {Herrero-Illana},
  {Mattila}, \& {Ryder}}]{RomAlb2017}
{Romero-Ca{\~n}izales}, C., {Alberdi}, A., {Ricci}, C., {et~al.} 2017, \mnras,
  467, 2504

\bibitem[{{Romero-Ca{\~n}izales} {et~al.}(2012){Romero-Ca{\~n}izales},
  {P{\'e}rez-Torres}, {Alberdi}, {Argo}, {Beswick}, {Kankare}, {Batejat},
  {Efstathiou}, {Mattila}, {Conway}, {Garrington}, {Muxlow}, {Ryder}, \&
  {V{\"a}is{\"a}nen}}]{RomPer2012}
{Romero-Ca{\~n}izales}, C., {P{\'e}rez-Torres}, M.~A., {Alberdi}, A., {et~al.}
  2012, \aap, 543, A72

\bibitem[{{Roussel} {et~al.}(2003){Roussel}, {Helou}, {Beck}, {Condon},
  {Bosma}, {Matthews}, \& {Jarrett}}]{RouHel2003}
{Roussel}, H., {Helou}, G., {Beck}, R., {et~al.} 2003, \apj, 593, 733

\bibitem[{{Rudnick} {et~al.}(2000){Rudnick}, {Rix}, \&
  {Kennicutt}}]{RudRix2000}
{Rudnick}, G., {Rix}, H.-W., \& {Kennicutt}, Jr., R.~C. 2000, \apj, 538, 569

\bibitem[{{Sales} {et~al.}(2015){Sales}, {Robinson}, {Axon}, {Gallimore},
  {Kharb}, {Curran}, {O'Dea}, {Baum}, {Elitzur}, \& {Mittal}}]{SalRob2015}
{Sales}, D.~A., {Robinson}, A., {Axon}, D.~J., {et~al.} 2015, \apj, 799, 25

\bibitem[{{Sanders}(1999)}]{San1999}
{Sanders}, D.~B. 1999, Astrophysics and Space Science, 266, 331

\bibitem[{{Sanders} {et~al.}(2003){Sanders}, {Mazzarella}, {Kim}, {Surace}, \&
  {Soifer}}]{SanMaz2003}
{Sanders}, D.~B., {Mazzarella}, J.~M., {Kim}, D.-C., {Surace}, J.~A., \&
  {Soifer}, B.~T. 2003, \aj, 126, 1607

\bibitem[{{Sanders} {et~al.}(1988){Sanders}, {Soifer}, {Elias}, {Madore},
  {Matthews}, {Neugebauer}, \& {Scoville}}]{SanSoi1988}
{Sanders}, D.~B., {Soifer}, B.~T., {Elias}, J.~H., {et~al.} 1988, \apj, 325, 74

\bibitem[{{Satyapal} {et~al.}(2014){Satyapal}, {Ellison}, {McAlpine}, {Hickox},
  {Patton}, \& {Mendel}}]{SatEll2014}
{Satyapal}, S., {Ellison}, S.~L., {McAlpine}, W., {et~al.} 2014, \mnras, 441,
  1297

\bibitem[{{Satyapal} {et~al.}(2017){Satyapal}, {Secrest}, {Ricci}, {Ellison},
  {Rothberg}, {Blecha}, {Constantin}, {Gliozzi}, {McNulty}, \&
  {Ferguson}}]{SatSec2017}
{Satyapal}, S., {Secrest}, N.~J., {Ricci}, C., {et~al.} 2017, \apj, 848, 126

\bibitem[{{Scoville} {et~al.}(2017){Scoville}, {Murchikova}, {Walter},
  {Vlahakis}, {Koda}, {Vanden Bout}, {Barnes}, {Hernquist}, {Sheth}, {Yun},
  {Sanders}, {Armus}, {Cox}, {Thompson}, {Robertson}, {Zschaechner}, {Tacconi},
  {Torrey}, {Hayward}, {Genzel}, {Hopkins}, {van der Werf}, \&
  {Decarli}}]{ScoMur2017}
{Scoville}, N., {Murchikova}, L., {Walter}, F., {et~al.} 2017, \apj, 836, 66

\bibitem[{{Shalyapina} {et~al.}(2004){Shalyapina}, {Moiseev}, {Yakovleva},
  {Hagen-Thorn}, \& {Burenkov}}]{ShaMoi2004}
{Shalyapina}, L.~V., {Moiseev}, A.~V., {Yakovleva}, V.~A., {Hagen-Thorn},
  V.~A., \& {Burenkov}, A.~N. 2004, Astronomy Letters, 30, 1

\bibitem[{{Shu} {et~al.}(2007){Shu}, {Wang}, {Jiang}, {Fan}, \&
  {Wang}}]{ShuWan2007}
{Shu}, X.~W., {Wang}, J.~X., {Jiang}, P., {Fan}, L.~L., \& {Wang}, T.~G. 2007,
  \apj, 657, 167

\bibitem[{{Silverman} {et~al.}(2011){Silverman}, {Kampczyk}, {Jahnke},
  {Andrae}, {Lilly}, {Elvis}, {Civano}, {Mainieri}, {Vignali}, {Zamorani},
  {Nair}, {Le F{\`e}vre}, {de Ravel}, {Bardelli}, {Bongiorno}, {Bolzonella},
  {Cappi}, {Caputi}, {Carollo}, {Contini}, {Coppa}, {Cucciati}, {de la Torre},
  {Franzetti}, {Garilli}, {Halliday}, {Hasinger}, {Iovino}, {Knobel},
  {Koekemoer}, {Kova{\v c}}, {Lamareille}, {Le Borgne}, {Le Brun}, {Maier},
  {Mignoli}, {Pello}, {P{\'e}rez-Montero}, {Ricciardelli}, {Peng}, {Scodeggio},
  {Tanaka}, {Tasca}, {Tresse}, {Vergani}, {Zucca}, {Brusa}, {Cappelluti},
  {Comastri}, {Finoguenov}, {Fu}, {Gilli}, {Hao}, {Ho}, \&
  {Salvato}}]{SilKam2011}
{Silverman}, J.~D., {Kampczyk}, P., {Jahnke}, K., {et~al.} 2011, \apj, 743, 2

\bibitem[{{Singh} {et~al.}(2012){Singh}, {Risaliti}, {Braito}, \&
  {Shastri}}]{SinRis2012}
{Singh}, V., {Risaliti}, G., {Braito}, V., \& {Shastri}, P. 2012, \mnras, 419,
  2089

\bibitem[{{Smith} {et~al.}(1996){Smith}, {Herter}, {Haynes}, {Beichman}, \&
  {Gautier}}]{SmiHer1996}
{Smith}, D.~A., {Herter}, T., {Haynes}, M.~P., {Beichman}, C.~A., \& {Gautier},
  III, T.~N. 1996, \apjs, 104, 217

\bibitem[{{Snyder} {et~al.}(2013){Snyder}, {Hayward}, {Sajina}, {Jonsson},
  {Cox}, {Hernquist}, {Hopkins}, \& {Yan}}]{SnyHay2013}
{Snyder}, G.~F., {Hayward}, C.~C., {Sajina}, A., {et~al.} 2013, \apj, 768, 168

\bibitem[{{Stierwalt} {et~al.}(2013){Stierwalt}, {Armus}, {Surace}, {Inami},
  {Petric}, {Diaz-Santos}, {Haan}, {Charmandaris}, {Howell}, {Kim}, {Marshall},
  {Mazzarella}, {Spoon}, {Veilleux}, {Evans}, {Sanders}, {Appleton}, {Bothun},
  {Bridge}, {Chan}, {Frayer}, {Iwasawa}, {Kewley}, {Lord}, {Madore},
  {Melbourne}, {Murphy}, {Rich}, {Schulz}, {Sturm}, {Vavilkin}, \&
  {Xu}}]{StiArm2013}
{Stierwalt}, S., {Armus}, L., {Surace}, J.~A., {et~al.} 2013, \apjs, 206, 1

\bibitem[{{Strickland} \& {Stevens}(2000)}]{StrSte2000}
{Strickland}, D.~K. \& {Stevens}, I.~R. 2000, \mnras, 314, 511

\bibitem[{{Surace} {et~al.}(2004){Surace}, {Sanders}, \&
  {Mazzarella}}]{SurSan2004}
{Surace}, J.~A., {Sanders}, D.~B., \& {Mazzarella}, J.~M. 2004, \aj, 127, 3235

\bibitem[{{Teng} \& {Veilleux}(2010)}]{TenVei2010}
{Teng}, S.~H. \& {Veilleux}, S. 2010, \apj, 725, 1848

\bibitem[{{Teng} {et~al.}(2005){Teng}, {Wilson}, {Veilleux}, {Young},
  {Sanders}, \& {Nagar}}]{TenWil2005}
{Teng}, S.~H., {Wilson}, A.~S., {Veilleux}, S., {et~al.} 2005, \apj, 633, 664

\bibitem[{{Terashima} {et~al.}(2015){Terashima}, {Hirata}, {Awaki}, {Oyabu},
  {Gandhi}, {Toba}, \& {Matsuhara}}]{TerHir2015}
{Terashima}, Y., {Hirata}, Y., {Awaki}, H., {et~al.} 2015, \apj, 814, 11

\bibitem[{{Turner} {et~al.}(2001){Turner}, {Reeves}, {Ponman}, {Arnaud},
  {Barbera}, {Bennie}, {Boer}, {Briel}, {Butler}, {Clavel}, {Dhez}, {Cordova},
  {Dos Santos}, {Ferrando}, {Ghizzardi}, {Goodall}, {Griffiths}, {Hochedez},
  {Holland}, {Jansen}, {Kendziorra}, {Lagostina}, {Laine}, {La Palombara},
  {Lortholary}, {Mason}, {Molendi}, {Pigot}, {Priedhorsky}, {Reppin},
  {Rothenflug}, {Salvetat}, {Sauvageot}, {Schmitt}, {Sembay}, {Short},
  {Str{\"u}der}, {Trifoglio}, {Tr{\"u}mper}, {Vercellone}, {Vigroux}, {Villa},
  \& {Ward}}]{TurRee2001}
{Turner}, M.~J.~L., {Reeves}, J.~N., {Ponman}, T.~J., {et~al.} 2001, \aap, 365,
  L110

\bibitem[{{Turner} \& {Miller}(2009)}]{TurMil2009}
{Turner}, T.~J. \& {Miller}, L. 2009, \aapr, 17, 47

\bibitem[{{Valiante} {et~al.}(2009){Valiante}, {Lutz}, {Sturm}, {Genzel}, \&
  {Chapin}}]{ValLut2009}
{Valiante}, E., {Lutz}, D., {Sturm}, E., {Genzel}, R., \& {Chapin}, E.~L. 2009,
  \apj, 701, 1814

\bibitem[{{Vega} {et~al.}(2008){Vega}, {Clemens}, {Bressan}, {Granato},
  {Silva}, \& {Panuzzo}}]{VegCle2008}
{Vega}, O., {Clemens}, M.~S., {Bressan}, A., {et~al.} 2008, \aap, 484, 631

\bibitem[{{Veilleux} {et~al.}(1999){Veilleux}, {Kim}, \&
  {Sanders}}]{VeiKim1999}
{Veilleux}, S., {Kim}, D.-C., \& {Sanders}, D.~B. 1999, \apj, 522, 113

\bibitem[{{Veilleux} {et~al.}(1995){Veilleux}, {Kim}, {Sanders}, {Mazzarella},
  \& {Soifer}}]{VeiKim1995}
{Veilleux}, S., {Kim}, D.-C., {Sanders}, D.~B., {Mazzarella}, J.~M., \&
  {Soifer}, B.~T. 1995, \apjs, 98, 171

\bibitem[{{Veilleux} {et~al.}(2009){Veilleux}, {Rupke}, {Kim}, {Genzel},
  {Sturm}, {Lutz}, {Contursi}, {Schweitzer}, {Tacconi}, {Netzer}, {Sternberg},
  {Mihos}, {Baker}, {Mazzarella}, {Lord}, {Sanders}, {Stockton}, {Joseph}, \&
  {Barnes}}]{VeiRup2009}
{Veilleux}, S., {Rupke}, D.~S.~N., {Kim}, D.-C., {et~al.} 2009, \apjs, 182, 628

\bibitem[{{Wang} {et~al.}(2009){Wang}, {Fabbiano}, {Elvis}, {Risaliti},
  {Mazzarella}, {Howell}, \& {Lord}}]{WanFab2009}
{Wang}, J., {Fabbiano}, G., {Elvis}, M., {et~al.} 2009, \apj, 694, 718

\bibitem[{{Weisskopf} {et~al.}(2000){Weisskopf}, {Tananbaum}, {Van Speybroeck},
  \& {O'Dell}}]{WeiTan2000}
{Weisskopf}, M.~C., {Tananbaum}, H.~D., {Van Speybroeck}, L.~P., \& {O'Dell},
  S.~L. 2000, in \procspie, Vol. 4012, X-Ray Optics, Instruments, and Missions
  III, ed. J.~E. {Truemper} \& B.~{Aschenbach}, 2--16

\bibitem[{{West}(1976)}]{Wes1976}
{West}, R.~M. 1976, \aap, 46, 327

\bibitem[{{Xu} {et~al.}(2015){Xu}, {Cao}, {Lu}, {Gao}, {Diaz-Santos},
  {Herrero-Illana}, {Meijerink}, {Privon}, {Zhao}, {Evans}, {K{\"o}nig},
  {Mazzarella}, {Aalto}, {Appleton}, {Armus}, {Charmandaris}, {Chu}, {Haan},
  {Inami}, {Murphy}, {Sanders}, {Schulz}, \& {van der Werf}}]{XuCao2015}
{Xu}, C.~K., {Cao}, C., {Lu}, N., {et~al.} 2015, \apj, 799, 11

\bibitem[{{Yeghiazaryan} {et~al.}(2016){Yeghiazaryan}, {Nazaryan}, \&
  {Hakobyan}}]{YegNaz2016}
{Yeghiazaryan}, A.~A., {Nazaryan}, T.~A., \& {Hakobyan}, A.~A. 2016, Journal of
  Astrophysics and Astronomy, 37, 1

\bibitem[{{Yuan} {et~al.}(2010){Yuan}, {Kewley}, \& {Sanders}}]{YuaKew2010}
{Yuan}, T.-T., {Kewley}, L.~J., \& {Sanders}, D.~B. 2010, \apj, 709, 884

\bibitem[{{Zhou} {et~al.}(2014){Zhou}, {Wu}, {Huang}, {Li}, {Zhou}, {Jia},
  {Lam}, \& {Zhu}}]{ZhoWu2014}
{Zhou}, Z.-M., {Wu}, H., {Huang}, L., {et~al.} 2014, Research in Astronomy and
  Astrophysics, 14, 1393

\bibitem[{{Zink} {et~al.}(2000){Zink}, {Lester}, {Doppmann}, \&
  {Harvey}}]{ZinLes2000}
{Zink}, E.~C., {Lester}, D.~F., {Doppmann}, G., \& {Harvey}, P.~M. 2000, \apjs,
  131, 413

\end{thebibliography}

\setcounter{table}{3}
\onecolumn
\afterpage{
\tabcolsep=0.15cm
\setlength{\LTleft}{-20cm plus -1fill}
\setlength{\LTright}{\LTleft}
\setcounter{table}{1}
\longtab{}{
\label{table:results}
\footnotesize
\centering
\begin{ThreePartTable}
\begin{longtable}[t]{clccccccccc}
\caption{X-ray spectral properties for the sample. } \\
\hline \hline
\begin{tabular}{c} No.\\ $\ $ \end{tabular}  & \begin{tabular}{c} Galaxy \\ $\ $ \end{tabular} & \begin{tabular}{c} $SX$\\ (1) \end{tabular} & \begin{tabular}{c} $HX$ \\ (2) \end{tabular} & \begin{tabular}{c} $HR$ \\ (3) \end{tabular}& \begin{tabular}{c} $F_{SX}$ \\ (4) \end{tabular} & \begin{tabular}{c} $F_{HX}$\\ (5) \end{tabular}  & \begin{tabular}{c} $L_{SX}$\\ (6) \end{tabular}  & \begin{tabular}{c} $L_{HX}$\\ (7) \end{tabular}  & \begin{tabular}{c} $SX/IR$\\ (8) \end{tabular}  & \begin{tabular}{c} $HX/IR$\\ (9) \end{tabular}  \\ \hline
\endfirsthead

\multicolumn{3}{c}%
{{\bfseries \tablename\ \thetable{} -- continued.}} \\
\hline \hline
\begin{tabular}{c} No.\\ $\ $ \end{tabular}  & \begin{tabular}{c} Galaxy \\ $\ $ \end{tabular} & \begin{tabular}{c} $SX$\\ (1) \end{tabular} & \begin{tabular}{c} $HX$ \\ (2) \end{tabular} & \begin{tabular}{c} $HR$ \\ (3) \end{tabular}& \begin{tabular}{c} $F_{SX}$ \\ (4) \end{tabular} & \begin{tabular}{c} $F_{HX}$\\ (5) \end{tabular}  & \begin{tabular}{c} $L_{SX}$\\ (6) \end{tabular}  & \begin{tabular}{c} $L_{HX}$\\ (7) \end{tabular}  & \begin{tabular}{c} $SX/IR$\\ (8) \end{tabular}  & \begin{tabular}{c} $HX/IR$\\ (9) \end{tabular}  \\ \hline
\endhead

\hline
\endfoot

\endlastfoot

45 & UGC 08387 & 15.47 $\pm$ 1.06 & 4.28 $\pm$ 0.59 & $-$0.57 $\pm$ 0.04 & 5.29 & 5.57 & 7.79 & 9.71 & $-$4.44 & $-$4.33 \\ 
47 & CGCG 436$-$030 & 9.01 $\pm$ 0.84 & 3.18 $\pm$ 0.54 & $-$0.48 $\pm$ 0.04 & 3.78 & 5.48 & 8.75 & 16.07 & $-$4.33 & $-$4.07 \\ 
49 & NGC 0695 & 15.58 $\pm$ 1.06 & 5.59 $\pm$ 0.66 & $-$0.47 $\pm$ 0.03 & 6.00 & 11.63 & 15.78 & 51.19 & $-$4.08 & $-$3.57 \\ 
50 & CGCG 043$-$099 & 2.60 $\pm$ 0.44 & 0.92 $\pm$ 0.32 & $-$0.48 $\pm$ 0.09 & 1.07 & 1.35 & 3.98 & 6.39 & $-$4.67 & $-$4.46 \\ 
51 & MCG+07$-$23$-$019 & 8.51 $\pm$ 0.45 & 1.17 $\pm$ 0.24 & $-$0.76 $\pm$ 0.04 & 3.55 & 1.93 & 10.91 & 7.10 & $-$4.17 & $-$4.35 \\ 
52 & NGC 6670 (E) & 5.76 $\pm$ 0.64 & 1.21 $\pm$ 0.35 & $-$0.65 $\pm$ 0.08 & 2.58 & 2.15 & 5.60 & 5.56 & $-$4.27 & $-$4.27 \\ 
52 & NGC 6670 (W) & 7.21 $\pm$ 0.72 & 2.93 $\pm$ 0.50 & $-$0.42 $\pm$ 0.04 & 3.31 & 6.18 & 7.24 & 22.73 & $-$3.95 & $-$3.45 \\ 
53 & UGC 02369 (S) & 11.16 $\pm$ 1.14 & 1.30 $\pm$ 0.48 & $-$0.79 $\pm$ 0.08 & 3.71 & 1.71 & 9.53 & 4.90 & $-$4.28 & $-$4.57 \\ 
54 & NGC 1614 & 38.57 $\pm$ 1.59 & 12.19 $\pm$ 0.91 & $-$0.52 $\pm$ 0.02 & 16.62 & 18.51 & 10.03 & 12.36 & $-$4.23 & $-$4.14 \\ 
56 & NGC 5331 (N) & 1.79 $\pm$ 0.38 & 0.73 $\pm$ 0.26 & $-$0.42 $\pm$ 0.10 & 0.95 & 1.31 & 2.89 & 4.86 & $-$4.06 & $-$3.84 \\ 
56 & NGC 5331 (S) & 4.37 $\pm$ 0.59 & 1.36 $\pm$ 0.36 & $-$0.53 $\pm$ 0.07 & 2.13 & 1.97 & 6.46 & 7.10 & $-$4.35 & $-$4.30 \\ 
57 & IRAS F06076$-$2139(N) & 1.96 $\pm$  0.37 & 0.97 $\pm$ 0.27 & $-$0.34 $\pm$ 0.08 & 0.93 & 1.48 & 3.60 & 6.26 & $-$4.68 & $-$4.44 \\ 
57 & IRAS F06076$-$2139(S) & 0.37 $\pm$ 0.17 & 0.35 $\pm$ 0.17 & $-$0.03 $\pm$ 0.17 &$-^{(a)}$ &$-^{(a)}$ &$-^{(a)}$ &$-^{(a)}$ &$-^{(a)}$ &$-^{(a)}$ \\ 
60 & IC 2810(NW) & 3.61 $\pm$ 0.50 & 0.18 $\pm$ 0.15 & $-$0.91 $\pm$ 0.13 & 1.67 & 0.55 & 4.27 & 2.10 & $-$4.33 & $-$4.73 \\ 
60 & IC 2810 (SE) & 1.75 $\pm$ 0.39 & 0.12 $\pm$ 0.19 & $-$0.87 $\pm$ 0.22 & 0.70 & 0.01 & 2.20 & 0.30 & $-$4.39 & $-$5.26 \\ 
63 & IRAS 18090+0130 (E) & 4.01 $\pm$ 0.55 & 1.27 $\pm$ 0.36 & $-$0.52 $\pm$ 0.08 & 1.59 & 2.50 & 4.08 & 6.39 & $-$4.53 & $-$4.34 \\ 
63 & IRAS 18090+0130(W) & 1.08 $\pm$ 0.30 & 0.33 $\pm$ 0.23 & $-$0.54 $\pm$ 0.18 &$-^{(a)}$ &$-^{(a)}$ &$-^{(a)}$ &$-^{(a)}$ &$-^{(a)}$ &$-^{(a)}$ \\ 
64 & III Zw 035 (N) & 2.34 $\pm$ 0.41 & 0.84 $\pm$ 0.26 & $-$0.47 $\pm$ 0.09 & 0.96 & 1.32 & 1.81 & 2.89 & $-$4.97 & $-$4.76 \\ 
65 & NGC 3256 & 263.36 $\pm$ 3.28 & 28.62 $\pm$ 1.67 & $-$0.80 $\pm$ 0.01 & 77.43 & 43.02 & 16.29 & 9.21 & $-$4.01 & $-$4.26 \\ 
67 & IRAS F16399$-$0937(N) & 4.02 $\pm$ 0.53 & 1.6 $\pm$ 0.34 & $-$0.43 $\pm$ 0.06 & 1.79 & 2.52 & 4.03 & 6.45 & $-$4.56 & $-$4.36 \\ 
67 & IRAS F16399$-$0937(S) & 2.06 $\pm$ 0.38 & 1.25 $\pm$ 0.31 & $-$0.24 $\pm$ 0.07 & 1.00 & 2.07 & 2.79 & 5.30 & $-$3.77 & $-$3.49 \\ 
68 & IRAS F16164$-$0746 & 3.91 $\pm$ 0.55 & 2.8 $\pm$ 0.52 & $-$0.17 $\pm$ 0.04 & 1.84 & 6.61 & 4.12 & 15.38 & $-$4.59 & $-$4.02 \\ 
69 & IC 4687 & 20.68 $\pm$ 1.21 & 4.27 $\pm$ 0.59 & $-$0.66 $\pm$ 0.04 & 8.60 & 5.97 & 7.97 & 5.30 & $-$4.11 & $-$4.30 \\
69 & IC 4686 & 4.87 $\pm$ 0.59 & 0.90 $\pm$ 0.28 & $-$0.69 $\pm$ 0.09 & 2.07 & 1.49 & 2.12 & 1.56 & $-$3.95 & $-$4.09 \\
69 & IC 4689 & 5.17 $\pm$ 0.63 & 0.00 $\pm$ 0.21 & $-$1.00 $\pm$ 0.14 & 2.45 & $-^{(a)}$ & 2.72 & $-^{(a)}$ & $-$4.11 & $-^{(a)}$ \\  
71 & NGC 2623 & 4.79 $\pm$ 0.52 & 3.85 $\pm$ 0.49 & $-$0.11 $\pm$ 0.02 & 1.74 & 7.93 & 1.49 & 12.85 & $-$5.01 & $-$4.08 \\ 
72 & IC 5298 & 11.76 $\pm$ 0.92 & 3.31 $\pm$ 0.56 & $-$0.56 $\pm$ 0.04 & 5.17 & 9.66 & 10.03 & 36.65 & $-$4.18 & $-$3.62 \\ 
73 & IRAS 20351+2521 & 9.14 $\pm$ 0.89 & 1.69 $\pm$ 0.52 & $-$0.69 $\pm$ 0.07 & 3.93 & 2.55 & 14.02 & 8.33 & $-$4.05 & $-$4.27 \\ 
75 & NGC 6090 (NE) & 16.71 $\pm$ 1.07 & 1.84 $\pm$ 0.37 & $-$0.80 $\pm$ 0.05 & 5.89 & 2.88 & 13.63 & 7.81 & $-$3.99 & $-$4.23 \\ 
75 & NGC6090 (SW) & 4.68 $\pm$ 0.57 & 0.56 $\pm$ 0.21 & $-$0.79 $\pm$ 0.10 & 1.71 & 3.11 & 4.00 & 9.01 & $-$3.55 & $-$3.20 \\ 
79 & NGC 5256 (NE) & 32.39 $\pm$ 1.31 & 7.97 $\pm$ 0.69 & $-$0.60 $\pm$ 0.02 & 11.62 & 18.06 & 24.23 & 55.81 & $-$3.33 & $-$2.97 \\ 
79 & NGC 5256 (SW) & 20.47 $\pm$ 1.02 & 1.7 $\pm$ 0.32 & $-$0.85 $\pm$ 0.04 & 7.98 & 3.30 & 16.56 & 7.72 & $-$3.72 & $-$4.05 \\ 
80 & IRAS F03359+1523(E) & 5.92 $\pm$ 0.67 & 1.41 $\pm$ 0.38 & $-$0.62 $\pm$ 0.07 & 2.22 & 2.30 & 8.68 & 8.32 & $-$4.20 & $-$4.21 \\ 
81 & ESO 550$-$IG025 (N) & 2.23 $\pm$ 0.41 & 0.55 $\pm$ 0.24 & $-$0.60 $\pm$ 0.12 & 0.73 & 1.17 & 1.78 & 3.58 & $-$4.51 & $-$4.22 \\ 
81 & ESO 550$-$IG025 (S) & 1.55 $\pm$ 0.35 & 0.76 $\pm$ 0.26 & $-$0.34 $\pm$ 0.10 & 0.69 & 1.51 & 1.65 & 4.29 & $-$4.46 & $-$4.16 \\ 
82 & NGC 0034 & 15.58 $\pm$ 1.08 & 6.68 $\pm$ 0.76 & $-$0.40 $\pm$ 0.03 & 7.23 & 12.97 & 6.40 & 17.78 & $-$4.27 & $-$3.83 \\ 
83 & MCG+12$-$02$-$001 (E) & 11.12 $\pm$ 0.89 & 3.77 $\pm$ 0.53 & $-$0.49 $\pm$ 0.04 & 4.66 & 5.59 & 3.01 & 3.93 & $-$4.56 & $-$4.44 \\ 
83 & MCG+12$-$02$-$001 (W) & 3.02 $\pm$ 0.46 & 1.49 $\pm$ 0.34 & $-$0.34 $\pm$ 0.06 & 1.29 & 2.92 & 1.26 & 2.39 & $-$3.98 & $-$3.71 \\ 
85 & IRAS F17138$-$1017 & 7.84 $\pm$ 0.76 & 6.19 $\pm$ 0.71 & $-$0.12 $\pm$ 0.02 & 3.67 & 12.68 & 3.94 & 17.64 & $-$4.48 & $-$3.83 \\ 
95 & ESO 440$-$IG058 (N) & 2.12 $\pm$ 0.38 & 1.02 $\pm$ 0.27 & $-$0.35 $\pm$ 0.08 & 0.99 & 1.27 & 1.72 & 2.22 & $-$3.74 & $-$3.63 \\ 
95 & ESO 440$-$IG058 (S) & 7.16 $\pm$ 0.70 & 0.97 $\pm$ 0.28 & $-$0.76 $\pm$ 0.08 & 3.40 & 1.53 & 5.60 & 2.97 & $-$4.23 & $-$4.50 \\ 
100 & NGC 7130 & 78.11 $\pm$ 1.44 & 8 $\pm$ 0.53 & $-$0.81 $\pm$ 0.01 & 25.37 & 13.28 & 16.89 & 13.29 & $-$3.78 & $-$3.88 \\ 
104 & NGC 7771 & 33.16 $\pm$ 1.54 & 12.55 $\pm$ 1.14 & $-$0.45 $\pm$ 0.02 & 12.87 & 20.19 & 6.26 & 12.35 & $-$4.15 & $-$3.85 \\ 
104 & NGC 7770 & 8.08 $\pm$0.73 & 0.34 $\pm$ 0.29 & $-$0.92 $\pm$ 0.09 & 3.15 & 1.27 & 1.59 & 0.74 & $-$3.77 & $-$4.11 \\ 
105 & NGC 7592 (E) & 8.42 $\pm$ 0.76 & 1.11 $\pm$ 0.32 & $-$0.77 $\pm$ 0.07 & 3.35 & 1.80 & 4.94 & 3.13 & $-$4.09 & $-$4.29 \\ 
105 & NGC 7592 (W) & 11.29 $\pm$ 0.88 & 2.04 $\pm$ 0.41 & $-$0.69 $\pm$ 0.05 & 4.21 & 3.15 & 6.23 & 5.87 & $-$3.75 & $-$3.78 \\ 
106 & NGC 6285 & 5.04 $\pm$ 0.64 & 0.45 $\pm$ 0.33 & $-$0.83 $\pm$ 0.12 & 1.94 & 0.85 & 1.73 & 0.97 & $-$3.83 & $-$4.08 \\ 
106 & NGC 6286 & 32.15 $\pm$ 1.64 & 1.27 $\pm$ 0.87 & $-$0.92 $\pm$ 0.05 & 11.96 & 2.58 & 10.93 & 2.92 & $-$3.86 & $-$4.43 \\ 
107 & NGC 4922 (N) & 10.93 $\pm$ 0.97 & 2.73 $\pm$ 0.62 & $-$0.60 $\pm$ 0.05 & 5.00 & 5.93 & 7.49 & 17.28 & $-$4.09 & $-$3.73 \\ 
110 & NGC 3110 & 21.92 $\pm$ 1.28 & 4.73 $\pm$ 0.77 & $-$0.65 $\pm$ 0.04 & 10.32 & 8.40 & 8.54 & 8.70 & $-$3.98 & $-$3.97 \\ 
114 & NGC 0232 & 10.91 $\pm$ 0.91 & 2.19 $\pm$ 0.55 & $-$0.67 $\pm$ 0.06 & 4.85 & 4.34 & 5.37 & 6.58 & $-$4.29 & $-$4.21 \\ 
117 & MCG+08$-$18$-$013(E) & 5.27 $\pm$ 0.67 & 2.1 $\pm$ 0.45 & $-$0.43 $\pm$ 0.06 & 2.43 & 3.56 & 4.07 & 7.84 & $-$4.32 & $-$4.03 \\ 
120 & CGCG 049$-$057 & 0.83 $\pm$ 0.28 & 0.76 $\pm$ 0.29 & $-$0.04 $\pm$ 0.09 & 0.37 & 1.26 & 0.20 & 1.02 & $-$5.63 & $-$4.93 \\ 
121 & NGC 1068 & 1975.11 $\pm$ 6.58 & 150.27 $\pm$ 2.03 & $-$0.86 $\pm$ 0.00 & 503.35 & 223.51 & 16.31 & 10.15 & $-$3.77 & $-$3.98 \\ 
123 & UGC 02238 & 6.94 $\pm$ 0.74 & 1.94 $\pm$ 0.52 & $-$0.56 $\pm$ 0.06 & 2.89 & 4.33 & 3.77 & 5.88 & $-$4.34 & $-$4.15 \\ 
127 & MCG$-$03$-$34$-$064 & 95.67 $\pm$ 3.73 & 49.5 $\pm$ 2.75 & $-$0.32 $\pm$ 0.01 & 37.92 & 158.12 & 35.11 & 289.53 & $-$3.32 & $-$2.40 \\ 
134 & ESO 350$-$IG038 & 26.85 $\pm$ 0.73 & 6.38 $\pm$ 0.43 & $-$0.62 $\pm$ 0.02 & 10.38 & 10.03 & 10.34 & 12.12 & $-$3.85 & $-$3.78 \\ 
136 & MCG$-$01$-$60$-$022 & 10.52 $\pm$ 0.84 & 6.7 $\pm$ 0.78 & $-$0.22 $\pm$ 0.02 & 4.30 & 16.83 & 5.42 & 33.62 & $-$4.12 & $-$3.33 \\ 
141 & IC 0564 & 7.14 $\pm$ 0.77 & 2.45 $\pm$ 0.53 & $-$0.50 $\pm$ 0.06 & 3.61 & 5.18 & 3.73 & 9.29 & $-$3.97 & $-$3.57 \\ 
141 & IC 0563 & 4.93 $\pm$ 0.64 & 2.36 $\pm$ 0.57 & $-$0.34 $\pm$ 0.05 & 2.56 & 4.74 & 2.64 & 6.90 & $-$4.05 & $-$3.63 \\ 
142 & NGC 5135 & 123.31 $\pm$ 2.14 & 12.39 $\pm$ 0.9 & $-$0.82 $\pm$ 0.01 & 39.38 & 26.95 & 19.69 & 42.37 & $-$3.59 & $-$3.26 \\ 
144 & IC 0860 & 0.68 $\pm$ 0.26 & 0.37 $\pm$ 0.27 & $-$0.29 $\pm$ 0.17 &$-^{(a)}$ &$-^{(a)}$ &$-^{(a)}$ &$-^{(a)}$ &$-^{(a)}$ &$-^{(a)}$ \\ 
147 & IC 5179 & 39.74 $\pm$ 2.14 & 6.71 $\pm$ 1.62 & $-$0.71 $\pm$ 0.04 & 15.51 & 14.38 & 4.99 & 6.35 & $-$4.13 & $-$4.02 \\ 
148 & CGCG 465$-$012 & 7.13 $\pm$ 0.77 & 1.52 $\pm$ 0.58 & $-$0.65 $\pm$ 0.08 & 2.63 & 4.01 & 3.65 & 5.64 & $-$4.22 & $-$4.04 \\ 
163 & NGC 4418 & 2.63 $\pm$ 0.59 & 0.38 $\pm$ 0.32 & $-$0.75 $\pm$ 0.18 & 0.87 & 0.91 & 0.15 & 0.19 & $-$5.61 & $-$5.50 \\ 
157 & MCG$-$02$-$33$-$098 & 2.97 $\pm$ 0.46 & 0.77 $\pm$ 0.26 & $-$0.59 $\pm$ 0.10 & 1.39 & 1.21 & 1.11 & 1.16 & $-$4.31 & $-$3.41 \\ 
157 & MCG$-$02$-$33$-$099 & 4.42 $\pm$ 0.56 & 1.32 $\pm$ 0.33 & $-$0.54 $\pm$ 0.07 & 2.19 & 2.25 & 1.76 & 2.15 & $-$4.37 & $-$4.44 \\ 
169 & ESO 343$-$IG013 (N) & 2.02 $\pm$ 0.38 & 2.05 $\pm$ 0.38 & 0.01 $\pm$ 0.05 & 0.91 & 4.45 & 0.85 & 6.74 & $-$4.70 & $-$3.80 \\ 
169 & ESO 343$-$IG013 (S) & 3.15 $\pm$ 0.47 & 1.02 $\pm$ 0.3 & $-$0.51 $\pm$ 0.08 & 1.67 & 1.24 & 1.62 & 1.36 & $-$3.86 & $-$3.93 \\ 
170 & NGC 2146 & 170.25 $\pm$ 4.38 & 43.74 $\pm$ 2.47 & $-$0.59 $\pm$ 0.01 & 59.70 & 75.77 & 2.50 & 3.76 & $-$4.31 & $-$4.13 \\ 
174 & NGC 5653 & 20.26 $\pm$ 1.20 & 3.18 $\pm$ 0.7 & $-$0.73 $\pm$ 0.04 & 8.00 & 5.37 & 3.59 & 3.44 & $-$4.16 & $-$4.18 \\ 
178 & NGC 4194 & 59.99 $\pm$ 1.33 & 7.91 $\pm$ 0.57 & $-$0.77 $\pm$ 0.02 & 22.87 & 12.40 & 5.23 & 3.37 & $-$3.97 & $-$4.16 \\ 
179 & NGC 7591 & 4.29 $\pm$ 1.01 & 1.09 $\pm$ 0.74 & $-$0.59 $\pm$ 0.17 &$-^{(a)}$ &$-^{(a)}$ &$-^{(a)}$ &$-^{(a)}$ &$-^{(a)}$ &$-^{(a)}$ \\ 
182 & NGC 0023 & 34.84 $\pm$ 1.45 & 3.88 $\pm$ 0.77 & $-$0.80 $\pm$ 0.03 & 12.89 & 6.04 & 7.12 & 3.47 & $-$3.85 & $-$4.17 \\ 
188 & NGC 7552 & 148.31 $\pm$ 5.56 & 15.63 $\pm$ 2.09 & $-$0.81 $\pm$ 0.03 & 55.45 & 24.24 & 3.76 & 1.92 & $-$4.12 & $-$4.41 \\ 
191 & ESO 420$-$G013 & 56.94 $\pm$ 2.19 & 4.17 $\pm$ 0.83 & $-$0.86 $\pm$ 0.03 & 23.59 & 5.81 & 7.86 & 2.04 & $-$3.76 & $-$4.34 \\ 
194 & ESO 432$-$IG006 (NE) & 8.09 $\pm$ 0.76 & 1.75 $\pm$ 0.48 & $-$0.64 $\pm$ 0.06 & 3.69 & 4.46 & 3.82 & 6.41 & $-$3.65 & $-$3.43 \\ 
194 & ESO 432$-$IG006 (SW) & 6.04 $\pm$ 0.66 & 1.81 $\pm$ 0.46 & $-$0.54 $\pm$ 0.06 & 2.79 & 4.03 & 2.87 & 4.53 & $-$4.01 & $-$3.81 \\ 
195 & NGC 1961 & 17.55 $\pm$ 0.89 & 4.48 $\pm$ 0.83 & $-$0.59 $\pm$ 0.03 & 12.45 & 15.65 & 6.22 & 10.63 & $-$3.85 & $-$3.62 \\ 
196 & NGC 7753 & 15.00 $\pm$ 1.65 & 4.43 $\pm$ 1.67 & $-$0.54 $\pm$ 0.07 & 4.52 & 7.75 & 3.36 & 5.88 & $-$3.94 & $-$3.70 \\ 
196 & NGC 7752 & 7.18 $\pm$ 0.83 & 1.21 $\pm$ 0.53 & $-$0.71 $\pm$ 0.09 & 2.72 & 1.83 & 1.97 & 1.53 & $-$3.92 & $-$4.03 \\ 
198 & NGC 1365 & 177.35 $\pm$ 3.63 & 121.5 $\pm$ 2.95 & $-$0.19 $\pm$ 0.00 & 66.03 & 374.52 & 2.62 & 28.89 & $-$4.17 & $-$3.12 \\ 
199 & NGC 3221 & 12.47 $\pm$ 1.08 & 4.53 $\pm$ 1.02 & $-$0.47 $\pm$ 0.04 & 4.87 & 12.20 & 2.59 & 12.26 & $-$4.26 & $-$3.59 \\ 
201 & NGC 0838 & 30.31 $\pm$ 0.75 & 3.91 $\pm$ 0.39 & $-$0.77 $\pm$ 0.02 & 14.91 & 6.50 & 5.50 & 2.70 & $-$3.89 & $-$4.20 \\ 
\hline \\
\end{longtable}
\begin{tablenotes}
\item \textbf{Notes:} Column (1): background corrected count rate in the 0.5$-$2 keV band in units of 10$^{-3}$ ct s$^{-1}$. Column (2): background corrected count rate in the 2$-$7 keV band in units of 10$^{-3}$ ct s$^{-1}$. Column (3): X-ray colour as defined by $HR = (H-S)/(H+S)$. Column (4): observed 0.5$-$2 keV band flux in units of 10$^{14}$ erg s$^{-1}$ cm$^{-2}$. Column (5): observed 2$-$7 keV band flux in units of 10$^{14}$ erg s$^{-1}$ cm$^{-2}$. Column (6): 0.5$-$2 keV band luminosity corrected for Galactic absorption in units of 10$^{40}$ erg s$^{-1}$. Column (7): 2$-$10 keV band luminosity corrected for Galactic absorption in units of 10$^{40}$ erg s$^{-1}$. Column (8): logarithmic luminosity ratio of the 0.5$-$2 keV and 8$-$1000 $\mu$m bands. Column (9): logarithmic luminosity ratio of the 2$-$10 keV and 8$-$1000 $\mu$m bands. $^{(a)}$  Missing values due to inability to fit X-ray spectra, not enough source counts. \newline
 \end{tablenotes}\end{ThreePartTable}
}
} 

\setcounter{table}{7}
\afterpage{
\tabcolsep=0.15cm
\setlength{\LTleft}{-20cm plus -1fill}
\setlength{\LTright}{\LTleft}
\setcounter{table}{5}
\longtab{}{
\label{table:parameters}
\footnotesize
\centering
\begin{ThreePartTable}
\begin{longtable}[t]{rlcc|ccc}
\caption{X-ray spectral properties for the sample. } \\
\hline \hline
\begin{tabular}{c} No.\\ $\ $ \end{tabular}  & \begin{tabular}{c} Galaxy \\ $\ $ \end{tabular} & \begin{tabular}{c} $\Gamma_{\rm H}$\\ (1) \end{tabular} & \begin{tabular}{c} $N_{\rm H}$ \\ (2) \end{tabular} & \begin{tabular}{c} $T_1$ \\ (3) \end{tabular}& \begin{tabular}{c} $T_2$ \\ (4) \end{tabular} & \begin{tabular}{c} $N_{\rm H}$\\ (5) \end{tabular}  \\ \hline
\endfirsthead

\multicolumn{3}{c}%
{{\bfseries \tablename\ \thetable{} -- continued.}} \\
\hline \hline
\begin{tabular}{c} No.\\ $\ $ \end{tabular}  & \begin{tabular}{c} Galaxy \\ $\ $ \end{tabular} & \begin{tabular}{c} $\Gamma_{\rm H}$\\ (1) \end{tabular} & \begin{tabular}{c} $N_{\rm H}$ \\ (2) \end{tabular} & \begin{tabular}{c} $T_1$ \\ (3) \end{tabular}& \begin{tabular}{c} $T_2$ \\ (4) \end{tabular} & \begin{tabular}{c} $N_{\rm H}$\\ (5) \end{tabular}  \\ \hline
\endhead

\hline
\endfoot

\endlastfoot

45 & UGC 08387 & 2.4$_{-0.5}^{+0.6}$ & & 0.35$_{-0.13}^{+0.29}$ &0.72$_{-0.14}^{+0.21}$ &1.09$_{-0.21}^{+0.21}$ \\ 
47 & CGCG 436$-$030 & 1.7$_{-0.6}^{+0.6}$ & & 0.68$_{-0.07}^{+0.12}$ &1.42$_{-0.33}^{+0.4}$ &1.5$_{-0.42}^{+0.62}$ \\ 
49 & NGC 0695 & 0.4$_{-0.4}^{+0.4}$ & & 0.36$_{-0.04}^{+0.10}$ &1.06$_{-0.12}^{+0.16}$ &1.25$_{-0.18}^{+0.2}$ \\ 
50 & CGCG 043$-$099 & 2.0 & & 0.75$_{-0.53}^{+0.59}$ &1.46$_{-0.99}^{+78.44}$ &0.98$_{-0.98}^{+2.11}$ \\ 
51 & MCG+07$-$23$-$019 & 2.2$_{-0.7}^{+0.8}$ & & 0.51$_{-0.04}^{+0.07}$ &1.03$_{-0.15}^{+0.27}$ &1.36$_{-0.23}^{+0.28}$ \\ 
52 & NGC 6670 (E) & 2.0 & & 0.46$_{-0.08}^{+0.11}$ & 0.90$_{-0.51}^{+0.16}$ & 1.82$_{-0.31}^{+0.42}$ \\ 
52 & NGC 6670 (W) & 0.5$_{-0.6}^{+0.5}$ & &  0.29$_{-0.06}^{+0.06}$ & 0.55$_{-0.15}^{+0.11}$& 1.15$_{-0.26}^{+0.39}$ \\ 
53 & UGC 02369 (S) & 2.0 & & 0.54$_{-0.18}^{+0.15}$ &1.12$_{-0.23}^{+0.33}$ &0.78$_{-0.27}^{+0.45}$ \\ 
54 & NGC 1614 & 2.4$_{-0.2}^{+0.3}$ & & 0.63$_{-0.05}^{+0.06}$ &1.12$_{-0.08}^{+0.16}$ &1.57$_{-0.18}^{+0.13}$ \\ 
56 & NGC 5331 (N) & 2.1$_{-0.6}^{+0.7}$ & & $-^{(a)}$ & $-^{(a)}$ & $-^{(a)}$\\ 
56 & NGC 5331 (S) & 2.2$_{-1.0}^{+0.8}$ & & 0.33$_{-0.05}^{+0.07}$ &1.37$_{-0.42}^{+78.53}$ &1.18$_{-1.14}^{+0.57}$ \\ 
57 & IRAS F06076$-$2139(N) & 2.0 & & 0.27$_{-0.10}^{+0.39}$ &0.90$_{-0.34}^{+0.84}$ &0.84$_{-0.82}^{+0.64}$ \\ 
57 & IRAS F06076$-$2139(S) & $-^{(d)}$ & & $-^{(d)}$ & $-^{(d)}$ & $-^{(d)}$ \\ 
60 & IC 2810(NW) & 2.0 & & 0.44$_{-0.07}^{+0.07}$ &1.14$_{-0.17}^{+0.27}$ &0.93$_{-0.2}^{+0.19}$ \\ 
60 & IC 2810 (SE) & 2.0 & & 1.23$_{-0.27}^{+0.27}$ & $-^{(b)}$ & $-^{(b)}$\\ 
63 & IRAS 18090+0130 (E) & 2.7$_{-1.0}^{+1.0}$ & & 0.70$_{-0.52}^{+0.36}$ &4.18$_{-1.54}^{+14.52}$ &0.39$_{-0.39}^{+0.56}$ \\ 
63 & IRAS 18090+0130(W) & $-^{(d)}$ & & $-^{(d)}$ & $-^{(d)}$ & $-^{(d)}$ \\ 
64 & III Zw 035 (N) & 2.0 & & 0.63$_{-0.32}^{+0.19}$ &1.06$_{-0.29}^{+0.37}$ &2.44$_{-0.77}^{+1.3}$ \\ 
65 & NGC 3256 & 2.6$_{-0.2}^{+0.2}$ & & 0.40$_{-0.01}^{+0.02}$ &0.84$_{-0.03}^{+0.026}$ &1.04$_{-0.02}^{+0.03}$ \\ 
67 & IRAS F16399$-$0937(N) & 1.6$_{-1.2}^{+1.2}$ & & 0.33$_{-0.15}^{+0.77}$ & 0.97$_{-0.23}^{+0.20}$ & 1.44$_{-0.39}^{+3.64}$  \\ 
67 & IRAS F16399$-$0937(S) & 1.8$_{-0.3}^{+0.3}$ & & $-^{(a)}$ & $-^{(a)}$ & $-^{(a)}$ \\ 
68 & IRAS F16164$-$0746 & 2.8$_{-1.5}^{+1.4}$ & & 0.71$_{-0.71}^{+0.40}$ &0.90$_{-0.27}^{+0.24}$ &1.68$_{-0.55}^{+0.73}$ \\ 
69 & IC 4687 & 3.4$_{-0.5}^{+0.5}$ & & 0.96$_{-0.14}^{+0.11}$ & 5.84$_{-2.61}^{+23.86}$ & 0.17$_{-0.08}^{+0.11}$ \\ 
69 & IC 4686 & 2.0 & & 0.49$_{-0.19}^{+0.15}$ & 1.11$_{-0.17}^{+0.66}$ & 1.58$_{-0.48}^{+0.38}$ \\ 
69 & IC 4689 & 2.0 & & 0.34$_{-0.06}^{+0.12}$ & $-^{(b)}$ & $-^{(b)}$ \\ 
71 & NGC 2623 & 0.3$_{-0.4}^{+0.4}$ & & 0.66$_{-0.28}^{+2.99}$ & 4.45$_{-2.22}^{+75.45}$ &  0.35$_{-0.23}^{+0.34}$\\ 
72 & IC 5298 & 1.8 & 4.80$_{-1.34}^{+1.45}$ & 0.63$_{-0.49}^{+0.16}$ &0.76$_{-0.15}^{+0.22}$ &1.43$_{-0.39}^{+0.48}$ \\ 
73 & IRAS 20351+2521 & 2.3$_{-1.9}^{+1.7}$ & & 0.60$_{-0.11}^{+0.08}$ &1.37$_{-0.45}^{+0.91}$ &1.03$_{-0.57}^{+0.51}$ \\ 
75 & NGC 6090 (NE) & 2.4$_{-0.6}^{+0.7}$ & & 0.66$_{-0.05}^{+0.07}$ &1.20$_{-0.16}^{+0.21}$ &1.18$_{-0.25}^{+0.28}$ \\ 
75 & NGC6090 (SW) & 2.0 & & 0.35$_{-0.09}^{+0.29}$ &0.71$_{-0.39}^{+0.59}$ &1.72$_{-0.41}^{+3.04}$ \\ 
79 & NGC 5256 (NE) & 1.8 & 0.72$_{-0.17}^{+0.18}$ & 0.29$_{-0.02}^{+0.02}$ &0.82$_{-0.08}^{+0.22}$ &0.96$_{-0.11}^{+0.12}$ \\ 
79 & NGC 5256 (SW) & 2.3$_{-0.9}^{+1.1}$ & & 0.24$_{-0.05}^{+0.02}$ &0.74$_{-0.08}^{+0.08}$ &0.63$_{-0.1}^{+0.1}$ \\ 
80 & IRAS F03359+1523(E) & 2.0 & & 0.45$_{-0.16}^{+0.40}$ &4.63$_{-3.69}^{+75.27}$ &0.25$_{-0.25}^{+1.28}$ \\ 
81 & ESO 550$-$IG025 (N) & 2.0 & & 0.78$_{-0.27}^{+0.34}$ &2.38$_{-0.95}^{+6.07}$ &0.83$_{-0.83}^{+3.84}$ \\ 
81 & ESO 550$-$IG025 (S) & 1.6$_{-0.5}^{+0.5}$ & & $-^{(a)}$ & $-^{(a)}$ & $-^{(a)}$ \\ 
82 & NGC 0034 & 1.8 & 1.05$_{-0.71}^{+0.73}$ & 0.38$_{-0.05}^{+0.06}$ &1.74$_{-0.34}^{+0.62}$ &0.79$_{-0.32}^{+0.3}$ \\ 
83 & MCG+12$-$02$-$001 (E) & 2.3$_{-0.4}^{+0.4}$ & & 0.71$_{-0.12}^{+0.14}$ &1.21$_{-0.22}^{+0.15}$ &1.3$_{-0.18}^{+0.23}$ \\ 
83 & MCG+12$-$02$-$001 (W) & 1.7$_{-0.3}^{+0.3}$ & & $-^{(a)}$ & $-^{(a)}$ & $-^{(a)}$ \\ 
85 & IRAS F17138$-$1017 & 1.1$_{-0.3}^{+0.3}$ & & 0.31$_{-0.07}^{+0.17}$ &1.00$_{-0.13}^{+0.14}$ &1.71$_{-0.24}^{+0.27}$ \\ 
95 & ESO 440$-$IG058 (N) & $-^{(c)}$  & & 3.91$_{-1.63}^{+8.85}$ & $-^{(c)}$ & $-^{(c)}$ \\ 
95 & ESO 440$-$IG058 (S) & 2.0 & & 0.25$_{-0.07}^{+0.08}$ &0.77$_{-0.09}^{+0.14}$ &0.81$_{-0.16}^{+0.17}$ \\ 
100 & NGC 7130 & 1.8 & 3.47$_{-1.84}^{+1.40}$& 0.30$_{-0.01}^{+0.01}$ &0.78$_{-0.03}^{+0.04}$ &0.89$_{-0.05}^{+0.04}$ \\ 
104 & NGC 7771 & 1.6$_{-0.3}^{+0.3}$ & & 0.52$_{-0.04}^{+0.08}$ &1.04$_{-0.09}^{+0.16}$ &1.7$_{-0.18}^{+0.21}$ \\ 
104 & NGC 7770 & 2.0 & & 0.36$_{-0.04}^{+0.06}$ &1.07$_{-0.24}^{+0.25}$ &1.1$_{-0.32}^{+0.43}$ \\ 
105 & NGC 7592 (E) & 2.0 & & 0.31$_{-0.06}^{+0.04}$ &0.72$_{-0.11}^{+0.14}$ &1.22$_{-0.26}^{+0.3}$ \\ 
105 & NGC 7592 (W) & 1.6$_{-0.8}^{+0.7}$ & & 0.36$_{-0.12}^{+0.30}$ &0.75$_{-0.27}^{+0.42}$ &1.35$_{-0.54}^{+0.98}$ \\ 
106 & NGC 6285 & 2.0 & & 2.96$_{-1.00}^{+3.74}$ & $-^{(b)}$ & $-^{(b)}$ \\ 
106 & NGC 6286 & 2.0 & & 0.37$_{-0.08}^{+0.05}$ &0.78$_{-0.16}^{+0.13}$ &0.57$_{-0.13}^{+0.13}$ \\ 
107 & NGC 4922 (N) & 1.8 & 2.94$_{-1.15}^{+1.41}$ & 0.60$_{-0.35}^{+0.10}$ & 0.58$_{-0.08}^{+0.16}$ & 1.34$_{-0.35}^{+2.04}$\\ 
110 & NGC 3110 & 1.7$_{-0.5}^{+0.5}$ & & 0.34$_{-0.03}^{+0.06}$ &0.89$_{-0.10}^{+0.09}$ &0.95$_{-0.14}^{+0.17}$ \\ 
114 & NGC 0232 & 1.6$_{-0.6}^{+0.7}$ & & 0.37$_{-0.08}^{+0.15}$ &0.91$_{-0.11}^{+0.12}$ &0.77$_{-0.18}^{+0.16}$ \\ 
117 & MCG+08$-$18$-$013(E) & 1.8$_{-0.8}^{+0.8}$ & & 0.43$_{-0.33}^{+2.16}$ &2.93$_{-1.45}^{+8.25}$ &0.63$_{-0.6}^{+0.84}$ \\ 
120 & CGCG 049$-$057 & 1.0$_{-0.6}^{+0.7}$ & & $-^{(a)}$ & $-^{(a)}$ & $-^{(a)}$ \\ 
121 & NGC 1068 & 2.0$_{-0.1}^{+0.1}$ & & 0.23$_{-0.00}^{+0.00}$ &0.66$_{-0.00}^{+0.01}$ &0.77$_{-0.01}^{+0.01}$ \\ 
123 & UGC 02238 & 1.8$_{-0.2}^{+0.2}$ & & $-^{(a)}$ & $-^{(a)}$ & $-^{(a)}$ \\ 
127 & MCG$-$03$-$34$-$064 & 1.8 & 5.02$_{-0.39}^{+0.40}$ & 0.28$_{-0.02}^{+0.02}$ &1.02$_{-0.06}^{+0.07}$ &0.9$_{-0.09}^{+0.1}$ \\ 
134 & ESO 350$-$IG038 & 2.0$_{-0.2}^{+0.2}$ && 0.33$_{-0.01}^{+0.02}$ &0.97$_{-0.05}^{+0.04}$ &1.19$_{-0.07}^{+0.07}$ \\ 
136 & MCG$-$01$-$60$-$022 & 1.8 & 1.10$_{-0.25}^{+0.27}$ & 0.30$_{-0.07}^{+0.09}$ &0.68$_{-0.10}^{+0.06}$ &1.08$_{-0.18}^{+0.21}$ \\ 
141 & IC 0564 & 0.7$_{-0.8}^{+0.9}$ & & 0.29$_{-0.06}^{+0.07}$ & 0.56$_{-0.08}^{+0.12}$ & 1.42$_{-0.32}^{+0.31}$ \\ 
141 & IC 0563 & 1.5$_{-0.6}^{+0.6}$ & & 0.65$_{-0.07}^{+0.11}$ & 1.93$_{-0.74}^{+77.97}$ & 3.28$_{-2.73}^{+2.54}$ \\ 
142 & NGC 5135 & 2.0 & & 0.25$_{-0.02}^{+0.02}$ &0.68$_{-0.04}^{+0.09}$ &0.73$_{-0.06}^{+0.06}$ \\ 
144 & IC 0860 & $-^{(d)}$ & & $-^{(d)}$ & $-^{(d)}$ & $-^{(d)}$ \\ 
147 & IC 5179 & 1.6$_{-0.5}^{+0.5}$ & & 0.67$_{-0.05}^{+0.06}$ &4.28$_{-1.12}^{+2.38}$ &0.01$_{-0.01}^{+0.04}$ \\ 
148 & CGCG 465$-$012 & 2.0$_{-1.1}^{+0.9}$ & & 0.38$_{-0.12}^{+0.22}$ &3.75$_{-2.15}^{+12.13}$ &0.44$_{-0.37}^{+0.76}$ \\ 
157 & MCG$-$02$-$33$-$098 & 2.0 & & 0.32$_{-0.08}^{+0.38}$ & 0.99$_{-0.22}^{+0.36}$ & 1.42$_{-0.35}^{+0.49}$ \\ 
157 & MCG$-$02$-$33$-$099 & 2.0 & & 0.25$_{-0.04}^{+0.05}$ & 1.02$_{-0.17}^{+0.38}$ & 1.25$_{-0.31}^{+0.28}$ \\ 
163 & NGC 4418 & 1.9$_{-0.5}^{+0.5}$ & & $-^{(a)}$ & $-^{(a)}$ & $-^{(a)}$ \\ 
169 & ESO 343$-$IG013 (N) & 0.9$_{-0.3}^{+0.4}$ & & $-^{(a)}$ & $-^{(a)}$ & $-^{(a)}$ \\ 
169 & ESO 343$-$IG013 (S) & 2.2$_{-0.4}^{+0.5}$ & & $-^{(a)}$ & $-^{(a)}$ & $-^{(a)}$ \\ 
170 & NGC 2146 & 1.7$_{-0.2}^{+0.2}$ & & 0.38$_{-0.02}^{+0.03}$ &0.72$_{-0.05}^{+0.06}$ &1.18$_{-0.07}^{+0.07}$ \\ 
174 & NGC 5653 & 1.4$_{-0.9}^{+0.8}$ & & 0.32$_{-0.02}^{+0.03}$ &0.80$_{-0.12}^{+0.15}$ &0.96$_{-0.14}^{+0.18}$ \\ 
178 & NGC 4194 & 2.3$_{-0.3}^{+0.3}$ & & 0.34$_{-0.01}^{+0.01}$ &0.87$_{-0.04}^{+0.1}$ &0.87$_{-0.06}^{+0.06}$ \\ 
179 & NGC 7591 & $-^{(d)}$ & & $-^{(d)}$ & $-^{(d)}$ & $-^{(d)}$ \\ 
182 & NGC 0023 & 3.0$_{-0.7}^{+0.6}$ & & 0.30$_{-0.07}^{+0.04}$ &0.84$_{-0.07}^{+0.05}$ &0.69$_{-0.06}^{+0.12}$ \\ 
188 & NGC 7552 & 2.4$_{-0.4}^{+0.4}$ & & 0.49$_{-0.05}^{+0.04}$ &0.87$_{-0.11}^{+0.1}$ &1.23$_{-0.13}^{+0.15}$ \\ 
191 & ESO 420$-$G013 & 2.9$_{-0.3}^{+0.4}$ & & 0.25$_{-0.02}^{+0.01}$ &0.77$_{-0.09}^{+0.06}$ &0.58$_{-0.09}^{+0.09}$ \\ 
194 & ESO 432$-$IG006 (NE) & 1.8 & 4.32$_{-1.78}^{+1.96}$& 0.65$_{-0.14}^{+0.10}$ &0.73$_{-0.51}^{+0.89}$ &1.99$_{-0.97}^{+6.05}$ \\ 
194 & ESO 432$-$IG006 (SW) & 1.8 & 1.22$_{-0.81}^{+0.76}$& 0.27$_{-0.07}^{+0.11}$ &0.75$_{-0.15}^{+0.35}$ &0.96$_{-0.2}^{+0.25}$ \\ 
195 & NGC 1961 & 1.0$_{-0.3}^{+0.4}$ & & 0.64$_{-0.04}^{+0.04}$ &4.70$_{-1.74}^{+5.6}$ &0.23$_{-0.15}^{+0.25}$ \\ 
196 & NGC 7753 & $-^{(c)}$ & & 6.48$_{-2.70}^{+19.82}$ & $-^{(c)}$ & $-^{(c)}$ \\ 
196 & NGC 7752 & 2.0 & & 0.49$_{-0.16}^{+0.12}$ & 0.70$_{-0.12}^{+0.28}$ & 1.46$_{-0.47}^{+0.49}$ \\ 
198 & NGC 1365 & 1.8 & 3.42$_{-0.13}^{+0.13}$ & 0.33$_{-0.01}^{+0.01}$ &0.96$_{-0.05}^{+0.04}$ &0.94$_{-0.06}^{+0.06}$ \\ 
199 & NGC 3221 & 0.3$_{-0.6}^{+0.6}$ & & 0.43$_{-0.09}^{+0.16}$ &0.70$_{-0.10}^{+0.15}$ &1.44$_{-0.18}^{+0.3}$ \\ 
201 & NGC 0838 & 2.5$_{-0.3}^{+0.3}$ & & 0.36$_{-0.03}^{+0.02}$ &1.03$_{-0.06}^{+0.05}$ &0.87$_{-0.07}^{+0.07}$ \\ 

\hline \\
\end{longtable}
\begin{tablenotes}
\item \textbf{Notes:} Column (1): Spectral power-law slope in the $2-7$ keV range. Column (2): Obscuring column density for galaxies with an absorbed AGN model fit in the $2-7$ keV range, in units of $10^{23}$ cm$^{-2}$. Column (3): external \textit{mekal} model temperature. Column (4): Internal, absorbed \textit{mekal} model temperature. Column (5): Obscuring column density associated with the internal \textit{mekal} model, in units of $10^{22}$ cm$^{-2}$. $^{(a)}$ Full spectrum fitted with a single power-law. $^{(b)}$ Soft band spectrum fitted with a single \textit{mekal} component. $^{(c)}$ Full spectrum fitted with  a single \textit{mekal} component.  $^{(d)}$ No fit. \newline
Values without errors are imposed, as described in section \ref{HBfit}. \newline
Errors reported correspond to $1\sigma$ for one parameter of interest, leaving 5 parameters free. 
 \end{tablenotes}
\end{ThreePartTable}
}
} 

\begin{appendix}

\twocolumn

\section{Notes on individual objects}\label{NotesIO} 

\textbf{[45] UGC 08387}: This sources meets our [Ne v] 14.32 $\mu$m line selection criterion, and is thus classified as an AGN, though there is no hint of its presence in the X-ray \textit{Chandra} data. As already seen in \citet{IwaSan2011}, a soft X-ray nebulae extends perpendicular to the plane of the galaxy; most likely associated with a galactic-scale outflow. 

Previous evidence of the AGN presence has come from detection of compact radio sources at mili-arcsec resolution \citep[e.g.][]{LonSmi1993,ParCon2010}, which \citet{RomPer2012} attributed to the presence of various supernovae in coexistence with a low-luminosity AGN. Using VLBI data, \citet{RomAlb2017} provide evidence for the presence of a pc-scale radio jet.

\textbf{[47] CGCG 436$-$030}: This galaxy shows three bright X-ray peaks in the soft band, with only the central one corresponding to a hard-band peak. The other two, placed following the spiral arm structure in the optical images, most likely correspond to star-forming regions. 

In the DSS image there is a hint of a bridge of material between the galaxy and a fainter galaxy $\sim 1$' to the east, with which it seem to be interacting \citep{MirSan1988,ZinLes2000}. This other galaxy is not detected in the \textit{\textit{Chandra}} X-ray data nor the MIPS data, and is only visible in near-infrared observations such as the largest-wavelength IRAC channels. Therefore, we have not considered any contribution to the \textit{IRAS} flux originating from this nearby companion. 

\textbf{[49] NGC 0695}: This source has a rather flat spectrum in the $0.4-7$ keV range. There is considerable extended emission in soft band, and both bands present a very intense emission from the central region. 

\textbf{[51] MCG+07$-$23$-$019}: This ring galaxy is composed of an elongated main body with double components separated by $\sim 5$" ($\sim 4$ kpc) and an oval ring with a diameter of ~16" to the west of the main body \citep{HatYos2004}. As already suggested by JHKL-band mapping, the nucleus of the galaxy lies between the two optical components and is heavily obscured in optical images \citep{JoyHar1987}. \textit{Chandra} data show clear emission coming from the elongated disk of the galaxy. The X-ray emission is more intense in the centre, and unobscured both in hard and soft bands. There is extended soft X-ray emission around the nucleus, which partly follows the oval ring, most likely tracing star formation. 

\textbf{[52] NGC 6670}: This closely interacting merger is composed of two sources, NGC 6670A (or East) and NGC 6670B (or West), separated $\sim 0.5$'. Both galaxies contribute to the \textit{IRAS} flux \citep{ChuSan2017}, with the western component being slightly brighter. 

X-ray emission from the eastern component is mostly observed in soft band, and concentrated around the nucleus. The western source, however, shows extended diffuse emission, particularly along the plane of the galaxy. The emission near the centre is more intense, both in hard and soft band, at both sides of the nucleus. This morphology suggests high absorption in the innermost region. The spectrum is also suggestive of a hard excess, and a simple power-law fit results in a photon index of $\Gamma = 0.5 \pm 0.5$. Fitting an absorbed power-law of fixed photon index $\Gamma = 1.8$ results in a moderate absorbing column density of $N_{\rm H} \sim 4 \times 10^{22}$ cm$^{-2}$, and no significant improvement on the fit. The X-ray luminosity of the source is $L_X \sim 10^{41}$ cm$^{-2}$. The excess at $\sim 6.4$ keV, if interpreted as a possible Fe K$\alpha$ line, is not significant to the 1$\sigma$ level. 

\textit{XMM-Newton} data for both sources, resolved, in this double system was analyzed by \citet{MudMat2014}, with no hint of AGN presence detected. However, their short exposure implied a detection of $\sim 100$ cts per source, lower than the counts detected in our \textit{Chandra} data.  

We consider that, even though we cannot rule out the possibility of the western source containing an AGN, we have no strong evidence to claim its presence. 

\textbf{[53] UGC 02369}: This double system, separated by $\sim 0.4$', is clearly dominated in X-rays by the southern component which, as shown in Table \ref{table:IRfrac}, is also responsible for $\sim 98$\% of the infrared emission. Due to the negligible contribution to the \textit{IRAS} flux originating in the northern galaxy, we do not present any results for this component. An X-ray analysis would also not be possible, as only $\sim 5$ cts are detected for this source. 

The southern source is compact in X-rays, with emission coming both from the nucleus and from a star forming region in the spiral arm, at south, both in soft and hard bands. 

\textbf{[54] NGC 1614}: This source has been classified as a possible obscured AGN through X-ray spectroscopy \citep{RisGil2000}, though VLBI studies with a sensitivity limit of 0.9 mJy do not detect a compact radio core in it \citep{HilHei2001}. Recent studies in sub-arcsec mid-IR observations do not completely rule out a possible (weak) AGN scenario, but they constrain that the nuclear luminosity is $< 5$\% of the overall bolometric luminosity of the galaxy \citep{PerCol2015}. ALMA observations do not detect the nucleus in either the CO (6-5) line emission and the 435 $\mu$m continuum, ruling out with relatively high confidence a Compton-thick AGN \citep{XuCao2015} .

In any case, this source also does not meet any of our AGN selection criteria, and we do not see any signs of AGN presence in the \textit{Chandra} spectrum.

Emission both in hard and soft band is peaked in the nucleus, and the soft band emission also shows elongated extension in the E-W direction, as opposed to the optical edge-on disk, which is elongated toward the N-S direction. Intense star formation, very compact in the nucleus, is the most likely origin of the X-ray emission. 

\textbf{[56] NGC 5331}: Both galaxies in this system, separated by $\sim 0.4$', contribute to the \textit{IRAS} flux, though the southern component is responsible for $\sim 80$\% of the emission, as shown in Table \ref{table:IRfrac}. However, since their X-ray luminosity is comparable, the northern galaxy has a much higher logarithmic ratio $(HX/IR)=-3.84$ (as defined in Table 3). This value is close to the expected given the correlation derived by \citet{RanCom2003}, but high when compared to the characteristic X-ray faintness of the GOALS sample. 

\textbf{[57] IRAS F06076$-$2139}: This closely interacting merger is clearly dominated by the northern source both in infrared and X-rays. However, with only $\sim 10$ cts, the southern source meets our HR criterion for AGN selection ($HR=-0.03 \pm 0.17$). The spectrum also shows an increase in flux toward higher energies, despite the significant error bars. Only the hardness ratio is computed as part of the analysis of this source, due to the low number of counts. For the same reason, a radial profile is not provided for this component. 

The northern source comes close to meeting the same AGN selection criteria, with $HR=-0.34 \pm 0.08$. The spectrum might hint toward an excess in hard band, though an absorbed power-law fit with a fixed photon index of 1.8 yields a column density of only $N_{\rm H} \sim 1.9 \times 10^{22}$ cm$^{-2}$. With a full-band X-ray luminosity of $L_{\rm X} \sim 10^{41}$ erg s$^{-1}$, and fitting statistics also favoring a non-absorbed power-law, we opt not to consider this source an AGN. 

\textbf{[60] IC 2810}: Both galaxies in this system, separated by $\sim 1.1$', contribute to the \textit{IRAS} flux as shown in Table \ref{table:IRfrac}, with the north-western source contributing $\sim 70$\% of the infrared luminosity.

\textbf{[63] IRAS 18090+0130}: Both galaxies in this system, separated by $\sim 1.3$', contribute to the \textit{IRAS} flux, with the eastern component being responsible for $\sim 80$\% of the emission. The western component has a low X-ray flux, and not enough counts to provide reliable data for any analysis further than computing a hardness ratio. 

\textbf{[64] III Zw 035}: This closely interacting double system is completely unresolved in both \textit{Herschel} and MIPS images used to derive the contribution of each galaxy into the \textit{IRAS} source, as shown in Table \ref{table:IRfrac}. However, \citet{ChaSta1990} mention that the majority of the radio continuum (and also probably FIR) emission originates in the northern galaxy. High angular resolution radio continuum observations from \citep{BarLer2017} indicate the northern component is the most compact source of the brightest and closest ULIRGs from the GOALS sample, while they don't detect the southern component at 33 GHz. Also, when considering IRAC channel 1 to 4 images (at 3.6, 4.5 ,5.8 and 8.0 $\mu m$), it is possible to see that the northern source clearly dominates and the southern source fades with increasing wavelength. Thus, we assign a contribution to the \textit{IRAS} flux of 100\% to the northern source. 

Due to the lack of \textit{IRAS} flux originating in the southern component, we do not present results for its X-ray analysis in this work, and we do not consider it a source within our sample. The total X-ray counts for this galaxy are $\sim 25$ cts in the full $0.5-7$ keV range, which also does not allow for a detailed X-ray analysis either, though a simple power-law fit gives an estimated $L_{\rm X} \sim 2 \times 10^{40}$ erg s$^{-1}$. However, while the soft band X-ray flux is dominated by the northern source, the hard band X-ray flux is very similar for both, and the southern source is optically classified as a Seyfert 2 \citep{YuaKew2010}. It is then possible that the northern source is responsible for the high \textit{IRAS} flux, most likely having a burst of star formation, while the NIR and MIR contribution from the southern source might be due to the presence of an AGN.

\textbf{[65] NGC 3256}: This source is assumed to be in an advanced merger stage, with a northern brighter component (the central peak) and a southern component at about $\sim 10$'', elongated in the E$-$W direction. The possibility of this being a merging, obscured, companion galaxy was first suggested by \citet{MooOli1994}, and radio observations by \citet{NorFor1995} supported this theory by detecting two equally bright knots. However, high-resolution MIR imaging shows that the northern peak is actually $\sim 20$ times brighter than the southern region, suggesting that most of the star formation in the galaxy is originated there \citep{LirWar2002}. They also find the northern peak to be brighter in X-rays, using \textit{Chandra} data. The images shown in this work mark the two hard-band peaks mentioned by them, the southern clearly falling in an obscured region, with dimmer soft band emission. 

The very advanced merger stage of this source makes it hard to determine how much of the surrounding extended emission was initially associated to any of the cores. Therefore, we analyze it as a single source in order to avoid introducing errors into the determination of its Infrared and X-ray emission. 

\textbf{[67] IRAS F16399$-$0937}: This closely interacting pair is unresolved both in \textit{Herschel} and MIPS data, though \citet{HaaSur2011} derive that most of the MIR emission ($>$90\%) is coming from the northern source. We use this value to correct for the fraction contributed to the \textit{IRAS} flux by each galaxy. 

The northern source shows two intense hard X-ray peaks, both corresponding to soft X-ray emitting regions, the more southern of which is the nucleus. The other, as well as the less-intense knots seen in both sources, probably correspond to star forming regions. The spectrum of this source shows an excess at $> 4$ keV, with a few hard counts coming from the nuclear region. Fitting an absorbed power-law with a photon index of 1.8 yields an absorbing column  density of $N_{\rm H} \sim 2 \times 10^{23}$ cm$^{-2}$, though an unabsorbed power-law of photon index $\sim 1.6$ is an equally good fit. With a net count number of $\sim 23$ cts, we do not believe we can confidently classify this source as an AGN. 

\citet{SalRob2015} also consider the possibility of the northern source containing an embedded AGN, fitting the $0.435-500$ $\mu$m SED with a model that includes an AGN torus component. The fit suggests the presence of an AGN with bolometric luminosity $L_{\rm bol} \sim 10^{44}$ erg s$^{-1}$, though the spectrum is also consistent with shocks ($v \sim 100-200$ km s$^{-1}$). This bolometric luminosity would imply a fraction $L_{\rm X}^{AGN} \sim 10^{42}-10^{43}$ erg s$^{-1}$, much larger than the $L_{\rm X} =10^{41}$ erg s$^{-1}$ detected in the \textit{Chandra} data. 

The southern source does not have a clear centre in X-rays or MIR and FIR, and so the centre for the radial profile is determined using the brightest region in the optical HST image. Because the source is clearly elongated, annuli centered on the eastern edge will obviously end up including photons from the northern source, and so to avoid interference, it is removed in the computation of radial profiles. This component meets our $HR$ AGN selection criterion, though the Appendix \ref{MWimages} image shows that no strong hard X-ray peak is coming from the nucleus of the source; and the origin of the hard counts is concentrated in two point-sources at the west of the nucleus. Therefore, we do not classify this source as an AGN. 

\textbf{[68] IRAS F16164$-$0746}: This source meets two of our AGN selection criteria, the $HR$ and the [Ne v] line, and is also classified as an optical Seyfert 2 in \citet{YuaKew2010}. The X-ray source is elongated in the soft band, in the direction perpendicular to the disk of the galaxy, which could be interpreted as an outflow. There is also a secondary point source $\sim 3$'' from the nucleus, both in soft and hard band, without any obvious overlap with a star-forming knot. With an associated X-ray luminosity in the $2-10$ keV range of $\sim 3 \times 10^{40}$ erg s$^{-1}$, it could be classified as a ULX. 

\textbf{[69] IC 4686/7}: This source is part of a triple merger system, formed by IC 4687 at north, closely interacting with the central galaxy, IC 4686, at $\sim 0.5$', and IC 4689 $\sim 1$' south of IC 4686 \citep{Wes1976}. All three of these galaxies contribute in more than 10\% to the \textit{IRAS} flux \citep{ChuSan2017}, and are therefore analyzed in this work. 

\textbf{[71] NGC 2623}: With a spectrum clearly raising toward higher energies, giving a hardness ratio of $HR=-0.11 \pm 0.02$, and also meeting the [Ne v] line criterion, this source is classified as an AGN. This source has been classified as an AGN previously in radio \citep{LonSmi1993} and X-rays \citep{MaiCom2003}.

Optical HST images show extended tidal tails, approximately $20-25$ kpc in length, with a southern region rich in bright star clusters \citep{EvaVav2008}, though no X-ray emission is detected with \textit{Chandra} in the region. 

\textbf{[72] IC 5298}: This source is a clear absorbed AGN, seen both in the \textit{Chandra} spectrum presented in this work, and through XMM-Newton data analysis. Using data from both telescopes, when a photon index of 1.8 is assumed, a column density of $N_{\rm H} \sim 4 \times 10^{23}$ is obtained. There is a hint of a line at 6.4 keV in the \textit{Chandra} data, with significance lower than 1$\sigma$. The presence of the line can be confirmed with a significance of $\sim 2\sigma$ when checking XMM-Newton EPIC data, with a fit that is also consistent with the derived absorbing column density. The AGN diagnostics is  also confirmed through the presence of the [Ne v] line, and the optical S2 classification \citep{VeiKim1995}. 

\textbf{[73] IRAS 20351+2521}: This galaxy shows strong central emission in X-rays, originating in the nucleus, with extended emission and point-sources along the spiral arms, tracing star-forming knots. 

\textbf{[75] NGC 6090}: This closely interacting system is completely unresolved in both \textit{Herschel} and MIPS data. Therefore, we resort to the analysis performed by \citet{HatYos2004} to derive the contribution of each component to the \textit{IRAS} flux, listed in Table \ref{table:IRfrac}. 

The north-eastern source shows hard-band diffuse emission corresponding to the optical central region of the galaxy, and a peak $\sim 3$'' north of the centre. It corresponds to a particularly bright region, both in optical and infrared, on one of the spiral arms. 

The sources are interacting so closely that the radial profiles interfere with each other past $~4-5$'' from each nucleus, and have therefore been limited to this radius.

\textbf{[79] NGC 5256}: This closely interacting system is surrounded by diffuse soft-band emission in X-rays, part of which extends toward the northern direction, following a blue tidal stream seen in optical images. Between the two sources, a slightly curved excess is visible, which can be interpreted as a shock between colliding winds from both galaxies \citep[see][]{MazIwa2012}. This excess is the reason the radial profile in Appendix \ref{MWimages} for the NE source shows an increase of soft-band surface brightness with distance at ~5$-$6''.

This source has been detected in the [Ne v] line at kpc scales, meeting our AGN selection criteria. However, due to the close proximity of both nuclei, it is not possible to know which (or if both) is responsible for this emission. Both optical classifications used in this work mark the NE source as a Seyfert 2, while leaving the SW source as LINER or composite. However, \citet{MazIwa2012} have the opposite optical classification for the sources: NE as LINER and SW as S2. 

Looking at the X-ray spectra, the NE source can be best fitted with an absorbed AGN model, fixing a spectral index of 1.8 and obtaining a column density of $N_{\rm H} \sim 8 \times 10^{22}$cm$^{-2}$, which is interpreted as a mildly absorbed AGN. 

The SW source shows an excess that can be fitted as an iron 6.4 keV line with a confidence of $\sim 2.1 \sigma$, which meets one of our X-ray AGN selection criteria.

As mentioned by \citet{MazIwa2012}, XMM-Newton EPIC data only marginally resolves the two sources, and the spectrum is presented for the whole system. However, given the spectra resolved by \textit{Chandra}, the iron line seen in the EPIC data is most likely originating in the south-western source. Their combined analysis also results in a Compton-Thick classification of the south-western source. 

\textbf{[80] IRAS F03359+1523}: Only one of the two sources present in this system is observed in X-rays, the eastern source, with an elongated morphology that corresponds to the length of the edge-on disk in the optical data. The sources are unresolved in the \textit{Herschel} data, and only one source is visible in the MIPS data, which is centred at the position of the eastern one. It is not possible to affirm if this is due to lack of resolution, or if the western source has no contribution to the MIPS flux. However, observing the IRAC images from channel 1 through 4 (at 3.6, 4.5 ,5.8 and 8.0 $\mu m$), it is possible to see that the eastern source clearly dominates and the western source fades with increasing wavelength. Also, \citet{GolJos1997} mention that only one (believed to be the eastern source) is prominent at radio wavelengths. This, together with the complete lack of X-ray emission originated in this companion source, leads us to believe that the western galaxy is not contributing to the \textit{IRAS} flux.

There is another source, prominent in radio wavelengths, $\sim 1.5$' to the south of IRAS F03359+1523.  \citet{CleVeg2008} use NVSS radio data to extrapolate that this nearby galaxy could be responsible for about $\sim 50$\% of the \textit{IRAS} flux. However, images at 8 and 24 microns show a weak source that fades completely at 70 micron, leading us to believe that its contribution to the far-infrared luminosity is most likely negligible. 

\textbf{[81] ESO 550$-$IG025}: Both sources in this system, with a separation of $\sim 0.3$', contribute to the \textit{IRAS} flux. The southern source has a rather flat spectrum, partly due to the contribution of the hard X-ray peak placed at about $\sim 4$'' west of the nucleus. This source cannot be easily interpreted as the X-ray counterpart to any star-forming regions in the galaxy. If it is associated to this galaxy, its X-ray luminosity in the $2-10$ keV range is of $\sim 3 \times 10^{40}$ erg s$^{-1}$, implying it could be classified as a ULX.

\textbf{[82] NGC 0034}: This source, optically classified as a Seyfert 2 \citep[e.g.][]{VeiKim1999, YuaKew2010}, has an X-ray spectrum that shows a hard band excess. Fitting an absorbed AGN with a fixed photon index of 1.8 gives an absorbing column density of $N_{\rm H} \sim 1 \times 10^{23}$ cm$^{-2}$. Previous analyses of XMM-Newton data confirm the presence of an AGN, be it through marginal detection of the Fe K$\alpha$ line or modeling of an absorption or reflection component \citep[e.g.][]{ShuWan2007, BriNan2011}.

\citet{RicBau2017} use joint data from \textit{Chandra}, XMM-Newton and NuSTAR and find a clear Fe K$\alpha$ feature at $6.48_{-0.05}^{+0.06}$ keV, and their spectral analysis shows a heavily obscured AGN with a column density of  $N_{\rm H} = 5.3 \pm 1.1 \times 10^{23}$ cm$^{-2}$. Their results certainly confirm the presence of the AGN, and their derived column density differs from the one derived with only \textit{Chandra}, most likely due to the latter's much lower sensitivity at high energies. 

\textbf{[83] MCG+12$-$02$-$001}: We consider this system to be composed of three individual sources: a northern component and a main pair, separated by $\sim 0.3$'. The western source in the pair is considered an individual galaxy in close interaction with the eastern source, though there is also the possibility of it being an extended star-forming region. Given the fact that we see an X-ray peak at its centre, and its X-ray luminosities of $L_{\rm SX}=1.3 \times 10^{40}$ erg s$^{-1}$ and $L_{\rm HX}=2.4 \times 10^{40}$ erg s$^{-1}$ are comparable to those of the eastern galaxy, we are inclined to treat this whole system as a triple.

The northern source does not contribute to the \textit{IRAS} flux \citep{DiaCha2010}, as specified in Table \ref{table:IRfrac}, and therefore it is not analyzed. It is detected with \textit{Chandra}, with $\sim 9$ cts in the full $0.5-7$ keV range.

\textbf{[85] IRAS F17138$-$1017}: This source has a rather flat spectrum in X-rays, with flux slightly increasing towards higher energies. It meets the $HR$ criterion for AGN selection, though no [Ne v] line is observed. Fitting with an absorbed AGN model, fixing a spectral index of 1.8, a low column density of $N_{\rm H} \sim 2 \times 10^{22}$cm$^{-2}$ is obtained.

Morphologically, \textit{\textit{Chandra}} data shows a soft X-ray deficit at the optical centre of the galaxy, which could be caused by absorption. The hard X-ray image does not show a clear peak of emission, but a rather homogeneous flux around a larger circular region. X-ray contours on the HST image show very prominent dust lanes close to the nucleus of the galaxy, to which the obtained column density could belong. These dust lanes are most likely absorbing an important part of the soft-band X-ray emission, and could be responsible for the hardness of the spectrum. Given all these facts, and the clear lack of observation of a hard-band peak in the nucleus, we opt not to classify this source as an AGN. 

\citet{RicBau2017} fit a combined \textit{Chandra} and NuSTAR spectrum with a simple power-law model, obtaining a photon index of $\sim 1.1$, harder than the typical X-ray emission expected of a star-forming region, but still consistent with this hypothesis. 

The hard-band X-ray luminosity of $L_{\rm HX} \sim 1.8 \times 10^{41}$ erg s$^{-1}$ we derive is high, but not incompatible with being originated by a strong starburst, as this source falls within the uncertainties of the correlation derived by \citet{RanCom2003}. 

\textbf{[95] ESO 440$-$IG058}: Both galaxies, with an angular separation of $\sim 2$', contribute to the \textit{IRAS} flux in this source; though the southern component dominates at almost $\sim 90$\% \citep{DiaCha2010}. Soft X-ray emission from the southern component is extended, most likely an outflow with its origin in a starburst wind.

\textbf{[100] NGC 7130}: This galaxy shows clear extended emission in soft X-rays around a strong peak that follows the disk of the face-on optical galaxy, tracing the spiral arms. The spectrum shows a hard excess due to absorption and an iron 6.4 keV line at high energies, which could be due to absorption in the soft band, or due to reflection. A reflection component fitting, using a \textit{pexrav} model \citep{MagZdz1995} results in an iron line with an equivalent width of $\sim 0.6$ keV, which is too low for a reflection-originated line. Therefore, our data favors an absorption model; which, when imposing a photon index of 1.8, results in a column density of $N_{\rm H}=3 \times 10^{23}$ cm$^{-2}$ and an iron line equivalent width of 0.8 keV; detected with a significance of $\sim 2.5 \sigma$. 

Based only on the \textit{Chandra} data, we find it difficult to distinguish between this scenario and a Compton-Thick source with an imposed photon index of $\Gamma=0.0$, as modeled by \citet{LevWea2005}. \citet{RicBau2017} confirm the presence of a Compton-Thick AGN using a combined analysis with \textit{NuSTAR} data.

\textbf{[104] NGC 7771}: This galaxy is part of an interacting quartet of galaxies, along with close companion NGC 7770 at an angular distance of $\sim 1.1$', NGC 7771A at $\sim 2.8$' and NGC 7769 at $\sim 5.4$' \citep[e.g.][]{YegNaz2016}.

About 90\% of the \textit{IRAS} flux originates in NGC 7771,  with NGC 7770 being responsible for the remaining $\sim 10$\% and NGC 7769 being resolved as a separate source by \textit{IRAS}. NGC 7771A is faint in the infrared, remaining undetected at 8 $\mu$m and above. There is also no detection for this small component in the \textit{Chandra} data. 

Of the many point sources seen along the disk of NGC 7771, \citet{LuaRob2015} classify $4^{+4}_{-0}$ as ULXs. 

\textbf{[105] NGC 7592}: This source is a closely interacting triple system, formed by two main infrared and X-ray sources (East and West) and a smaller, southern source. This third source, seen in the optical SDSS images, is undetected in X-rays, and also does not contribute to the \textit{IRAS} flux. 

There is unresolved detection of the [Ne v] line for this triple source that meets our AGN selection criterion. However, as the western source is classified as an optical Seyfert 2, it is likely that it is the origin of the infrared line. The spectrum of the western source shows an excess between $6-7$ keV, originating in the nucleus, that can be fitted as a Gaussian line with an energy of $6.7_{-0.3}^{0.1}$ keV. The significance of this line is at the $1 \sigma$ level, and thus we do not use this hint as any selection criterion; though given the uncertainties, a 6.4 keV line cannot be ruled out completely, specially when combined with the continuum. 

The western source presents very compact X-ray emission, as derived from its radial profile, compared to its extended infrared emission (as plotted in Fig. \ref{fig:RSXvsRIR}).

\textbf{[106] NGC 6286}: This source is interacting with NGC 6285, $\sim 1.5$' to the northwest, showing very extended tidal disruption features. Both sources contribute to the \textit{IRAS} flux \citep{ChuSan2017}.

The X-ray spectrum of NGC 6286 shows hard excess emission above 5 keV. With less than 20 cts in the $5-8$ keV range, the excess is difficult to fit as an absorbed AGN using only \textit{Chandra} data. MIR studies find possible hints of an AGN presence \citep[e.g.][]{VegCle2008,DudSat2009}, which are confirmed by hard X-ray \textit{NuSTAR} observations. \citet{RicBau2016} find compelling evidence of a Compton-thick, low-luminosity AGN ($N_{\rm H} \simeq (0.95-1.32)\times 10^{24}$ cm$^{-2}$). We thus classify this source as an AGN.

This galaxy shows a very extended soft X-ray emission, spreading perpendicular to the optical edge-on disk up to a distance of $\sim 5 -7$ kpc, depending on direction. Presence of a super-wind outflow generated by a strong starburst was already suggested by \citet{ShaMoi2004} through detection of an increase of $[NII]\lambda 6583 / H_{\alpha}$ ratios; and the presence of an emission nebula extending up to $\sim 9$ kpc from the galactic plane. 

\textbf{[107] NGC 4922}: This system contains two galaxies, separated by distance of $\sim 0.4$', with the northern one being brighter in both X-rays and infrared \citep{DiaCha2010} . The southern source contributes $\sim 1$\% of the \textit{IRAS} flux, and therefore its analysis is not included in this work. With only $\sim 40$ cts, all in the $0.4-2$ keV range, it is also a weak X-ray source.  There is also another source (2MASX J13012200+2920231) $\sim 1.7$' to the north, which is undetected at 8 $\mu$m and above, and most likely does not contribute to the \textit{IRAS} flux. 

The northern source is selected as an AGN through the presence of the [Ne v] line and the pair (unresolved) is also classified as a Seyfert 2 in \citet{YuaKew2010}. Our X-ray analysis also classifies it as an absorbed AGN, with a column density of $N_{\rm H} \sim 3 \times 10^{23}$ cm$^{-2}$ when fixing the photon index to 1.8. There is a hint of the presence of an iron 6.4 keV line, though only at a significance of about $1 \sigma$. 

\citet{RicBau2017} analyse NuSTAR observations and conclude, based on their similar \textit{Chandra} results, that the source detected at high energies must correspond to NGC 4922 (N), given the companion's non-detection in the $2-7$ keV range. They detect a prominent Fe K$\alpha$ line at $6.48_{-0.07}^{+0.07}$ keV, and find that the source is actually Compton-thick, with $N_{\rm H} \geq 4.27 \times 10^{24}$ cm$^{-2}$; which is more than one order of magnitude higher than our best \textit{Chandra} fit. 

\textbf{[110] NGC 3110}: This source is in interaction with nearby galaxy MCG$-$01$-$26$-$013 at its south-west, separated $\sim 1.8$', which is not detected in X-rays in the \textit{Chandra} data. Both sources are contributing to the \textit{IRAS} flux, though $\sim 90$\% of the infrared emission has its origin in NGC 3110 \citep{DiaCha2010}. The companion galaxy is not analyzed due to lack of significant X-ray emission and low infrared luminosity, though the \textit{IRAS} flux associated to NGC 3110 is corrected for the pair's contribution.

This source has diffuse soft X-ray emission along the spiral arms, which also contain strong hard X-ray peaks that are most likely associated with star-forming knots. The nucleus of the galaxy, however, does not show peaked emission in the $2-7$ keV band. Due to this particular morphology, HST optical and IRAC channel 1 images are used to centre the derived radial profiles.

\textbf{[114] NGC 0232}: This source is paired with NGC 0235, at a distance of $\sim 2$', which is a known Seyfert 2 galaxy. Despite having previously been classified as non-interacting, a faint tidal bridge has been observed connecting the two galaxies \citep{DopPer2002}. NGC 0235 has two nuclei, and is classified as a minor interaction \citep{LarSan2016}. However, as this companion galaxy is resolved by \textit{IRAS} \citep{SurSan2004} as an individual source, we do not include it in our analysis. 

\textbf{[117] MCG+08$-$18$-$013}: This galaxy is paired with MCG+08$-$18$-$012 at $\sim 1$' to its west. MCG+08$-$18$-$013 is clearly dominant in the infrared and the origin of the \textit{IRAS} flux \citep{ChuSan2017}. This component's X-ray emission originates from two point sources close to the nucleus of the galaxy, one of which is bright in soft-band X-rays, and could be associated to a star-forming region, and the other in hard-band X-rays. We consider this hard-band peak to originate from the nucleus of the galaxy, and use it to center the radial profiles. 

MCG+08$-$18$-$013 is undetected in X-rays and therefore not included in the analysis.

\textbf{[120] CGCG 049$-$057}: This source, despite only having a total of $\sim 30$ cts in the $0.5-7$ keV band of the available \textit{\textit{Chandra}} observation, has a hardness ratio of $HR=-0.04 \pm 0.09$, and so meets one of the X-ray AGN selection criteria. The spectrum of the source shows, despite the large error bars, a tendency to rising flux toward higher energies. The X-ray image in the $2-7$ keV band shows about $\sim 5$ cts originating from the innermost region of the source, and thus we classify it as an AGN.

\citet{BaaKlo2006} also classified it as an AGN based on radio observations. Though optical and MIR observations \citep[e.g.][]{VeiKim1995, StiArm2013, MelHec2014} classify it as a starburst, \textit{Herschel} spectroscopic data analysed by \citet{FalGon2015} detect very high column densities in the nucleus ($N_{\rm H} = 0.3 -1.0 \times 10^{25}$ cm$^{-2}$), meaning that a Compton-thick AGN could be present. This would explain the X-ray weakness we observe, which was already reported by \citet{LehAle2010}. 

\textbf{[121] NGC 1068}: This well-known AGN meets our selection criteria in all bands: Seyfert 2 in both of the used optical classifications, presence of the [Ne v] line in infrared and clear detection of the Fe K$\alpha$ line at 6.4 keV in X-rays with a significance of $\sim 3.6 \sigma$. Note that the equivalent width of the 6.2 $\mu$m PAH feature is not presented in \citet{StiArm2013} due to saturation of the spectrograph. \citet{HowMaz2007} analyze PAH and warm dust emission in NGC 1068 in detail. Their 6.2 $\mu$m images are saturated within the inner $r\sim 500$ pc, though they measure an equivalent width of the PAH feature right outside the region of saturation of $\sim 0.1$. This is suggestive that the value of of the equivalent width would drop below 0.1 further in.  

Diffuse X-ray emission is clearly observed in this source, following the optical spiral arms and star-forming regions. In order to outline all features, X-ray contours to a low enough level were necessary, which result also in the clear contours around the saturated feature that diagonally crosses the image.

Individual point sources can be seen spreading along the disk of the galaxy in both the soft and hard bands, which most likely correspond to X-ray binaries. We note that we have not marked these as hard X-ray peaks in Appendix \ref{MWimages} images, as they are numerous and clearly not originating from a region near the nucleus of the galaxy. None of these point sources were removed in order to derive radial profiles. \citet{LuaRob2015} classify 3 of them as ULXs. 

\textbf{[123] UGC 02238}: This source presents a rather diffuse emission, showing three main X-ray peaks near the nucleus, only one of which (the westernmost) is also peaked in the $2-7$ keV band. However, as shown by the contours over the IRAC channel 1 image, this region is outside of the nucleus of the galaxy. We consider this emission to most likely originate from different intense starburst regions, given how there is no clear hard-band central peak. Optical and infrared imaging data show a highly disturbed disk and tidal tails, and classify this galaxy as a post-merger stage \citep[e.g.][]{SmiHer1996, LarSan2016}, which is consistent with the described X-ray morphology.  

We fit the overall X-ray $0.5-7$ keV emission of this galaxy with a single power law. Attempts to fit a one or two component \textit{mekal} model produce unsatisfactory results. Two strong point sources can be seen in the $2-7$ keV band image presented in Appendix \ref{MWimages} which, if associated with the galaxy, would be classified as ULXs. The point-source at the easternmost edge of the disk of the galaxy would have an estimated luminosity of  $\sim 4 \times 10^{40}$ erg s$^{-1}$, and the strong point-source right south of the nucleus of the galaxy would be emitting $\sim 1 \times 10^{40}$ erg s$^{-1}$; in the $2-10$ keV range. Note, however, that due to the very low number of counts these are very rough estimations. 

\textbf{[127] MCG$-$03$-$34$-$064}: This source has a north-eastern companion at $\sim 1.8$', MCG$-$03$-$34$-$063, which is responsible for about $\sim 25$\% of the \textit{IRAS} flux \citep{ChuSan2017}. The analysis of this companion source is not included in this work because it is undetected in the \textit{Chandra} data. A correction to the IR luminosity for the contribution of MCG$-$02$-$34$-$063 has been considered for this source. 

MCG-03-34-064 presents a very peaked central emission in all X-ray bands, and is a clear absorbed AGN, as seen from the spectrum. Fitting with a fixed photon index of 1.8 yields an absorbing column density of $N_{\rm H} \sim 5 \times 10^{23}$ cm$^{-2}$. The iron K$\alpha$ line at 6.4 keV is detected, with a $\sim 3 \sigma$ significance.  With a $HR=-0.32 \pm 0.01$ our other X-ray AGN selection criterion is almost also met. 

\citet{RicBau2017} perform a NuSTAR analysis of this source, combined with XMM-Newton EPIC data, and derive an absorbing column density of $N_H = 5.42_{-0.09}^{+0.07} \times 10^{23}$ cm$^{-2}$, which is compatible with our derived result within the errors. They also detect the Fe K$_{\alpha}$ line, and a Gaussian line at $6.62_{-0.01}^{+0.01}$ keV with $EW \sim 0.2$ keV, which is not detected in the \textit{Chandra} data. 

This source meets also the [Ne v] line and 6.2 $\mu$m PAH feature AGN selection criteria, though the same cannot be said about the optical ones. \citet{YuaKew2010} classify it as a star-forming galaxy and \citet{VeiKim1995,VeiKim1999} do not analyze it, although other works classify it as a Seyfert galaxy \citep[e.g.][]{LipNei1988, CorNor2002}.

\textbf{[134] ESO 350$-$IG038}: This galaxy presents three main star-forming condensations \citep{KunLei2003, AteKun2008}. Only two of these knots, the eastern and western, are clearly resolved as X-ray sources in the \textit{\textit{Chandra}} data, both presenting peaked emission in both the soft and hard bands. The region is surrounded by diffuse, soft X-ray emission. These knots, separated by $\sim 4$'', are analyzed together as the X-ray source corresponding to the \textit{IRAS} source, and not separated as two individual galaxies, as there is no clear evidence of them being individual galaxy nuclei in a state of closely interacting merger. 

\textbf{[136] MCG$-$01$-$60$-$022}: This source is near galaxies MCG$-$01$-$60$-$021 and Mrk 0399, at $\sim 4.4$' interacting with the former \citep[e.g.][]{DopPer2002}, connected through thin and long tidal bridges. Both nearby galaxies are undetected in the \textit{Chandra} data, and are detected together as another \textit{IRAS} source, resolved from MCG$-$01$-$60$-$022 \citep{DiaCha2010}. 

This source presents diffuse soft band emission surrounding the central X-ray peak, which has its origin in an absorbed AGN. This sources meets the $HR$ X-ray criterion, and spectral fitting of an absorbed power-law with a fixed photon index of 1.8 yields a an absorbing column density of $N_{\rm H} \sim 1 \times 10^{23}$ cm$^{-2}$.

\textbf{[141] IC 0563/4}: This source is a double system, composed of IC 0564 at north and IC 0563 at south, separated by $\sim 1.6$'. Both contribute similarly to the \textit{IRAS} flux, and to the overall X-ray luminosity. Morphologically, both galaxies have faint emission originating in the nucleus and various point-sources spreading along the spiral disks. 

IC 0563 has a hardness ratio of $-0.34 \pm 0.05$, which exceeds our AGN selection threshold. However, the origin of the hardness (also seen as an excess at $3-5$ keV in the spectrum shown in Fig. \ref{fig:XraySpectra}) is not the nucleus of the galaxy, but a point-source located north of it. If the source is associated with the galaxy, with a roughly estimated luminosity of $\sim 3 \times 10^{40}$ erg s$^{-1}$, it could be classified as a ULX. Interestingly, the point-source spectrum shows a hint of a line at $\sim 1.50 \pm 0.03$ keV, with a significance of $\sim 2 \sigma$. If this source is a ULX within the galaxy, this excess cannot be easily explained as an emission line. If this source is a background quasar, for which we are detecting a redshifted 6.4 keV iron line, a high $z \sim 3.3$ would be necessary. If it were an object at $z=3.3$, the X-ray spectrum would suggest that its origin is in reflected light form a Compton-thick AGN. This scenario, however, leads to an unreasonably luminous quasar. 

Similarly, IC 0564 shows a spectrum with a large flux at high energies, though the errorbars are significative, which is also emitted by a northern point-source. Also with a roughly estimated luminosity of $\sim 3 \times 10^{40}$ erg s$^{-1}$, it could be classified as a ULX, if it is associated to the galaxy. 

Both point-sources are marked with green crosses in the images shown in Appendix \ref{MWimages}. 

\textbf{[142] NGC 5135}: This galaxy is classified as an optical Seyfert 2 \citep[e.g.][]{YuaKew2010}, and meets our infrared [Ne v] line criterion for AGN selection \citep{PetArm2011}. 

It shows an excess in hard X-rays, with a Fe $K_{\alpha}$ line at 6.4 keV with a $\sim 2.9 \sigma$ significance, which could be the result of either absorption or a reflection component. Fitting an absorbed power-law with fixed photon index of 1.8 yields an absorbing column density of $N_H \sim 4 \times 10^{23}$ cm$^{-2}$, which is not large enough to produce the equivalent width of the iron line of $EW \sim 0.9$ keV obtained through the same model. Fitting a \textit{pexrav} model \citep{MagZdz1995} with a fixed photon index of 2.0 yields a believable $EW \sim 1.1$ keV, which means our data favour a reflection dominated AGN. \textit{Suzaku} observations extending up to 50 keV allow to better estimate the absorbing column density, of$\sim 2.5 \times 10^{24}$ cm$^{-2}$, classifying this source as Compton-thick and providing a good estimate of the strength of the reflection component \citep{SinRis2012}. 

Morphologically, this source presents a very extended soft band emission, with two central X-ray peaks, visible only when the smoothing in the image is set to 0.5'' or less. The northern peak is responsible for the iron emission line, which indicates it's associated to the nucleus of the galaxy. The southern peak is brighter in the $0.5-2$ keV band, and most likely associated with a star-formation region. From the many point-sources seen in the full band image, up to $6^{+1}_{-2}$ are classified as ULXs \citep{LuaRob2015}.

\textbf{[144] IC 0860}: With only $\sim 25$ cts, no X-ray analysis beyond the calculation of the HR and the extraction of the radial profile has been performed for this source. However, despite the low count-rate,  this galaxy is classified as an AGN with a value of $HR=-0.29 \pm 0.17$. The spectrum obtained also shows a rising tendency toward higher energies. However, due to the small number of counts detected, classification of this X-ray source remains ambiguous.

\textbf{[147] IC 5179}: This galaxy shows dim soft-band extended emission near the nucleus and many X-ray point-sources spreading along the optical disk, which most likely correspond to X-ray binaries, $8^{+0}_{-3}$ of which are classified as ULXs \citep{LuaRob2015}. The presence of these sources makes the radial profile seem rather irregular, especially in hard band (see Appendix \ref{MWimages}).

\textbf{[148] CGCG 465$-$012}: This galaxy is paired with UGC 02894 at its north-west, at a distance of $\sim 4.2$', which is resolved as a separate source by \textit{IRAS}.

CGCG 465$-$012 shows diffuse soft X-ray emission along its optical disk, concentrated in the nucleus and in a north-eastern region $\sim 5$'' from it, most likely a star-forming region. The hard-band emission is very dim and non-peaked. 

\textbf{[157] MCG$-$02$-$33$-$098/9}: This system is composed of two very closely interacting galaxies (separated by $\sim 14$''), with the western one contributing $\sim 70$\% of the \textit{IRAS} flux \citep{DiaCha2010}. Two nearby galaxies, at $\sim 0.7$' north-west and $\sim 2$' south-east, most likely do not contribute to the \textit{IRAS} flux, as they are not detected in MIR wavelengths. 

\cite{TerHir2015}, using \textit{XMM-Newton EPIC-PN} data, report the detection of a hint of an iron line at 6.97 keV. The \textit{Chandra} spectrum of MCG$-$02$-$33$-$098 shows a slight increase at $\sim 7$ keV, which is not significant enough to claim the presence of an excess. 

\textbf{[163] NGC 4418}: This galaxy is paired with MCG+00$-$32$-$013 at its south-east, at a distance of $\sim 3$'. Using infrared photometry, it can be determined that more than 99\% of the \textit{IRAS} flux originates in NGC 4418 \citep{ChuSan2017}, therefore this nearby galaxy is not considered in the analysis. It also meets our PAH EW selection criteria \citep{StiArm2013}, and thus we classify it as AGN, despite the current debate regarding its nature. 

Only two central peaks can be seen in the \textit{Chandra} data; the eastern one is brighter in the hard band, and thus is used to center the radial profiles. However, this source is known as a possible candidate for containing a heavily obscured AGN, and it is possible that we are resolving the non-absorbed emission at both sides of the nucleus. While some studies in radio and infrared seemed to favor a compact starburst as a central source \citep[e.g.][]{RouHel2003, LahSpo2007}, it is at least clear that the nucleus is extremely Compton-thick, and could host either an AGN, or a starburst of obscuration as extreme as the one in Arp 220 \citep[see][and references therein]{CosAal2013,CosSak2015}. 

\textbf{[169] ESO 343$-$IG013}: Both galaxies in this closely interacting merger, separated by only $\sim 0.2$', contribute to the \textit{IRAS} flux, with the northern component dominant in both infrared \citep{DiaCha2010} and X-rays. 

The X-ray emission is diffuse in the southern source and between sources. The northern source shows a bright X-ray peak in both the soft and hard bands, which meets one of our AGN selection criteria, having $HR=0.01 \pm 0.05$. The X-ray spectrum of this component also shows an increase of flux toward higher energies.

The radial profile of the southern source has been centered using the brightest peak in optical and infrared images, which corresponds to the dimmer X-ray peak of the $0.5-7$ keV image shown in Appendix \ref{MWimages}. The northern source presents very compact X-ray emission, as derived from its radial profile, compared to its extended infrared emission (as plotted in Fig. \ref{fig:RSXvsRIR}).

\textbf{[170] NGC 2146}: This galaxy, most likely a post-merger object \citep{HutLo1990}, shows very extended soft-band X-ray emission in the direction perpendicular to the plane of the optical disk, originating in a super-wind driven by the central starburst \citep[see][]{KreArm2014}. The hard-band emission is limited to the region encompassed by the galaxy disk, which presents a lack of soft-band emission. This lack is most likely a result of absorption in the plane of the galaxy. The radial profiles have been centered using the brightest peak in the NIR IRAC channel 1 image.  

A detailed \textit{Chandra} analysis of point sources in the galaxy, including 7 ULXs, and the extended emission can be found in \citet{InuMat2005}. 

\textbf{[174] NGC 5653}: This source presents a diffuse X-ray emission along the spiral arms seen in the optical images, with a bright X-ray knot at a distance of $\sim 15$'' from the nucleus. This knot falls on one of the spiral arms and appears very blue in optical images, also being the strongest X-ray and infrared source in the galaxy \citep{DiaCha2010}. It could be argued that this source is actually a second galaxy, merging with the bigger NGC 5653. This source has been classified as a lopsided galaxy \citep{RudRix2000}, which is usually assumed to be an indicator of weak tidal interaction. However, as we have no clear evidence of this, and do not see a central point-source in hard X-rays, we opt to consider it a particularly active star-forming region.  \citet{LuaRob2015} find $1 \pm 1$ ULXs in this source. 

The NIR IRAC channel 1 image was used to centre the radial profile, due to the difficulty of finding a clear nucleus in the X-ray data. 

\textbf{[178] NGC 4194}: This source, commonly known as the Medusa, is the result of a merger with very particular tidal features, as seen in optical images. The X-ray morphology of the source is also particular, showing a very extended emission in soft X-rays, especially toward the north-west, most likely indicative of a strong starburst-driven wind. \citet{LuaRob2015} find $1 \pm 1$ ULXs in this source, while a detailed study of all X-ray point sources was performed by \citet{KaaAlo2008}.

\textbf{[179] NGC 7591}: This galaxy is interacting with PGC 214933, at $\sim 1.8$' southwest; and a dimmer galaxy lying $\sim 3.6$' to the east is detected in HI \citep{KuoLim2008}. PGC 214933 contributes $\sim 6$\% of the \textit{IRAS} flux \citep{ChuSan2017}, and is undetected in the \textit{Chandra} data. 

With only $\sim 26$ cts in the $0.5-7$ keV band, no X-ray analysis beyond the calculation of the $HR$ and the extraction of the radial profile has been performed for NGC 7591.

NGC 7591 is the only source in our sample that is optically classified as a Seyfert 2 galaxy \citep{YuaKew2010}, while being classified as a LINER by \citet{VeiKim1999}, and does not meet any of our AGN selection criteria. The contribution of an AGN component to the the bolometric luminosity of the galaxy, as estimated by \citep{DiaArm2017}, is low: $AGN_{bol}=0.09\pm0.02$. With a flux estimation using CIAO's tool \textit{srcflux}, the obtained luminosity of this source in the $0.5-7$ keV band is $L_X= 2.5_{-1.3}^{+1.8} \times 10^{40}$ erg s$^{-1}$. This X-ray luminosity is low for an AGN, and only possible for this type of source if it's Compton-thick. Even if we cannot rule out this possibility, we consider that a single optical classification as Seyfert 2 is not a strong enough criterion to classify this source as an AGN. 

\textbf{[182] NGC 0023}: This source is paired with NGC 0026, at a distance of $\sim 9.2$' \citep{HatYos2004}, which is far enough to guarantee no contribution to the \textit{IRAS} flux. 

This source shows central extended emission surrounding the nucleus, which is not a strongly peaked hard X-ray source. The source spectrum shows a lack of emission at $> 3$ keV, and a small excess at higher energies, which is not significant enough to hint toward the presence of an AGN. \citet{LuaRob2015} classify $2 \pm 2$ of the galaxy's point sources as ULXs. 

\textbf{[188] NGC 7552}: This source shows extended soft-band X-ray emission in the inner region of the galaxy, surrounded by numerous point-sources, which most likely correspond to X-ray binaries. Of these, $2^{+3}_{-1}$ are classified as ULXs \citep{LuaRob2015}.

\textbf{[191] ESO 420$-$G013}: This source meets our infrared [Ne v] line criterion for AGN selection, and shows a slight excess at around 6.4 keV, which is not significant enough in the \textit{Chandra} data to confirm the presence of an X-ray AGN. Though \citet{YuaKew2010} classify is as HII dominated, other optical classifications have pointed toward a Seyfert nature previously \citep[e.g.][]{MaiSuz1996}

\textbf{[194] ESO 432$-$IG006}: Both galaxies in this system, separated by $\sim 0.5$', contribute similarly to the \textit{IRAS} flux, and have significant X-ray emission. Also both galaxies present signs of the presence of an absorbed AGN. Fitting such a model, with a fixed photon index of 1.8, on the north-eastern source yields an absorbing column density of $N_H \sim 4 \times 10^{23}$ cm$^{-2}$; and on the south-western source it yields $N_{\rm H} \sim 1 \times 10^{23}$ cm$^{-2}$. 

\textbf{[195] NGC 1961}: This source has most of its X-ray emission concentrated on the nucleus, with a few point sources spreading along the spiral arms. Another peak of emission is found in a region about $\sim 20$'' west of the nucleus. It is unclear if this source is truly overlapping with emission regions in infrared images (IRAC channels 1$-$4, MIPS24/70), or actually has no counterpart.

The spectrum presented in Fig. \ref{fig:XraySpectra} includes the full region, and the hard X-ray excess actually originates in this outer source. Therefore, we see no spectral traces of the presence of an AGN. 

\textbf{[196] NGC 7752/3}: This double system is composed by NGC 7752 at south-west and NGC 7753 $\sim 2$' to the north-east, the latter being the dominant source in the \textit{IRAS} flux \citep{ChuSan2017}.

X-ray emission in the north-eastern source is point-like in the nucleus, with a few other point sources spreading along the galaxy's spiral arms. The X-ray best spectral fit for this source is performed using a single \textit{mekal} component for the full spectrum, though no fit is truly satisfactory.

The south-western source presents much more diffuse emission, with no clear central point source. The most concentrated emission, in soft band, is coming from a region just west of the infrared nucleus of the galaxy, which is the one used to centre the radial profile. This region seems to correspond to a slight increase of infrared emission in the IRAC channel 1 image, which means it is most likely a star-forming region. Using optical diagnostic diagrams \citep{KewDop2001, KewGro2006} this source can be classified either as an AGN or composite of AGN and starburst \citep{ZhoWu2014,PerAlo2015}, though we see no X-ray signs of activity. 

\textbf{[198] NGC 1365}: This very bright source meets all three of our X-ray AGN selection criteria, with $HR=-0.19 \pm 0.00$, an absorbing column density of $N_{\rm H} \sim 3 \times 10^{23}$ cm$^{-2}$ when fitting an absorbed AGN model with fixed photon index of 1.8, and a 6.4 keV iron $K_\alpha$ line at a significance of $\sim 2.9 \sigma$. This galaxy hosts a very well-known AGN, with frequent dramatic spectral variability \citep[e.g.][]{RisElv2005, RisElv2007}, attributable to variations in the column density along the line of sight. 

The \textit{Chandra} data show bright and extended diffuse emission in soft band, spreading along the central region of the galaxy, and a strong central point source \citep[see e.g.][]{WanFab2009}. \citet{LuaRob2015} find $6^{+2}_{-1}$ ULXs in this galaxy. Note that we have chosen to show only the central emission in the galaxy in Appendix \ref{MWimages}, and not the open spiral arms around it, in order to better distinguish the X-ray morphology of the region of interest. 

\textbf{[199] NGC 3221}: This source consists of, mostly, point-like sources scattering along the optical edge-on disk of the galaxy, $6^{+0}_{-1}$ of which are classified as ULXs \citep{LuaRob2015}. The hard-band spectrum can be fitted with a power-law of index $\Gamma=0.3 \pm 0.6$, with the nucleus being a stronger hard X-ray source than the other point sources. Even though such a flat spectrum could be indicative of an obscured AGN, fitting an absorbed power-law with a fixed photon index of 1.8 yields an absorbing column density of only  $N_{\rm H} \sim 5 \times 10^{22}$ cm$^{-2}$. We conclude that we cannot confirm the presence of an absorbed AGN, even though we cannot rule out the possibility either.

\textbf{[201] NGC 0838}: This galaxy is in a complex system. NGC 0839 is placed $\sim 2.4$' to the south-east, and the centre of the closely interacting system formed by NGC 0833 and NGC 0835 is found at $\sim 4$' to its west. \textit{IRAS} resolves three of these four sources, being NGC 0833/0835 too closely interacting to derive their fluxes separately. However, because NGC 0838 is resolved, we do not include the rest of the components of the complex system in this work.

\citet{OdaUed2018} analyze $3-50$ keV \textit{NuSTAR} data of this compact group. NGC 0838 is not detected above 8 keV, showing no evidence of an obscured AGN. The conclusion that NGC 0838 is a starburst-dominated galaxy is also reached in the detailed works by \citet{OsuZez2014,TurRee2001}.

This source is very bright in X-rays, and has a complex morphology of diffuse soft-band emission surrounding the nucleus, a clear example of a strong starburst wind. Among the point sources, $2^{+1}_{-0}$ are classified as ULXs \citep{LuaRob2015}.

\section{Observations}\label{MWimages}

\begin{figure*}
\centering
\includegraphics[width=\textwidth,keepaspectratio]{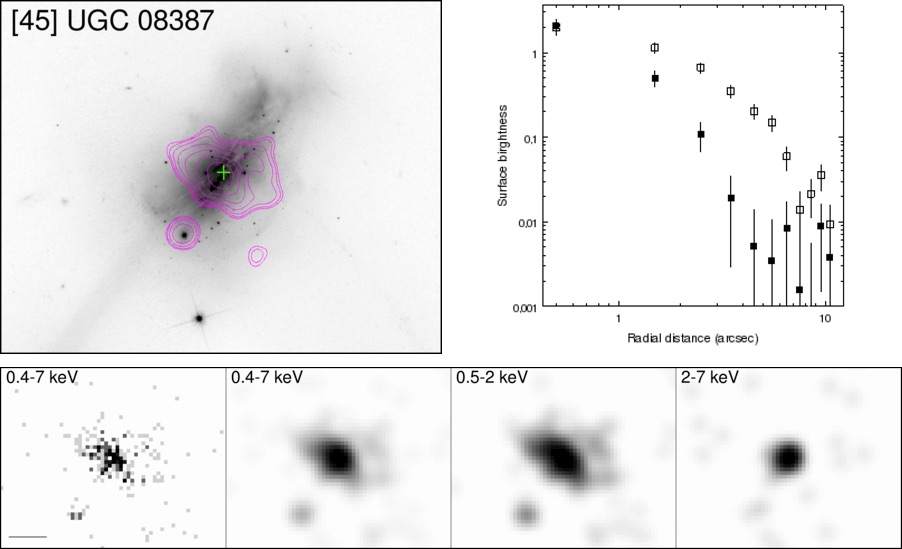}
\caption{\textit{Overlay on:} HST-ACS F814W. \textit{Contours:} Interval 1.}
\end{figure*}

\begin{figure*}
\centering
\includegraphics[width=\textwidth,keepaspectratio]{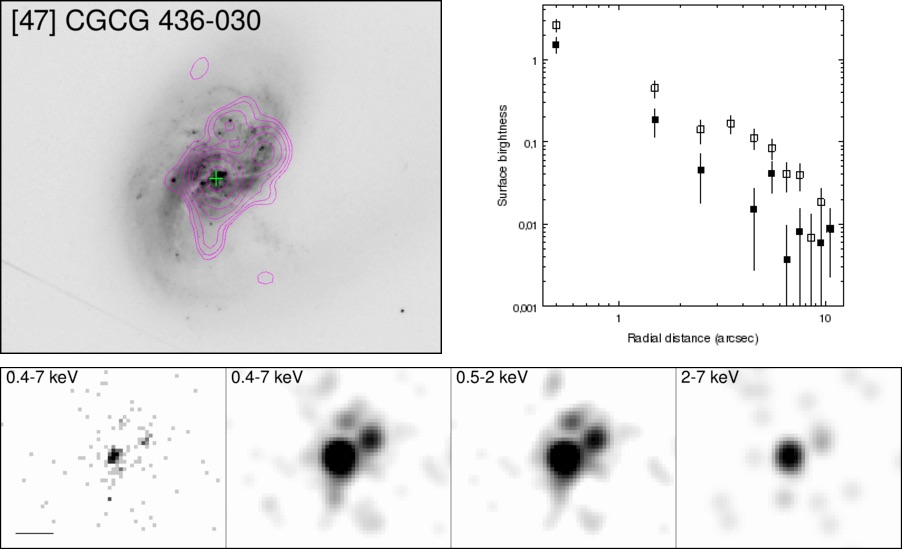}
\caption{\textit{Overlay on:} HST-ACS F814W. \textit{Contours:} Interval 1.}
\end{figure*}

\begin{figure*}
\centering
\includegraphics[width=\textwidth,keepaspectratio]{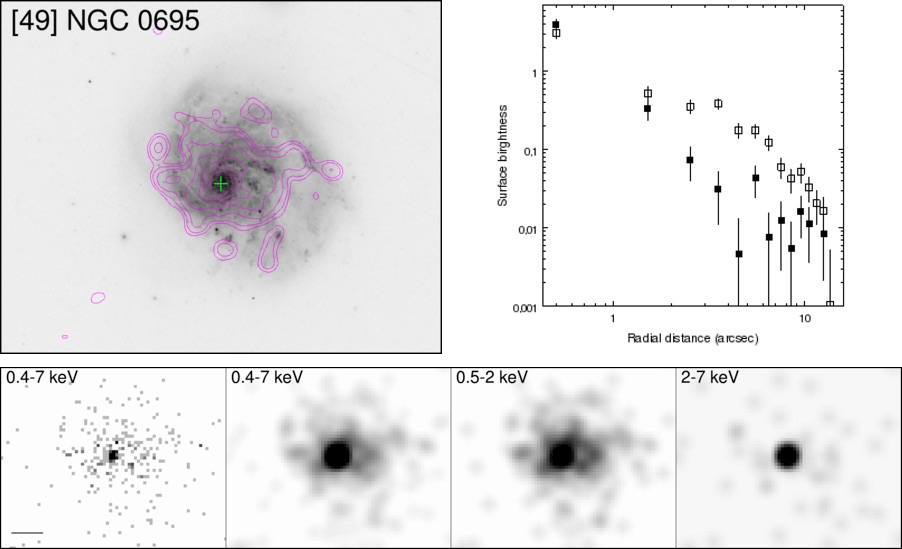}
\caption{\textit{Overlay on:} HST-ACS F814W. \textit{Contours:} Interval 1.}
\end{figure*}

\begin{figure*}
\centering
\includegraphics[width=\textwidth,keepaspectratio]{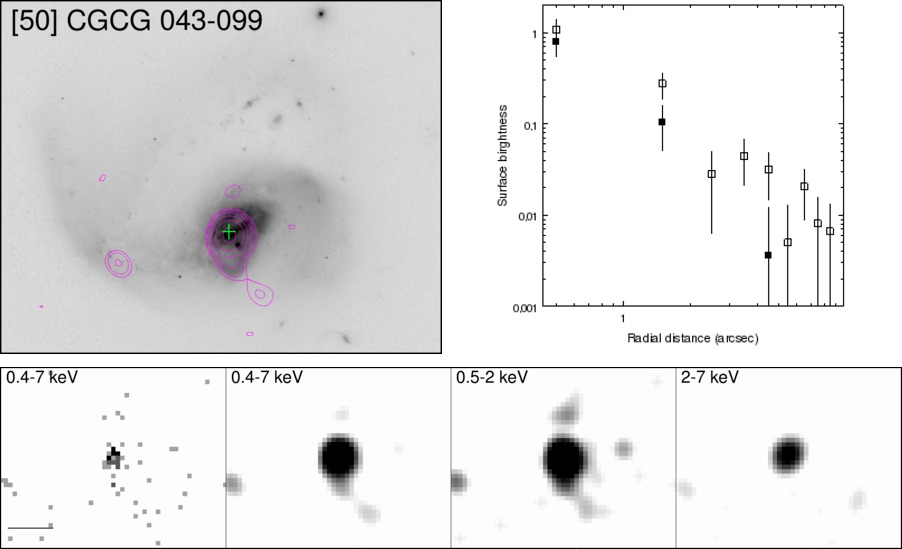}
\caption{\textit{Overlay on:} HST-ACS F814W. \textit{Contours:} Interval 1.}
\end{figure*}

\begin{figure*}
\centering
\includegraphics[width=\textwidth,keepaspectratio]{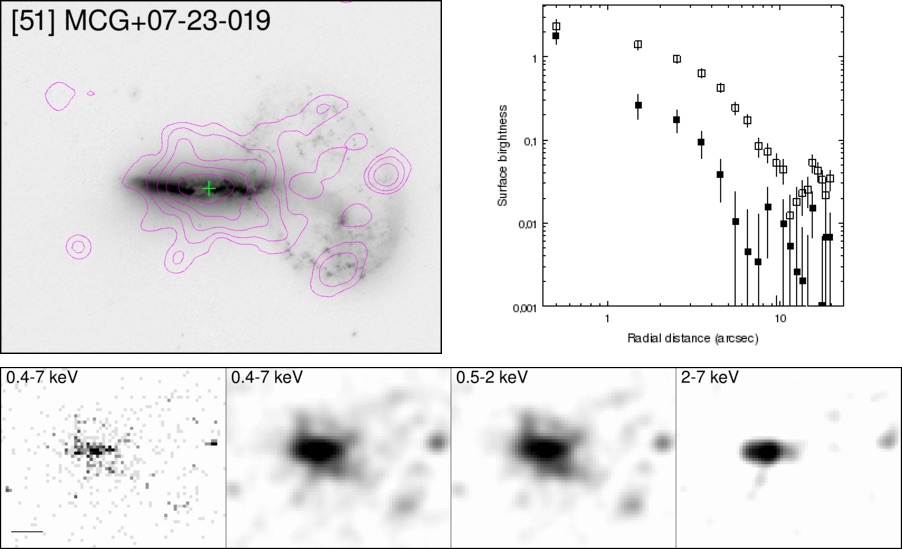}
\caption{\textit{Overlay on:} HST-ACS F814W. \textit{Contours:} Interval 3.}
\end{figure*}

\begin{figure*}
\centering
\includegraphics[width=\textwidth,keepaspectratio]{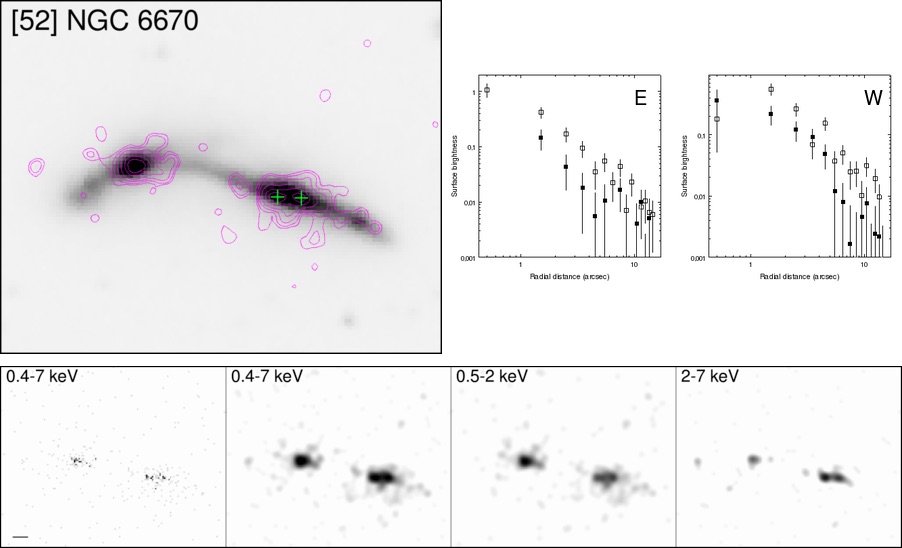}
\caption{\textit{Overlay on:} HST-ACS F814W. \textit{Contours:} Interval 1.}
\end{figure*}

\begin{figure*}
\centering
\includegraphics[width=\textwidth,keepaspectratio]{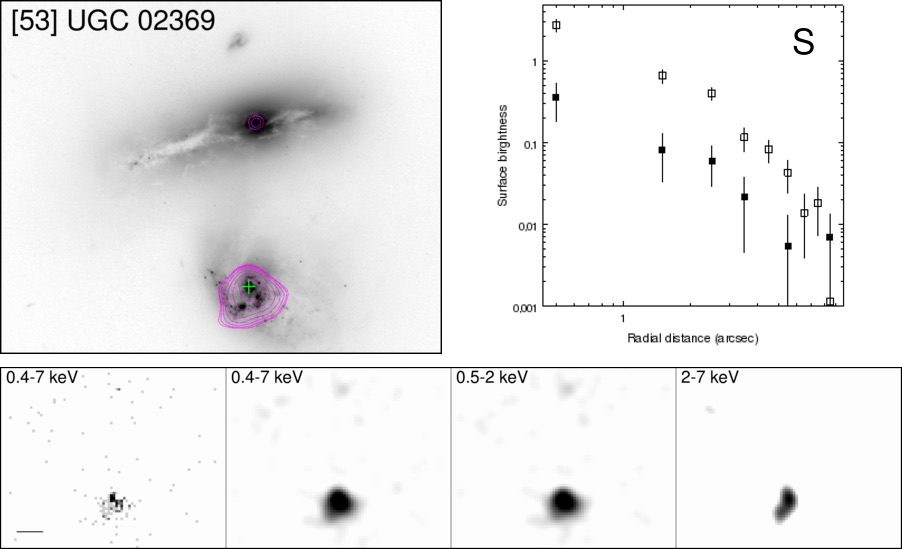}
\caption{\textit{Overlay on:} HST-ACS F814W. \textit{Contours:} Interval 4.}
\end{figure*}

\begin{figure*}
\centering
\includegraphics[width=\textwidth,keepaspectratio]{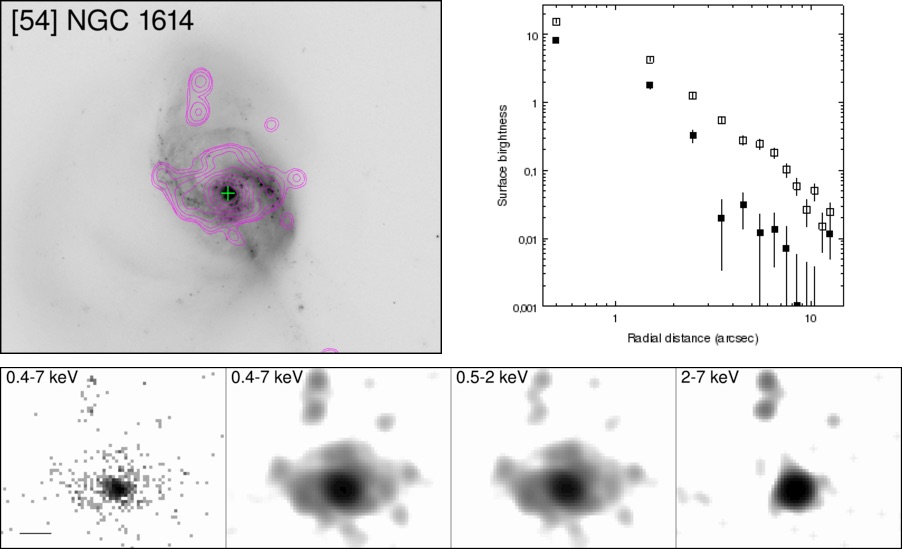}
\caption{\textit{Overlay on:} HST-ACS F814W. \textit{Contours:} Interval 1.}
\end{figure*}

\begin{figure*}
\centering
\includegraphics[width=\textwidth,keepaspectratio]{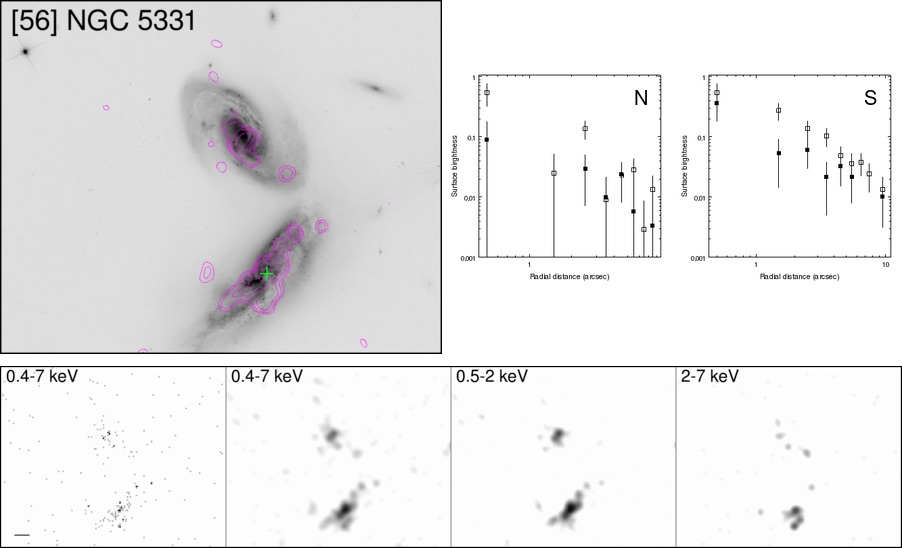}
\caption{\textit{Overlay on:} HST-ACS F814W. \textit{Contours:} Interval 2.}
\end{figure*}

\begin{figure*}
\centering
\includegraphics[width=\textwidth,keepaspectratio]{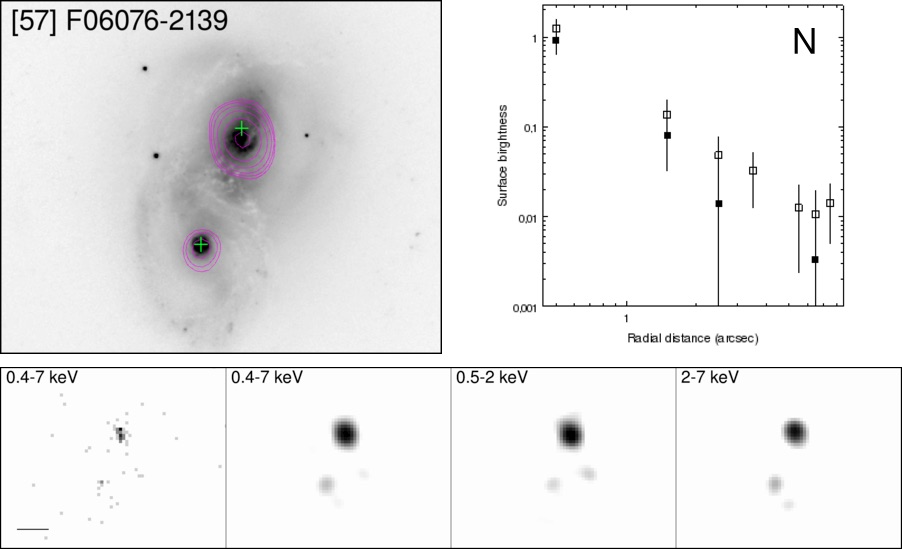}
\caption{\textit{Overlay on:} HST-ACS F814W. \textit{Contours:} Interval 1.}
\end{figure*}

\begin{figure*}
\centering
\includegraphics[width=\textwidth,keepaspectratio]{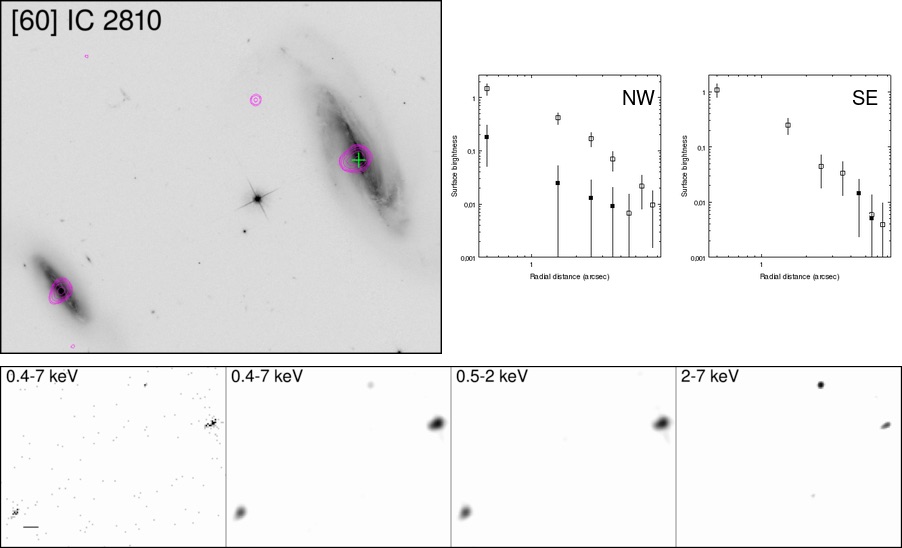}
\caption{\textit{Overlay on:} HST-ACS F814W. \textit{Contours:} Interval 1.}
\end{figure*}

\begin{figure*}
\centering
\includegraphics[width=\textwidth,keepaspectratio]{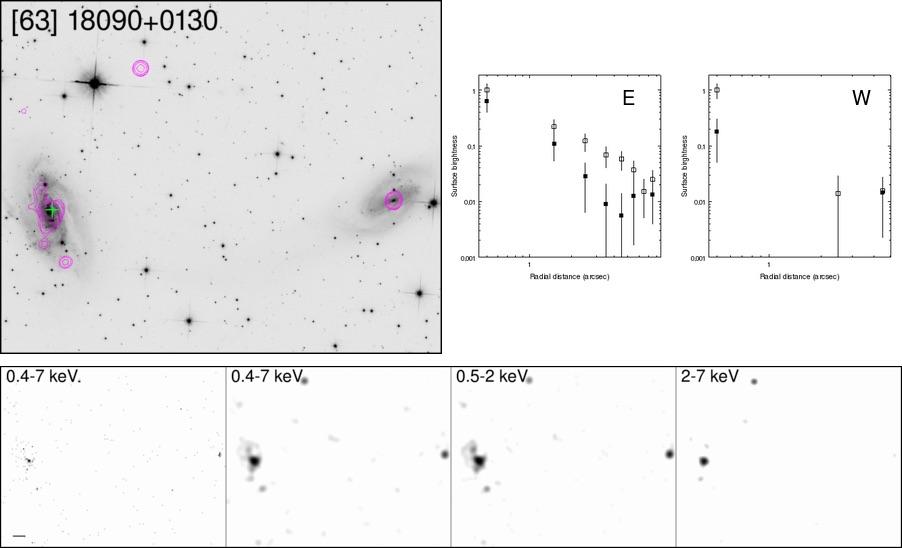}
\caption{\textit{Overlay on:} HST-ACS F814W. \textit{Contours:} Interval 1.}
\end{figure*}

\begin{figure*}
\centering
\includegraphics[width=\textwidth,keepaspectratio]{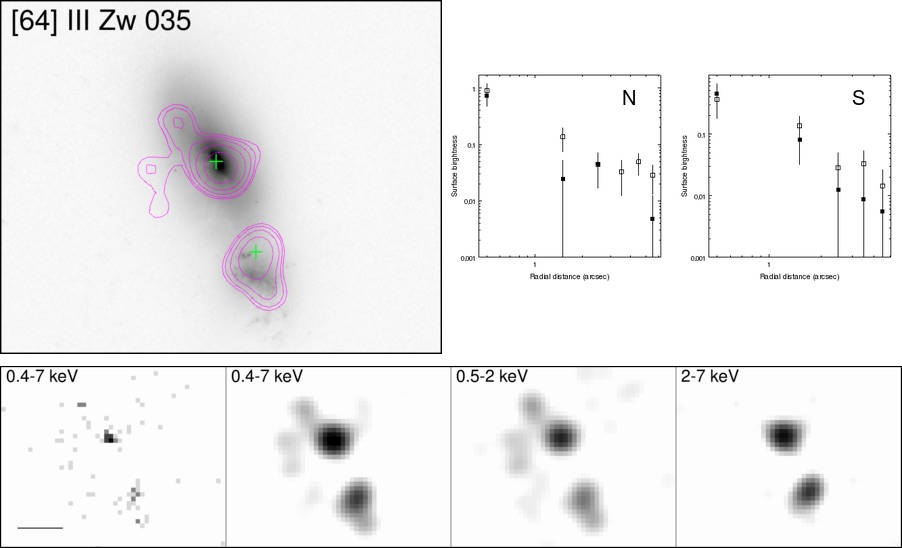}
\caption{\textit{Overlay on:} HST-ACS F814W. \textit{Contours:} Interval 1.}
\end{figure*}

\begin{figure*}
\centering
\includegraphics[width=\textwidth,keepaspectratio]{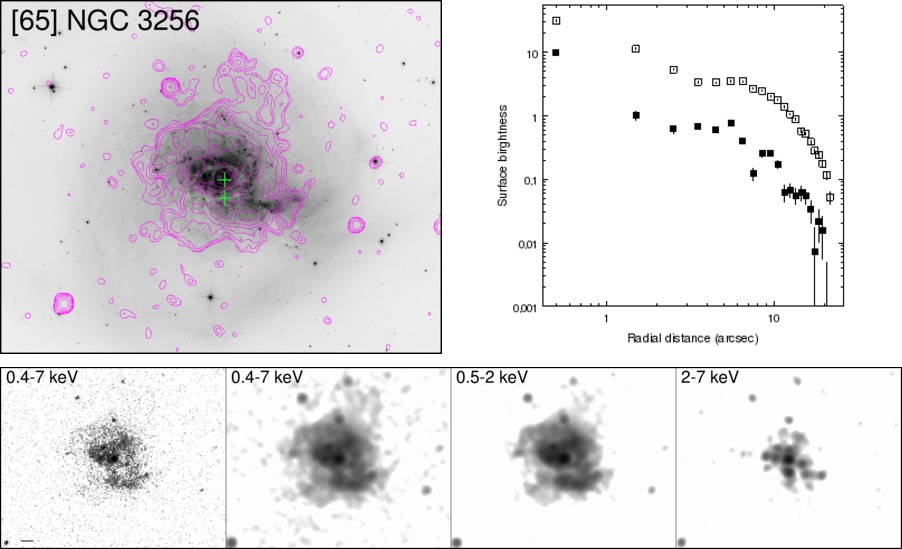}
\caption{\textit{Overlay on:} HST-ACS F814W. \textit{Contours:} Interval 1.}
\end{figure*}

\begin{figure*}
\centering
\includegraphics[width=\textwidth,keepaspectratio]{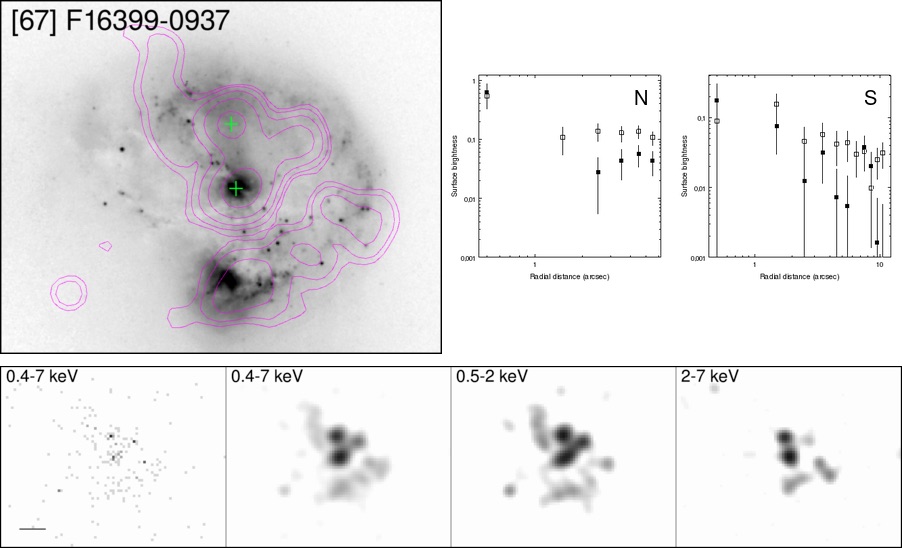}
\caption{\textit{Overlay on:} HST-ACS F814W. \textit{Contours:} Interval 1.}
\end{figure*}

\begin{figure*}
\centering
\includegraphics[width=\textwidth,keepaspectratio]{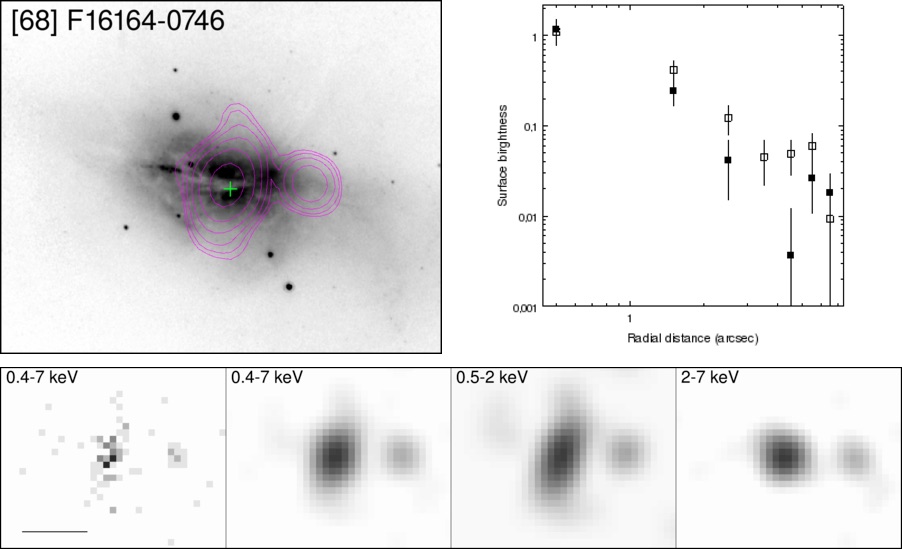}
\caption{\textit{Overlay on:} HST-ACS F814W. \textit{Contours:} Interval 1.}
\end{figure*}

\begin{figure*}
\centering
\includegraphics[width=\textwidth,keepaspectratio]{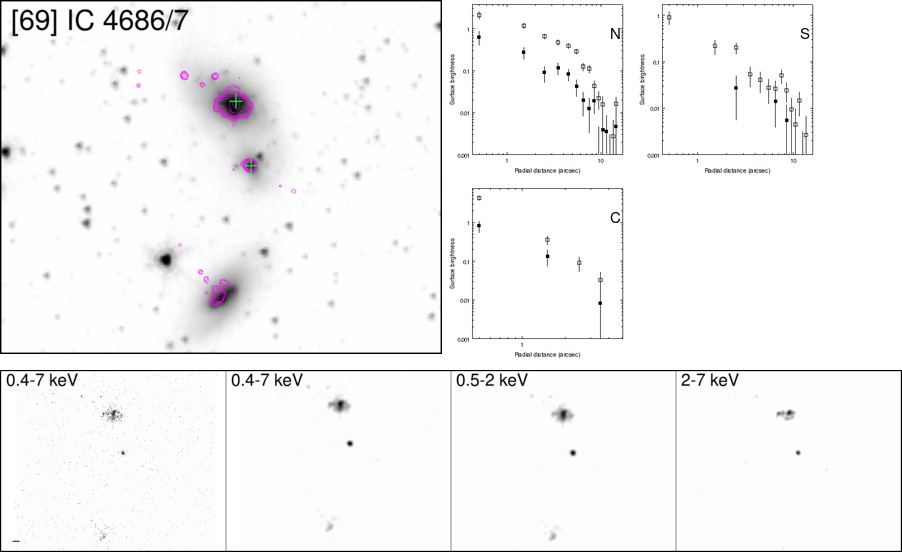}
\caption{\textit{Overlay on:} HST-ACS F814W. \textit{Contours:} Interval 1.}
\end{figure*}

\begin{figure*}
\centering
\includegraphics[width=\textwidth,keepaspectratio]{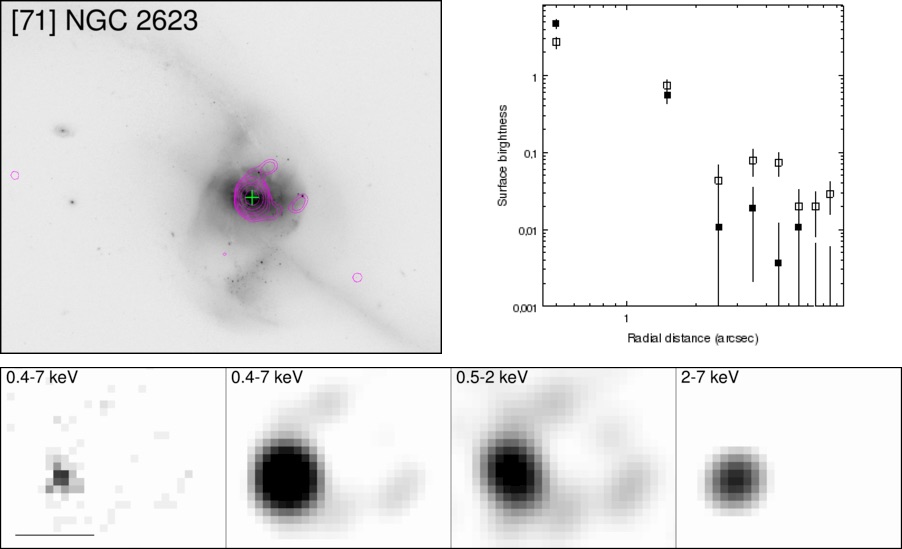}
\caption{\textit{Overlay on:} HST-ACS F814W. \textit{Contours:} Interval 1.}
\end{figure*}

\begin{figure*}
\centering
\includegraphics[width=\textwidth,keepaspectratio]{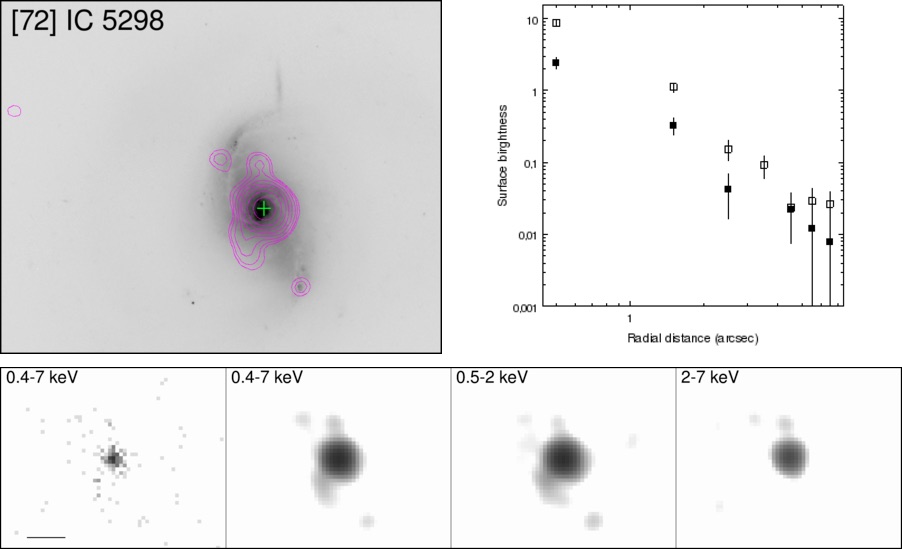}
\caption{\textit{Overlay on:} HST-ACS F814W. \textit{Contours:} Interval 1.}
\end{figure*}

\begin{figure*}
\centering
\includegraphics[width=\textwidth,keepaspectratio]{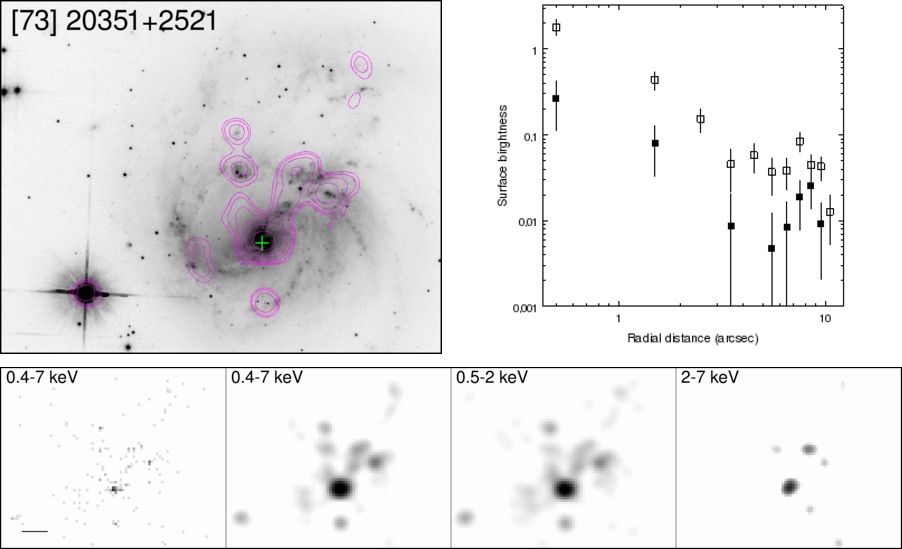}
\caption{\textit{Overlay on:} HST-ACS F814W. \textit{Contours:} Interval 1.}
\end{figure*}

\clearpage

\begin{figure*}
\centering
\includegraphics[width=\textwidth,keepaspectratio]{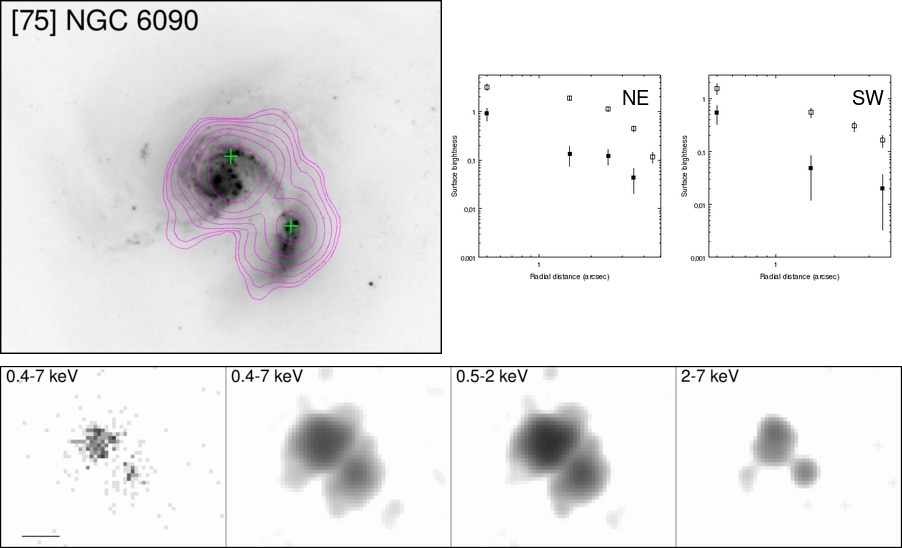}
\caption{\textit{Overlay on:} HST-ACS F814W. \textit{Contours:} Interval 1.}
\end{figure*}

\begin{figure*}
\centering
\includegraphics[width=\textwidth,keepaspectratio]{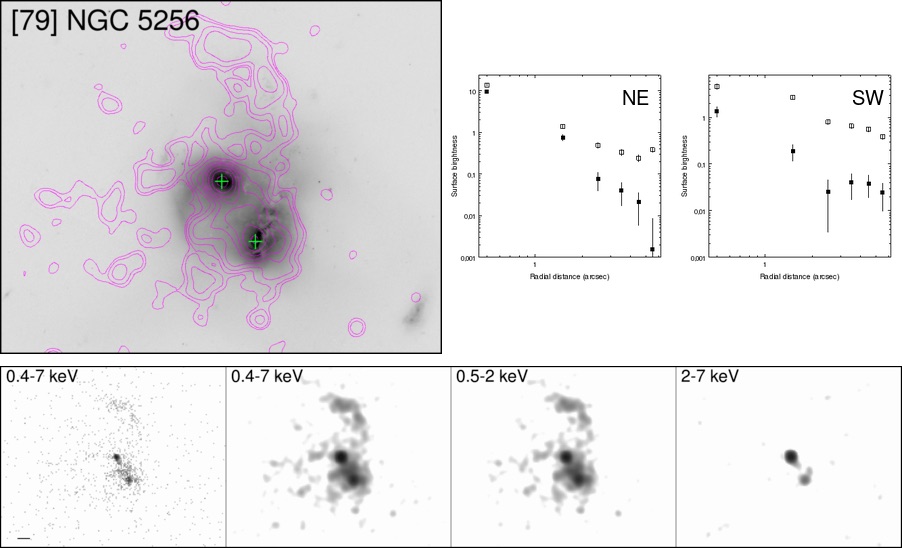}
\caption{\textit{Overlay on:} HST-ACS F814W. \textit{Contours:} Interval 2.}
\end{figure*}

\begin{figure*}
\centering
\includegraphics[width=\textwidth,keepaspectratio]{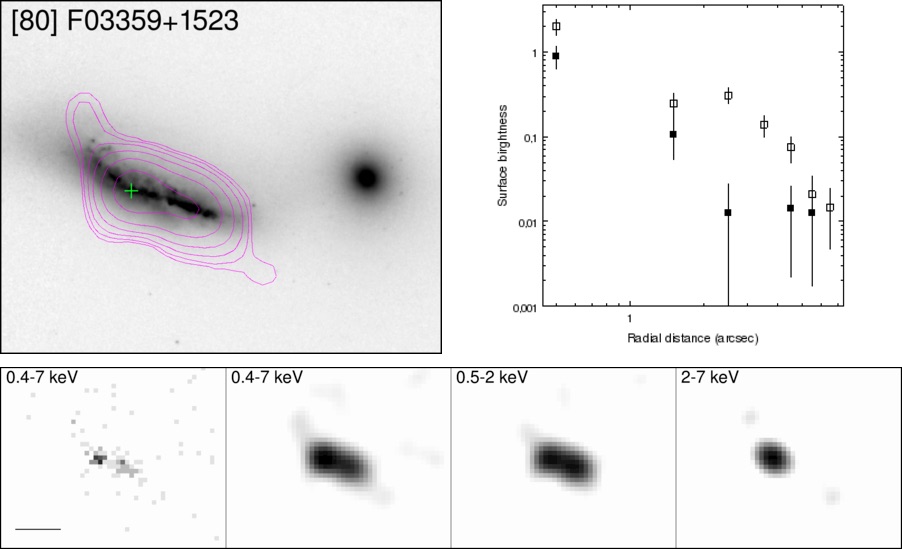}
\caption{\textit{Overlay on:} HST-ACS F814W. \textit{Contours:} Interval 1.}
\end{figure*}

\begin{figure*}
\centering
\includegraphics[width=\textwidth,keepaspectratio]{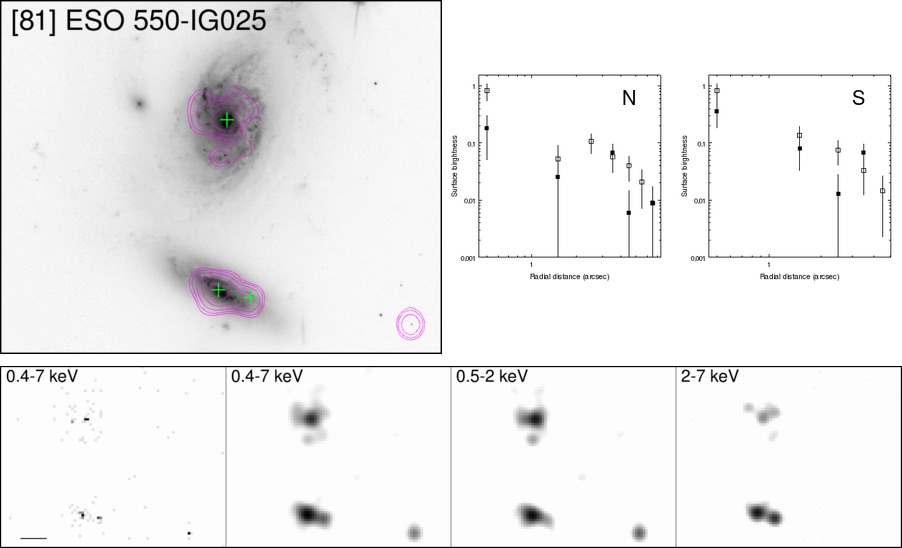}
\caption{\textit{Overlay on:} HST-ACS F814W. \textit{Contours:} Interval 1.}
\end{figure*}

\begin{figure*}
\centering
\includegraphics[width=\textwidth,keepaspectratio]{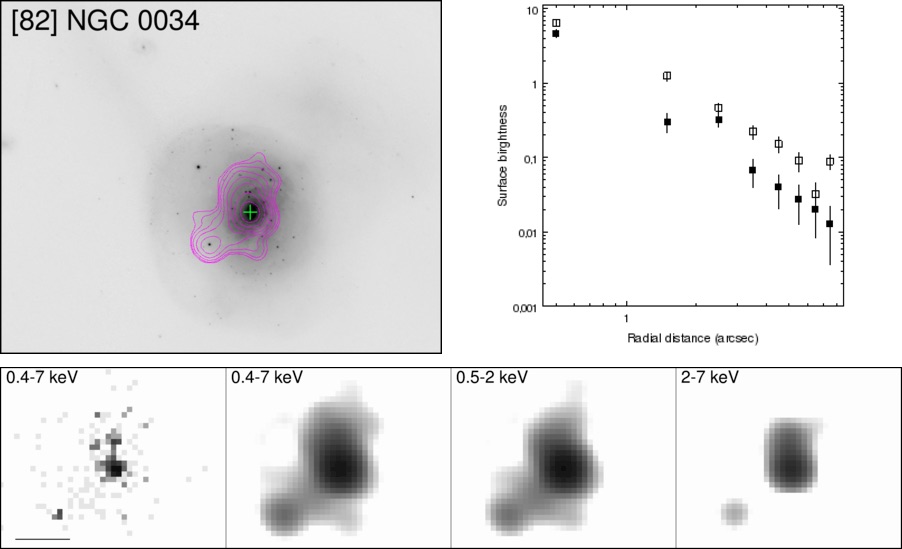}
\caption{\textit{Overlay on:} HST-ACS F814W. \textit{Contours:} Interval 1.}
\end{figure*}

\begin{figure*}
\centering
\includegraphics[width=\textwidth,keepaspectratio]{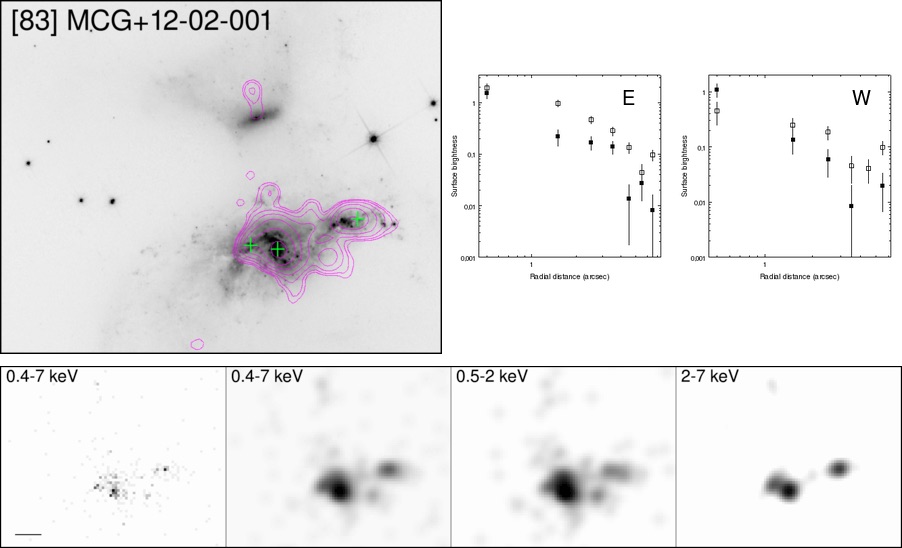}
\caption{\textit{Overlay on:} HST-ACS F814W. \textit{Contours:} Interval 1.}
\end{figure*}

\begin{figure*}
\centering
\includegraphics[width=\textwidth,keepaspectratio]{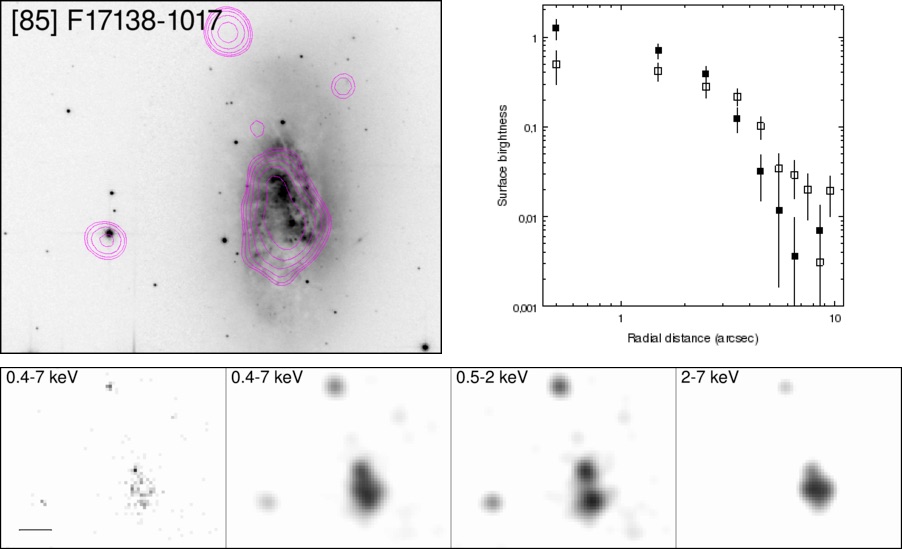}
\caption{\textit{Overlay on:} HST-ACS F814W. \textit{Contours:} Interval 1.}
\end{figure*}

\begin{figure*}
\centering
\includegraphics[width=\textwidth,keepaspectratio]{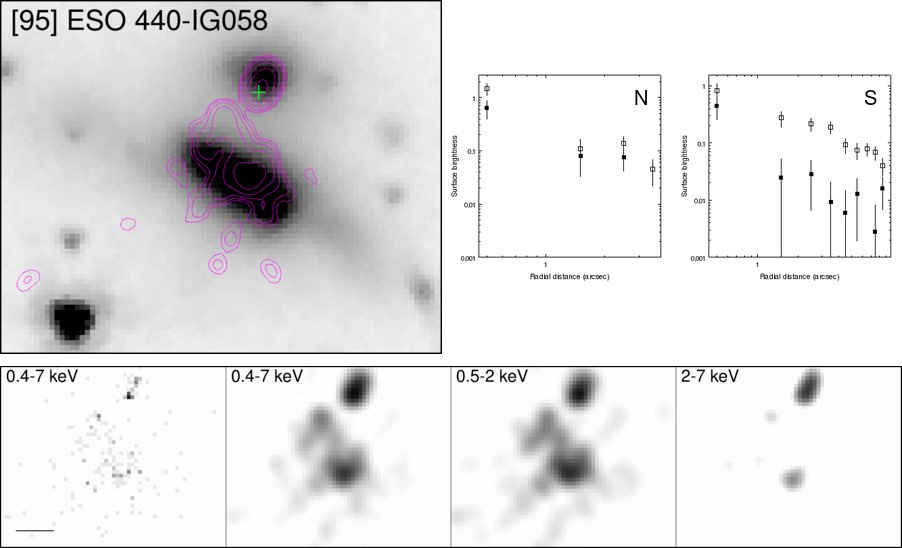}
\caption{\textit{Overlay on:} IRAC channel 1. \textit{Contours:} Interval 2.}
\end{figure*}

\begin{figure*}
\centering
\includegraphics[width=\textwidth,keepaspectratio]{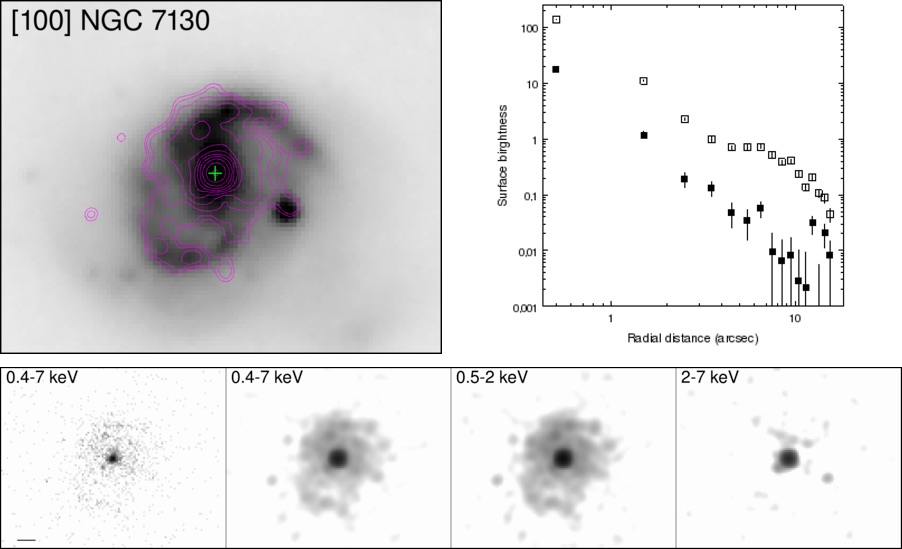}
\caption{\textit{Overlay on:} IRAC channel 1. \textit{Contours:} Interval 1.}
\end{figure*}

\begin{figure*}
\centering
\includegraphics[width=\textwidth,keepaspectratio]{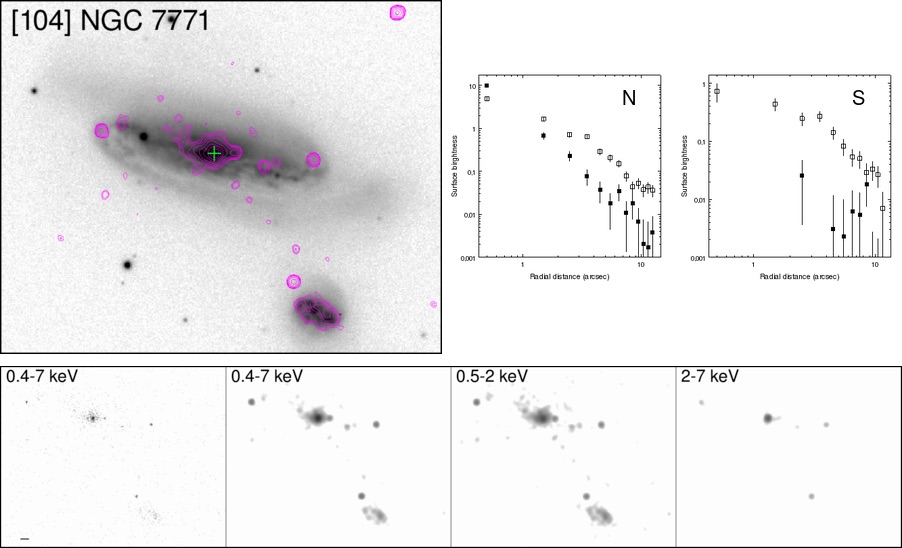}
\caption{\textit{Overlay on:} SDSS DR-12 i-band. \textit{Contours:} Interval 1.}
\end{figure*}

\begin{figure*}
\centering
\includegraphics[width=\textwidth,keepaspectratio]{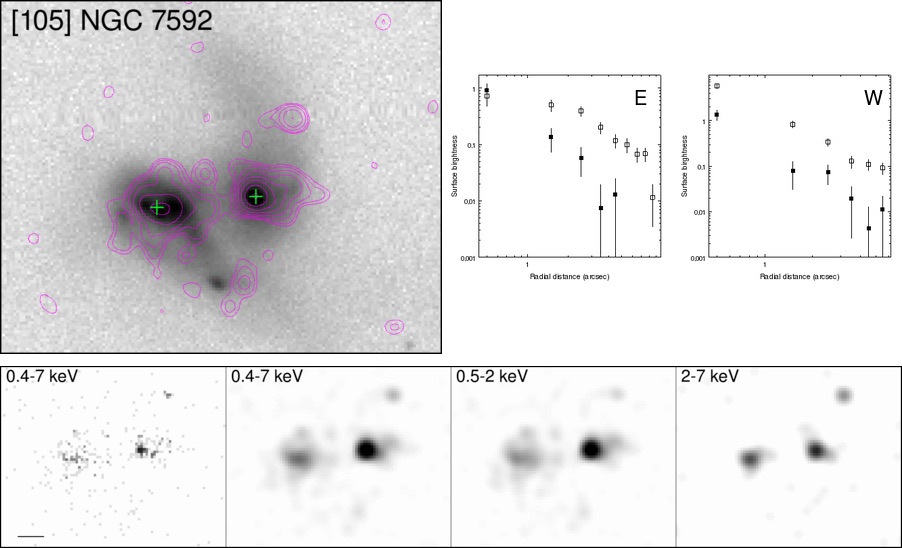}
\caption{\textit{Overlay on:} SDSS DR-12 i-band. \textit{Contours:} Interval 2.}
\end{figure*}

\begin{figure*}
\centering
\includegraphics[width=\textwidth,keepaspectratio]{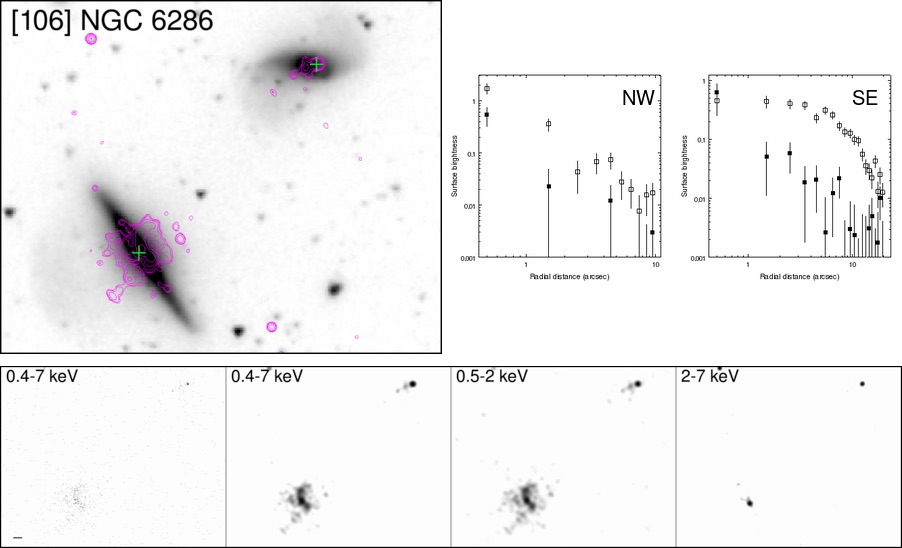}
\caption{\textit{Overlay on:} IRAC channel 1. \textit{Contours:} Interval 1.}
\end{figure*}

\begin{figure*}
\centering
\includegraphics[width=\textwidth,keepaspectratio]{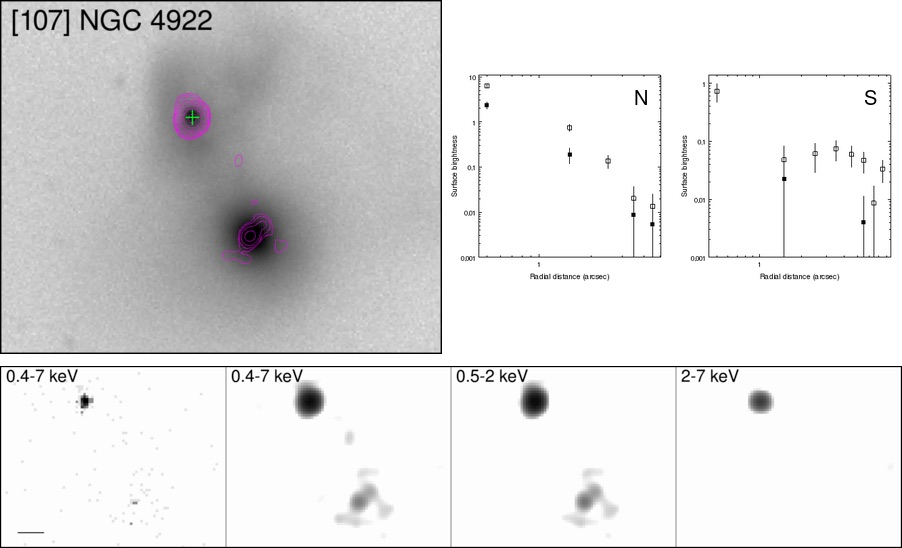}
\caption{\textit{Overlay on:} SDSS DR-12 i-band. \textit{Contours:} Interval 1.}
\end{figure*}

\begin{figure*}
\centering
\includegraphics[width=\textwidth,keepaspectratio]{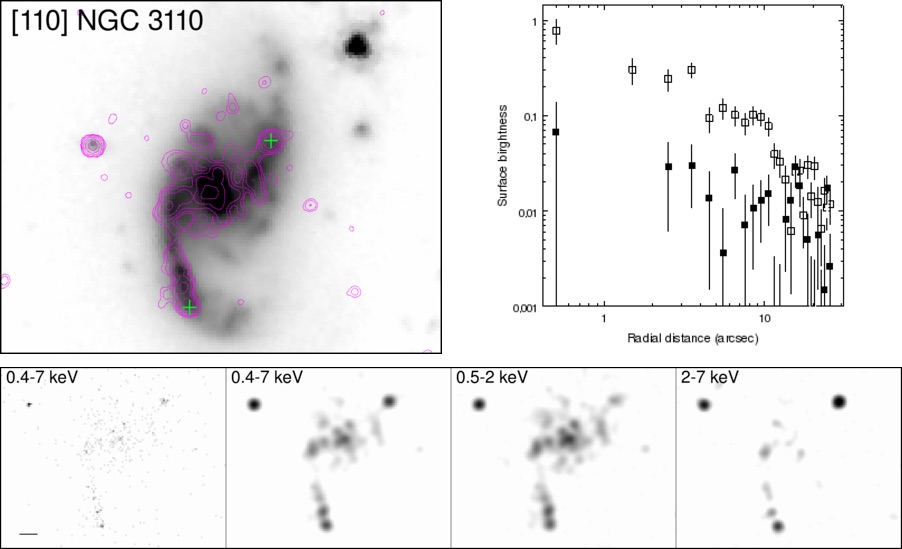}
\caption{\textit{Overlay on:} IRAC channel 1. \textit{Contours:} Interval 2.}
\end{figure*}

\begin{figure*}
\centering
\includegraphics[width=\textwidth,keepaspectratio]{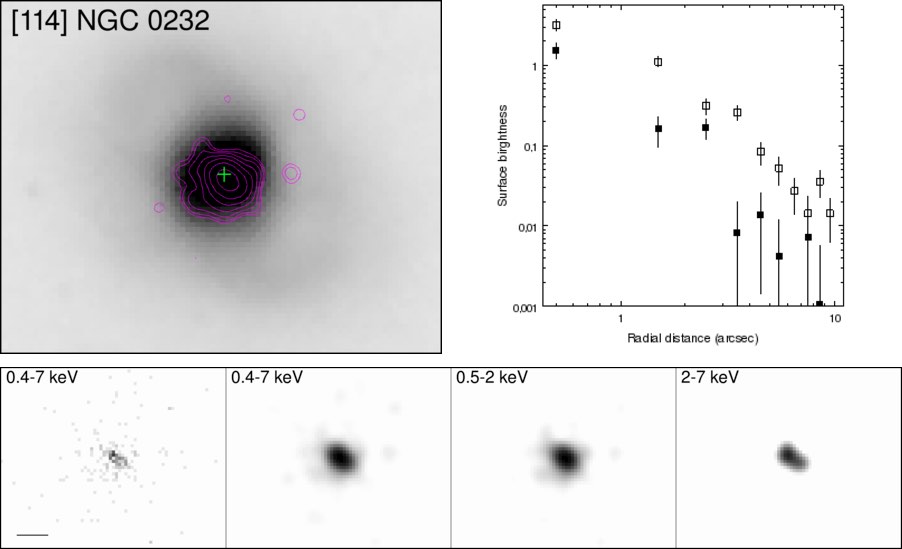}
\caption{\textit{Overlay on:} IRAC channel 1. \textit{Contours:} Interval 1.}
\end{figure*}

\begin{figure*}
\centering
\includegraphics[width=\textwidth,keepaspectratio]{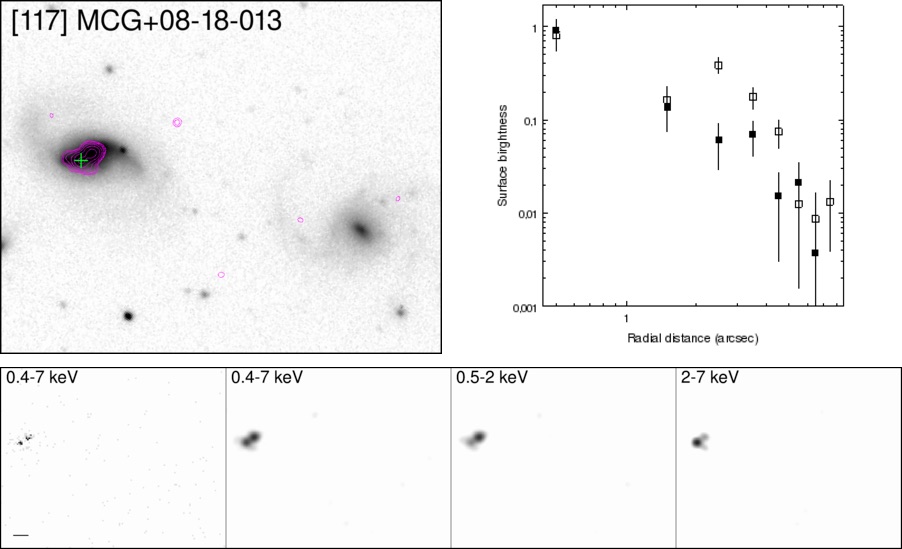}
\caption{\textit{Overlay on:} SDSS DR-12 i-band. \textit{Contours:} Interval 1.}
\end{figure*}

\begin{figure*}
\centering
\includegraphics[width=\textwidth,keepaspectratio]{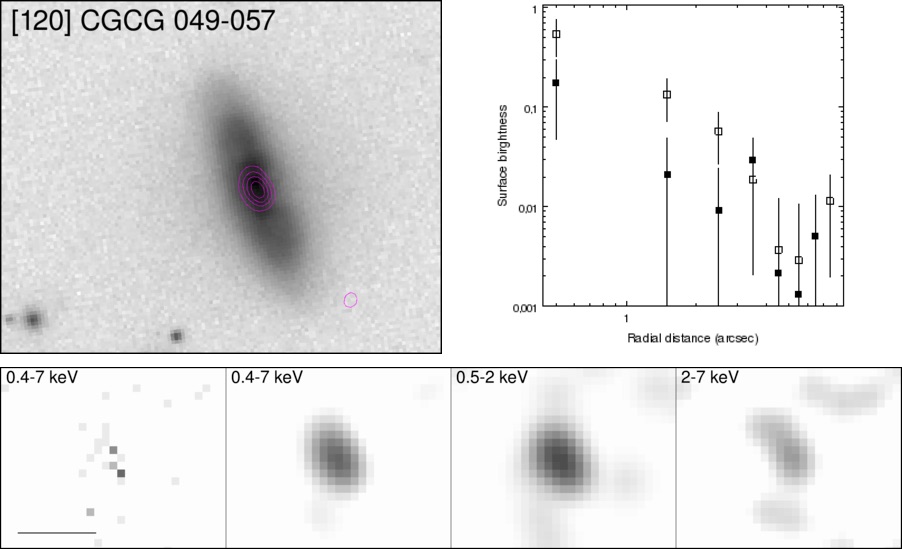}
\caption{\textit{Overlay on:} SDSS DR-12 i-band. \textit{Contours:} Custom.}
\end{figure*}

\begin{figure*}
\centering
\includegraphics[width=\textwidth,keepaspectratio]{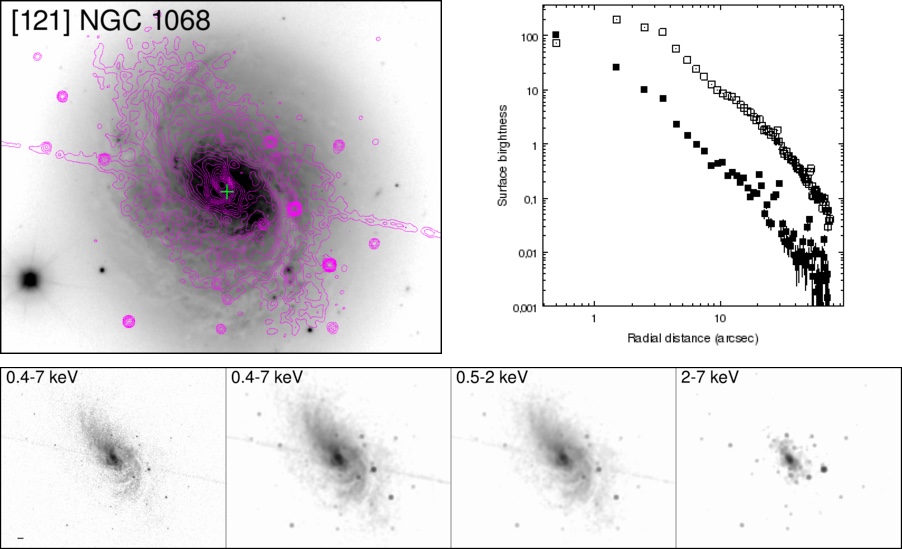}
\caption{\textit{Overlay on:} SDSS DR-12 i-band. \textit{Contours:} Custom.}
\end{figure*}

\clearpage

\begin{figure*}
\centering
\includegraphics[width=\textwidth,keepaspectratio]{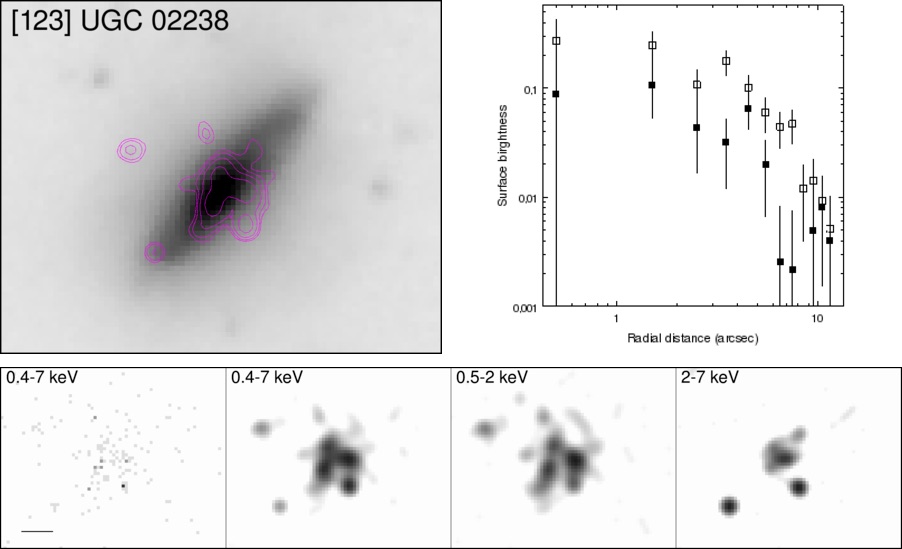}
\caption{\textit{Overlay on:} IRAC channel 1. \textit{Contours:} Interval 1.}
\end{figure*}

\begin{figure*}
\centering
\includegraphics[width=\textwidth,keepaspectratio]{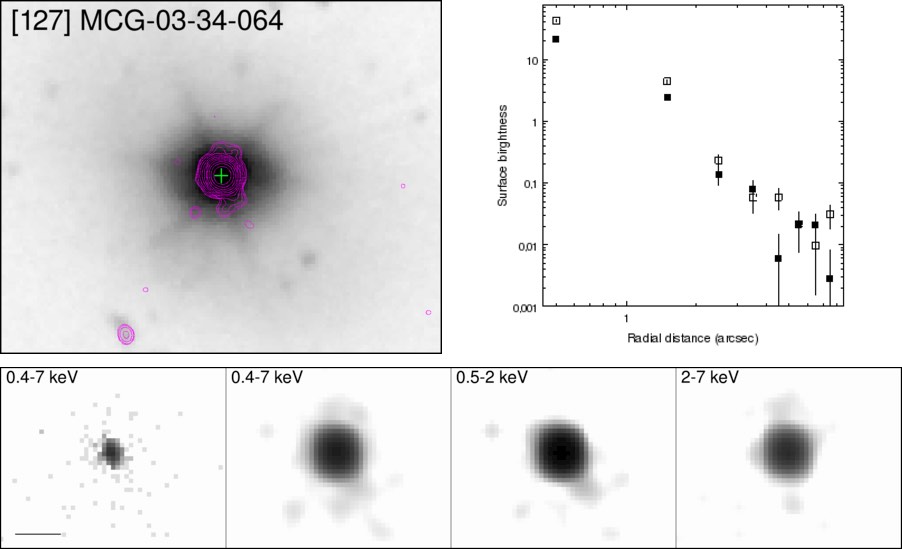}
\caption{\textit{Overlay on:} IRAC channel 1. \textit{Contours:} Interval 4.}
\end{figure*}

\begin{figure*}
\centering
\includegraphics[width=\textwidth,keepaspectratio]{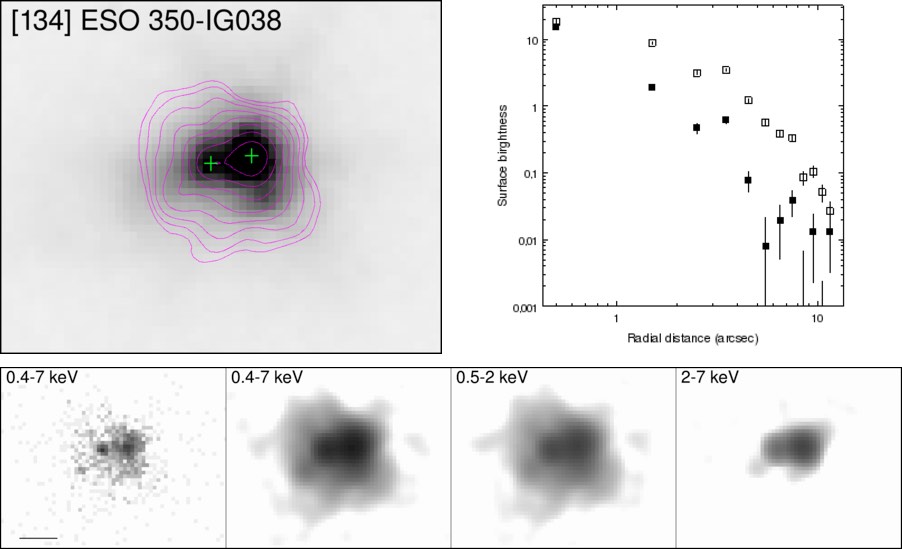}
\caption{\textit{Overlay on:} IRAC channel 1. \textit{Contours:} Interval 2.}
\end{figure*}

\begin{figure*}
\centering
\includegraphics[width=\textwidth,keepaspectratio]{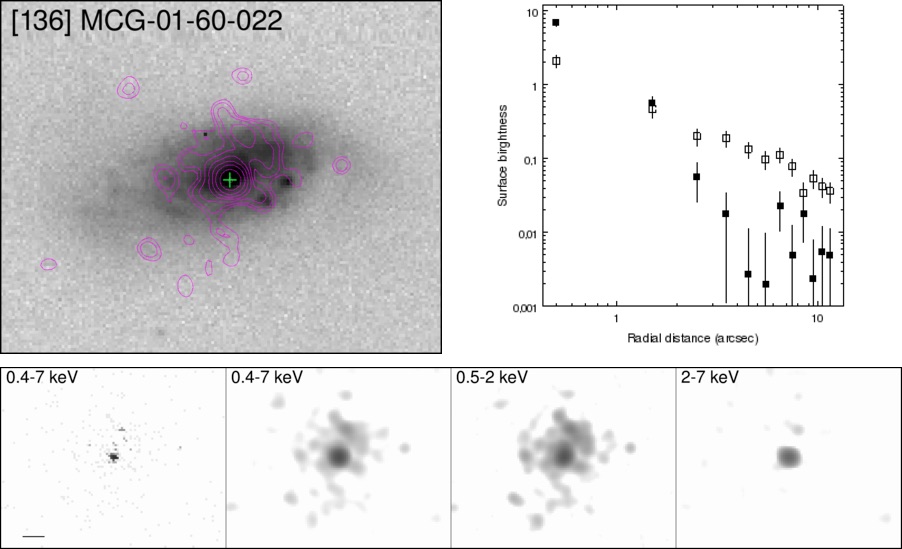}
\caption{\textit{Overlay on:} SDSS DR-12 i-band. \textit{Contours:} Interval 2.}
\end{figure*}

\begin{figure*}
\centering
\includegraphics[width=\textwidth,keepaspectratio]{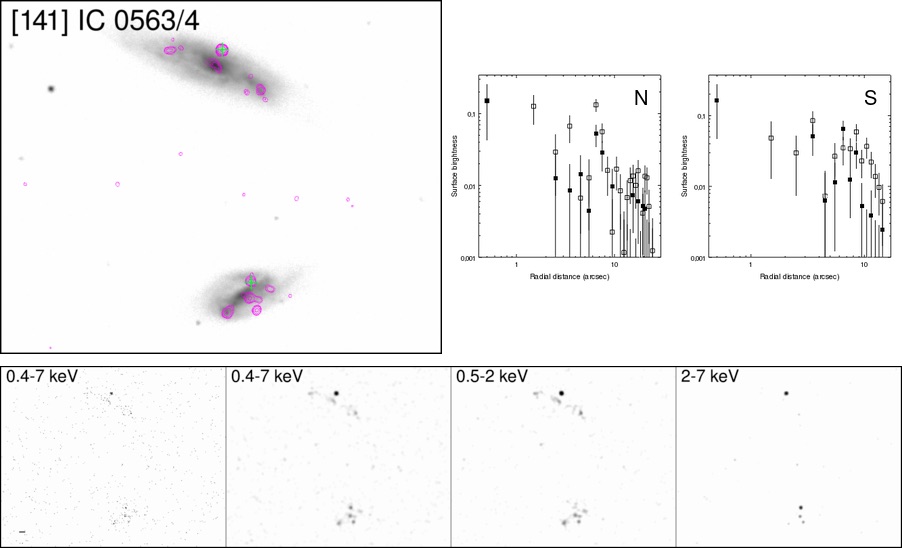}
\caption{\textit{Overlay on:} SDSS DR-12 i-band. \textit{Contours:} Interval 1.}
\end{figure*}

\begin{figure*}
\centering
\includegraphics[width=\textwidth,keepaspectratio]{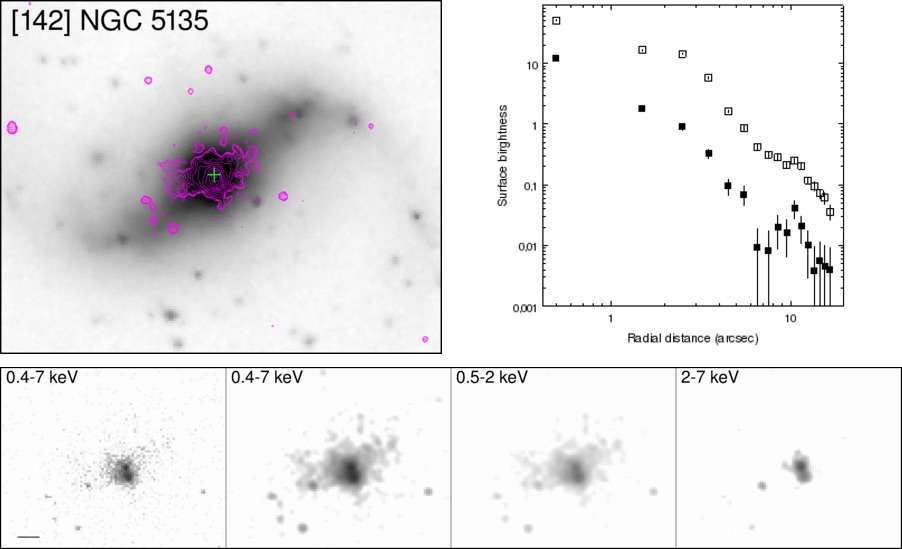}
\caption{\textit{Overlay on:} IRAC channel 1. \textit{Contours:} Custom.}
\end{figure*}

\begin{figure*}
\centering
\includegraphics[width=\textwidth,keepaspectratio]{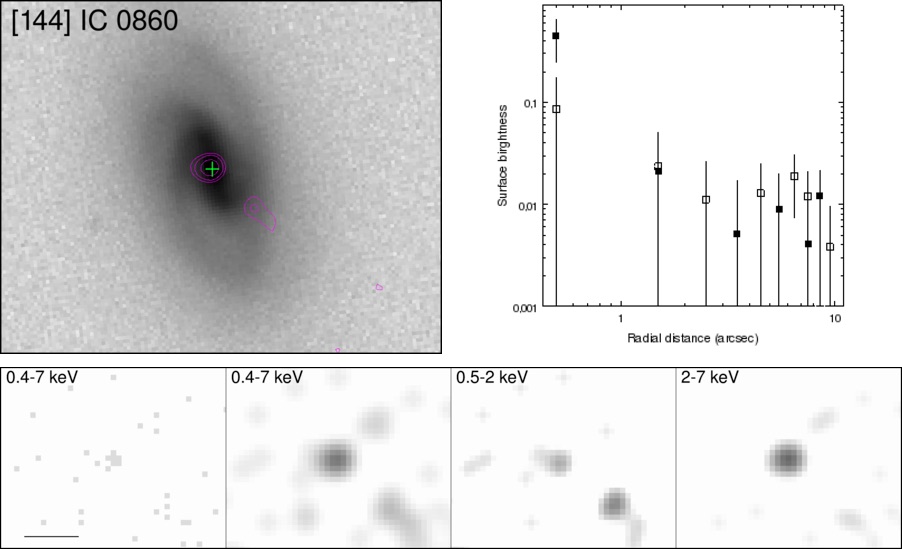}
\caption{\textit{Overlay on:} SDSS DR-12 i-band. \textit{Contours:} Interval 2.}
\end{figure*}

\begin{figure*}
\centering
\includegraphics[width=\textwidth,keepaspectratio]{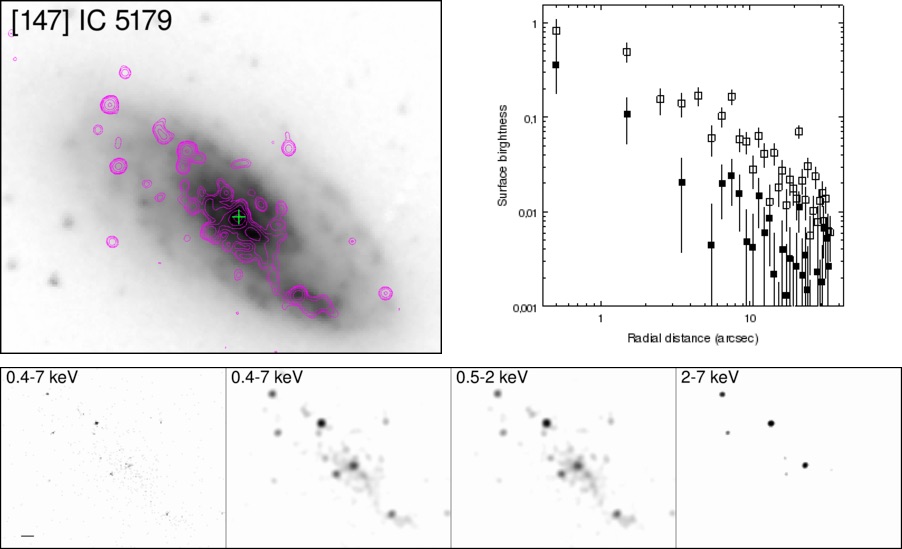}
\caption{\textit{Overlay on:} IRAC channel 1. \textit{Contours:} Interval 1.}
\end{figure*}

\begin{figure*}
\centering
\includegraphics[width=\textwidth,keepaspectratio]{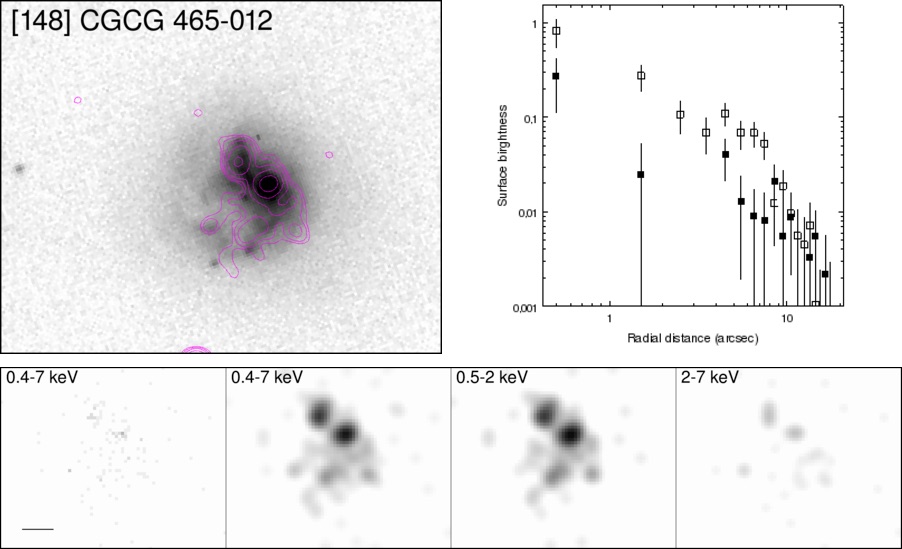}
\caption{\textit{Overlay on:} SDSS DR-12 i-band. \textit{Contours:} Interval 2.}
\end{figure*}

\begin{figure*}
\centering
\includegraphics[width=\textwidth,keepaspectratio]{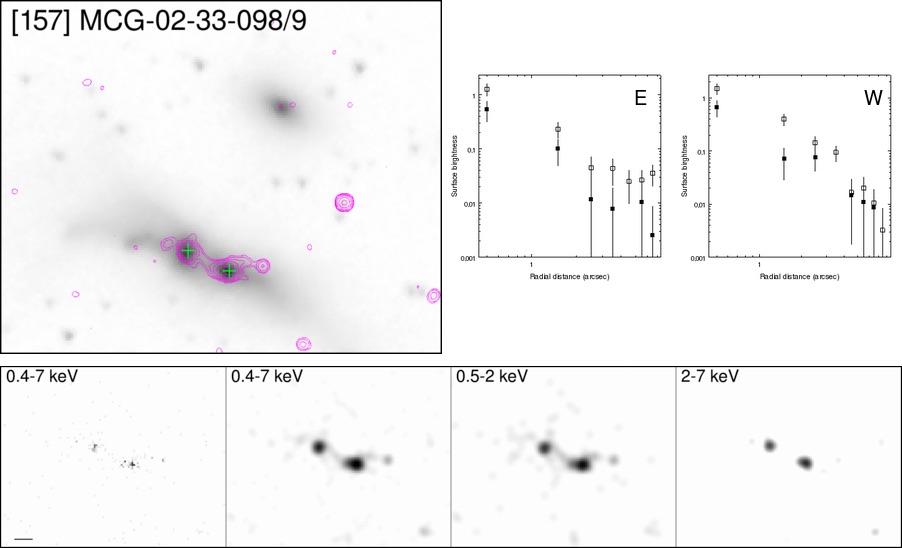}
\caption{\textit{Overlay on:} IRAC channel 1. \textit{Contours:} Interval 1.}
\end{figure*}

\begin{figure*}
\centering
\includegraphics[width=\textwidth,keepaspectratio]{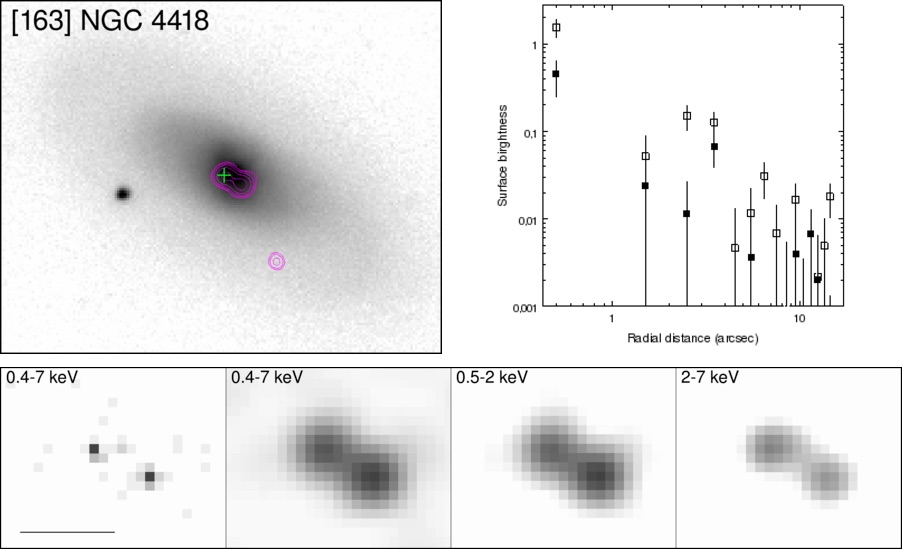}
\caption{\textit{Overlay on:} SDSS DR-12 i-band. \textit{Contours:} Interval 1.}
\end{figure*}

\begin{figure*}
\centering
\includegraphics[width=\textwidth,keepaspectratio]{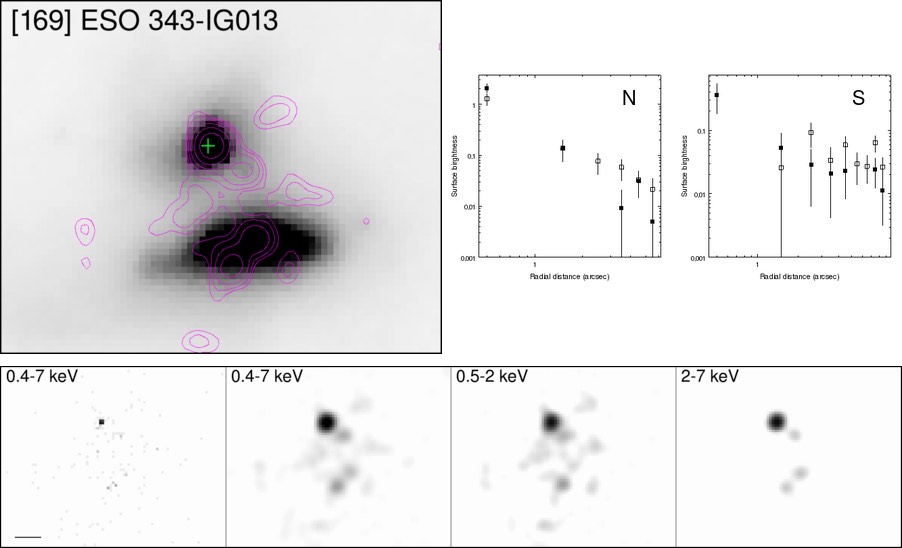}
\caption{\textit{Overlay on:} IRAC channel 1. \textit{Contours:} Interval 2.}
\end{figure*}

\begin{figure*}
\centering
\includegraphics[width=\textwidth,keepaspectratio]{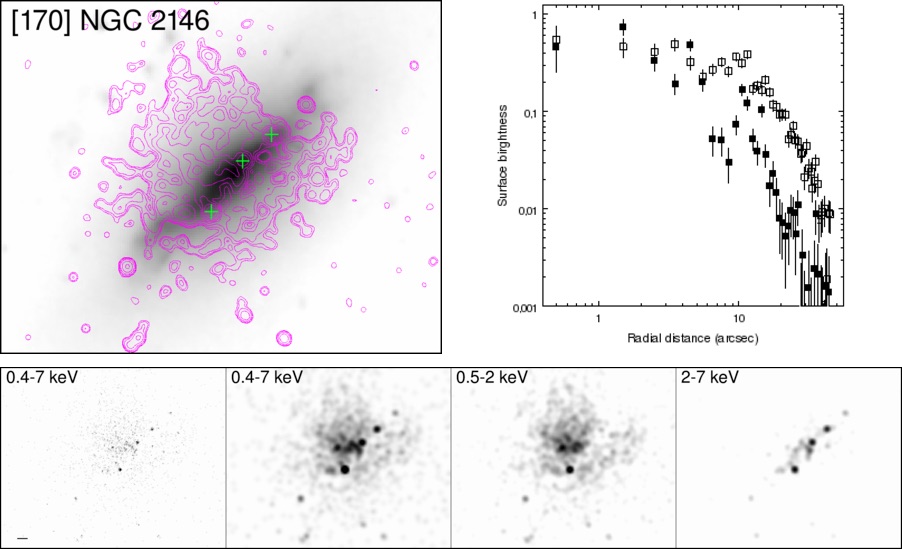}
\caption{\textit{Overlay on:} IRAC channel 1. \textit{Contours:} Interval 1.}
\end{figure*}

\begin{figure*}
\centering
\includegraphics[width=\textwidth,keepaspectratio]{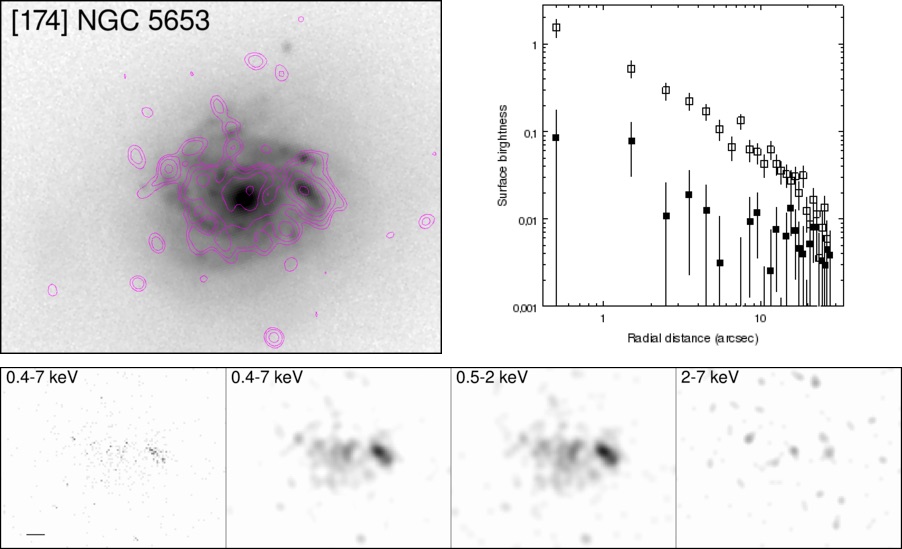}
\caption{\textit{Overlay on:} SDSS DR-12 i-band. \textit{Contours:} Interval 2.}
\end{figure*}

\begin{figure*}
\centering
\includegraphics[width=\textwidth,keepaspectratio]{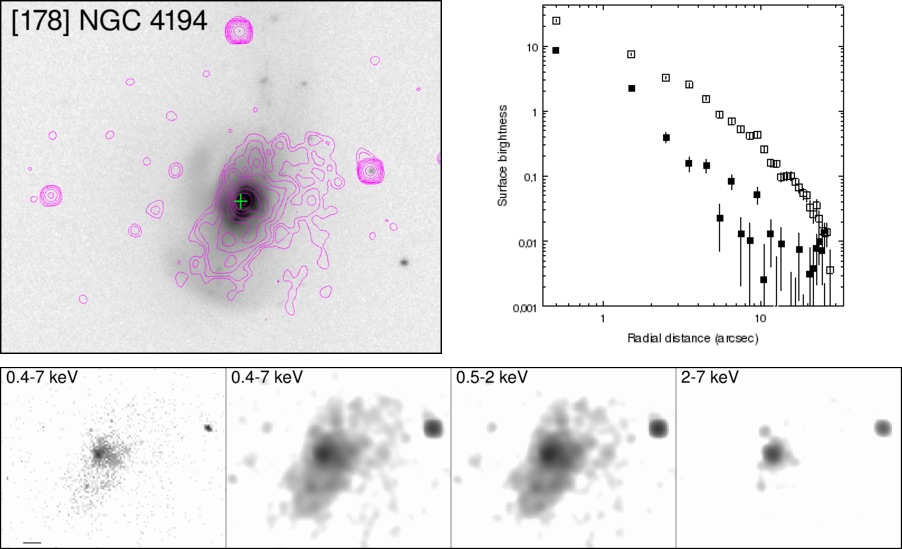}
\caption{\textit{Overlay on:} SDSS DR-12 i-band. \textit{Contours:} Interval 1.}
\end{figure*}

\begin{figure*}
\centering
\includegraphics[width=\textwidth,keepaspectratio]{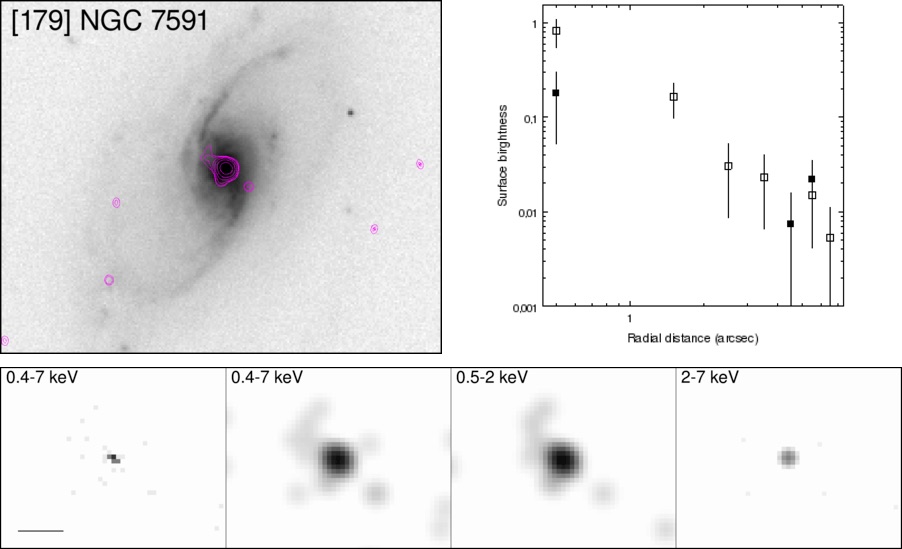}
\caption{\textit{Overlay on:} SDSS DR-12 i-band. \textit{Contours:} Interval 4.}
\end{figure*}

\begin{figure*}
\centering
\includegraphics[width=\textwidth,keepaspectratio]{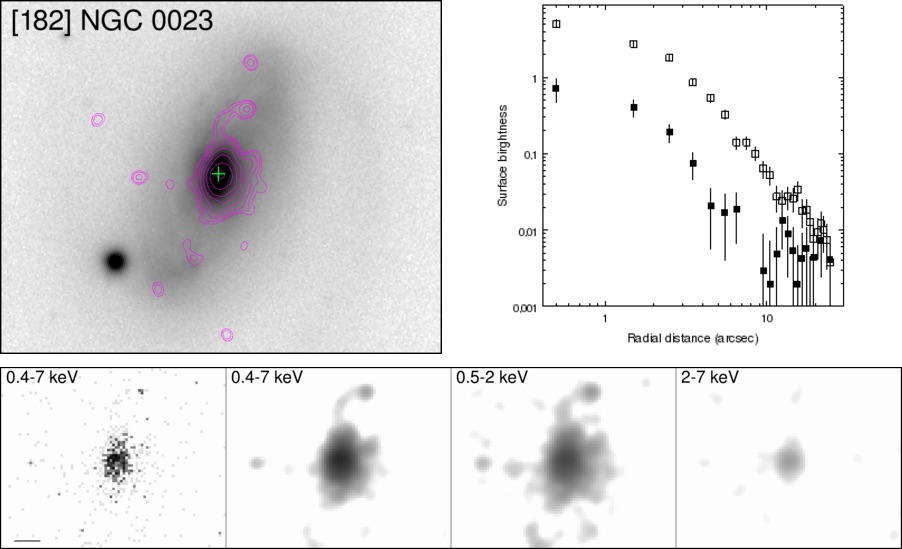}
\caption{\textit{Overlay on:} SDSS DR-12 i-band. \textit{Contours:} Interval 1.}
\end{figure*}

\begin{figure*}
\centering
\includegraphics[width=\textwidth,keepaspectratio]{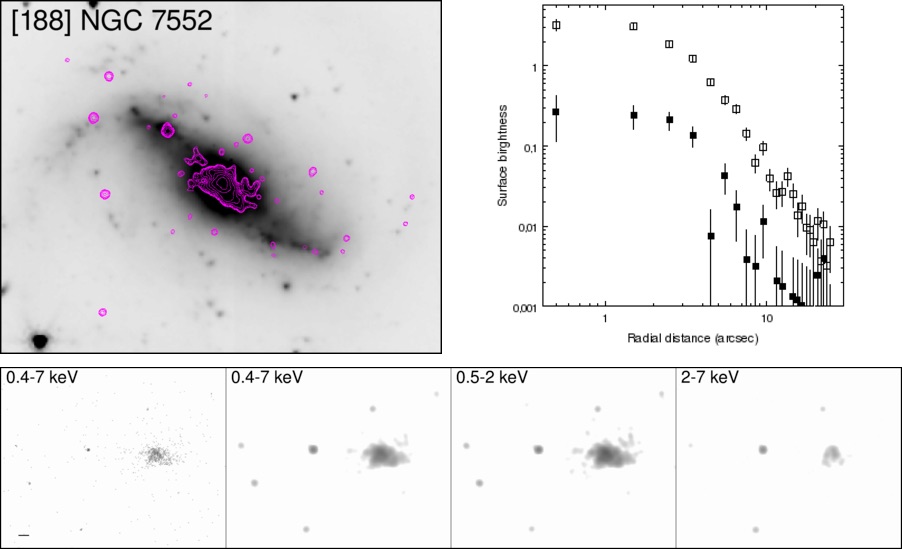}
\caption{\textit{Overlay on:} IRAC channel 1. \textit{Contours:} Interval 4.}
\end{figure*}

\begin{figure*}
\centering
\includegraphics[width=\textwidth,keepaspectratio]{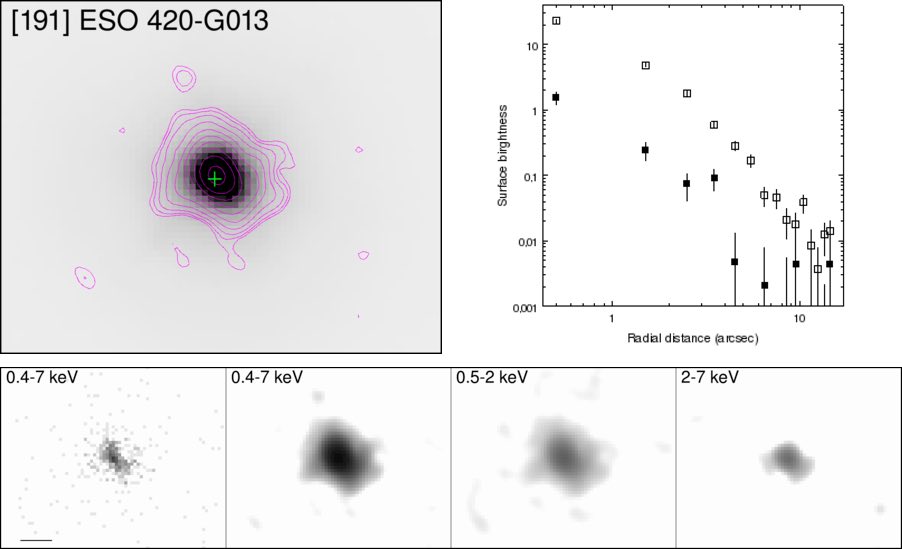}
\caption{\textit{Overlay on:} IRAC channel 1. \textit{Contours:} Interval 1.}
\end{figure*}

\begin{figure*}
\centering
\includegraphics[width=\textwidth,keepaspectratio]{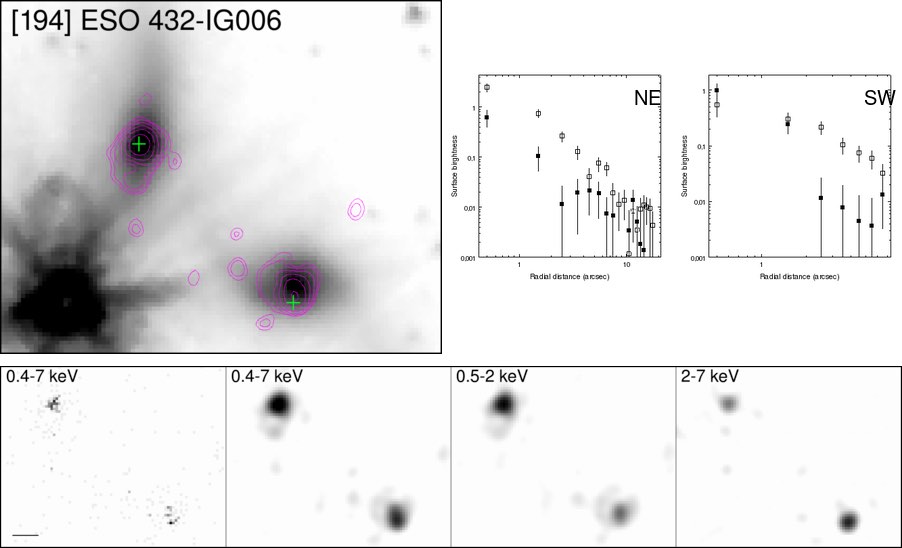}
\caption{\textit{Overlay on:} IRAC channel 1. \textit{Contours:} Interval 1.}
\end{figure*}

\clearpage

\begin{figure*}
\centering
\includegraphics[width=\textwidth,keepaspectratio]{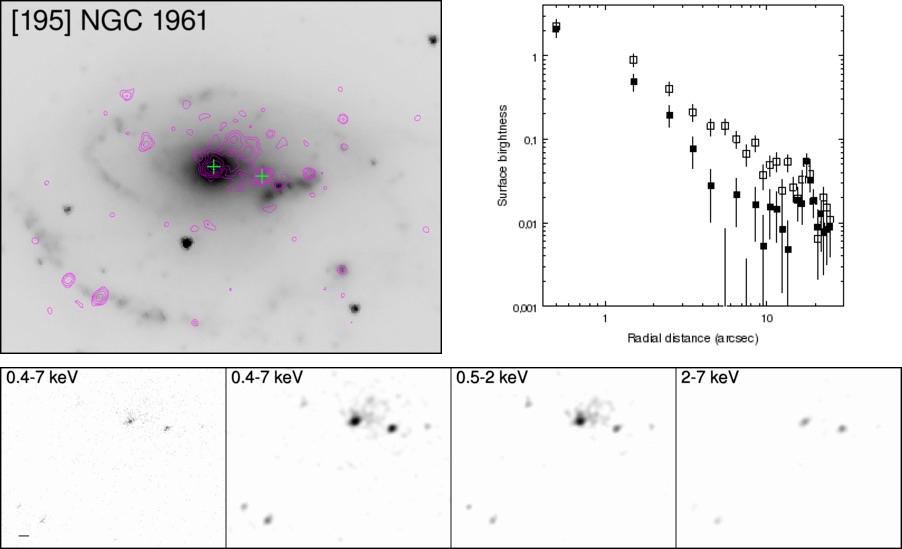}
\caption{\textit{Overlay on:} IRAC channel 1. \textit{Contours:} Interval 3.}
\end{figure*}

\begin{figure*}
\centering
\includegraphics[width=\textwidth,keepaspectratio]{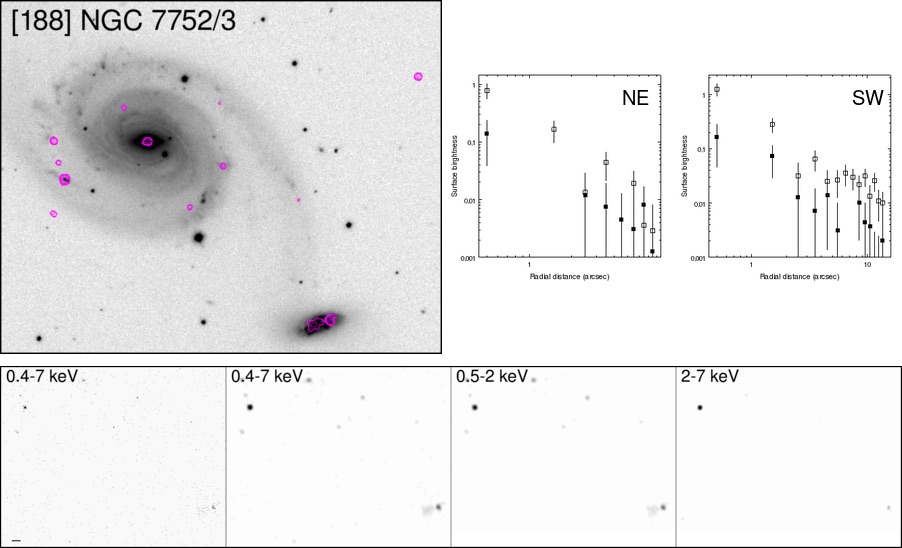}
\caption{\textit{Overlay on:} SDSS DR-12 i-band. \textit{Contours:} Interval 4.}
\end{figure*}

\begin{figure*}
\centering
\includegraphics[width=\textwidth,keepaspectratio]{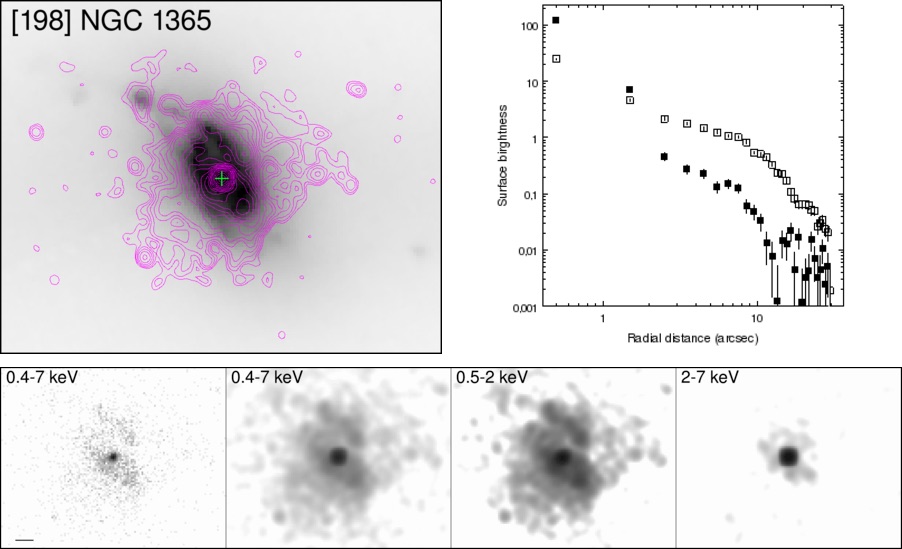}
\caption{\textit{Overlay on:} IRAC channel 1. \textit{Contours:} Custom.}
\end{figure*}

\begin{figure*}
\centering
\includegraphics[width=\textwidth,keepaspectratio]{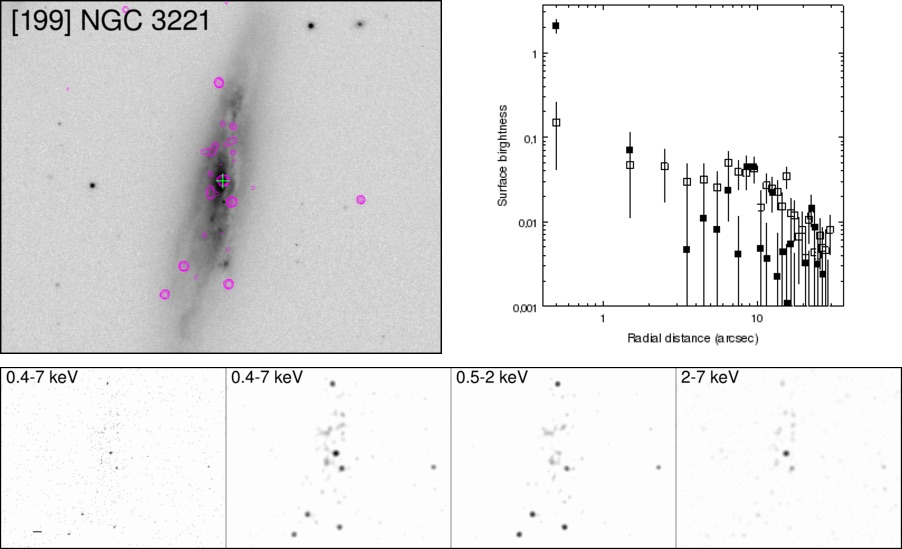}
\caption{\textit{Overlay on:} SDSS DR-12 i-band. \textit{Contours:} Interval 1.}
\end{figure*}

\begin{figure*}
\centering
\includegraphics[width=\textwidth,keepaspectratio]{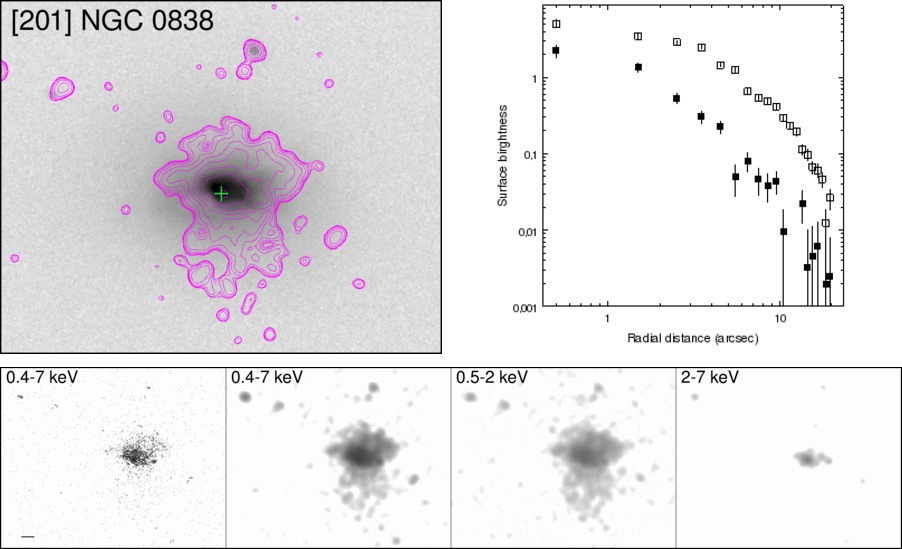}
\caption{\textit{Overlay on:} SDSS DR-12 i-band. \textit{Contours:} Custom.}
\end{figure*}

\clearpage

\section{X-ray Spectra}\label{AllSpectra}
\begin{figure*}
  \centering 
\includegraphics[width=0.93\textwidth,keepaspectratio]{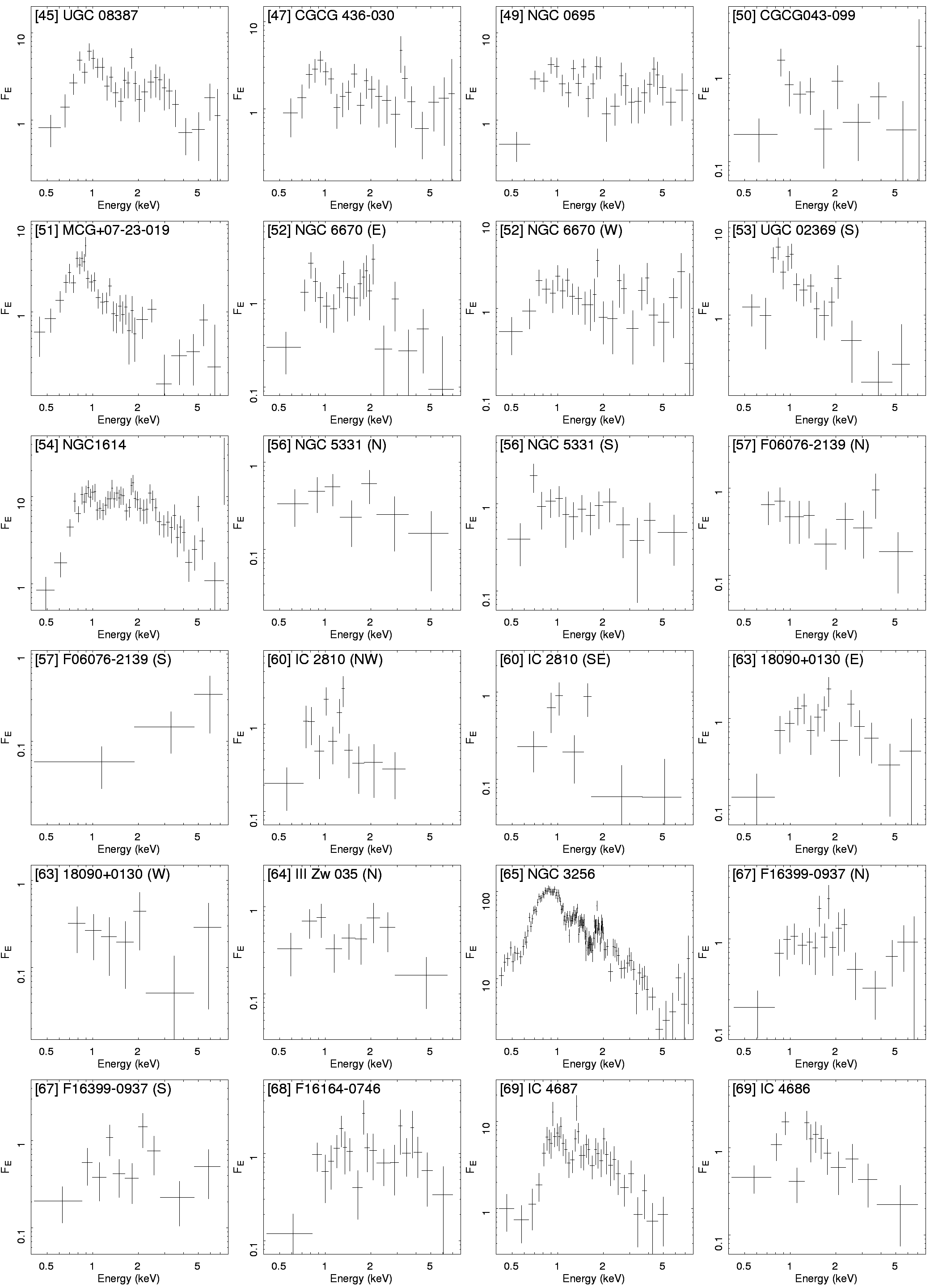}
  \caption{X-ray flux density spectra for the 84 individual galaxies of CGII, obtained from the \textit{Chandra} ACIS. Flux density in units of $10^{-14}$ erg s$^{-1}$ cm$^{-2}$ keV$^{-1}$.}
  \label{fig:XraySpectra}
\end{figure*}

\begin{figure*}
  \ContinuedFloat 
  \centering 
 \includegraphics[width=0.93\textwidth,keepaspectratio]{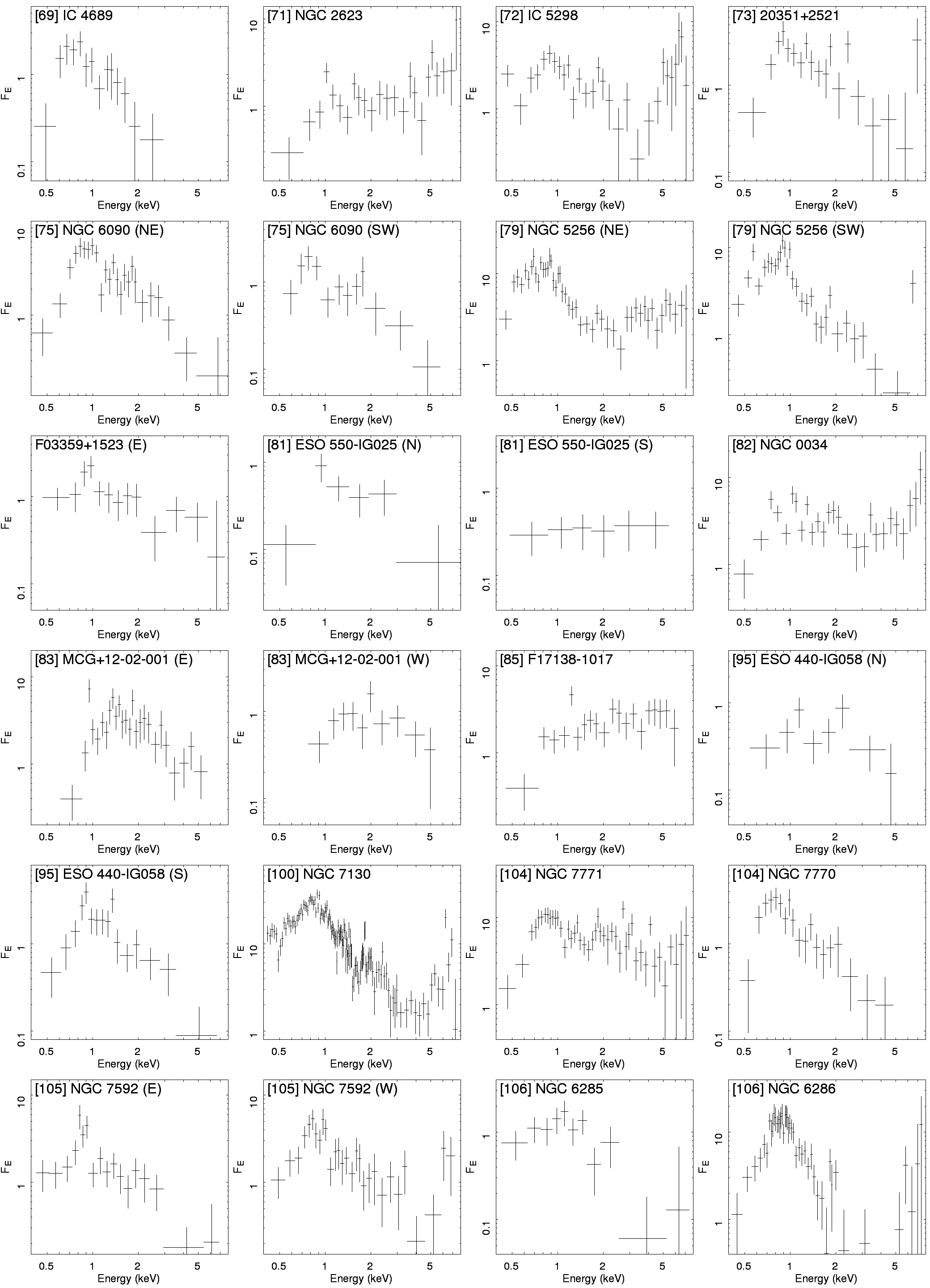}
  \caption[]{continued.}
  \label{fig:XraySpectra}
\end{figure*}

\begin{figure*}
  \ContinuedFloat 
  \centering 
 \includegraphics[width=0.93\textwidth,keepaspectratio]{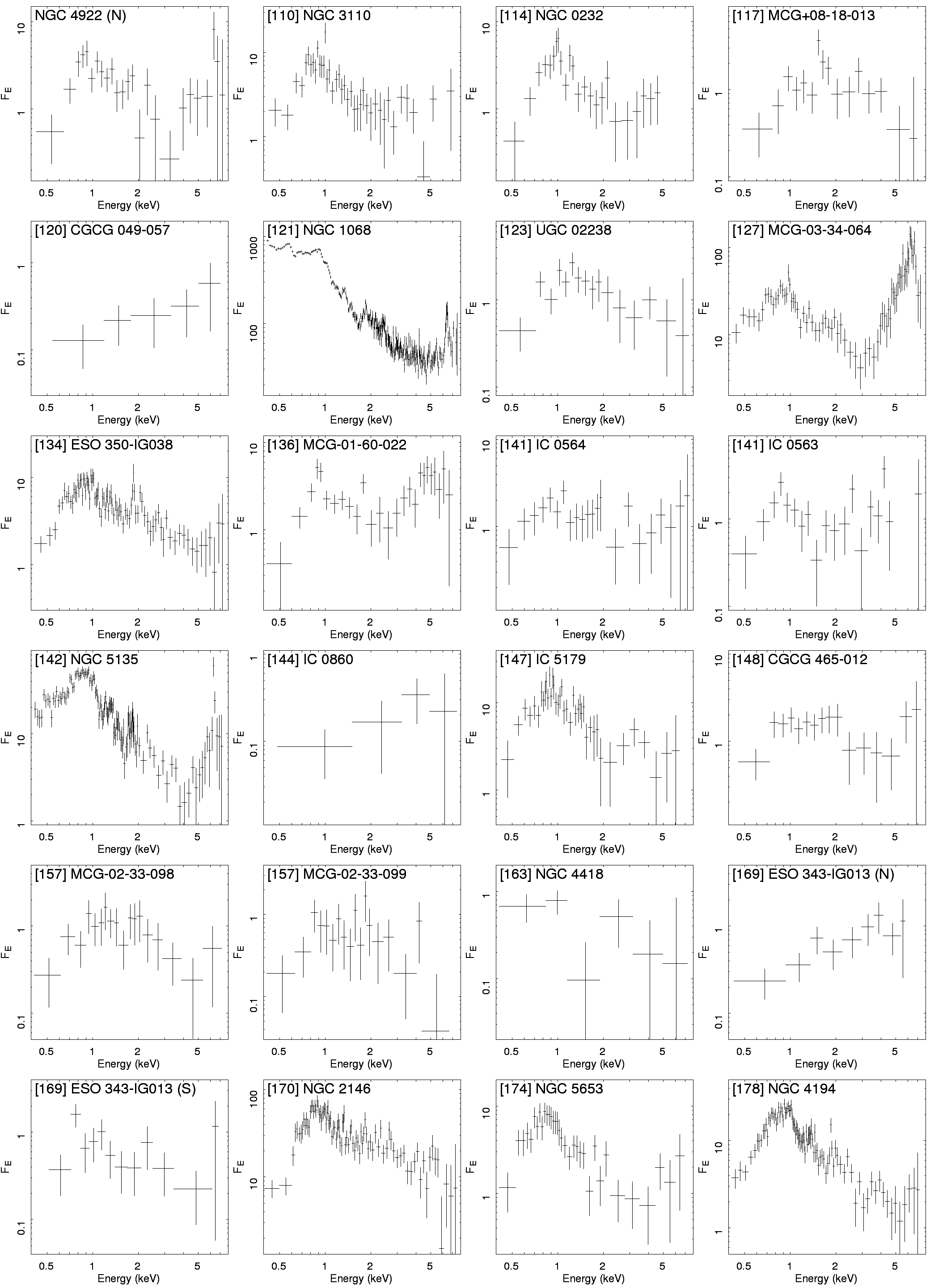}
  \caption[]{continued.}
  \label{fig:XraySpectra}
\end{figure*}

\begin{figure*}
  \ContinuedFloat 
  \centering 
 \includegraphics[width=0.93\textwidth,keepaspectratio]{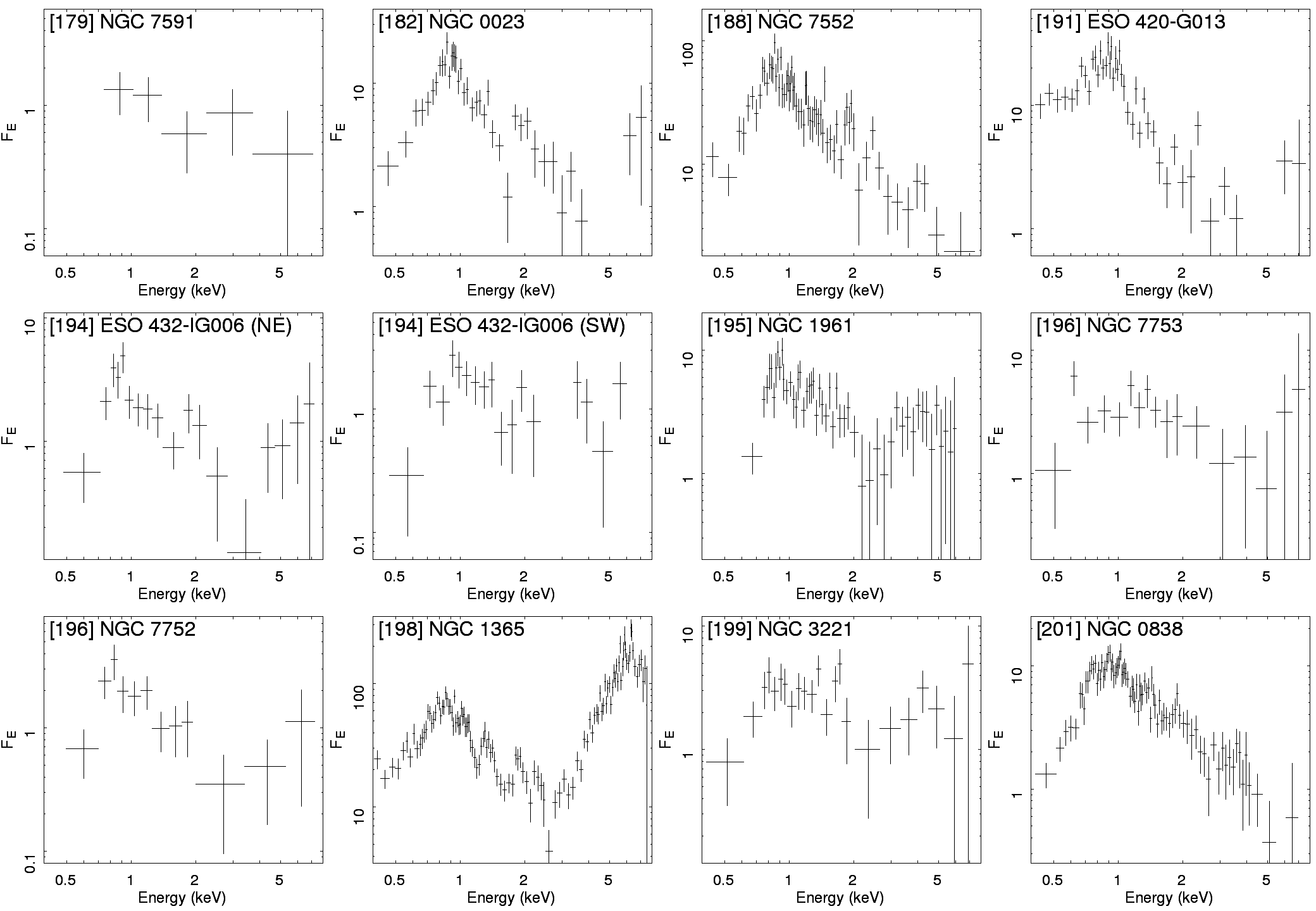}
  \caption[]{continued.}
  \label{fig:XraySpectra}
\end{figure*}

\end{appendix}
\end{document}